\documentclass[fleqn,usenatbib,useAMS]{mnras}

\usepackage{adjustbox}
\usepackage{multirow}
\usepackage{graphicx}	
\usepackage{amsmath}	
\usepackage{amssymb}	
\usepackage{multicol}        
\usepackage{bm}		
\usepackage{pdflscape}	
\usepackage{threeparttable}

\defcitealias{2020MNRAS.499.2173B}{Paper~I}
\newcommand\inlineeqno{\stepcounter{equation}\ (\theequation)}

\usepackage[T1]{fontenc}
\usepackage{ae,aecompl}

\usepackage{newtxtext,newtxmath}

\title[Shock--multicloud systems II]{Shock--multicloud interactions in galactic outflows - II. Radiative fractal clouds and cold gas thermodynamics}

\author[W.~E.~Banda-Barrag\'{a}n et al.]{W.~E.~Banda-Barrag\'{a}n,$^{1}$\thanks{E-mail: wlady.bsc@gmail.com (WBB)}
M.~Br\"uggen,$^{1}$
V.~Heesen,$^{1}$
E.~Scannapieco,$^{2}$
J.~Cottle,$^{2}$\newauthor
C.~Federrath,$^{3}$ and 
A.~Y.~Wagner$^{4}$\\
$^{1}$Hamburger Sternwarte, Universit\"{a}t Hamburg, Gojenbergsweg 112, D-21029 Hamburg, Germany\\
$^{2}$School of Earth and Space Exploration, Arizona State University, Tempe AZ, USA\\
$^{3}$Research School of Astronomy and Astrophysics, Australian National University, Canberra, ACT 2611, Australia\\
$^{4}$Center for Computational Sciences, University of Tsukuba, 1-1-1 Tennodai, Tsukuba, Ibaraki 305-8577, Japan\\
}

\date{Accepted XXX. Received YYY; in original form ZZZ}

\pubyear{2021}

\begin{document}
\label{firstpage}
\pagerange{\pageref{firstpage}--\pageref{lastpage}}
\maketitle

\begin{abstract}
Galactic winds are crucial to the cosmic cycle of matter, transporting material out of the dense regions of galaxies. Observations show the coexistence of different temperature phases in such winds, which is not easy to explain. We present a set of 3D shock-multicloud simulations that account for radiative heating and cooling at temperatures between $10^2\,\rm K$ and $10^7\,\rm K$. The interplay between shock heating, dynamical instabilities, turbulence, and radiative heating and cooling creates a complex multi-phase flow with a rain-like morphology. Cloud gas fragments and is continuously eroded, becoming efficiently mixed and mass loaded. The resulting warm mixed gas then cools down and precipitates into new dense cloudlets, which repeat the process. Thus, radiative cooling is able to sustain fast-moving dense gas by aiding condensation of gas from warm clouds and the hot wind. In the ensuing outflow, hot gas with temperatures $\gtrsim 10^6\,\rm K$ outruns the warm and cold phases, which reach thermal equilibrium near $\approx 10^4\,\rm K$ and $\approx 10^2\,\rm K$, respectively. Although the volume filling factor of hot gas is higher in the outflow, most of the mass is concentrated in dense gas cloudlets and filaments with these temperatures. More porous multicloud layers result in more vertically extended outflows, and dense gas is more efficiently produced in more compact layers. The cold phase is not accelerated by ram-pressure, but, instead, precipitates from warm and mixed gas out of thermal equilibrium. This cycle can explain the presence of high-velocity H\,{\sc i} gas with $N_{\rm H\,{\scriptstyle I}}=10^{19-21}\,\rm cm^{-2}$ and $\Delta v_{{\rm FWHM}}\lesssim37\,\rm km\,s^{-1}$ in the Galactic centre outflow.
\end{abstract}




\section{Introduction}
\label{sec:Introduction}
Radiative processes play a key role in the formation and evolution of interstellar clouds (e.g., see \citealt{2015ApJ...804..137P,2019ApJ...875..158W}; \citealt*{2020MNRAS.492.4484F}). In the context of galactic outflows, the ability of interstellar clouds to radiate away some of the energy injected by external shocks is crucial to understanding their survival (e.g., see \citealt{2008ApJ...674..157C,2016MNRAS.455.1830T,2017ApJ...834..144S,2017ApJ...837...28S}). Dense gas clouds (atomic and molecular) in such outflows, with temperatures between $\sim 10^2\,\rm K$ and $\sim 10^4\,\rm K$, are generally embedded in much hotter environments with temperatures $\gtrsim10^5\,\rm K$ (e.g., see \citealt*{1997A&A...320..378S}; \citealt{1998ApJ...493..129S,2016ApJ...826..215L,2017ApJ...843...18V,2017ApJ...835..265W,2019ApJ...881...43K,2020ApJ...901..151S, 2020A&ARv..28....2V}), which poses some serious constraints on how long they can live and how much they can travel under those conditions (see recent discussions in \citealt{2017MNRAS.468.4801Z} and \citealt{2019MNRAS.486.4526B}).\par

Adiabatic simulations of wind-cloud and shock-cloud interactions have shown that the hotter environment can very effectively remove material from the clouds, heat up their gas, and destroy them via dynamical instabilities (e.g., see \citealt*{1994ApJ...420..213K}; \citealt{1995ApJ...454..172X,2006ApJS..164..477N,2016MNRAS.455.1309B,2016MNRAS.457.4470P,2018MNRAS.476.2209G}). On the other hand, simulations that include the effect of radiative cooling have shown that the lifetime of clouds can be prolonged when cooling is efficient (e.g., see \citealt*{2002AA...395L..13M}; \citealt{2009ApJ...703..330C,2015ApJ...805..158S,2018ApJ...865...64G}; \citealt*{2019MNRAS.482.5401S}).\par

Even though radiative cooling can extend the lifetime of clouds, previous simulations of single clouds exposed to supersonic winds/shocks have shown that cooling can reduce their cross section and increase their density via contraction (e.g., see \citealt{2005ApJ...619..327F,2015ApJ...805..158S}), which, in some cases, can lead to shattering (e.g., see \citealt{2018MNRAS.473.5407M,2020MNRAS.492.1970G}) or splattering (see \citealt{2019ApJ...875..158W}). Since cloud acceleration occurs as a result of momentum transfer from the hot gas to the dense component\footnote{Note that cosmic-ray pressure forces can also transfer momentum to cold gas. For recent work on the driving of clouds by cosmic rays, see, e.g., \citet{2019MNRAS.489..205W} and \citet{2020ApJ...905...19B}.}, radiative cooling makes them harder to accelerate (e.g., see \citealt{2015ApJ...805..158S,2017ApJ...834..144S}). As a result, simulations of dense clouds in hot outflows show that they do not travel large distances even when cooling is very efficient. Despite this, it has recently been shown by \cite{2018MNRAS.480L.111G} that the warm, mixed gas that is removed from radiative clouds can condense back into dense cloudlets and filaments (see also \citealt{2010MNRAS.404.1464M}). This occurs as a result of a `focusing effect' that can lead to mass growth and sustained pulsations, i.e., to continuous contractions and expansions of dense gas driven by pressure gradients (see \citealt{2020MNRAS.494L..27G}). This process can entrain hot gas and continuously replenish dense gas in dynamic outflows. In addition, studies where non-radiative clouds are placed along a stream show that hydrodynamical shielding can be an effective mechanism for, both, prolonging the lifetimes of clouds and aiding their acceleration (see \citealt{2019AJ....158..124F}, and \cite{2020MNRAS.499.2173B}; hereafter \citetalias{2020MNRAS.499.2173B}). Thus, the effects of cooling in multi-phase outflows also depend on how high-density, cold gas is distributed in the outflow.\par

In this context, in \citetalias{2020MNRAS.499.2173B} we showed that in non-radiative shock-multicloud scenarios, ram-pressure-driven outflows carry some information of the density structure of the launching site. By comparing models with compact and porous cloud density distributions, i.e., with log-normal density fields characteristic of supersonic (${\cal M}_{\rm turb}\sim 5.5$) turbulence driven by solenoidal and compressive modes, respectively (e.g., see \citealt{2008ApJ...688L..79F,2010A&A...512A..81F}), we showed that the morphology, dynamics, vertical extent, and momentum distributions of entrained gas are different in both cases. We demonstrated that in compact solenoidal models, cloud destruction is very efficient and the dispersed and mixed gas forms a shell that can be effectively mass loaded into the flow (see also \citealt{2012MNRAS.425.2212A}). In porous compressive models, a few dense cores survive, but mixed gas does not form a coherent shell and entrained gas consists of mostly low-density material. In both cases, direct dense-gas entrainment is highly inefficient, but both hot and warm gas components become readily mass-loaded. In this paper, we study how radiative cooling and heating affect the evolution of mass-loaded gas in shock-swept multicloud systems with log-normal density distributions of the same type discussed in \citetalias{2020MNRAS.499.2173B}.\par

Including radiative cooling in numerical models increases the computational costs significantly, compared to their non-radiative counterparts. Dense gas is expensive to resolve by shock-capturing codes, so a commonly-used technique to reduce the time needed for the simulations has been to switch off cooling below a threshold temperature (usually $\sim 10^4\,\rm K$). Although this approach prevents runaway cooling and is justified as a first approximation, the heating rate of cold gas (with temperatures $\leq 10^4\,\rm K$) depends on the gas density, so the balance between heating and cooling will also depend on it for a given temperature. Thus, in this paper we relax this assumption by allowing gas to naturally cool down to $\sim 10^2\,\rm K$ (our new cooling floor) and be heated by a density-weighted heating rate, which represents an interstellar radiation field. In addition, capturing some radiative-driven processes requires very high resolutions (e.g., see \citealt*{2010ApJ...722..412Y}). Therefore, previous studies investigating, e.g., the cooling-induced shattering/splattering of dense gas have resorted to 2D and 1D models (e.g., see \citealt{2018MNRAS.473.5407M,2019ApJ...875..158W}) or to single-cloud scenarios in 3D models (e.g., see \citealt{2020MNRAS.492.1970G}). Thus, the parameter space to be explored in radiative models is still very broad, and studying different shock-cloud configurations and cooling regimes is needed as they may lead to distinct scenarios, e.g., shattering, pulsations/oscillations, or coagulation/coalescence (see \citealt{2019ApJ...876L...3W,2020MNRAS.494L..27G}).\par

In this paper we present the second part of a systematic study of the interaction between shocks and multicloud systems. In Section \ref{sec:Method}, we describe the equations, methods, simulation set-ups, diagnostics, and time-scales important for these systems. In Section \ref{sec:Results}, we discuss our results in two parts. In the first one we discuss the overall evolution of radiative shock-multicloud systems, and how this varies compared to non-radiative models. In the second one, we provide an overview of the dynamics and density distribution of gas in different temperature bins. In Section \ref{sec:Application}, we apply our results to observations of neutral hydrogen (H\,{\sc i}) gas in the Galactic Centre outflow (e.g., see \citealt{2013ApJ...770L...4M,2018ApJ...855...33D,2020ApJ...888...51L}). In Section \ref{sec:Resolution}, we discuss the effects of numerical resolution and the limitations of this work. In Section \ref{sec:Conclusions}, we summarise our main results.

\section{Method}
\label{sec:Method}

\subsection{Simulation code}
\label{subsec:SimulationCode}
For the numerical simulations of shock-multicloud systems reported in this paper we use the {\sevensize PLUTO v4.3} code (\citealt{2007ApJS..170..228M}) to solve the following equations of mass, momentum, and energy conservation:
\begin{equation}
\frac{\partial \rho}{\partial t}+\bm{\nabla\cdot}\left[{\rho \bm{v}}\right]=0,
\label{eq:MassConservation}
\end{equation}

\begin{equation}
\frac{\partial \left[\rho \bm{v}\right]}{\partial t}+\bm{\nabla\cdot}\left[{\rho\bm{v}\bm{v}}+{\bm{I}}P\right]=0,
\label{eq:MomentumConservation}
\end{equation}

\begin{equation}
\frac{\partial E}{\partial t}+\bm{\nabla\cdot}\left[\left(E+P\right)\bm{v}\right]=\Gamma-\Lambda,
\label{eq:EnergyConservation}
\end{equation}

\noindent where $\rho=\mu\,m_u n$ is the mass density, $\mu$ is the mean particle mass, $m_u$ is the atomic mass unit, $n$ is the gas number density, $\bm{v}$ is the velocity, $P=\left(\gamma-1\right)\rho\epsilon$ is the gas thermal pressure, $\gamma=5/3$ is the adiabatic index, $E=\rho\epsilon+\frac{1}{2}\rho\bm{v^2}$ is the total energy density, $\epsilon$ is the specific internal energy, $\Gamma$ is the volumetric heating rate, and $\Lambda$ is the volumetric cooling rate. In addition, we solve the following advection equation to track gas originally in the multicloud system:
\begin{equation}
\frac{\partial\left[\rho C\right]}{\partial t}+\bm{\nabla\cdot}\left[{\rho C \bm{v}}\right]=0,
\label{eq:tracer}
\end{equation}
\noindent where $C$ is a Lagrangian scalar defined as $C=1$ for gas inside the multicloud layer and $C=0$ everywhere else (see Section \ref{subsec:ComputationalSetup} for further details). The above system of equations is numerically solved using the \verb#HLLC# approximate Riemann solver \citep*{Toro:1994} with a Courant--Friedrichs--Lewy (CFL) number of $C_{\rm a}=0.3$. 

\subsection{Cooling and heating function}
\label{subsec:CoolingHeating}

\begin{figure}
\begin{center}
  \begin{tabular}{c} 
      (1a) Net cooling and heating function\\ 
      \resizebox{80mm}{!}{\includegraphics{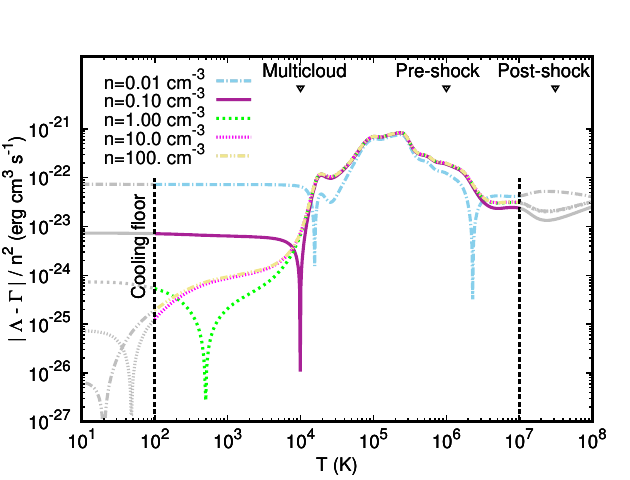}}\\
  \end{tabular}
  \caption{Net temperature- and density-dependent cooling and heating function, $|\Lambda-\Gamma|/n^2$, employed in our simulations. We show the curves for gas with five different number densities. The vertical dashed lines indicate the cooling limits (e.g. the cooling floor at $10^2\,\rm K$). The black triangles on the top indicate the initial average temperatures in the multicloud systems ($\langle T_{\rm mc} \rangle=10^4\,\rm K$), the pre-shock ambient medium ($T_{\rm ambient}=10^6\,\rm K$) and the post-shock ambient medium ($T_{\rm psh}\approx3\times10^7\,\rm K$).} 
  \label{Figure1}
\end{center}
\end{figure}

In our simulations, we include a temperature- and density-dependent, customised cooling and heating function. In the case of radiative cooling we use a tabulated cooling function, $\tilde{\Lambda}$, which includes radiative cooling from atomic species in the temperature range from $10\,\rm K$ to $10^{10}\,\rm K$ (see \citealt*{2008A&A...488..429T}, but note that we switch off cooling for $T<10^2\,\rm K$ and $T>10^7\,\rm K$). The cooling rates were pre-compiled with {\small CLOUDY} (see \citealt{1998PASP..110..761F}) for a solar mix at redshift zero and tabulated in units of $\rm erg\,cm^3\,s^{-1}$, so $\Lambda=[\rho/(\mu\,m_u)]^2\,\tilde{\Lambda}=n^2\tilde{\Lambda}$. In addition to this cooling function, we also add a density-weighted heating function, which is initially calculated from the cooling rates such that diffuse gas in the cloud with $n=0.1\,\rm cm^{-3}$ has an equilibrium temperature of $10^4\,\rm K$ at the beginning of the simulations, i.e., at $t_0$. This heating rate, $\tilde{\Gamma}$, is then applied to gas with different densities by weighting it with the local density, $\rho$, so that $\Gamma=[\rho/(\mu\,m_u)]\,\tilde{\Gamma}=n\tilde{\Gamma}$. By following this process, the thermal equilibrium equation is $n\tilde{\Lambda}=\tilde{\Gamma}$. This configuration provides a close approximation to cooling and heating functions (e.g. \citealt{2013MNRAS.434.1043O}) that include the redshift-zero Haardt \& Madau background (see \citealt{2001cghr.confE..64H,2012ApJ...746..125H}). In the end we obtain a net heating and cooling function, $|\Lambda-\Gamma|/n^2$, that depends on both density and temperature (see Figure \ref{Figure1}).\par

This function implies that low-density gas is not able to cool below $\sim 10^4\,\rm K$ as heating and cooling balance out near that temperature, but high-density gas can reach thermal equilibrium at lower temperatures because heating is not as efficient at balancing cooling in dense regions. This means that some high-density gas in the clouds can cool down to temperatures of $\sim 10^2\,\rm K$, which corresponds to the floor in our cooling function. Unlike in previous studies, we do not set a $\sim 10^4\,\rm K$ threshold, but allow our heating function to counteract cooling as a function of density, which is what we expect in realistic situations in the interstellar medium (ISM). Similarly, our recipe implies that hot, diffuse gas can efficiently cool down if it has temperatures between $10^4\,\rm K$ and a few $\times 10^6\,\rm K$, but cannot be cooled down or shock heated excessively above that temperature. Combined with the relatively long cooling time-scale of such diffuse gas, this ensures that the post-shock flow remains nearly isothermal at all times.

\subsection{Computational set-up}
\label{subsec:ComputationalSetup}
\begin{figure}
\begin{center}
  \begin{tabular}{c c c} 
      \hspace{-1.5cm}(2a) Compact solenoidal model & \hspace{-1cm}(2b) Porous compressive model\\
\hspace{-0.30cm}\resizebox{48mm}{!}{\includegraphics{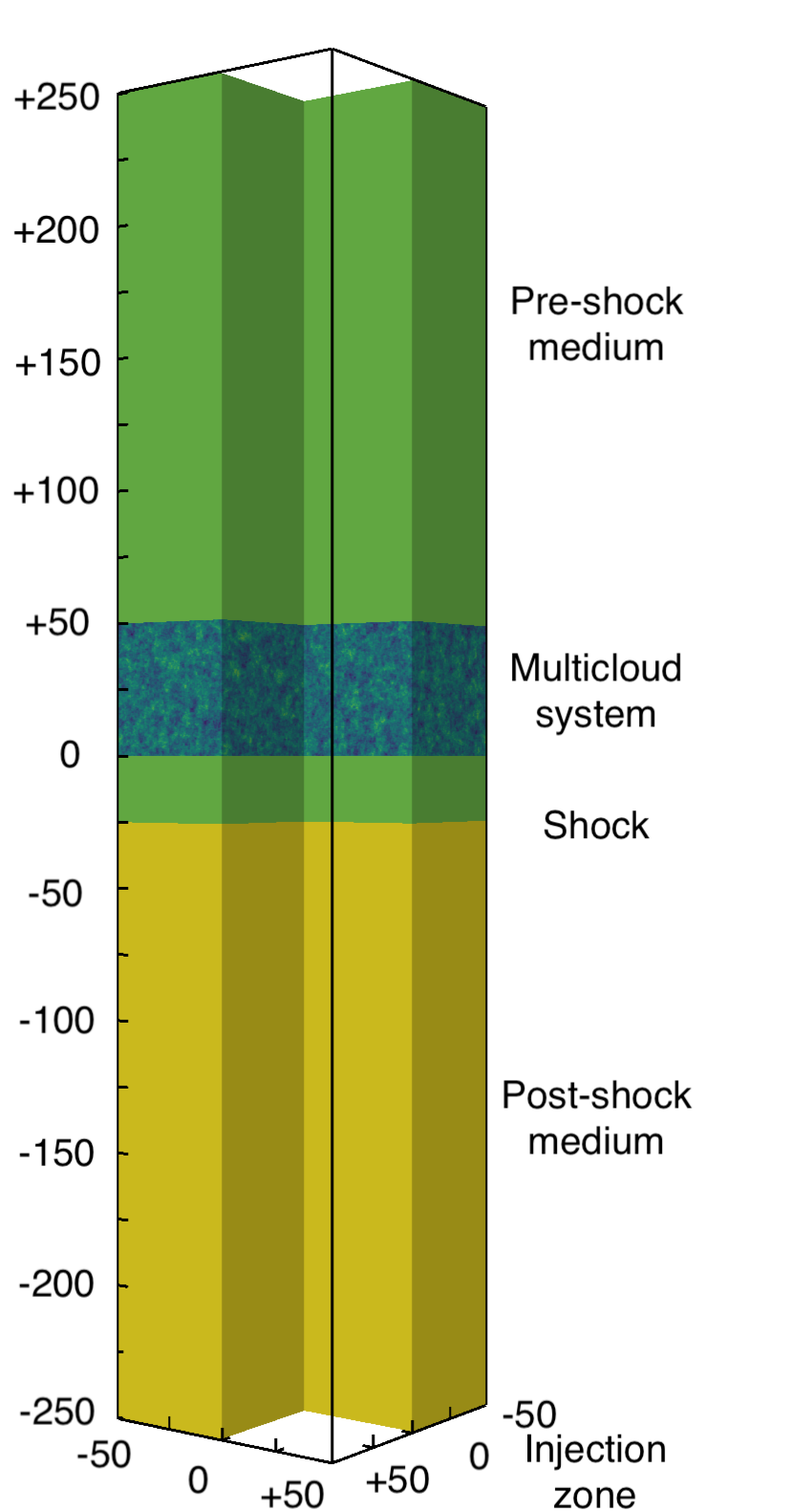}} & \hspace{-1.05cm}\resizebox{35.6mm}{!}{\includegraphics{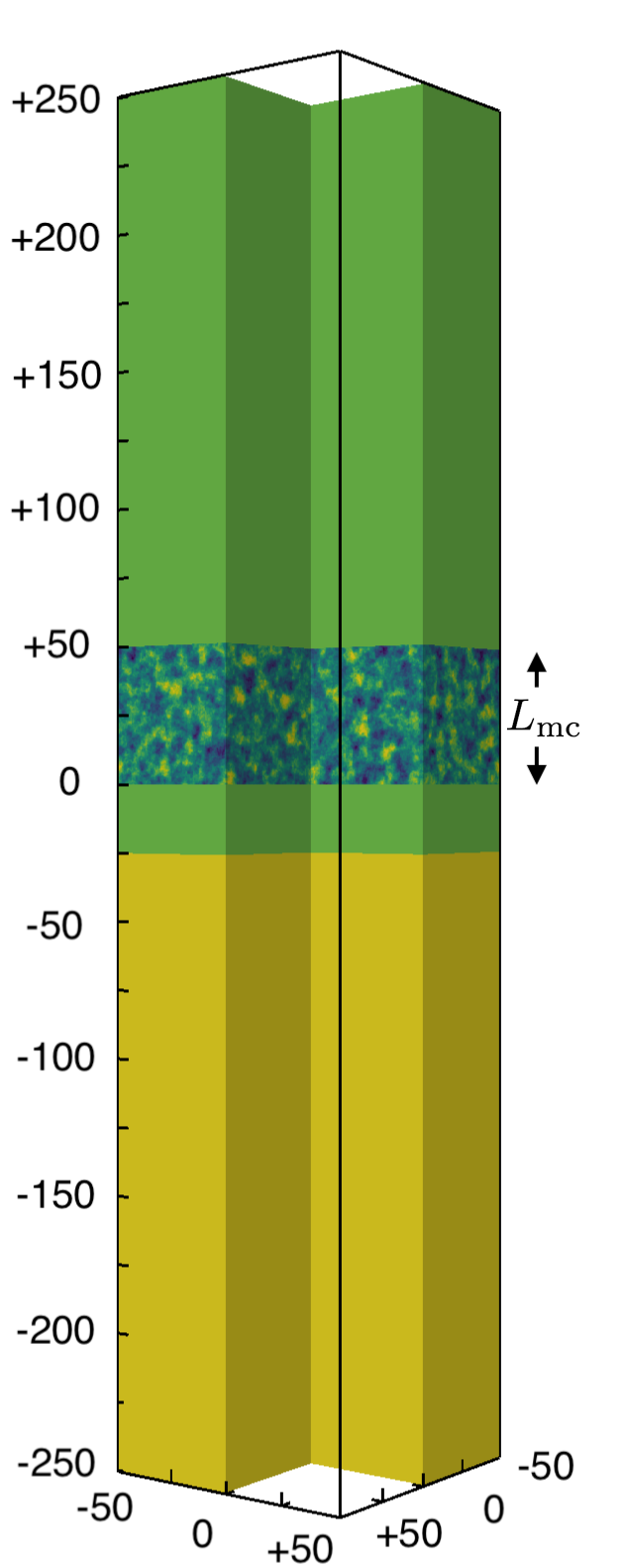}} & \hspace{-0.60cm}\resizebox{12mm}{!}{\includegraphics{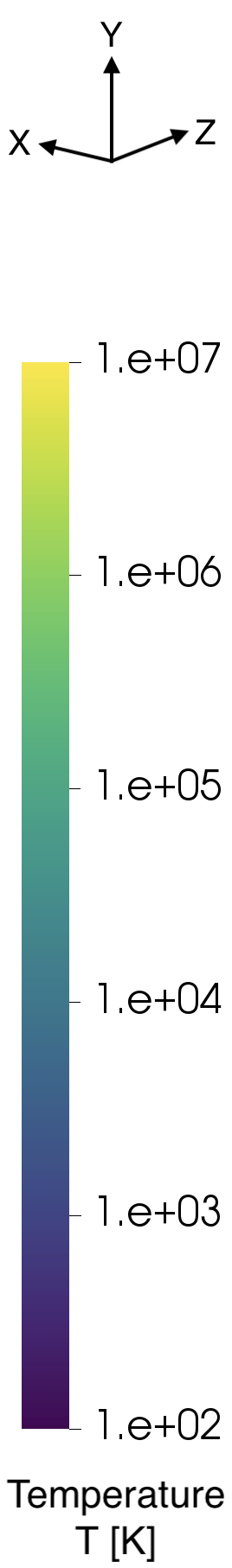}} \\
  \end{tabular}
  \caption{3D renderings of the initial temperature in logarithmic scale of a compact solenoidal cloud model (left panel) and a porous compressive cloud model (right panel). We have clipped a quarter of the volume in both cases to show the interior of the computational domains. The clump substructure can be viewed in Figure \ref{FigureA2}.} 
  \label{Figure2}
\end{center}
\end{figure}

Following \citetalias{2020MNRAS.499.2173B}, we use 3D rectangular prisms as computational domains (see Figure \ref{Figure2}). Each domain has dimensions of $L\times 5L\times L$ with $L=100\,\rm pc$, so that the physical size of the domain is $100\,\rm pc\times 500\,\rm pc\times100\,\rm pc$. The domain has a uniform grid with $(256\times1280\times256)$ cells in our standard-resolution models, $(512\times2560\times512)$ cells in our high-resolution models, and $(128\times640\times128)$ cells in our low-resolution models. Therefore, the numerical resolution is $\approx0.39\,\rm pc$ in our standard models, and $\approx0.20\,\rm pc$ and $\approx0.78\,\rm pc$ in high- and low-resolution models, respectively. These resolutions are in between the numerical resolution achieved in disc-scale simulations of starburst systems (e.g., $\Delta x,y,z\approx5\,\rm pc$; see \citealt{2020ApJ...895...43S}) and that in wind/shock-cloud models with isolated clouds (e.g., $\Delta x,y,z\approx0.08\,\rm pc$; see \citealt{2016MNRAS.455.1309B}). Thus, our shock-multicloud models allow us to resolve smaller scales than global outflow models, and, at the same time, capture the collective effects of cloud conglomerates.\par 

\subsubsection{Initial Conditions}
\label{subsec:InitialConditions}
The domain contains three media: a pre-shock multicloud layer with an initial average number density $\bar{n}_{\rm cloud,0}=\bar{\rho}_{\rm cloud,0}/{\mu m_u}=1\,\rm cm^{-3}$ and a tracer $C=1$ (see equation \ref{eq:tracer}), a pre-shock ambient medium with a constant number density $n_{\rm ambient}={\rho_{\rm ambient}}/{\mu m_u}=0.01\,\rm cm^{-3}$ and $C=0$, and a post-shock ambient medium with ${n_{\rm psh}}\approx 4\,{n_{\rm ambient}}=0.04\,\rm cm^{-3}$ and $C=0$. Thus, the initial density contrast between gas in the multicloud layer and the pre-shock ambient material is $\chi=\bar{\rho}_{\rm cloud,0}/{\rho_{\rm ambient}}=\bar{n}_{\rm cloud,0}/{n_{\rm ambient}}=10^2$, where $\bar{\rho}_{\rm cloud,0}$ is the mean of the initial density field, represented by a log-normal probability density function (PDF) with a standard deviation, $\sigma_{\rm cloud}=\sigma_{\rho_{\rm cloud,0}}/\bar{\rho}_{\rm cloud,0}$ (see Section 2.2.1 in \citetalias{2020MNRAS.499.2173B} for further details).\par

\begin{table*}\centering
\caption{Initial conditions for our shock-multicloud models. Column 1 indicates the model name. Columns 2 and 3 indicate the type of density field in the cloud layer and its normalised standard deviation, $\sigma_{\rm cloud,0}=\sigma_{\rho_{\rm cloud,0}}/\bar{\rho}_{\rm cloud,0}$, respectively. Columns 4, 5, and 6 show the $L_{\rm mc}$-normalised domain size, the number of grid cells in the computational volume, and the size of the domain in physical units, respectively. Column 7 indicates the length of the cloud layer in the streaming direction, $L_{\rm mc}$. Columns 8 and 9 report the cloudlet sizes, $r_{\rm cloudlet}$, in the multicloud system, and the number of grid cells covering a cloudlet radius, respectively. The displayed upper limits correspond to $r_{\rm cloudlet,k_{\rm min}}$, where $k_{\rm min}=8$ in all models. Column~10 indicates whether or not heating/cooling is used. In all models, the adiabatic index is $\gamma=\frac{5}{3}$, the density fields are consistent with those of ${\cal M}_{\rm turb}\approx 5.5$ turbulence, the $L$-normalised domain is $(L\times 5L \times L)$, and the initial density contrast between the multicloud system and the ambient medium is $\chi=10^2$.}
\begin{adjustbox}{max width=\textwidth}
 \hspace*{-0.35cm}\begin{tabular}{c c c c c c c c c c}
\hline
\textbf{(1)} & \textbf{(2)} & \textbf{(3)} & \textbf{(4)} & \textbf{(5)} & \textbf{(6)} & \textbf{(7)} & \textbf{(8)} & \textbf{(9)} & \textbf{(10)}\\
\textbf{Model} & \textbf{Density} & $\sigma_{\rm cloud}$ & \textbf{Domain} & \textbf{Number of cells} & \textbf{Fiducial domain}& $L_{\rm mc}$ & $r_{\rm cloudlet}$ & $\frac{{\rm cells}}{r_{\rm cloudlet}}$ & \textbf{Cooling \&}\\
 &  & & & & $[\rm pc^3]$ & $[\rm pc]$ & $[\rm pc]$ & & \textbf{Heating} \\ \hline
sole-k8-M10 & Compact solenoidal & $1.9$ & $(2\times 10 \times 2)\,L_{\rm mc}$ & $(256\times1280\times256)$ & $(100\times 500 \times 100)$ & $50$ & $\leq6.3$ & $\leq 16$ & No \\
sole-k8-M10-rad & Compact solenoidal& $1.9$ & $(2\times 10 \times 2)\,L_{\rm mc}$ & $(256\times1280\times256)$ & $(100\times 500 \times 100)$ & $50$ & $\leq6.3$ & $\leq 16$ & Yes\\
comp-k8-M10 & Porous compressive & $5.9$ & $(2\times 10 \times 2)\,L_{\rm mc}$ & $(256\times1280\times256)$ & $(100\times 500 \times 100)$ & $50$ & $\leq6.3$ & $\leq 16$ & No \\
comp-k8-M10-rad & Porous compressive & $5.9$ & $(2\times 10 \times 2)\,L_{\rm mc}$ & $(256\times1280\times256)$ & $(100\times 500 \times 100)$ & $50$ & $\leq6.3$ & $\leq 16$ & Yes\\\hline
sole-k8-M10-rad-hr & Compact solenoidal & $1.9$ & $(2\times 10 \times 2)\,L_{\rm mc}$ & $(512\times2560\times512)$ & $(100\times 500 \times 100)$ & $50$ & $\leq6.3$ & $\leq 32$ & Yes \\
sole-k8-M10-rad-lr & Compact solenoidal & $1.9$ & $(2\times 10 \times 2)\,L_{\rm mc}$ & $(128\times640\times128)$ & $(100\times 500 \times 100)$ & $50$ & $\leq6.3$ & $\leq 8$ & Yes \\
comp-k8-M10-rad-hr & Porous compressive & $5.9$ & $(2\times 10 \times 2)\,L_{\rm mc}$ & $(512\times2560\times512)$ & $(100\times 500 \times 100)$ & $50$ & $\leq6.3$ & $\leq 32$ &Yes \\
comp-k8-M10-rad-lr & Porous compressive & $5.9$ & $(2\times 10 \times 2)\,L_{\rm mc}$ & $(128\times640\times128)$ & $(100\times 500 \times 100)$ & $50$ & $\leq6.3$ & $\leq 8$ & Yes \\\hline
\end{tabular}
\end{adjustbox}
\label{Table1}
\end{table*} 

As in our previous study, we compare two types of cloud density distributions, consistent with supersonic (${\cal M}_{\rm turb}\approx 5.5$) turbulence (see Table \ref{Table1}; and \citealt{2008ApJ...688L..79F} for a description of the relevant parameters). One type is consistent with divergence-free driven turbulence (`compact solenoidal' clouds with $\sigma_{\rm cloud}=1.9$ and a power-law spectrum, $D(k)\propto k^{-0.78}$, in Fourier space), and the other one is consistent with curl-free driven turbulence (`porous compressive' clouds with $\sigma_{\rm cloud}=5.9$ and a smaller fractal dimension, i.e., $D(k)\propto k^{-1.44}$), but note that we do not drive turbulence explicitly, and we do not include turbulent velocity or magnetic fields in this paper. Instead, the log-normal density fields are generated with the pyFC library\footnote{Available at \url{https://bitbucket.org/pandante/pyfc}}, which produces scalar fields that are spatially correlated on a range of scales determined by a minimum wavenumber, $k_{\rm min}=8$ (which determines the number and size of the largest perturbations or `cloudlets') and the Nyquist limit, $k_{\rm max}$ (which corresponds to the grid resolution). To insert the pyFC-generated fractal clouds, we mask regions in the cloud layer outside a vertical length of $L_{\rm mc}$, scale the average density to $\bar{\rho}_{\rm cloud,0}$, and interpolate the resulting density data cube into the 3D domain. This process ensures that `compact solenoidal' and `porous compressive' multicloud models contain gas with the same initial average density.\par

Similarly to \citetalias{2020MNRAS.499.2173B}, we use the minimum wavenumber ($k_{\rm min}=8$) of the multicloud density distributions to calculate the minimum number of perturbations (`cloudlets') in the distribution as $N_{\rm cloudlet,k_{min}}\approx k_{\rm min}^3\,L_{\rm mc}/{L}=256$, and also their largest sizes as $r_{\rm cloudlet,k_{\rm min}}\approx {L}/({2\,k_{\rm min}})=6.3\,\rm pc$, so in our standard models there are up to $16$ grid cells covering a `cloudlet' radius (i.e. the numerical resolution is $\leq R_{16}$ in the conventional notation). However, we note that the fractal nature of the density fields implies that the number of cloudlets and their sizes depend on which $k$ we choose from the spectrum, so in general $N_{\rm cloudlet,k}\geq N_{\rm cloudlet,k_{min}}$ and $r_{\rm cloudlet,k}\leq r_{\rm cloudlet,k_{\rm min}}$ (see Appendix \ref{AppendixA}).\par

The top half of the domain contains a 3D multicloud layer spanning the region between the $Y=0$ and $Y=L_{\rm mc}=+50\,\rm pc$ planes. The bottom half of the domain is needed to maintain the properties of the post-shock flow intact throughout the simulation. It contains a shock with a Mach number ${\cal M_{\rm shock}}=v_{\rm shock}/{c_{\rm ambient}}=10$, where $v_{\rm shock}=1440\,\rm km\,s^{-1}$ and $c_{\rm ambient}=\sqrt{\gamma {P_{\rm ambient}}/{\rho_{\rm ambient}}}=144\,\rm km\,s^{-1}$ are the shock speed and the sound speed of the pre-shock ambient medium, respectively. The shock is defined as a discontinuity at $Y=-L/4=-25\,\rm pc$ in all models, and we use the Rankine-Hugoniot jump conditions (\citealt{1987flme.book.....L}) to calculate the density, pressure, and velocity jumps across the shock. For ${\cal M_{\rm shock}}=10$, ${P_{\rm psh}}\approx 125\,{P_{\rm ambient}}$ (with $P_{\rm ambient}/k_B\approx10^4\,\rm K\,cm^{-3}$), and ${v_{\rm psh}}\approx 0.75\,v_{\rm shock}=1080\,\rm km\,s^{-1}$. Also, the initial (mass-weighted) average temperatures in the multicloud systems is $10^4\,\rm K$, in the pre-shock ambient medium is $10^6\,\rm K$, and in the post-shock ambient medium is $3\times10^7\,\rm K$. We evolve the systems in the rest frame of the pre-shock media, so the pre-shock media are initially stationary ($v_{\rm ambient}=0$) and in thermal pressure equilibrium. Note also that since the shock is initially $-25\,\rm pc$ away from the multicloud layer, there is a delay, $\Delta t_{\rm ini}$, until the shock reaches the multicloud layer (which we define as $t=0$).




\subsubsection{Boundary conditions}
\label{subsec:Boundaries}
We set up diode boundary conditions on the top side of the simulation domain, periodic conditions on the lateral sides, and an inflow boundary condition on the bottom side. The inflow zone injects a constant supply of gas with properties of the post-shock gas into the computational domain, and it is located at the ghost zone that faces the bottom side of the multicloud system (see Figure \ref{Figure2}).

\subsubsection{Models}
\label{subsec:models}
In total, we present $8$ models (see Table \ref{Table1}). We use two of the non-radiative models presented in \citetalias{2020MNRAS.499.2173B} as control runs, one with a compact solenoidal cloud layer (sole-k8-M10) and one with a porous compressive cloud layer (comp-k8-M10). In addition, we include $6$ new models with radiative cloud layers (for which we add `-rad' termination at the end of the name), three of them are compact solenoidal and three of them are porous compressive. We keep the minimum wavenumber of the cloud distribution ($k_{\rm min}=8$), the cloud layer thickness ($L_{\rm mc}=50\,\rm pc$), the shock Mach number (${\cal M}_{\rm shock}=10$), and the cloud-generating seed ($S_d$) constant in all models, and only vary the numerical resolution within each of the new model samples. Models with higher numerical resolutions are labelled with `-hr' and models with lower numerical resolutions with `-lr'.\par

The initial conditions in our models represent the ISM conditions at the base of galactic winds driven by stellar feedback (e.g., see \citealt{2008ApJ...674..157C,2017ApJ...834..144S}). In particular, our models are relevant for the galactic outflows in galaxy M82 (e.g., see \citealt{1998ApJ...493..129S,2009ApJ...697.2030S}) and in our Galaxy (e.g., see \citealt{2003ApJ...582..246B,2013ApJ...770L...4M,2019Natur.573..235H}). Thus, similarly to \citetalias{2020MNRAS.499.2173B} the multicloud layer thickness, $L_{\rm mc}$, is selected so that its value is lower than disc scale heights of $\sim 300\,\rm pc$ of these systems. Our computational models can be interpreted as idealised vertical sections of a global 3D outflow (e.g., see \citealt{2020ApJ...895...43S}).

\subsection{Diagnostics and dynamical time-scales}
\label{subsec:Diagnostics}
To analyse the simulations presented in this paper, we use the same diagnostics introduced in \citetalias{2020MNRAS.499.2173B}, and we list them in Table \ref{Table2} in the order of appearance in this paper, jointly with their respective definition. The reader is referred to Section 2.4 of \citetalias{2020MNRAS.499.2173B} and references therein for additional details on how these diagnostics are computed. As in \citetalias{2020MNRAS.499.2173B}, in Sections \ref{sec:Results} and \ref{sec:Application} we show the evolution of these diagnostics only until they are not affected by the exit of the shock front or cloud gas from the simulation domain.\par

\begin{table}\centering
\caption{List of diagnostics used in the analysis of our simulations.}
\hspace{-0.2cm}\begin{adjustbox}{max width=86mm}
\begin{threeparttable}
\begin{tabular}{c c c}
\hline
\textbf{(1)} & \textbf{(2)} & \\
\textbf{Diagnostic} & \textbf{Definition} \\ \hline\vspace{+0.15cm}
x,y,z Mach number & ${\cal M}_{x,y,z}\approx \left|\frac{4}{3}\frac{\Delta v_{x,y,z}}{c_{\rm ambient}}\right|$ & $\inlineeqno$\\\vspace{+0.15cm}
rs/ts Mach number \tnote{(a,b)} & ${\cal M}_{\rm rs/ts}=\max\left(\frac{\int {\cal M}_i\,dxdz}{\int \,dxdz}\right)_{-\vec{y}/+\vec{y}}$ & $\inlineeqno$\\\vspace{+0.15cm}
Cloud gas / cloud material & $\rho C$ & $\inlineeqno$ \\\vspace{+0.15cm}
Cloud mass \tnote{(c)} & $M_{\rm mc}=\int \rho\,C\,dV$ & $\inlineeqno$ \\\vspace{+0.15cm}
Cloud volume \tnote{(c)} & $V_{\rm mc}=\int \,C\,dV$ & $\inlineeqno$ \\\vspace{+0.15cm}
Cloud column density \tnote{(d)} & $N_{\rm mc}=\int n\,C\,dZ$ & $\inlineeqno$ \\\vspace{+0.15cm}
Mean cloud pressure & $\left[~P_{{\rm cloud}}~\right]=\frac{\int P\,C\,dV}{\int C\,dV}$ & $\inlineeqno$ \\\vspace{+0.15cm}
Cloud volume filling factor & $F_{v}=\frac{\int C\,dV}{\int dV}$ & $\inlineeqno$\\\vspace{+0.15cm}
Mixing fraction \tnote{(e,f)} & $f_{{\rm mix}}=\frac{\int \rho\,C_{\rm mix}\,dV}{\int \rho_0\,C\,dV}$ & $\inlineeqno$ \\\vspace{+0.15cm}
j-Velocity dispersion \tnote{(c)} & $\delta_{{\rm v}_{{\rm j}}}=\sqrt{\langle~v^2_{{\rm j}}~\rangle-\langle~v_{{\rm j}}~\rangle^2}$; $\rm j=X,Z$ & $\inlineeqno$\\\vspace{+0.15cm}
Velocity dispersion & $\delta_{{\rm v}}=\sqrt{\sum_{\rm j}\delta_{{\rm v}_{{\rm j}}}^2}$; $\rm j=X,Z$ & $\inlineeqno$ \\\vspace{+0.15cm}
Cloud displacement along $Y$ \tnote{(c)} & $\langle~d_{{\rm y}}~\rangle=\frac{\int \rho\,Y CdV}{\int \rho\,C\,dV}$ & $\inlineeqno$\\\vspace{+0.15cm}
Cloud speed along $Y$ \tnote{(c)} & $\langle~v_{{\rm y}}~\rangle=\frac{\int \rho\,v_{{\rm y}}\,C\,dV}{\int \rho\,C\,dV}$ & $\inlineeqno$ \\\vspace{+0.15cm}
Cloud mass with $\rho C \geq \bar{\rho}_{\rm cloud, 0}/3$ \tnote{(g)} & $M_{\rm mc_{1/3}}=\int [\rho\,C]_{\rho C \geq \bar{\rho}_{\rm cloud, 0}/3}\,dV$ & $\inlineeqno$\\\vspace{+0.15cm}
Cloud mass with $\rho C \geq \bar{\rho}_{\rm cloud, 0}$ \tnote{(g)} & $M_{\rm mc_{1}}=\int [\rho\,C]_{\rho C \geq \bar{\rho}_{\rm cloud, 0}}\,dV$ & $\inlineeqno$\\ 
\vspace{+0.15cm}
Total mass with $\rho \geq \bar{\rho}_{\rm cloud, 0}/3$ & $M_{1/3}=\int [\rho]_{\rho \geq \bar{\rho}_{\rm cloud, 0}/3}\,dV$ & $\inlineeqno$\\\vspace{+0.15cm}
Total mass with $\rho \geq \bar{\rho}_{\rm cloud, 0}$ & $M_{1}=\int [\rho]_{\rho \geq \bar{\rho}_{\rm cloud, 0}}\,dV$ & $\inlineeqno$\\
\hline
\end{tabular}
\begin{tablenotes}
	\item[(a)] `rs'$\equiv$ reverse shock, `ts'$\equiv$ transmitted forward shock.
	\item[(b)] ${\cal M}_{i}=\sqrt{{\cal M}^2_{x_i}+{\cal M}^2_{y_i}+{\cal M}^2_{z_i}}$ is the Mach number in each cell, $i$.
	\item[(c)] $M_{{\rm mc}_{\rm T}}$, $V_{{\rm mc}_{\rm T}}$, ${\delta_{{\rm v}_{{\rm j}}}}_{\rm T}$, ${\langle~d_{{\rm y}}~\rangle}_{\rm T}$, ${\langle~v_{{\rm y}}~\rangle}_{\rm T}$ are measured for specific temperature  (T) bins.
	\item[(d)] $N_{\rm H I}$ is defined in a similar manner, but for H\,{\sc i}-emitting gas solely.
	\item[(e)] $C_{\rm mix}=C$ if $0.1\leq C \leq 0.9$, and $C_{\rm mix}=0$ otherwise.
	\item[(f)] $M_{\rm mc,0}=\int \rho_0\,C\,dV$ is the initial mass of the multicloud layer.
	\item[(g)]  ${\langle~d_{{\rm y}}~\rangle}_{\rm 1/3}$, ${\langle~v_{{\rm y}}~\rangle}_{\rm 1/3}$ are also calculated for $\rho C \geq \bar{\rho}_{\rm cloud, 0}/3$.
\end{tablenotes}
\end{threeparttable}
\end{adjustbox}
\label{Table2}
\end{table}

Similarly, we study the evolution of shock-multicloud models using the same set of pre-defined time-scales reported in Section 2.5 of \citetalias{2020MNRAS.499.2173B}, jointly with some additional time-scales related to radiative processes and dynamical instabilities. First, the approximate time for the transmitted shock to travel across the multicloud layer is the shock-passage time, 
\begin{equation}
t_{\rm sp}=\frac{L_{\rm mc}}{v_{\rm ts}}=0.200\,\rm Myr,
\label{eq:ShockPassageTime}
\end{equation}

\noindent where $v_{\rm ts}=\chi^{-\frac{1}{2}}\,(F_{\rm c1}F_{\rm st})^{\frac{1}{2}}{\cal M}_{\rm shock}\,c_{\rm ambient}\approx250\,\rm km\,s^{-1}$ is the approximate speed of the internal shock transmitted to the multicloud layer. Here, $F_{\rm st}$ and $F_{\rm c1}$ are dimensionless factors that relate the postshock ambient pressure to the stagnation pressure, and the stagnation pressure to the pressure behind the transmitted shock, respectively (see \citealt{2002ApJ...576..832P} and Section 2.5 in \citetalias{2020MNRAS.499.2173B}). We use the shock-passage time, $t_{\rm sp}$, as our normalisation time-scale for all our models. In addition, for individual cloudlets inside the multicloud layer, the relevant dynamical time-scale is the cloud-crushing time, defined in \cite*{1994ApJ...420..213K,2002ApJ...576..832P}
\begin{equation}
t_{\rm cc}=\frac{2\,r_{\rm cloudlet}}{v_{\rm ts}}=0.25\,t_{\rm sp}=0.050\,\rm Myr.
\label{eq:CloudCrushing}
\end{equation}

Time-scales associated with radiative processes are also important. The cloud cooling time is: 
\begin{equation}
t_{\rm cool,mc}=\frac{\frac{3}{2}\,n_{\rm cloud}\,k_B\,\langle T_{\rm mc} \rangle}{n_{\rm cloud}^2\,\tilde{\Lambda}_{\langle T_{\rm mc} \rangle}}=0.04\,t_{\rm sp}=0.008\,\rm Myr,
\label{eq:CoolingTime}
\end{equation}

\noindent which is associated with a cooling length (\citealt*{2010ApJ...722..412Y,2013ApJ...766...45J}) $l_{\rm cool, mc}=v_{\rm ts}\,t_{\rm cool, mc}=2.11\,\rm pc$, and a shattering length (\citealt{2018MNRAS.473.5407M,2019MNRAS.482.5401S}) $l_{\rm shatter, mc}=c_{\rm cloud}\,t_{\rm cool, mc}=0.12\,\rm pc$. Note that $t_{\rm cool,mc}$, $l_{\rm cool, mc}$, and $l_{\rm shatter, mc}$ are calculated using the initial average temperature of the multicloud system, $\langle T_{\rm mc} \rangle=10^4\,\rm K$, but gas in the cloud layers actually spans a broad range of temperatures owing to their inhomogeneous density fields. Thus, different parts of a multicloud layer have cooling times and cooling/shattering lengths of their own.\par

\noindent Similarly, the cooling time of mixed gas (\citealt{2018MNRAS.480L.111G}) with a temperature of $T_{\rm mix}\sim (T_{\rm psh}\,\langle T_{\rm mc} \rangle)^{0.5}=5.6\times10^5\,\rm K$ is 
\begin{equation}
t_{\rm cool, mix}=\chi\,\frac{\tilde{\Lambda}_{\langle T_{\rm mc} \rangle}}{\tilde{\Lambda}_{T_{\rm mix}}}\,t_{\rm cool, mc}=0.12\,t_{\rm sp}=0.024\,\rm Myr.
\label{eq:CoolingTime}
\end{equation}


In addition, the growth times of shear-driven Kelvin--Helmholtz (KH) and acceleration-driven Rayleigh--Taylor (RT) instabilities (see \citealt{1961hhs..book.....C}) with wavelengths similar to the cloud radius are
\begin{equation}
t_{\rm KH}\approx \frac{r_{\rm cloudlet} \chi^{0.5}}{2\pi\,(v_{\rm psh})}=0.05\,t_{\rm sp}=0.009\,\rm Myr,\: \rm and
\label{KHtime}
\end{equation}

\begin{equation}
t_{\rm RT}\approx\left[\frac{r_{\rm cloudlet}}{2\pi\,(a_{\rm eff})}\right]^{0.5}=0.18\,t_{\rm sp}=0.036\,\rm Myr,
\label{RTtime}
\end{equation}

\noindent respectively. In the latter equation $a_{\rm eff}\approx 0.4v_{\rm psh}^2/(\chi r_{\rm cloudlet})$ is the effective cloud acceleration (see \citealt{2019MNRAS.486.4526B}).\par

These estimates indicate that the cooling time and the KH and RT instability growth time-scales in our simulations are all smaller than the shock-passage time, which implies that KH and RT instabilities are dynamically important for the evolution of shocked multicloud models. These calculations also indicate that these systems are subjected to strong cooling, which also suppresses the growth of large-wavelength dynamical instabilities (\citealt{2009ApJ...703..330C,2015ApJ...805..158S}). Therefore, as we discuss below, the evolution of radiative systems is controlled by the interplay between thermodynamical processes and the generation of turbulence.\par

Similar to \citetalias{2020MNRAS.499.2173B}, simulations start at $t_0=-0.09\,t_{\rm sp}$, the shock reaches the multicloud layer at $t=0$, the shock-multicloud interaction time is $t_{\rm sim}=3\,t_{\rm sp}$, and the total simulation time is $t_{\rm totsim}=\Delta t_{\rm ini}+t_{\rm sim}$, where $\Delta t_{\rm ini}=0.09\,t_{\rm sp}$ is measured from $t_0$ to $t=0$. Also, the cloud destruction time \citep[see, e.g.][]{2015ApJ...805..158S,2019MNRAS.486.4526B} is defined as the time when only a quarter of the initial mass of the multicloud layer has densities above $1/3$ of the initial mean density, $\bar{\rho}_{\rm cloud,0}$.\par

\section{Results}
\label{sec:Results}

\subsection{Non-radiative versus radiative multicloud models}
\label{subsec:Evolution}

\begin{figure*}
\begin{center}
  \begin{tabular}{c c c c c c c}
       \multicolumn{1}{l}{\hspace{-2mm}3a) sole-k8-M10 \hspace{+0.2mm}$t_0$} & \multicolumn{1}{c}{$0.5\,t_{\rm sp}=0.10\,\rm Myr$} & \multicolumn{1}{c}{$1.1\,t_{\rm sp}=0.22\,\rm Myr$} & \multicolumn{1}{c}{$1.8\,t_{\rm sp}=0.36\,\rm Myr$} & \multicolumn{1}{c}{$2.4\,t_{\rm sp}=0.48\,\rm Myr$} & \multicolumn{1}{c}{$3.0\,t_{\rm sp}=0.60\,\rm Myr$} & $\frac{n}{n_{\rm ambient}}$\\   
       \hspace{-0.3cm}\resizebox{27mm}{!}{\includegraphics{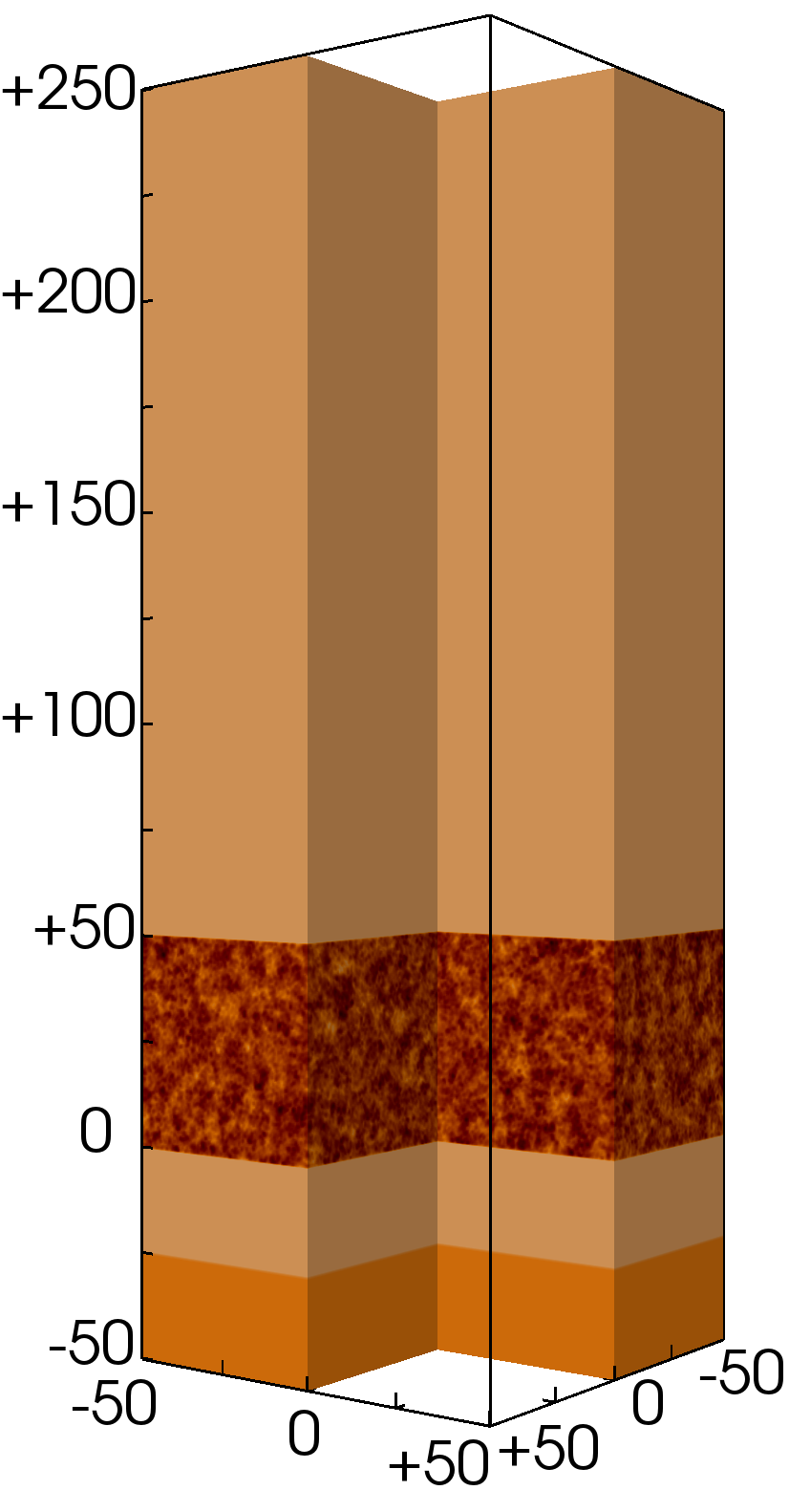}} & \hspace{-0.3cm}\resizebox{27mm}{!}{\includegraphics{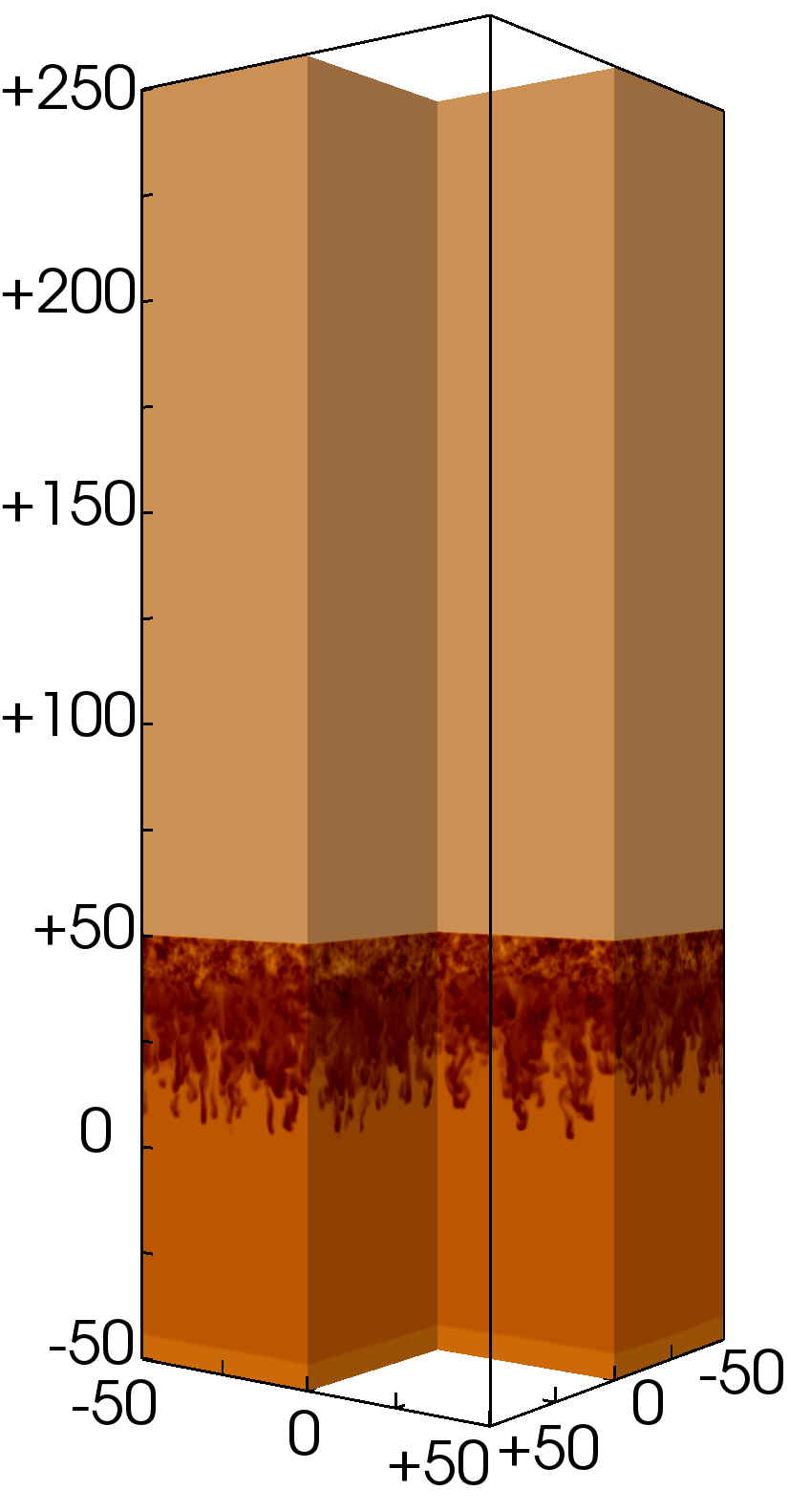}} & \hspace{-0.3cm}\resizebox{27mm}{!}{\includegraphics{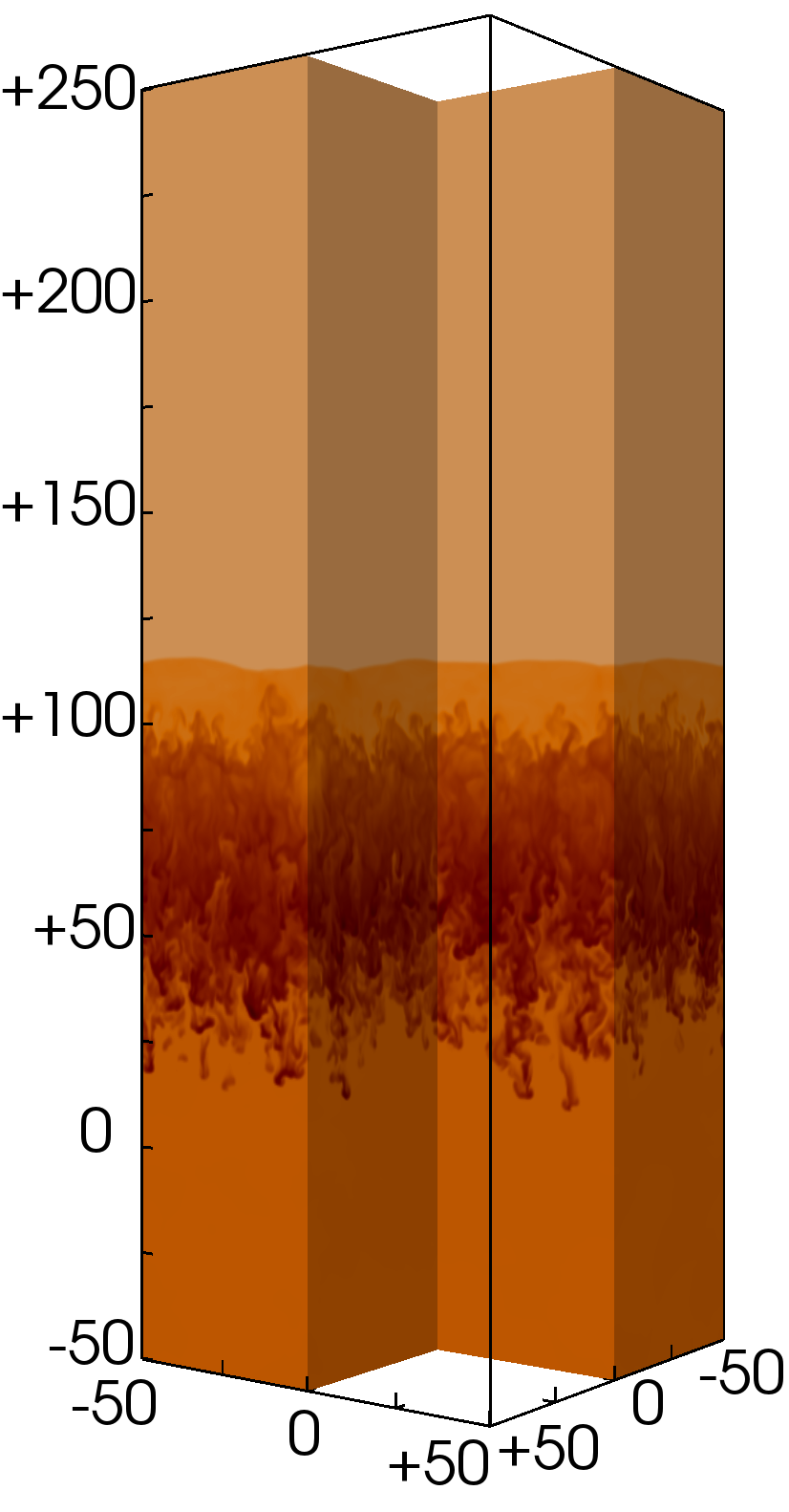}} & \hspace{-0.3cm}\resizebox{27mm}{!}{\includegraphics{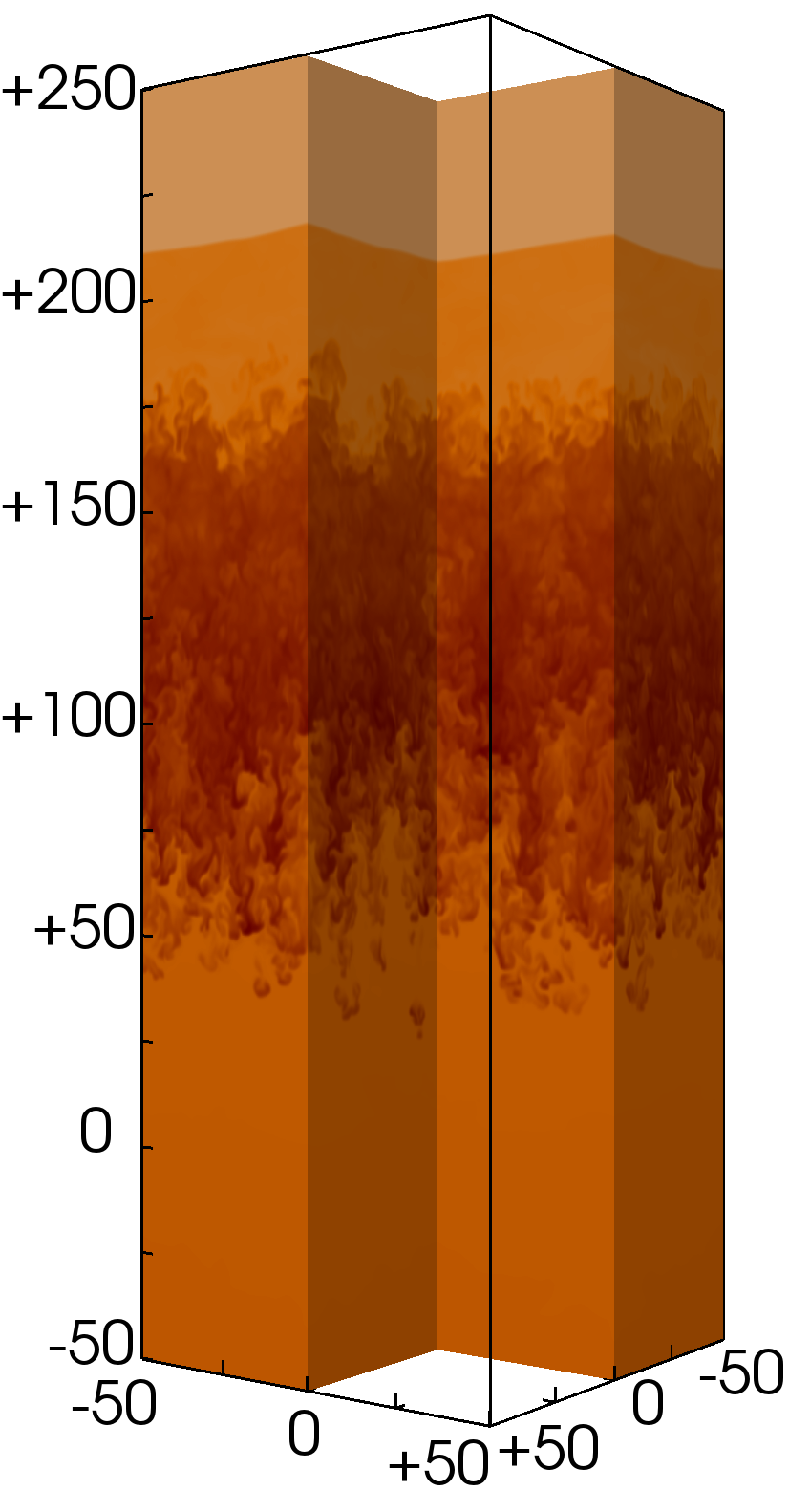}} & \hspace{-0.3cm}\resizebox{27mm}{!}{\includegraphics{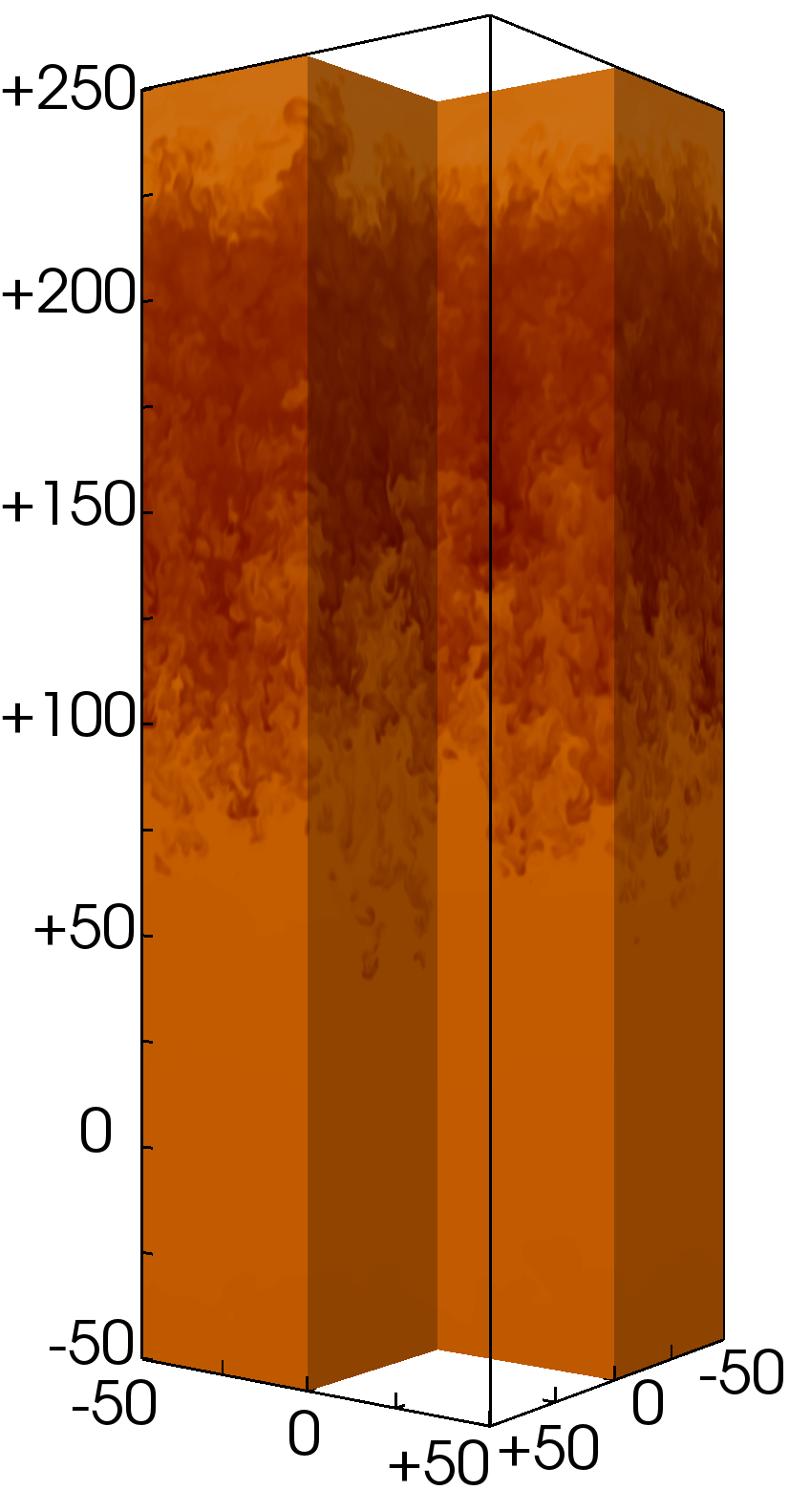}} & \hspace{-0.3cm}\resizebox{27mm}{!}{\includegraphics{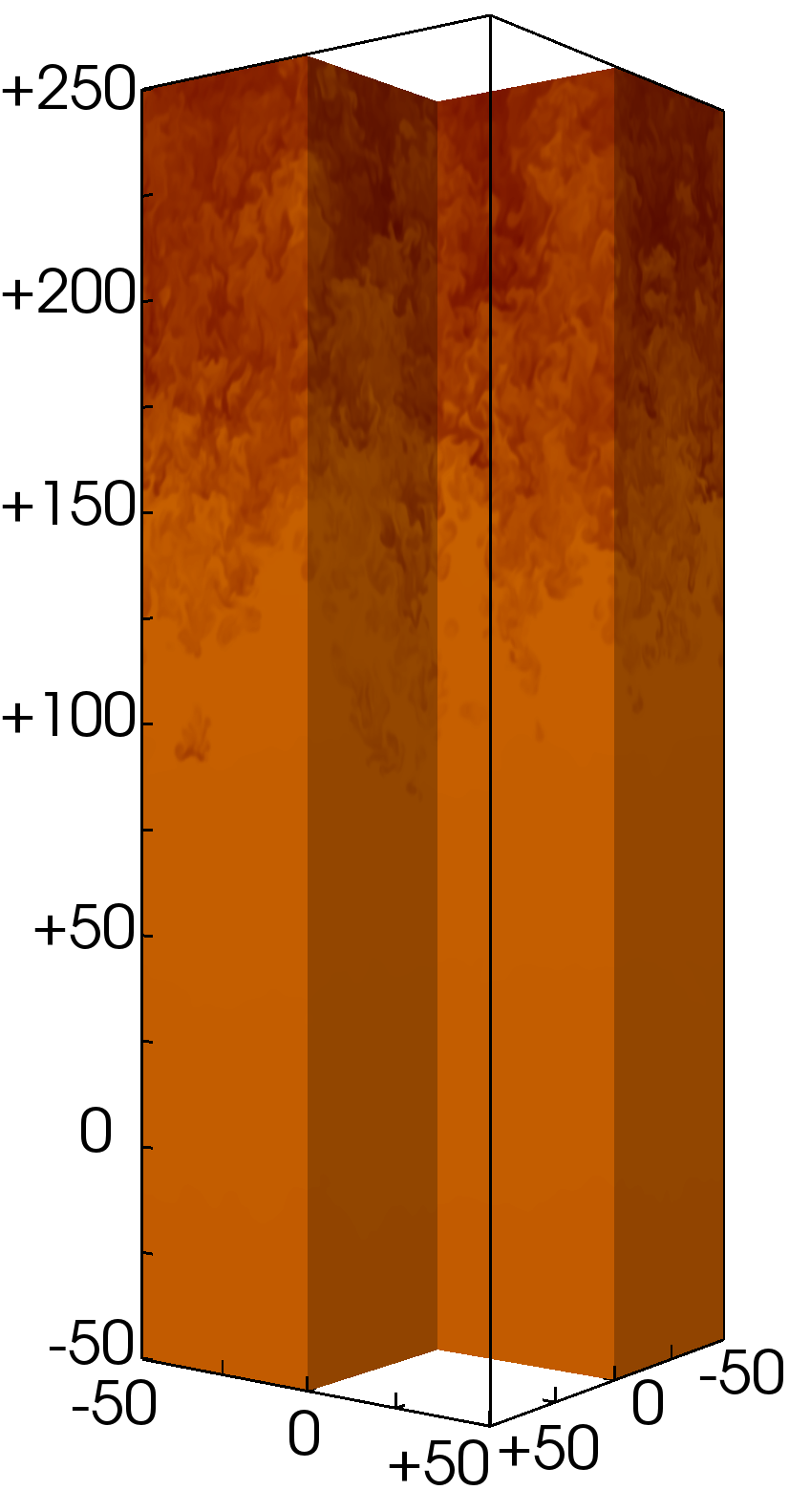}} &
\hspace{-0.3cm}\resizebox{9.8mm}{!}{\includegraphics{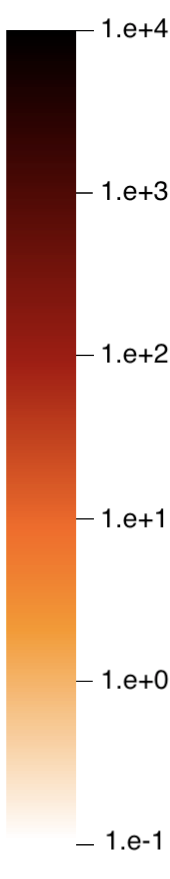}}\\
       \multicolumn{1}{l}{\hspace{-2mm}3b) sole-k8-M10-rad \hspace{+0.2mm}$t_0$} & \multicolumn{1}{c}{$0.5\,t_{\rm sp}=0.10\,\rm Myr$} & \multicolumn{1}{c}{$1.1\,t_{\rm sp}=0.22\,\rm Myr$} & \multicolumn{1}{c}{$1.8\,t_{\rm sp}=0.36\,\rm Myr$} & \multicolumn{1}{c}{$2.4\,t_{\rm sp}=0.48\,\rm Myr$} & \multicolumn{1}{c}{$3.0\,t_{\rm sp}=0.60\,\rm Myr$} & $\frac{n}{n_{\rm ambient}}$\\      
       \hspace{-0.3cm}\resizebox{27mm}{!}{\includegraphics{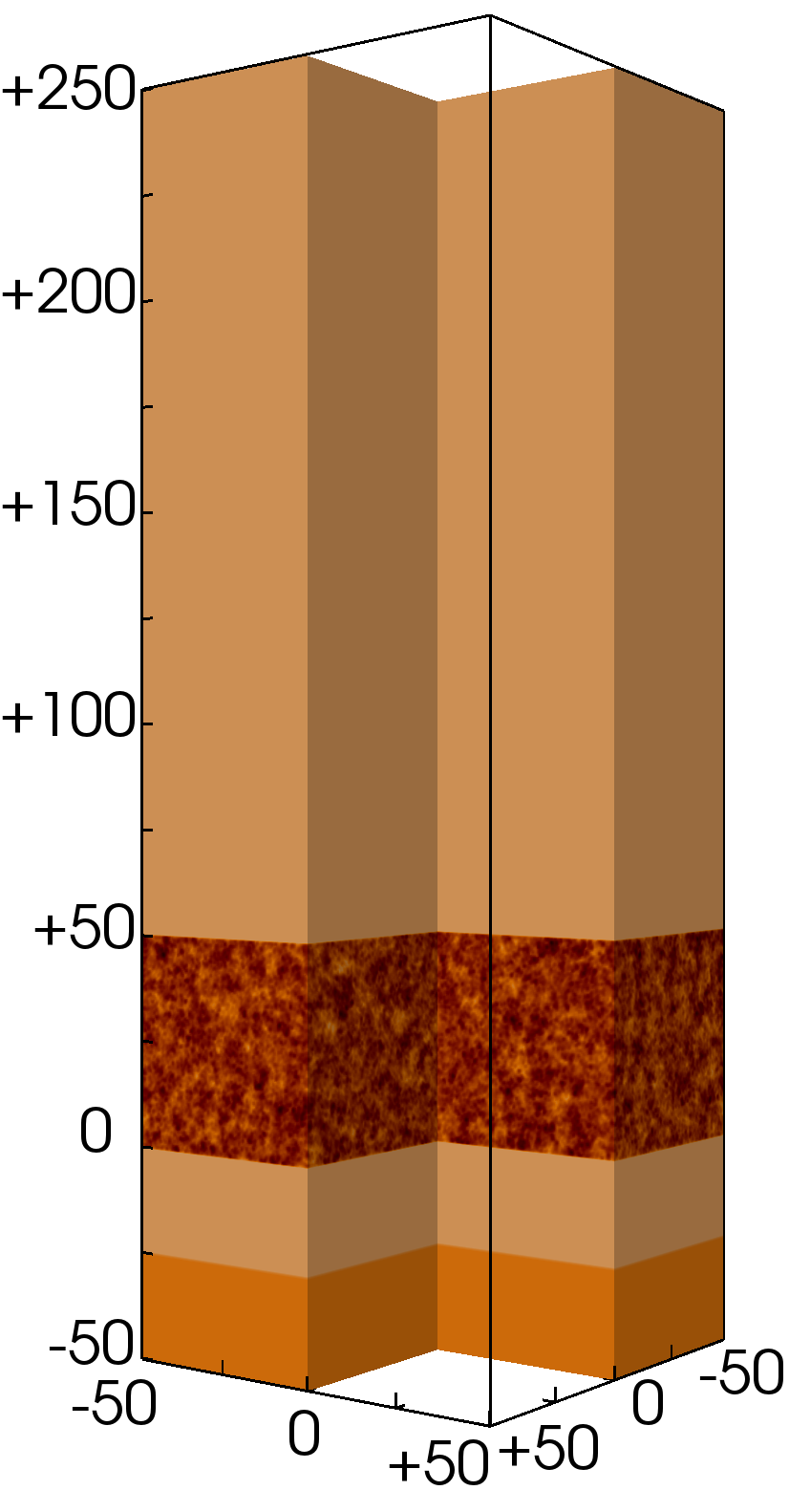}} & \hspace{-0.3cm}\resizebox{27mm}{!}{\includegraphics{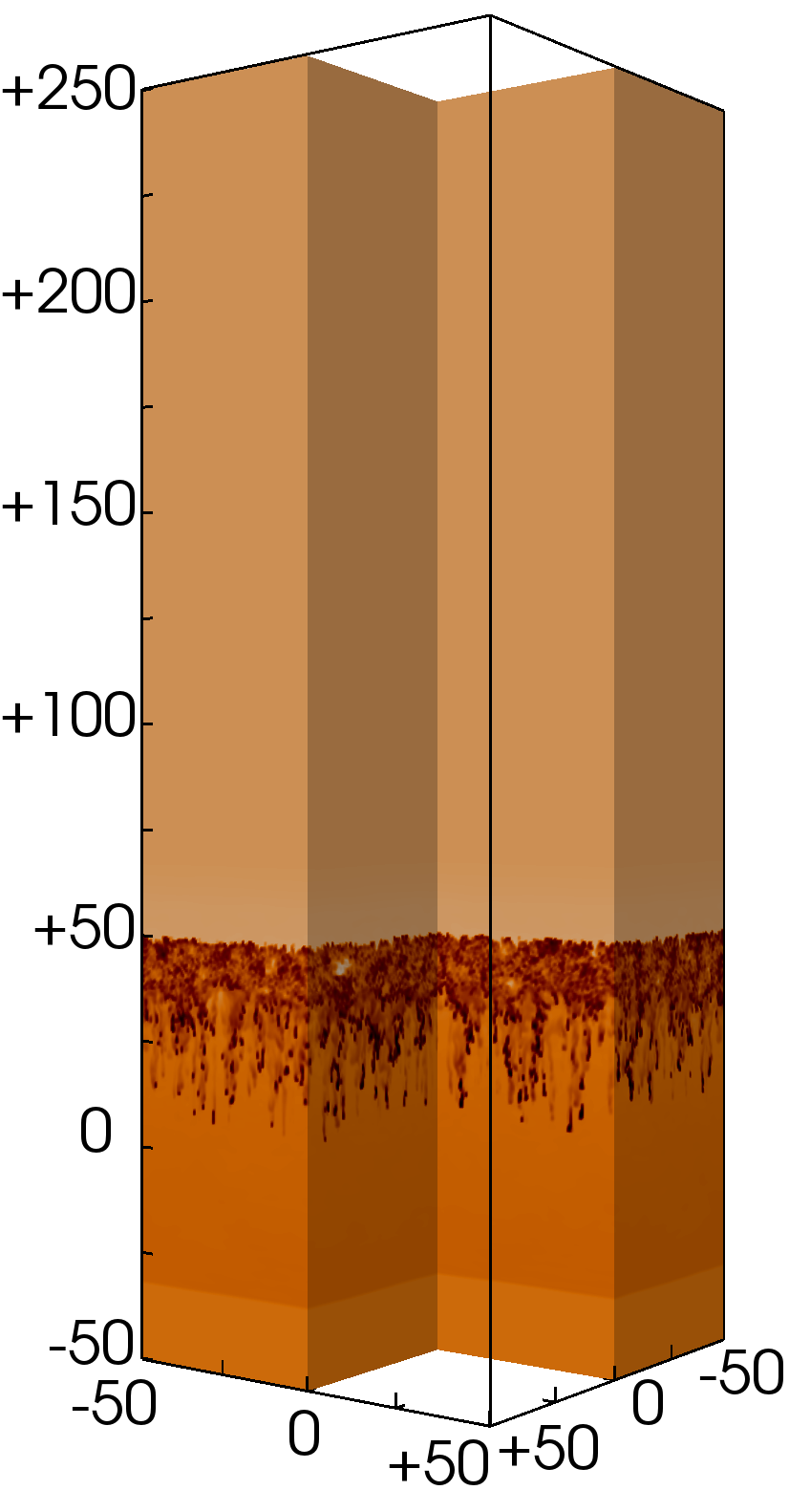}} & \hspace{-0.3cm}\resizebox{27mm}{!}{\includegraphics{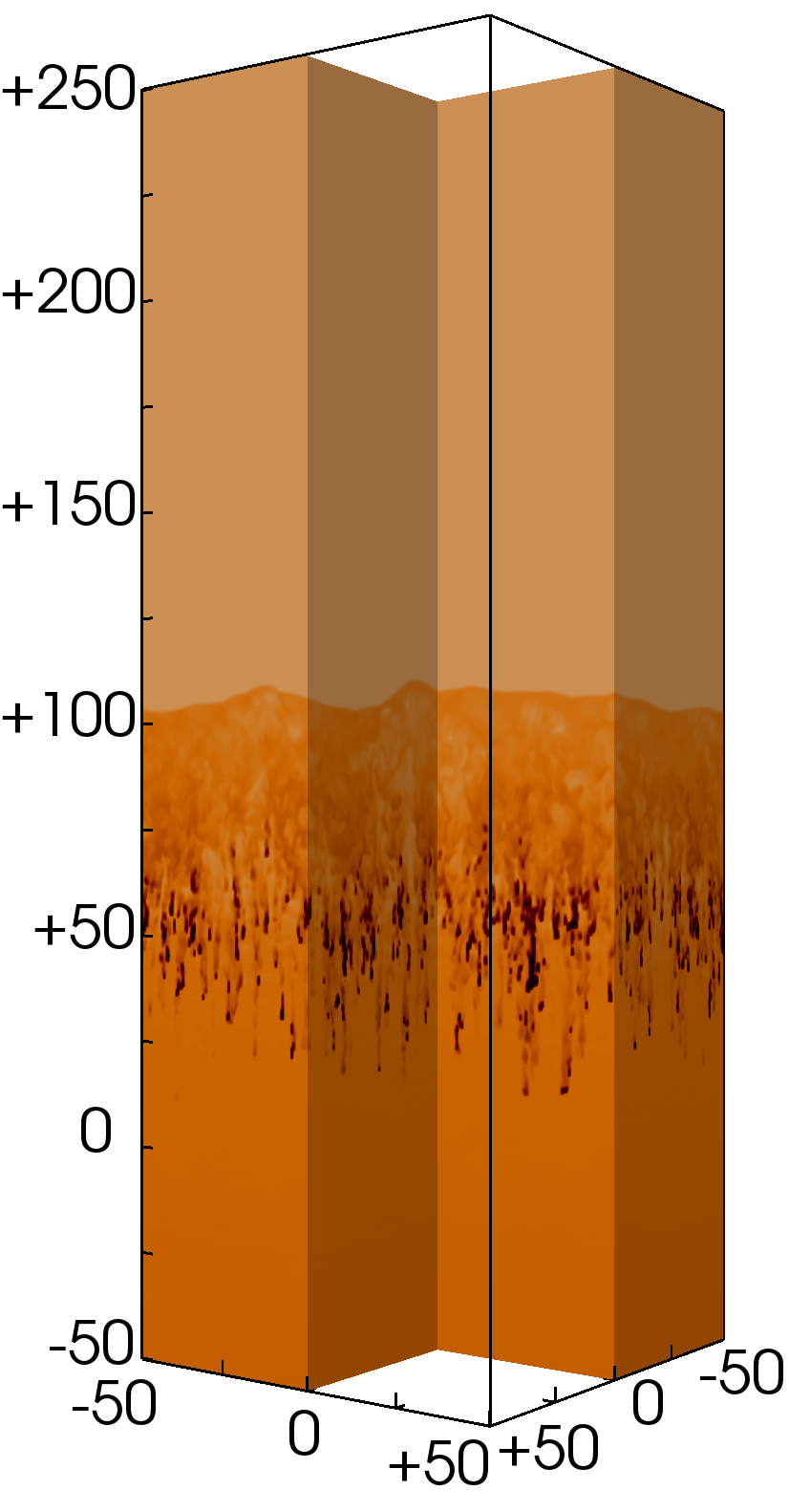}} & \hspace{-0.3cm}\resizebox{27mm}{!}{\includegraphics{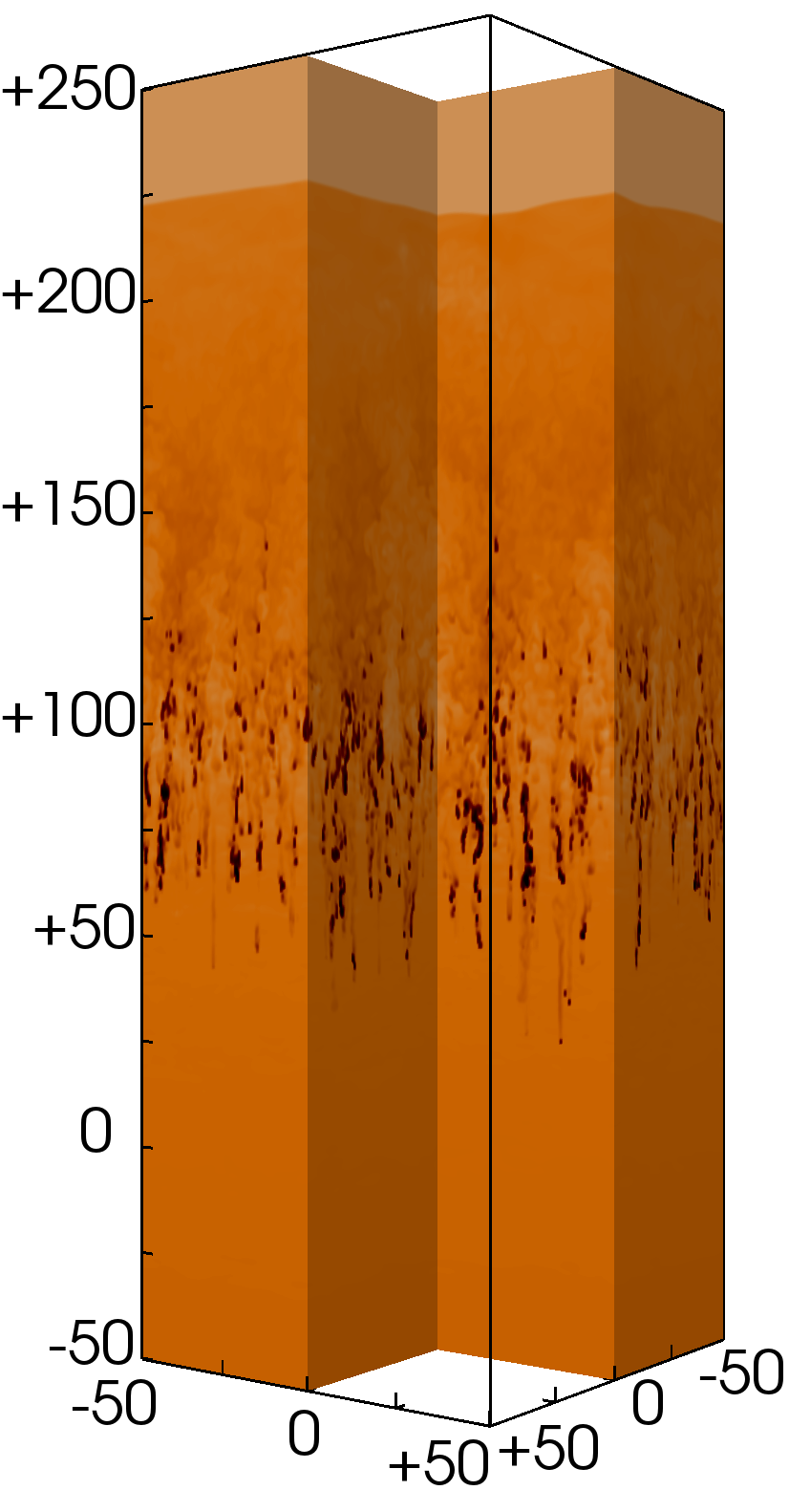}} & \hspace{-0.3cm}\resizebox{27mm}{!}{\includegraphics{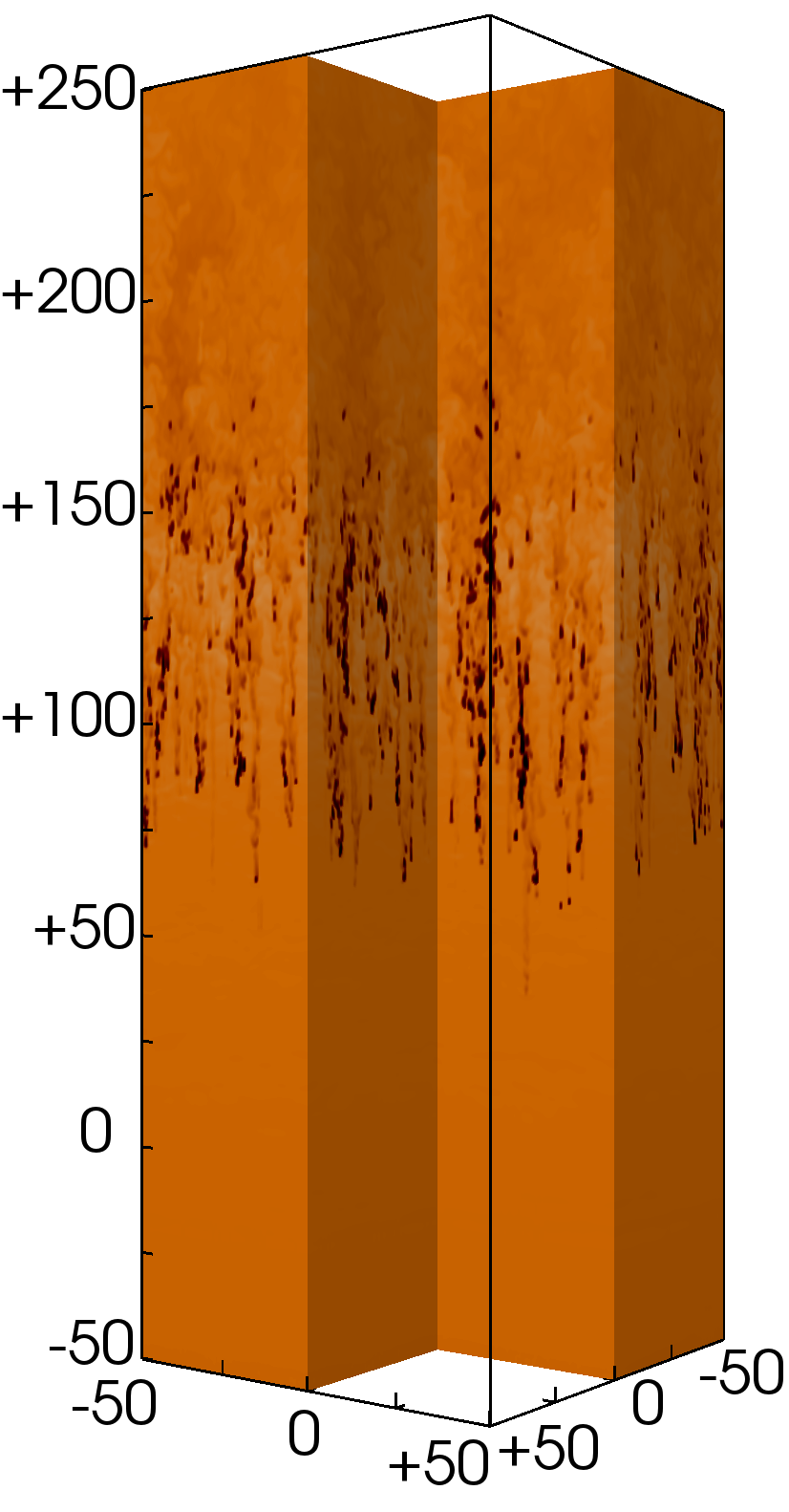}}  & \hspace{-0.3cm}\resizebox{27mm}{!}{\includegraphics{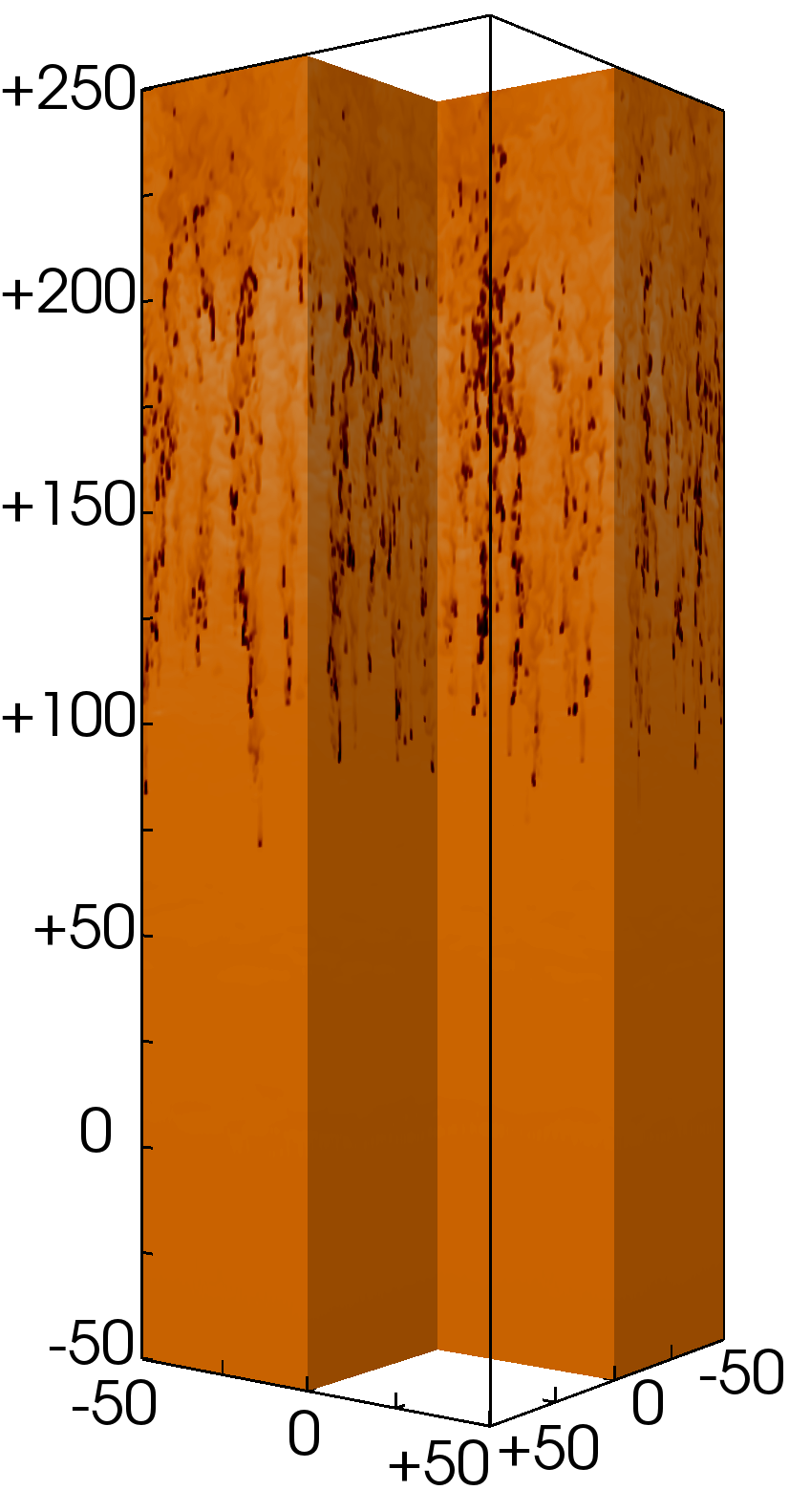}} &
\hspace{-0.3cm}\resizebox{9.8mm}{!}{\includegraphics{bar_vert_2.png}}\\
       \multicolumn{1}{l}{\hspace{-2mm}3c) comp-k8-M10 \hspace{+0.2mm}$t_0$} & \multicolumn{1}{c}{$0.5\,t_{\rm sp}=0.10\,\rm Myr$} & \multicolumn{1}{c}{$1.1\,t_{\rm sp}=0.22\,\rm Myr$} & \multicolumn{1}{c}{$1.8\,t_{\rm sp}=0.36\,\rm Myr$} & \multicolumn{1}{c}{$2.4\,t_{\rm sp}=0.48\,\rm Myr$} & \multicolumn{1}{c}{$3.0\,t_{\rm sp}=0.60\,\rm Myr$} & $\frac{n}{n_{\rm ambient}}$\\   
       \hspace{-0.3cm}\resizebox{27mm}{!}{\includegraphics{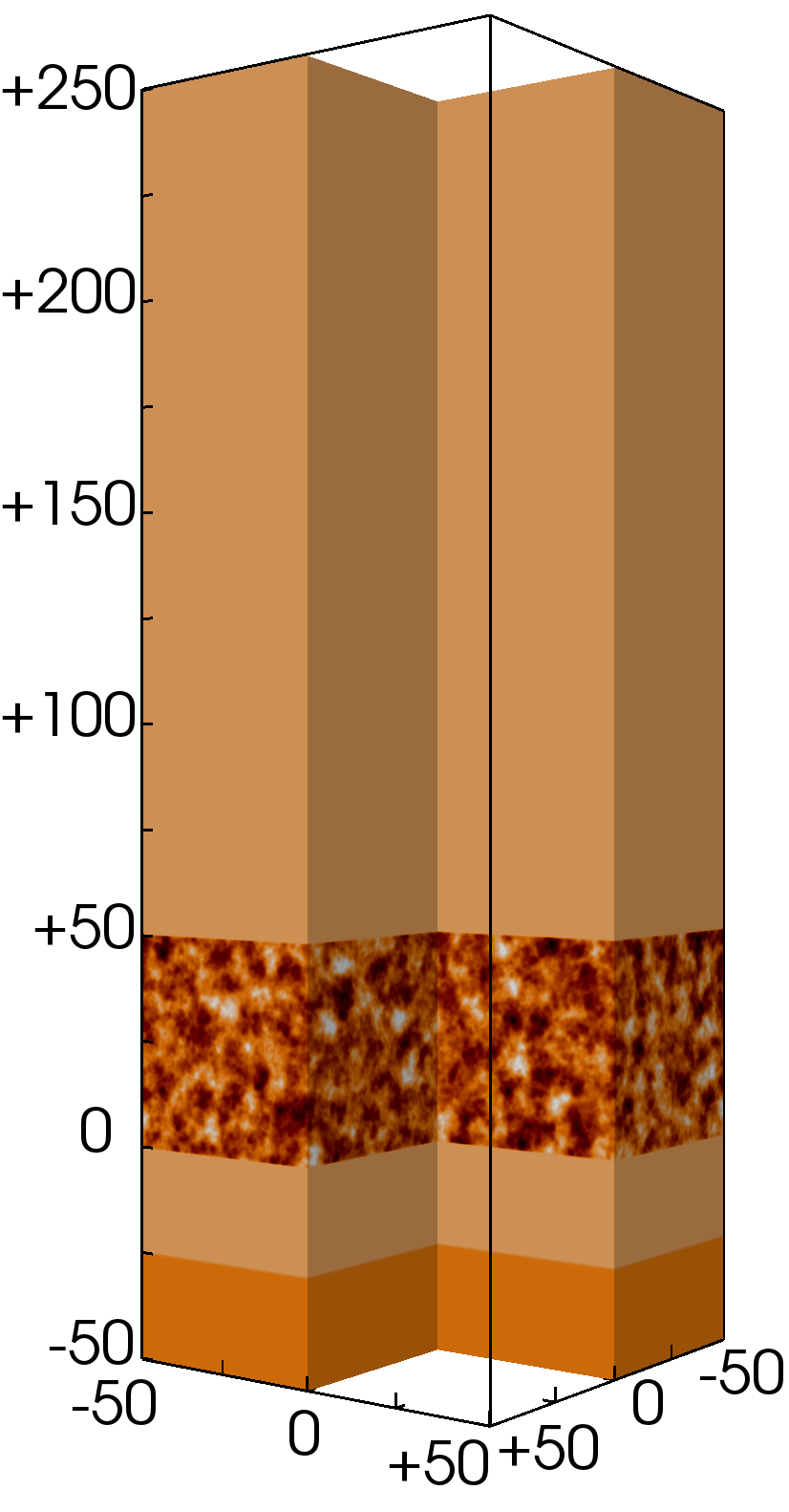}} & \hspace{-0.3cm}\resizebox{27mm}{!}{\includegraphics{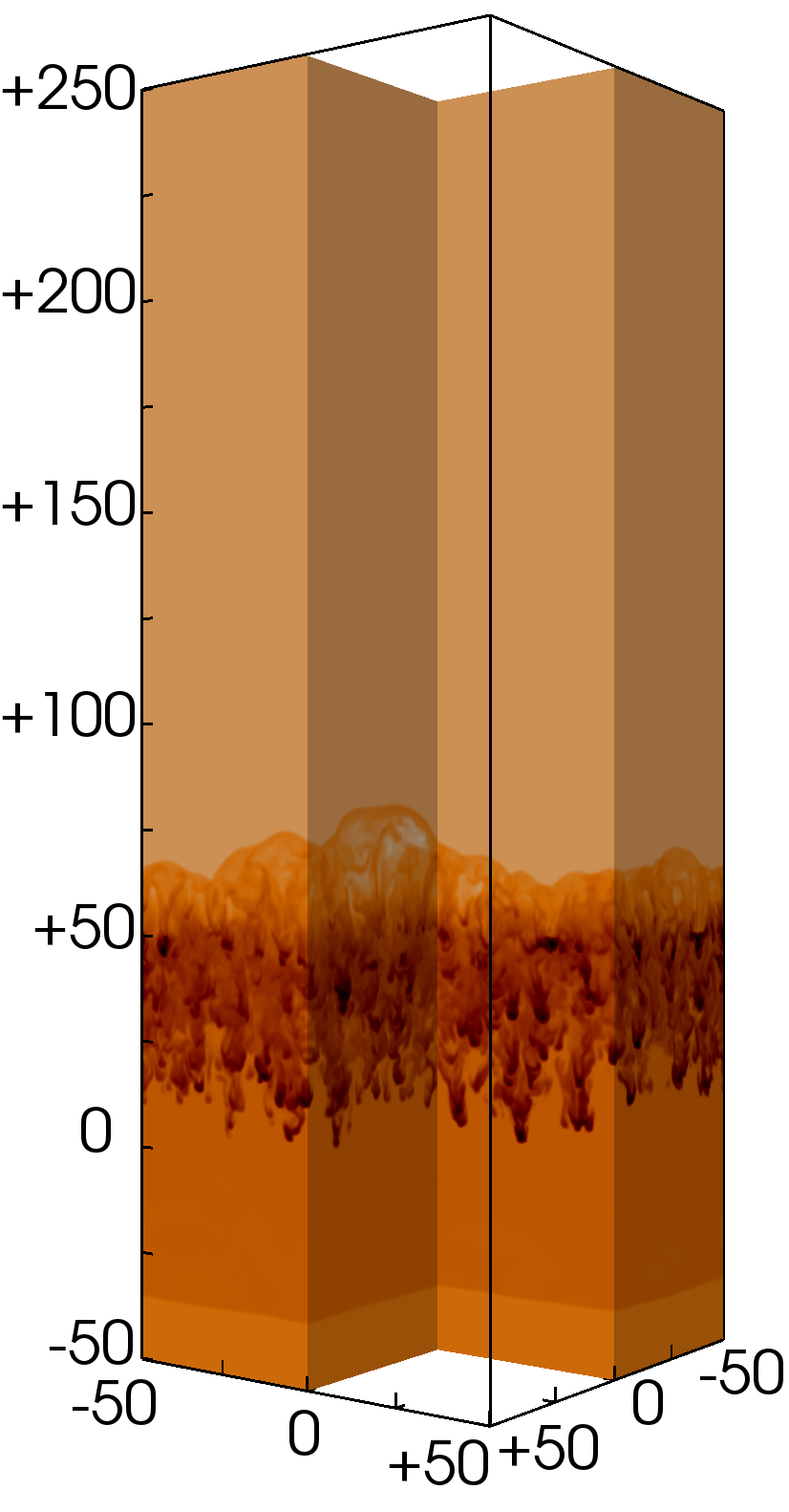}} & \hspace{-0.3cm}\resizebox{27mm}{!}{\includegraphics{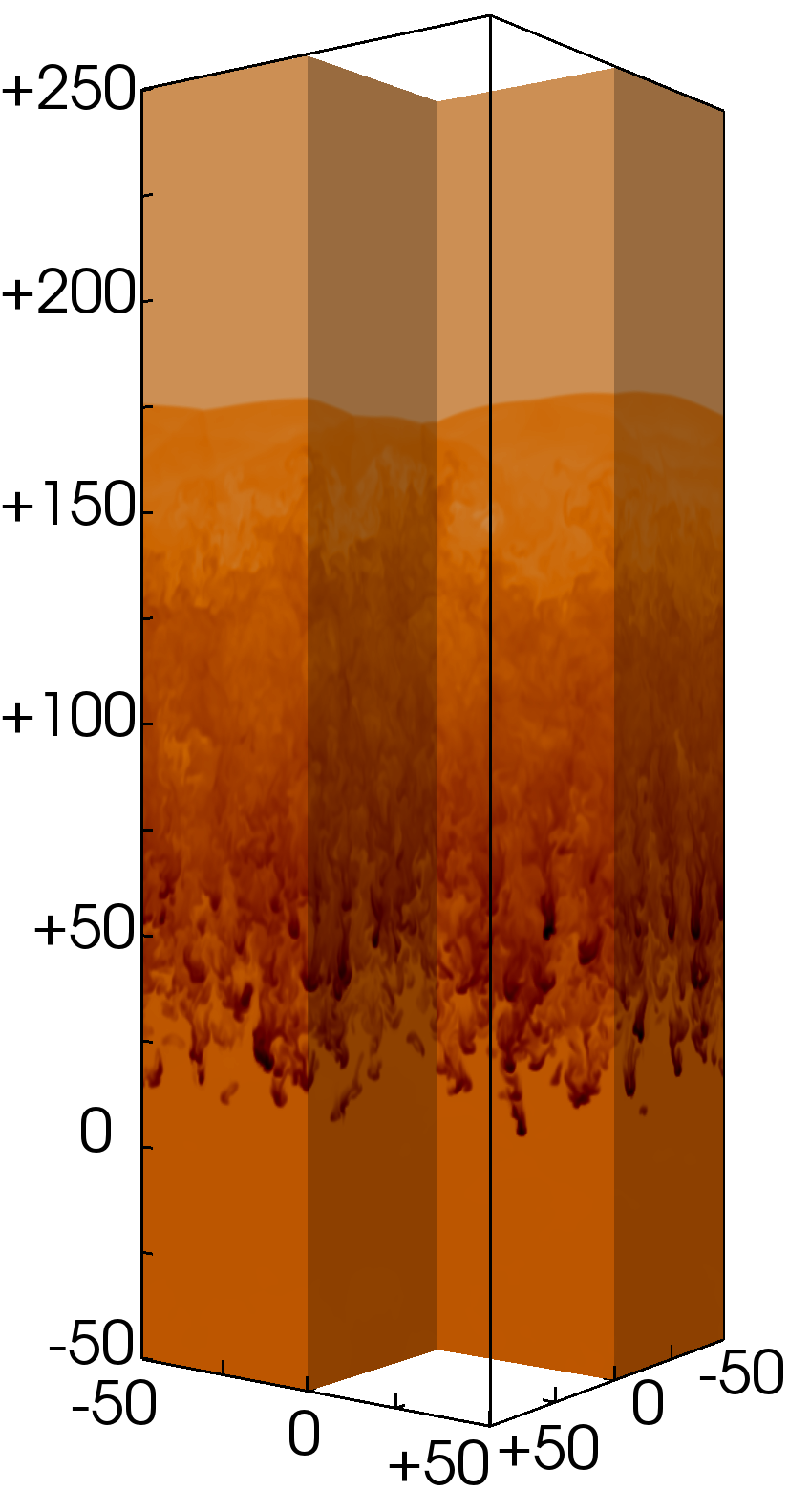}} & \hspace{-0.3cm}\resizebox{27mm}{!}{\includegraphics{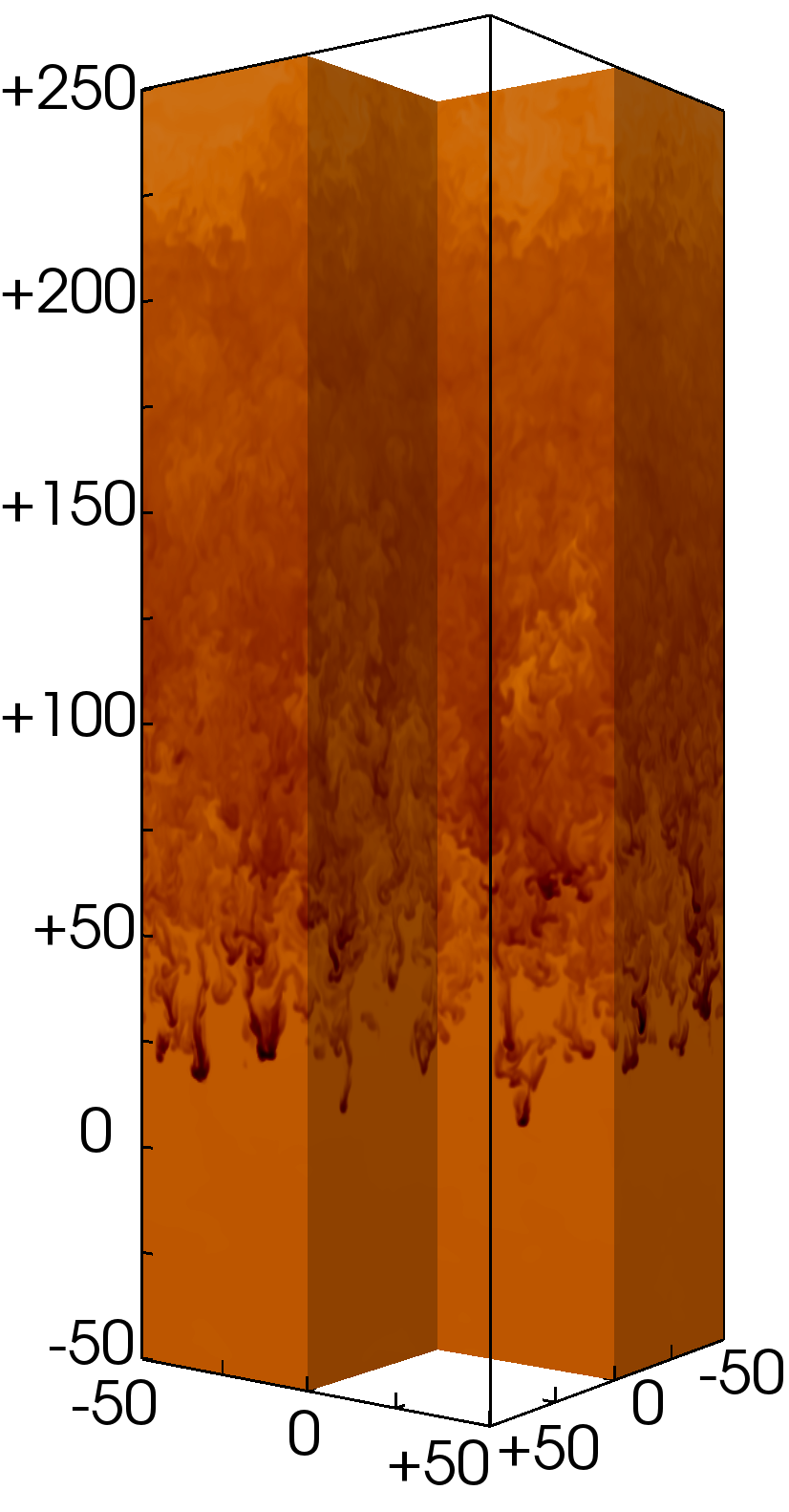}} & \hspace{-0.3cm}\resizebox{27mm}{!}{\includegraphics{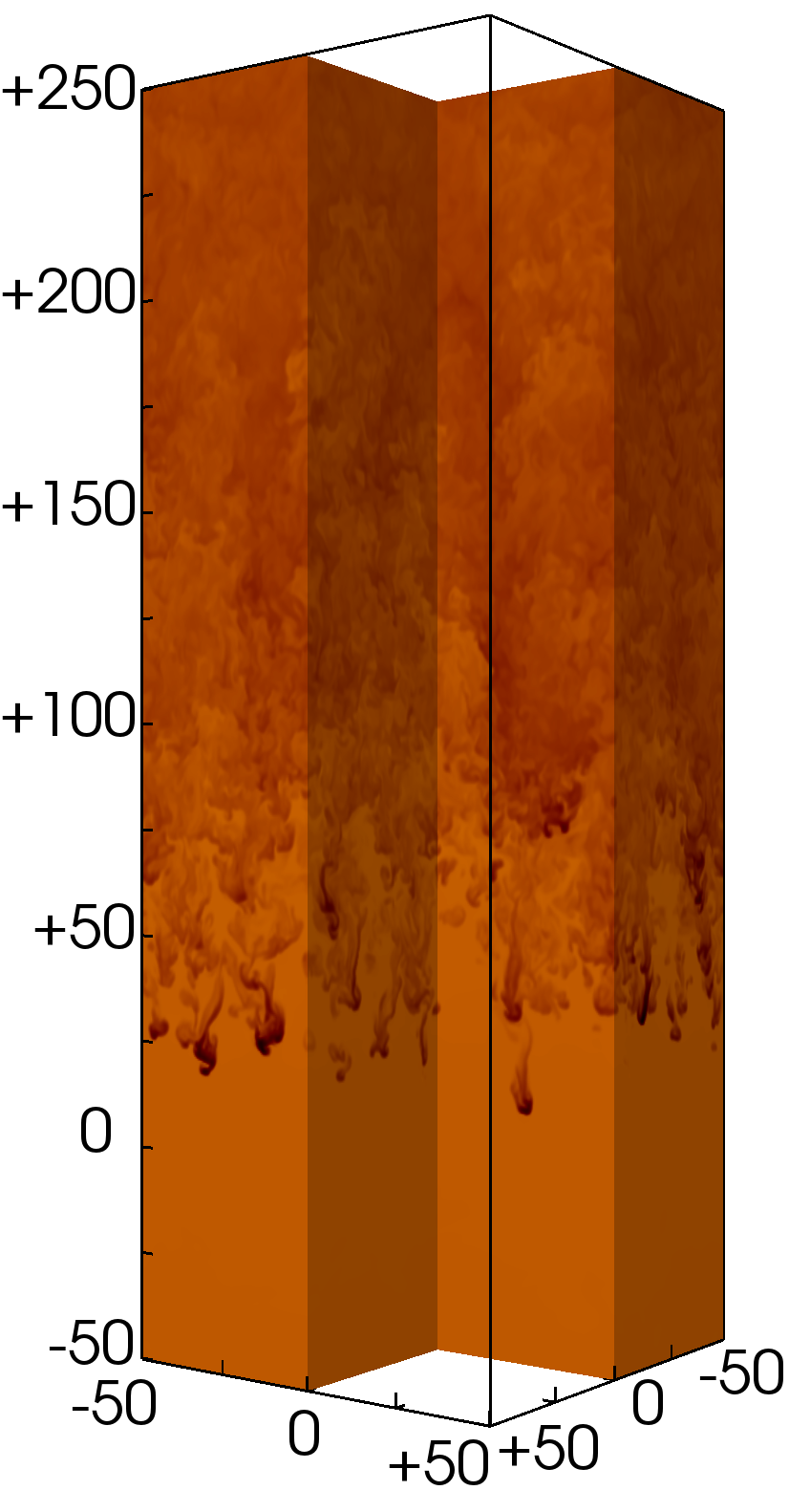}} & \hspace{-0.3cm}\resizebox{27mm}{!}{\includegraphics{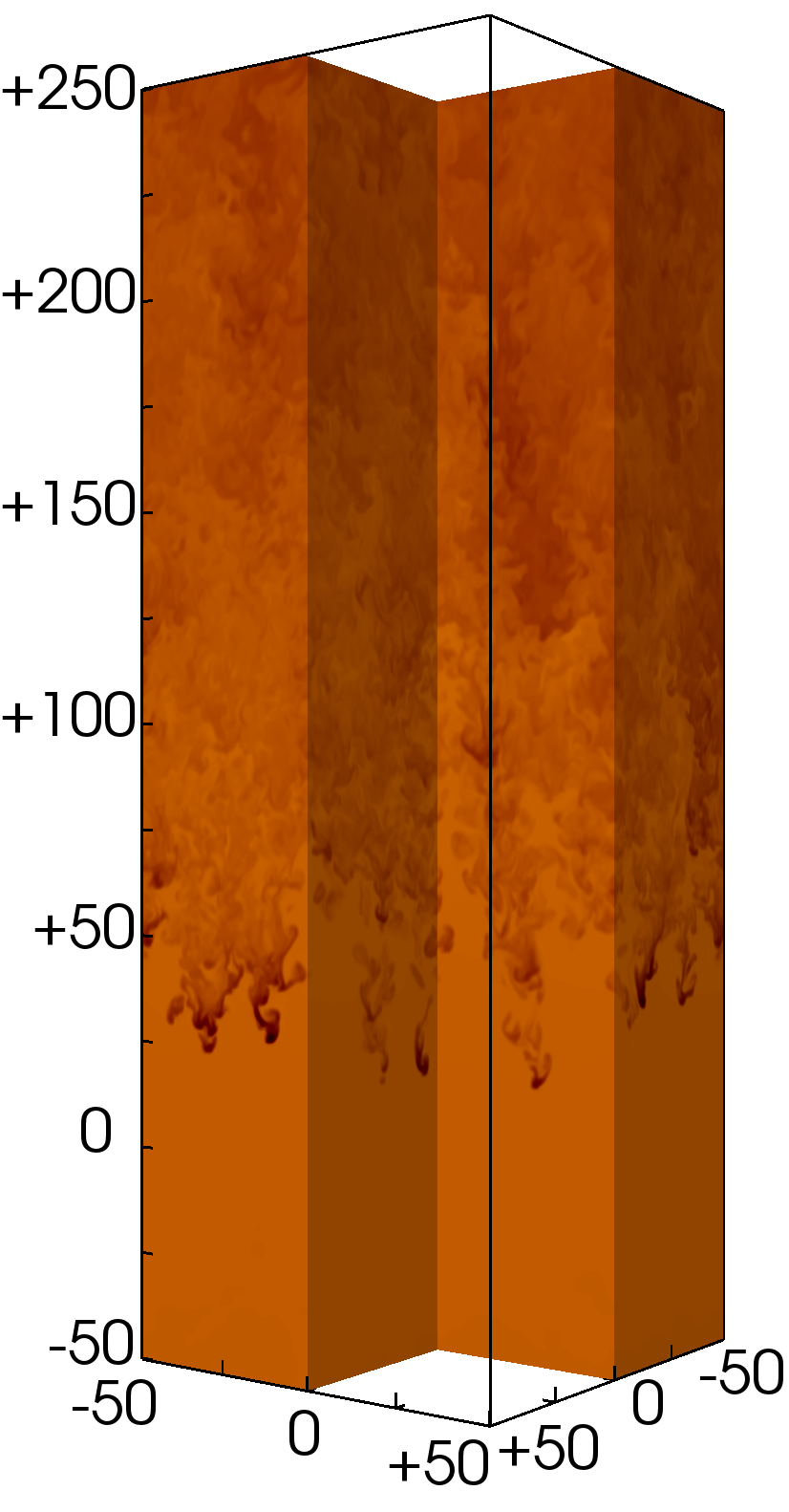}} &
\hspace{-0.3cm}\resizebox{9.8mm}{!}{\includegraphics{bar_vert_2.png}}\\
       \multicolumn{1}{l}{\hspace{-2mm}3d) comp-k8-M10-rad \hspace{+0.2mm}$t_0$} & \multicolumn{1}{c}{$0.5\,t_{\rm sp}=0.10\,\rm Myr$} & \multicolumn{1}{c}{$1.1\,t_{\rm sp}=0.22\,\rm Myr$} & \multicolumn{1}{c}{$1.8\,t_{\rm sp}=0.36\,\rm Myr$} & \multicolumn{1}{c}{$2.4\,t_{\rm sp}=0.48\,\rm Myr$} & \multicolumn{1}{c}{$3.0\,t_{\rm sp}=0.60\,\rm Myr$} & $\frac{n}{n_{\rm ambient}}$\\     
       \hspace{-0.3cm}\resizebox{27mm}{!}{\includegraphics{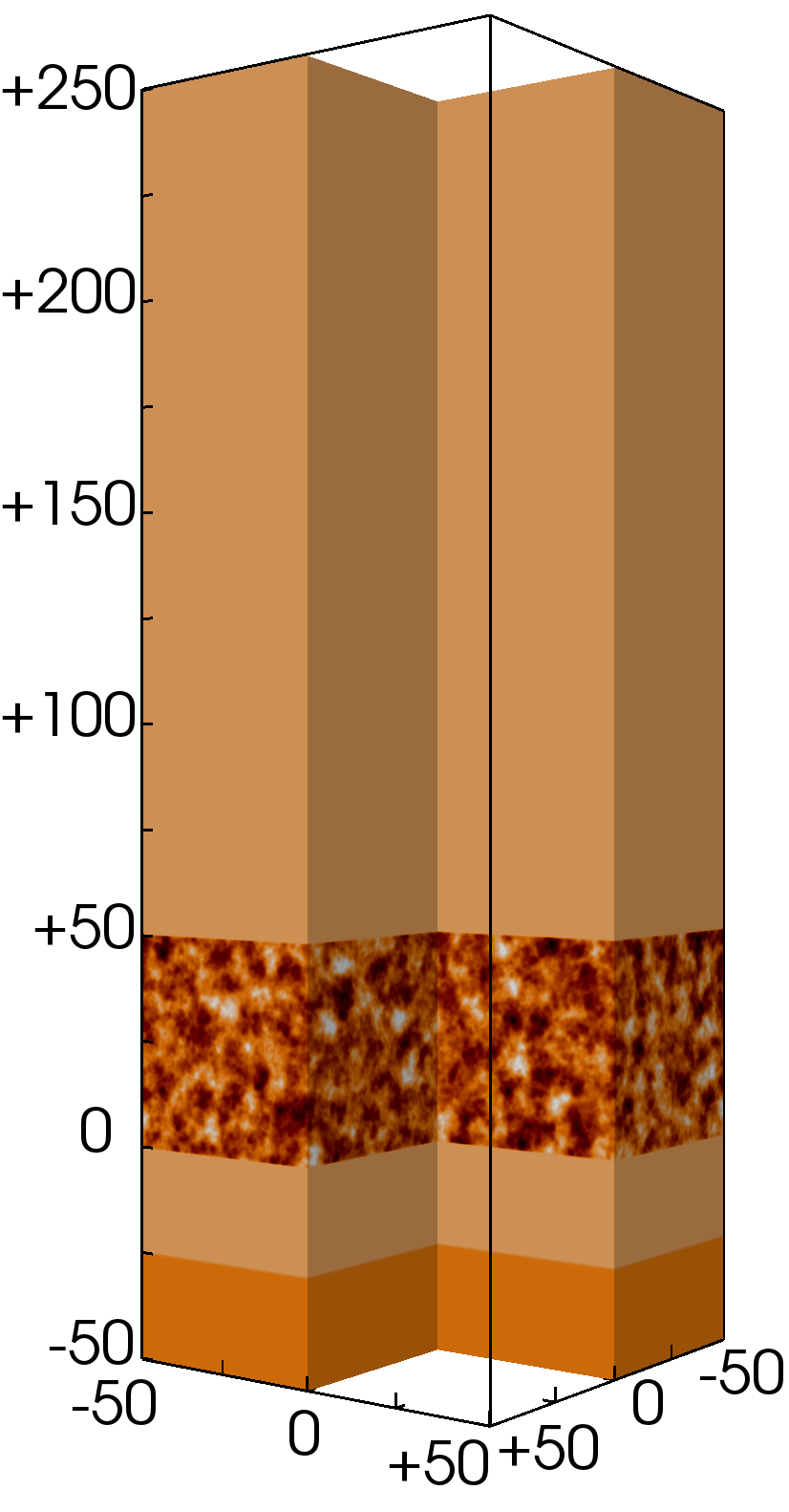}} & \hspace{-0.3cm}\resizebox{27mm}{!}{\includegraphics{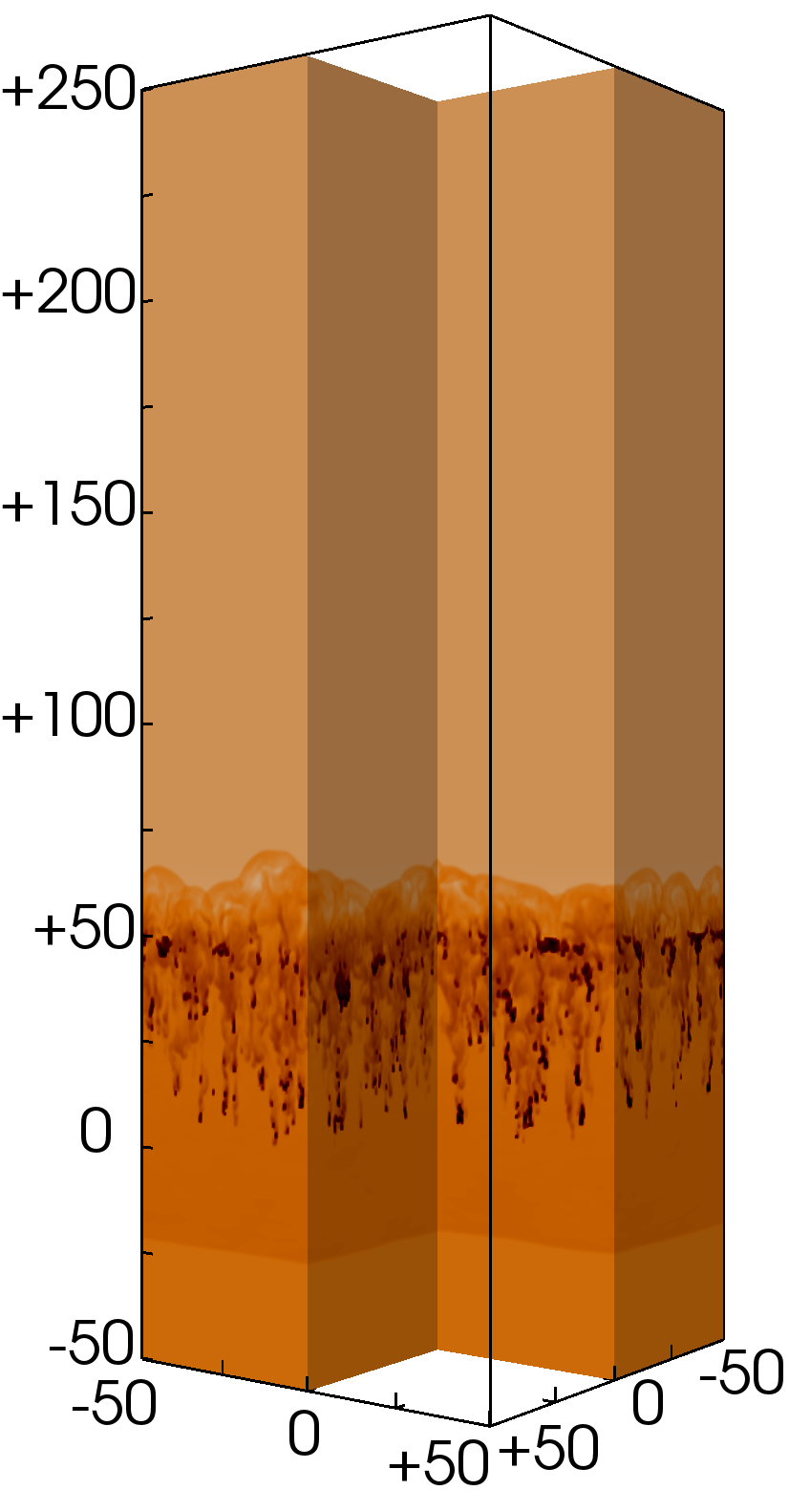}} & \hspace{-0.3cm}\resizebox{27mm}{!}{\includegraphics{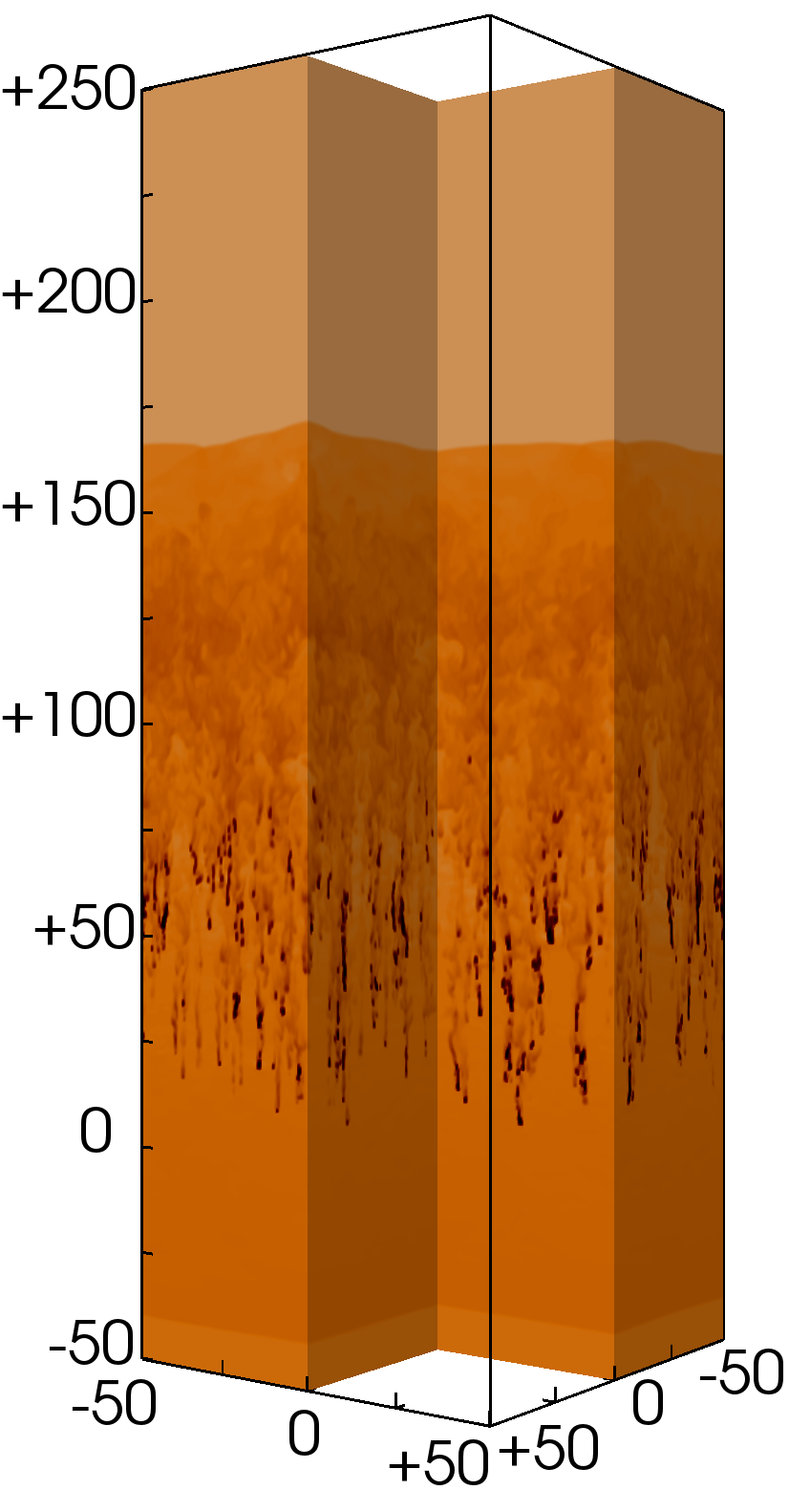}} & \hspace{-0.3cm}\resizebox{27mm}{!}{\includegraphics{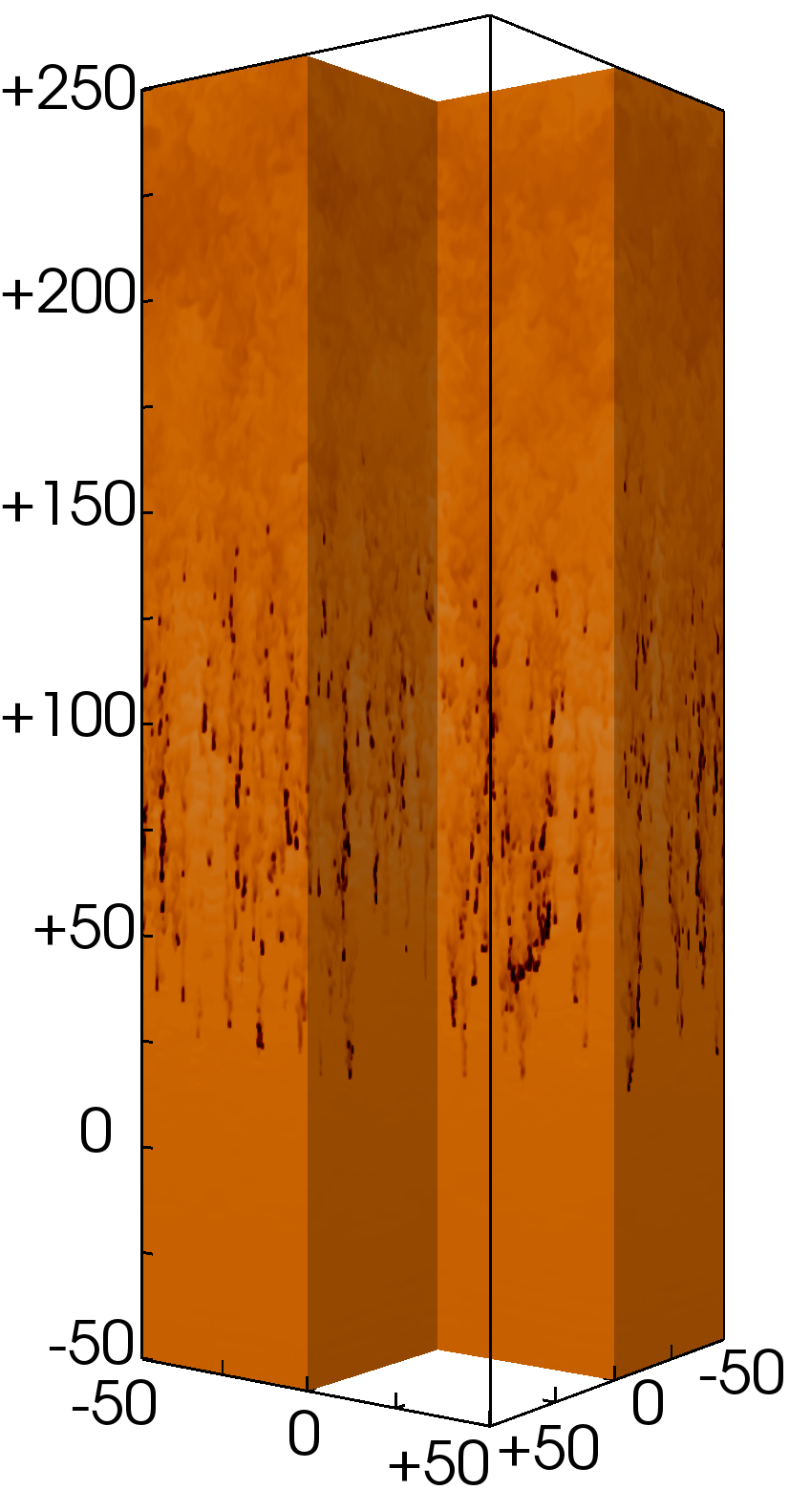}} & \hspace{-0.3cm}\resizebox{27mm}{!}{\includegraphics{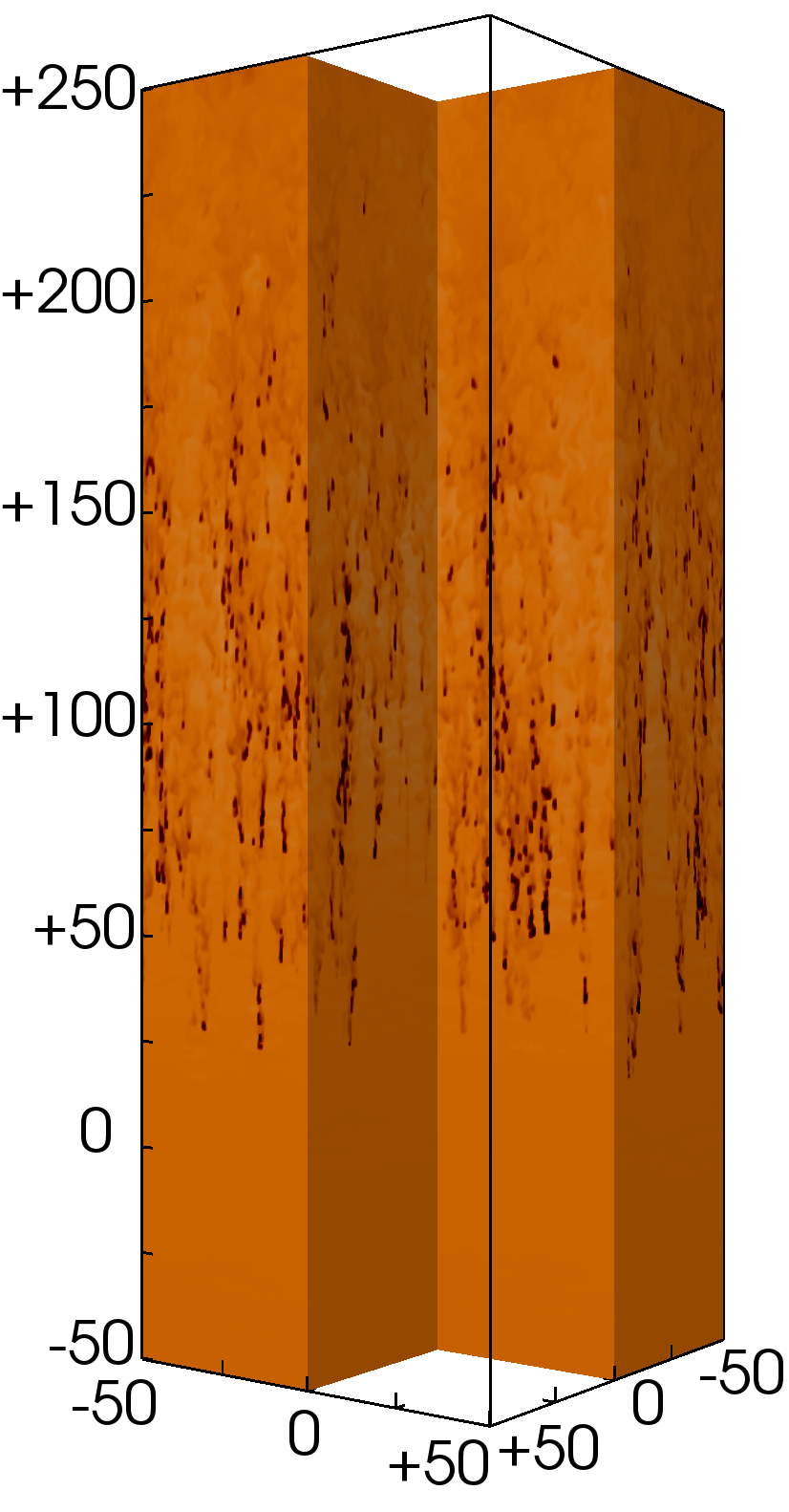}}  & \hspace{-0.3cm}\resizebox{27mm}{!}{\includegraphics{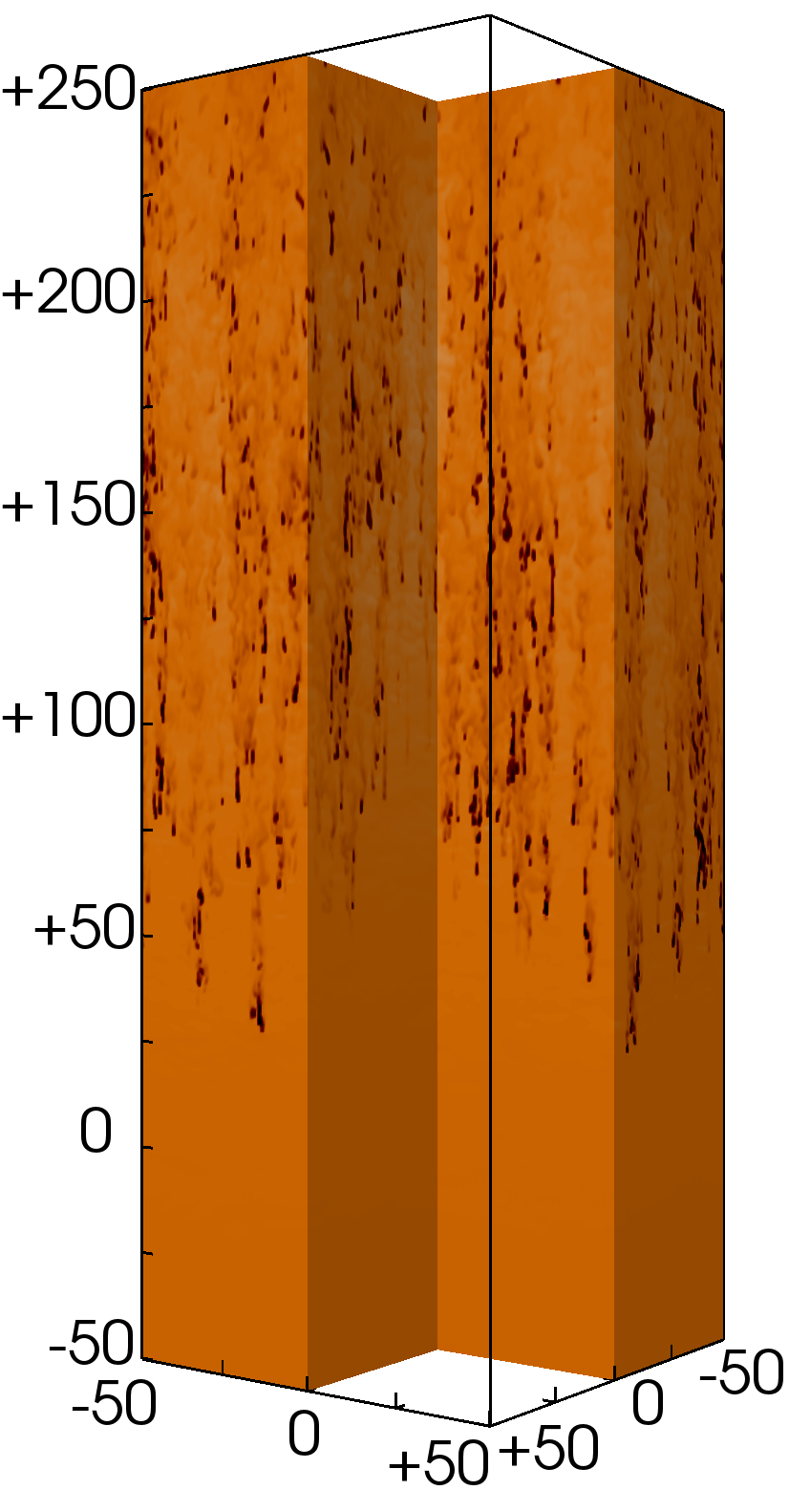}} &
\hspace{-0.3cm}\resizebox{9.8mm}{!}{\includegraphics{bar_vert_2.png}}\vspace{-0.4cm}\\
  \end{tabular}
  \caption{3D renderings showing the evolution for $t/t_{\rm sp}\leq3.0$ of the gas number density ($n$), normalised with respect to the ambient number density ($n_{\rm ambient}$). We show two compact solenoidal models, sole-k8-M10 (panel 3a) and sole-k8-M10-rad (panel 3b), and two porous compressive models, comp-k8-M10 (panel 3c) and comp-k8-M10-rad (panel 3d), in non-radiative and radiative configurations. The spatial ($X,Y,Z$) extent is ($L\times3L\times L$) as we cropped the bottom part of the domain to zoom into the multicloud region. In physical units the time range corresponds to $t\leq0.6\,\rm Myr$, and the $X$, $Y$, and $Z$ axes are given in $\rm pc$, so they cover a spatial extent of ($100\,\rm pc\times300\,\rm pc\times100\,\rm pc$).} 
  \label{Figure3}
\end{center}
\end{figure*}

Figure \ref{Figure3} shows a time sequence of 3D renderings of the gas number density for different models. Panels 3a and 3c show the non-radiative models (discussed in \citetalias{2020MNRAS.499.2173B}) for compact (solenoidal) and porous (compressive) multicloud systems, respectively. Panels 3b and 3d show the radiative models for the same compact and porous multicloud systems, respectively. These panels indicate that the evolution of shock-multicloud models is different in non-radiative and radiative models, and also different in compact and porous multicloud systems.\par

In non-radiative models the shock compresses the cloud layer and injects kinetic energy into it. This extra energy is converted into heat and is responsible for the expansion and mixing of cloud gas with ambient gas when the shock leaves the cloud layer. As explained in \citetalias{2020MNRAS.499.2173B}, this process leads to the formation of extended shells of mixed gas that is prone to dynamical (KH and RT) instabilities, which efficiently break up the densest cores in the layers. Then, the mixed gas is advected away with the post-shock flow, leaving behind only a few low-momentum cloudlets (particularly in porous compressive cases).\par

In radiative models we do not see the same behaviour, and the morphology of cloud gas depends not only on the growth rates of dynamical (KH and RT) instabilities that generate turbulence (e.g., see \citealt{2020ApJ...894L..24F}), but also on the heating and cooling rates as they generate pressure gradients that enhance mixing (e.g., see \citealt{2020MNRAS.492.1970G}). In our radiative models, the thermodynamical interplay between heating and cooling, which depends on the local density and temperature of the gas, creates a rain-like outflowing structure characterised by the presence of very dense cloudlets, surrounded by more diffuse envelopes and extended filamentary tails (see panels 3b and 3d in Figure \ref{Figure3}).\par

\begin{figure}
\begin{center}
  \begin{tabular}{l}
    \hspace{0cm}4a) Forward and reverse shock Mach numbers\\
    \hspace{-0.40cm}\resizebox{80mm}{!}{\includegraphics{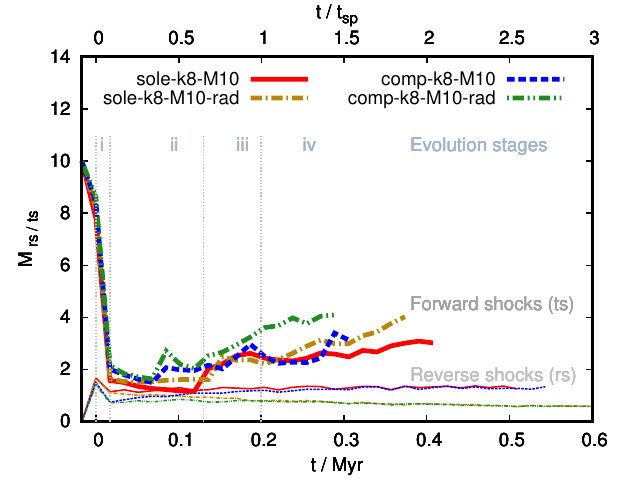}}\\
    \hspace{0cm}4b) Thermal pressure in the multicloud system\\
    \hspace{-0.40cm}\resizebox{80mm}{!}{\includegraphics{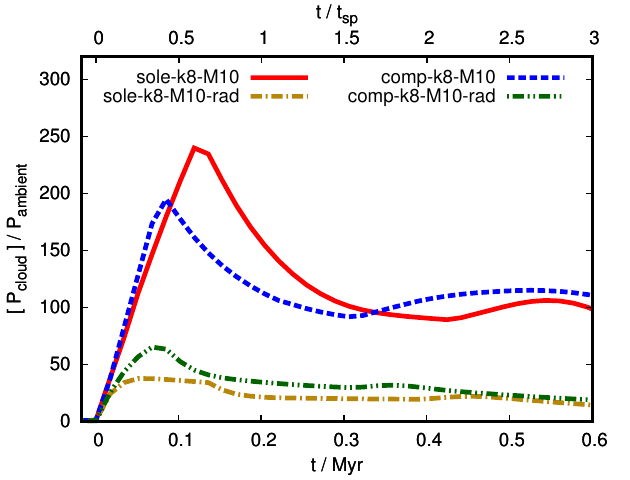}}\\
    \hspace{0cm}4c) Cloud volumetric filling factor\\
    \hspace{-0.40cm}\resizebox{80mm}{!}{\includegraphics{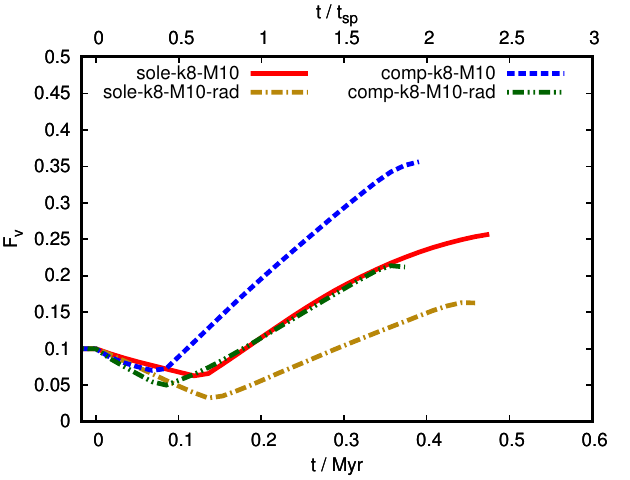}}\\
  \end{tabular}
  \caption{Evolution of the Mach numbers of forward (thick lines) and reverse (thin lines) shocks (panel 4a), the thermal pressure in the multicloud layer (panel 4b), and the volumetric filling factor of cloud material in the computational domain (panel 4c). The forward shocks in radiative models acquire higher velocities after leaving the cloud layer, while the reverse shocks (which are located upstream) are much weaker than in their non-radiative counterparts and degenerate into subsonic waves. Thermal energy is very efficiently radiated away in cooling multicloud systems, which prevents cloudlets from remaining shock heated and delays their disruption. Porous compressive cloud layers are more-vertically extended than their compact solenoidal counterparts, in both non-radiative and radiative models.} 
  \label{Figure4}
\end{center}
\end{figure}

In Figure \ref{Figure4} we show the evolution of the forward and reverse shock Mach numbers (panel 4a), the thermal pressure in cloud material (panel 4b), and the cloud volumetric filling factor (panel 4c). Panel 4a shows that the evolution in radiative models still consists of a four-stage process: (1) Initial contact and shock splitting; (2) Cloud layer compression and shock steady crossing; (3) Cloudlet expansion and shock re-acceleration; and (4) Cloud mixing and turbulence emergence (see Section 3.1 in \citetalias{2020MNRAS.499.2173B} for a full description of the interaction). However, forward and reverse shocks evolve differently. In radiative models, forward shocks approach ${\cal M}_{\rm rs}\approx 4$ and are stronger than in non-radiative models, while reverse shocks reach only half of the values reported for non-radiative cases, thus degenerating into subsonic waves with ${\cal M}_{\rm rs}\approx0.5$. The shock-heated shells formed in non-radiative models act as an effective barrier for the incoming main shock, which both slows the forward shocks and also reflects some hot gas further upstream. In radiative models, condensation prevents these barriers from forming by increasing the clumpiness inside the layer.\par


\begin{figure*}
\begin{center}
  \begin{tabular}{c c c c c}
       \multicolumn{1}{l}{\hspace{-2mm}5a) sole-k8-M10 \hspace{+2.5mm}$t_0$} & \multicolumn{1}{c}{$0.5\,t_{\rm sp}=0.10\,\rm Myr$} & \multicolumn{1}{c}{$1.8\,t_{\rm sp}=0.36\,\rm Myr$} & \multicolumn{1}{c}{$3.0\,t_{\rm sp}=0.60\,\rm Myr$} & $\frac{M_{\rm mc}}{M_{\rm mc,0}}$\\      
       \hspace{-0.40cm}\resizebox{40mm}{!}{\includegraphics{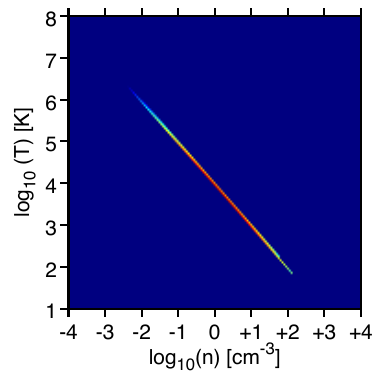}} & \hspace{-0.4cm}\resizebox{40mm}{!}{\includegraphics{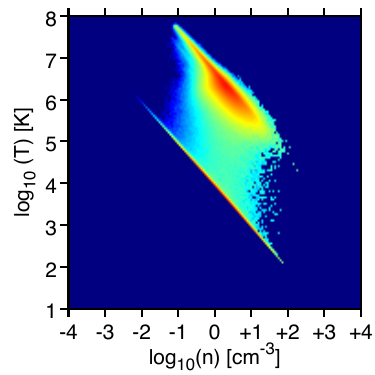}} & \hspace{-0.4cm}\resizebox{40mm}{!}{\includegraphics{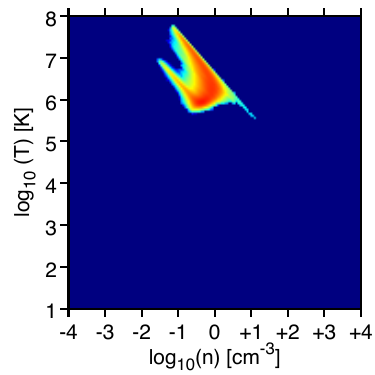}} & \hspace{-0.4cm}\resizebox{40mm}{!}{\includegraphics{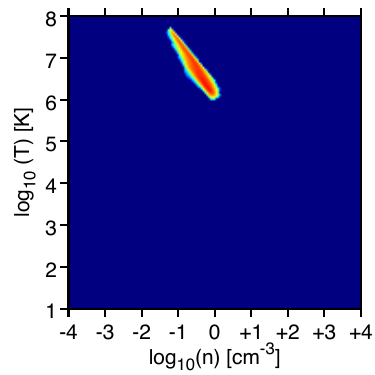}} &
\hspace{-0.2cm}\resizebox{9mm}{!}{\includegraphics{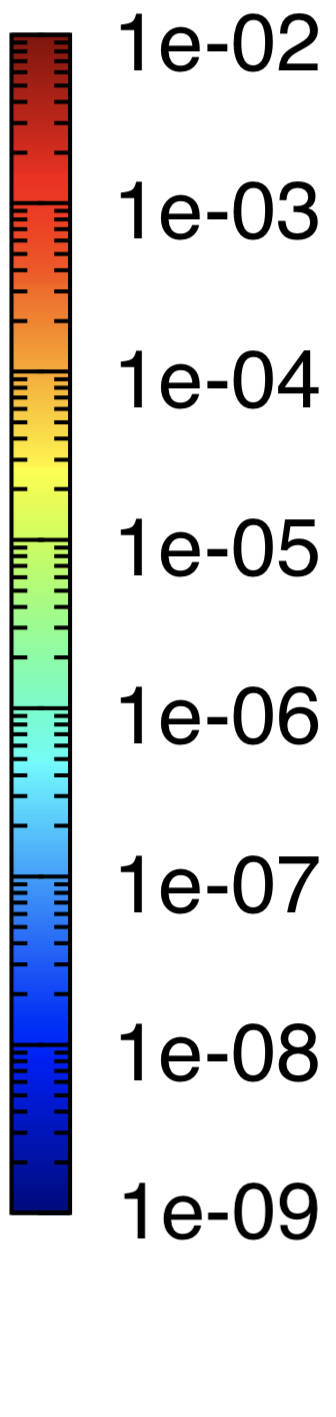}}\\
       \multicolumn{1}{l}{\hspace{-2mm}5b) sole-k8-M10-rad \hspace{+2.5mm}$t_0$} & \multicolumn{1}{c}{$0.5\,t_{\rm sp}=0.10\,\rm Myr$} & \multicolumn{1}{c}{$1.8\,t_{\rm sp}=0.36\,\rm Myr$} & \multicolumn{1}{c}{$3.0\,t_{\rm sp}=0.60\,\rm Myr$} & $\frac{M_{\rm mc}}{M_{\rm mc,0}}$\\       
       \hspace{-0.40cm}\resizebox{40mm}{!}{\includegraphics{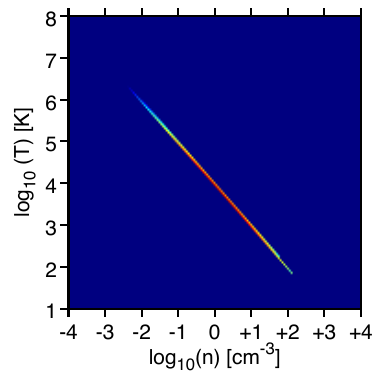}} & \hspace{-0.4cm}\resizebox{40mm}{!}{\includegraphics{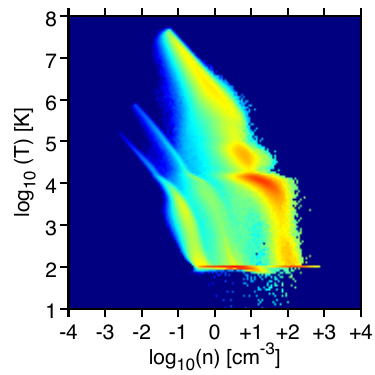}} & \hspace{-0.4cm}\resizebox{40mm}{!}{\includegraphics{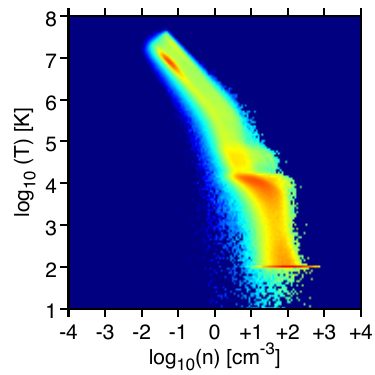}} & \hspace{-0.4cm}\resizebox{40mm}{!}{\includegraphics{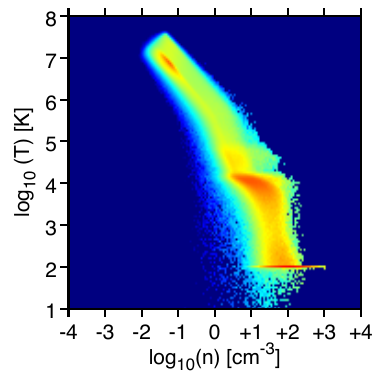}} &
\hspace{-0.2cm}\resizebox{9mm}{!}{\includegraphics{mass-bar.png}}\\
       \multicolumn{1}{l}{\hspace{-2mm}5c) comp-k8-M10 \hspace{+2.5mm}$t_0$} & \multicolumn{1}{c}{$0.5\,t_{\rm sp}=0.10\,\rm Myr$} & \multicolumn{1}{c}{$1.8\,t_{\rm sp}=0.36\,\rm Myr$} & \multicolumn{1}{c}{$3.0\,t_{\rm sp}=0.60\,\rm Myr$} & $\frac{M_{\rm mc}}{M_{\rm mc,0}}$\\           
       \hspace{-0.40cm}\resizebox{40mm}{!}{\includegraphics{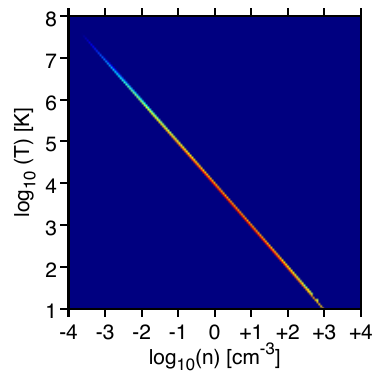}} & \hspace{-0.4cm}\resizebox{40mm}{!}{\includegraphics{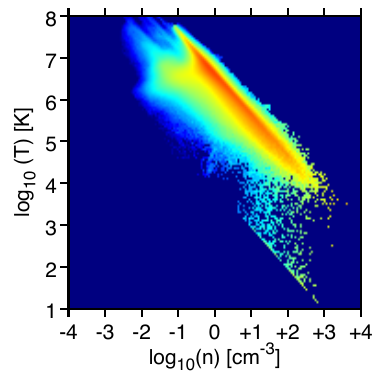}} & \hspace{-0.4cm}\resizebox{40mm}{!}{\includegraphics{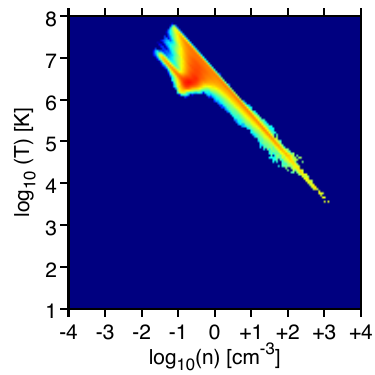}} & \hspace{-0.4cm}\resizebox{40mm}{!}{\includegraphics{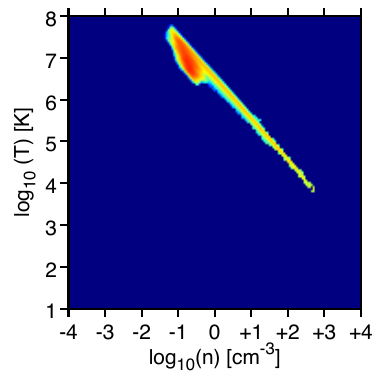}} &
\hspace{-0.2cm}\resizebox{9mm}{!}{\includegraphics{mass-bar.png}}\\
       \multicolumn{1}{l}{\hspace{-2mm}5d) comp-k8-M10-rad \hspace{+2.5mm}$t_0$} & \multicolumn{1}{c}{$0.5\,t_{\rm sp}=0.10\,\rm Myr$} & \multicolumn{1}{c}{$1.8\,t_{\rm sp}=0.36\,\rm Myr$} & \multicolumn{1}{c}{$3.0\,t_{\rm sp}=0.60\,\rm Myr$} & $\frac{M_{\rm mc}}{M_{\rm mc,0}}$\\        
       \hspace{-0.40cm}\resizebox{40mm}{!}{\includegraphics{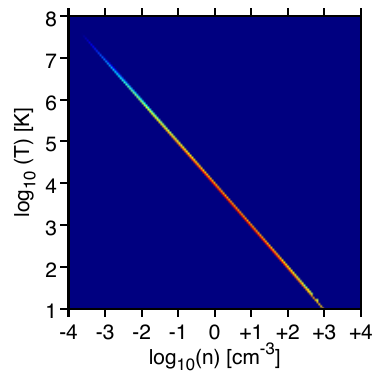}} & \hspace{-0.4cm}\resizebox{40mm}{!}{\includegraphics{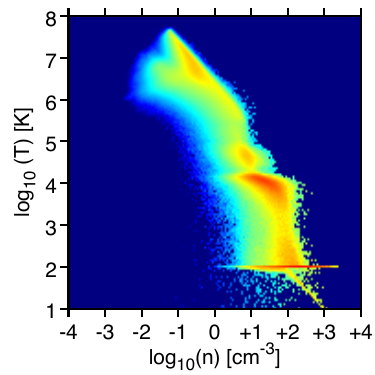}} & \hspace{-0.4cm}\resizebox{40mm}{!}{\includegraphics{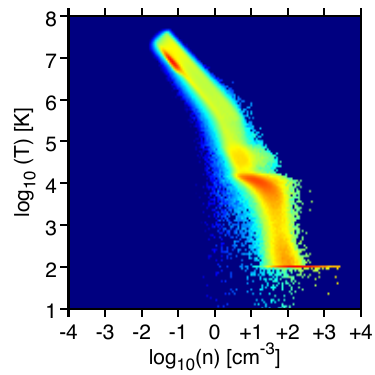}} & \hspace{-0.4cm}\resizebox{40mm}{!}{\includegraphics{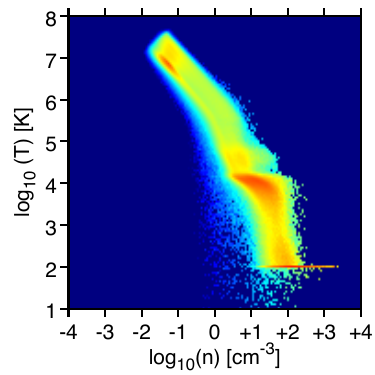}} &
\hspace{-0.2cm}\resizebox{9mm}{!}{\includegraphics{mass-bar.png}}\\
  \end{tabular}
  \caption{Mass-weighted phase diagrams showing the 2D temperature--number density distribution of cloud material in different models, sole-k8-M10 (panel 5a), sole-k8-M10-rad (panel 5b), comp-k8-M10 (panel 5c), and comp-k8-M10-rad (panel 5d), at four different times. While in non-radiative models most of the gas is at temperatures $>10^4\,\rm K$ at late times, in radiative models the balance between shock heating and radiative cooling and heating creates a multi-phase structure with most of the mass at three temperatures, $\sim 10^7\,\rm K$, $\sim 10^4\,\rm K$ and $\sim10^2\,\rm K$. The differences between compact solenoidal and porous compressive models, seen in non-radiative models, vanish with time in radiative models, which acquire a similar temperature-density distribution.} 
  \label{Figure5}
\end{center}
\end{figure*}

In radiative models, strong cooling removes energy that would otherwise be converted into heat. This prevents the pronounced increase in thermal pressure inside the multicloud system, which we see in non-radiative models (see panel 4b of Figure \ref{Figure4}), and also reduces cloud expansion, promoting the clumping and fragmentation of dense gas (see panel 4c of Figure \ref{Figure4}). Thus, radiative cooling leads to a denser and more clumpy medium, in which cooled, dense gas remains protected from instabilities enveloped by a warm radiative layer of mixed, medium-density gas (see also \citealt{2004ApJ...604...74F,2009ApJ...703..330C}). Compression and clumping triggered by cooling leaves a more porous and hollow system that over time occupies a smaller volume, so forward shocks can travel through the intercloud medium in the cloud layer more easily than in non-radiative models. Due to momentum conservation, this also means that reflected shocks/waves travelling upstream are weaker and spatially closer to the upstream edge of the multicloud system than in non-radiative models. Indeed, the 3D number density renderings at $t=0.5\,t_{\rm sp}=0.10\,\rm Myr$ in Figure \ref{Figure3} display larger standoff distances for reverse shocks in non-radiative models than in radiative models.\par

\subsection{The role of radiative cooling and heating}
\label{subsec:CoolingHeating}
As mentioned above, the evolution of radiative multicloud systems not only depends on the energy and momentum injected into them by the main shock, but also on the thermodynamical balance between cooling and heating. Figure \ref{Figure5} shows the evolution of mass-weighted phase diagrams of temperature versus number density for both non-radiative and radiative, and compact solenoidal and porous compressive multicloud systems. These panels allow us to study the thermodynamical path followed by cloud gas (see Table \ref{Table2}). In non-radiative models, most cloud material is rapidly shock heated to temperatures $>10^4\,\rm K$, with very few cloudlets remaining dense and cold. Without a mechanism to release the extra energy in the system, the leftover cold gas is also eventually shredded and heated up either by refracted shocks or by the post-shock flow. Thus, by $t=3.0\,t_{\rm sp}=0.6\,\rm Myr$ most of the cloud mass in non-radiative models corresponds to hot, shock-heated gas with temperatures $\gtrsim 10^6\,\rm K$.\par

In radiative models, cloud gas follows a different thermodynamical path, depending on its local density and temperature. Early on in the simulation ($t\leq0.5\,t_{\rm sp}=0.10\,\rm Myr$), there is shocked and unshocked gas in the multicloud layer. During this time, downstream unshocked cloud gas cools down, while upstream shocked gas is heated first and then cools down. Unshocked cloud gas with initial temperatures $>10^4\,\rm K$ and number densities $\sim 0.1-1\,\rm cm^{-3}$ cools down to temperatures $\lesssim 10^4\,\rm K$. Unshocked colder and denser cloud gas cools further reaching temperatures between $\sim 10^2$--$10^3\,\rm K$, depending on the local density. Concurrently, upstream cloud gas is rapidly shock heated by the internal forward shocks moving through the clouds. This creates a transient and broad 2D temperature-density distribution for $t\leq0.1\,\rm Myr$ (see the phase diagrams in the second column of Figure \ref{Figure5}). As time progresses, all cloud gas becomes shock heated and again prone to cooling. The more diffuse of such shocked gas with $n\sim 0.01-0.1\,\rm cm^{-3}$ remains hot at $\sim 10^7\,\rm K$ (hot phase) while gas with $0.1\lesssim n\lesssim 1\,\rm cm^{-3}$ transits between $10^4\,\rm K$ and a few $\times10^6\,\rm K$. Dense gas with $1\lesssim n\lesssim 10^2\,\rm cm^{-3}$ stays near $\sim 10^4\,\rm K$ (warm phase), but some very dense gas with $n\gtrsim 10^2\,\rm cm^{-3}$ is able to cool even further transiting between $\sim 10^4\,\rm K$ and the cooling floor temperature of $\sim10^2\,\rm K$ (cold phase). Our heating function is able to counteract the rapid cooling of warm gas keeping it at temperatures $>10^3\,\rm K$, but cannot prevent denser gas from further cooling and reaching values near the cooling floor temperature.\par 


Owing to cooling-driven fragmentation, the degree of porosity in the multicloud layer increases with time in radiative models. Dense cloudlets become smaller (with radii between $\sim 1$--$4\,\rm pc$ in the direction transverse to the shock normal) and more exposed to the fast-moving post-shock flow. As a result, these cloudlets are slowly eroded by dynamical KH instabilities and develop tails (with lengths between $\sim 4$--$10\,\rm pc$ in the direction parallel to the shock normal), which then generate a turbulent, multi-filamentary, rain-like 3D structure (see the third column in panels 3b and 3d of Figure \ref{Figure3} and Appendix \ref{AppendixA}). Dense gas is therefore continuously exposed to erosion and heating, but can rapidly cool down once it is in the intermediate phases. As cloud gas continues mixing with the post-shock flow and interacting with the forward refracted shocks, gas with temperatures $\gtrsim 10^4\,\rm K$ and number densities $\lesssim 10\,\rm cm^{-3}$ (i.e. gas that has the optimal conditions for fast cooling to act) continuously forms and also continuously precipitates back into the $10^3$--$10^4\,\rm K$ phase. Lighter gas escapes this fate and remains shock heated at higher temperatures ($\sim 10^6$--$10^7\,\rm K$), while denser gas continuously cycles between temperatures $\sim 10^2$--$10^3\,\rm K$.\par

As a result, the balance between shock heating, and radiative cooling and heating redistributes most of the cloud mass into a three-phase flow, akin to the flow structure produced by wind-cloud interactions with radiative fractal clouds (e.g., compare our Figure \ref{Figure5} with Figure 8 in \citealt{2017ApJ...834..144S}). Shock heating, dynamical instabilities, and cooling-induced pressure gradients promote gas mixing and the emergence of turbulence (in line with recent studies on mixing layers, e.g., see \citealt*{2019MNRAS.487..737J}; \citealt{2020MNRAS.494.2641M}), radiative cooling leads to the continuous replenishment of dense gas in the outflow (as suggested by \citealt{2016MNRAS.455.1830T} and shown by \citealt{2018MNRAS.480L.111G,2020MNRAS.492.1970G,2020MNRAS.492.1841L}), and radiative heating prevents runaway cooling in dense gas keeping warm cloud gas in thermal balance at temperatures $\sim 10^3$--$10^4\,\rm K$ and cold cloud gas at temperatures $\sim 10^2\,\rm K$. The presence of dense gas, produced by this mechanism and located $\gtrsim 100\,\rm pc$ away from the initial location of the multicloud layer, can be viewed in the last columns of Figure \ref{Figure3} (see also Section \ref{sec:tempbins} for a discussion on the evolution of gas in different temperature bins).

\subsection{Density PDF evolution}
\label{subsec:DensityPDF}
\begin{figure}
\begin{center}
  \begin{tabular}{l}
    6a) $t_0$\vspace{-0.2cm}\\ 
    \hspace{-0.60cm}\resizebox{80mm}{!}{\includegraphics{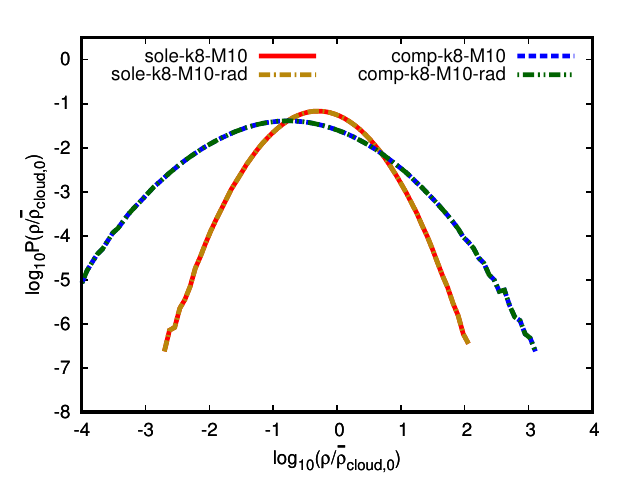}}\\
    6b) $t=1.1\,t_{\rm sp}=0.22\,\rm Myr$\\
    \hspace{-0.60cm}\resizebox{80mm}{!}{\includegraphics{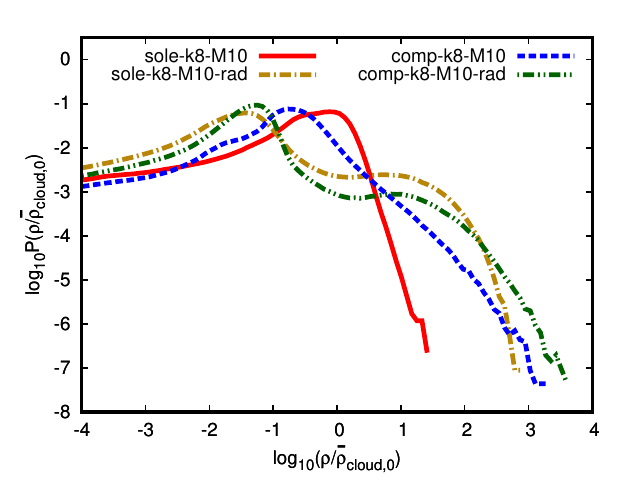}}\\
    6c) $t=3.0\,t_{\rm sp}=0.60\,\rm Myr$\\
    \hspace{-0.60cm}\resizebox{80mm}{!}{\includegraphics{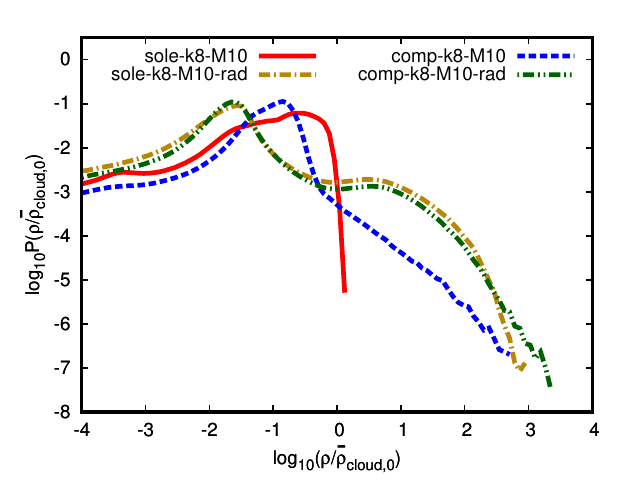}}\\
  \end{tabular}
  \caption{Volume-weighted density PDFs in non-radiative and radiative models with compact solenoidal and porous compressive multicloud systems, at three different times of their evolution. The inclusion of radiative cooling and heating produces bi-modal density distributions. The first peak corresponds to gas with densities between $\sim 0.01$--$0.1\,\rho_{\rm cloud,0}$ and the second peak corresponds to gas with densities between $\sim10$--$10^{2}\rho_{\rm cloud,0}$. Radiative models develop very similar density PDFs regardless of their initial density distribution, indicating that the interplay between heating and cooling is what ultimately shapes the outflow density structure in these cases. Despite this, more marked signatures of the initial density PDFs are still present in other diagnostics of radiative models until late times, e.g., see Figures \ref{Figure4} and \ref{Figure7}.} 
  \label{Figure6}
\end{center}
\end{figure}

The effects of our cooling and heating function and the role of the initial density distributions can also be studied by looking into the evolution of the density PDFs in compact solenoidal and porous compressive cases. Figure \ref{Figure6} shows the density PDFs of cloud material at three different stages of the evolution in both non-radiative and radiative models initialised with compact (solenoidal) and porous (compressive) density distributions.\par

While in the compact, non-radiative model the high-density end of the density PDF moves to lower densities rather quickly, in its radiative counterpart the evolution is different and the initially log-normal PDF degenerates into a bi-modal distribution with a first peak at densities $\sim 0.01$--$0.1\,\rho_{\rm cloud,0}$ and a second peak at densities $\sim10$--$10^{2}\rho_{\rm cloud,0}$. As mentioned in the previous section, the first peak corresponds to hot diffuse gas with temperatures $\sim 10^7\,\rm K$, while the second peak corresponds to dense warm and cold gas at temperatures between $\sim 10^2$--$10^4\,\rm K$. Some of the gas in between both peaks and around the second peak corresponds to the intermixed medium, which is either removed from cooled clouds or exists in between the clouds. A bi-modal distribution (with peaks at similar densities) also characterises the evolution of the porous radiative model, but it is entirely absent in non-radiative models, which are rather characterised by a uni-modal distribution.\par

Note also that unlike what we find for non-radiative models, in which porous models retain more high-density cores than solenoidal models, in radiative models the PDFs of both multicloud systems develop nearly identical morphologies at the end of the simulations (see panel 6c in Figure \ref{Figure6}). This signifies that the thermodynamical evolution of shocked gas steadily overrides the initial differences in their density distributions. Despite this, as we discuss below, we do find differences in the evolution of radiative compact and porous models that can be attributed to the initial density PDFs. We further note that while the non-radiative models have power-law tails at high densities, the high-density tails in the radiative models have a log-normal shape, as expected for turbulent, dense, cold gas (e.g., see \citealt{2020MNRAS.493.3098M}).\par

\subsection{Compact (solenoidal) versus porous (compressive) radiative multicloud systems}
\label{subsec:SolvsComp}
\subsubsection{Morphology}
The 3D density renderings in Figure \ref{Figure3} show that porous (compressive) cloud layers evolve into more vertically-extended distributions of cold gas than their compact (solenoidal) counterparts. This is true regardless of whether the models are non-radiative (panels 3a and 3c) or radiative (panels 3b and 3d). Similarly, panel 4c of Figure \ref{Figure4} shows that the volume occupied by porous cloud models is systematically larger by a factor of $1.5-2$ than in compact cloud models. In both cases the differences seen at late stages can be attributed to the initial density distribution.\par

Porous cloud layers have high-density cores that are already cold at the start of the evolution, so these cores function as natural, low-momentum footpoints (i.e., as dense gas reservoirs) for more diffuse gas because they are harder to disrupt. In porous cloud models (both non-radiative and radiative) the main shock can move more easily across the multicloud system, thus exiting it earlier and advecting warm, mixed gas farther away than in compact cloud models. Dense gas in cloudlets lags behind shock-heated, mixed, turbulent gas. In non-radiative models such warm gas keeps accelerating and expanding as warm/hot gas, whereas in radiative models the warm medium condenses back into the cold phase while expanding downstream, thus also spreading the cold component in the outflow over a larger vertical volume.\par

\subsubsection{Mixing}
\label{subsection:mixing}
\begin{figure}
\begin{center}
  \begin{tabular}{l}
    7a) Mixing fraction\\
    \hspace{-0.40cm}\resizebox{80mm}{!}{\includegraphics{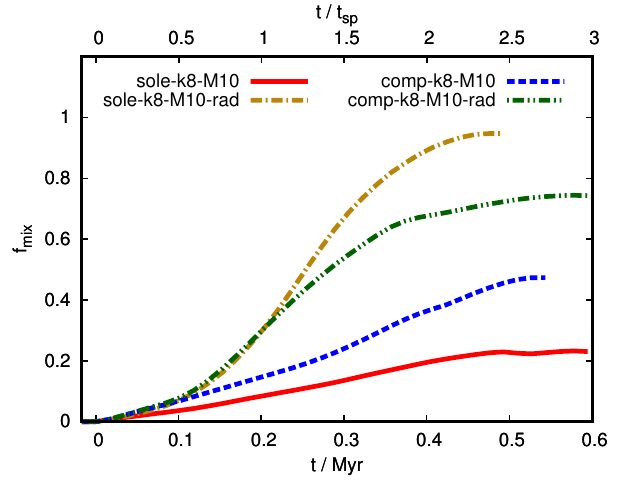}}\\
    7b) Mass fraction of gas denser than $\rho_{\rm cloud, 0}/3$ vs. time\\
    \hspace{-0.40cm}\resizebox{80mm}{!}{\includegraphics{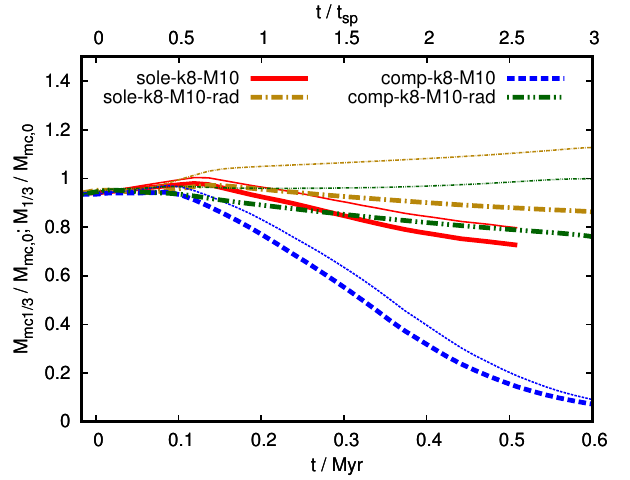}}\\
    7c) Mass fraction of gas denser than $\rho_{\rm cloud, 0}$ vs. time\\
    \hspace{-0.40cm}\resizebox{80mm}{!}{\includegraphics{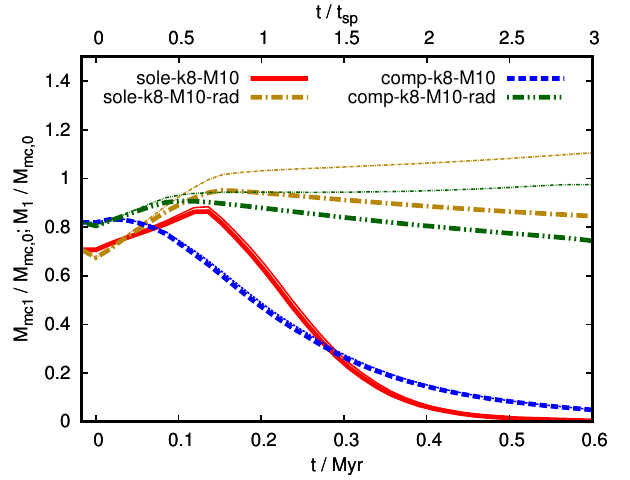}}\\
  \end{tabular}
  \caption{Evolution of the mixing fraction (panel 7a) and the mass fractions of gas with densities above $\rho_{\rm cloud, 0}/3$ (panel 7b) and $\rho_{\rm cloud, 0}$ (panel 7c) in compact solenoidal and porous compressive multicloud models. The thick lines show mass fractions including only cloud gas, and the thin lines include all the gas. In radiative models, mixing fractions are higher in compact cloud models than in porous models, and mass fractions of gas denser than $\rho_{\rm cloud, 0}/3$ and $\rho_{\rm cloud, 0}$ grow as a result of re-condensation of warm, mixed gas and entrainment from the hot flow. This effect is slightly more pronounced in the compact model owing to its higher mixing fraction. At least $\sim 20$ per cent of the dense gas at $t=3.0\,t_{\rm sp}=0.60\,\rm Myr$ is material entrained from the hot wind, while the rest is recycled from mass-loaded cloud gas.} 
  \label{Figure7}
\end{center}
\end{figure}

The mixing and disruption properties of cloud material are also different between compact and porous systems, regardless of whether or not radiative processes are included in the models. Figure \ref{Figure7} shows the evolution of three parameters (see their definitions in Table \ref{Table2}), the cloud gas mixing fraction (panel 7a), the mass fraction of gas above the conventional density threshold (e.g., see \citealt{2015ApJ...805..158S}) of $\rho_{\rm cloud,0}/3$ (panel 7b), considering only cloud gas (thick lines) and all the gas in the domain (thin lines), and the mass fraction of gas with densities above $\rho_{\rm cloud,0}$ (panel 7c), also for cloud gas (thick lines) and for all the gas in the domain (thin lines).\par

First, we find that mixing fractions are in general higher in radiative models than in non-radiative models. Mixing processes in radiative clouds are intrinsically more complex than in non-radiative clouds. In non-radiative cases, mixing is the result of KH and RT instabilities arising at shear layers at gas-gas boundaries (e.g., see \citealt{2015ApJS..217...24S,2016MNRAS.458.1139P}). On the other hand, in radiative cases, mixing via instabilities can be suppressed (e.g., see \citealt{2009ApJ...703..330C,2017ApJ...834..144S}), but additional mixing can occur as a result of cooling-driven pressure gradients (e.g., see \citealt{2020MNRAS.492.1970G}), gas heating, and precipitation from the warm phase (where cooling is very efficient). In addition, in our radiative models, where the global $t_{\rm cool, mix}<t_{\rm cc}$, entrainment of hot ambient gas onto the cold flow also occurs (see Section \ref{Ambient_entrainment} below, and Section 2.2. in \citealt{2020MNRAS.494.2641M} for further discussions). While the hot and warm mixed phases in non-radiative models are carried away by the post-shock flow, in radiative models such phases can precipitate again into colder and denser gas phases, thus increasing the amount of gas in the warm (mixed) and dense phases of the multi-phase flow. The efficiency of mixing in radiative models is, therefore, not solely tied to the development of dynamical instabilities, but also linked to the efficiency of cooling.\par

Second, we find that the behaviour of mixing fractions in compact and porous radiative layers is reversed at late times, compared to non-radiative models. In non-radiative models porous systems lead to higher mixing, while in radiative models the opposite occurs. Although the earlier departure of fast-moving gas in porous models may explain this discrepancy, the narrower density distributions of compact systems seem to facilitate very fast cycles of shock-heating, cooling, heating, and re-cooling, which characterise radiative models. If more warm, mixed gas is available downstream, then we would expect the amount of high-density gas to also be higher in compact systems as this precipitates from the warm component. Panels 7b and 7c of Figure \ref{Figure7} confirm this effect as both show that compact systems favour the regrowth of dense gas.\par

Panels 7b and 7c show that the fraction of dense cloud gas (thick lines) with densities above $\rho_{\rm cloud,0}/3$ and $\rho_{\rm cloud,0}$, respectively, increases during the shock-heating phase, and then it decreases steadily at a much slower rate than in the non-radiative scenarios. For instance, at $t=3.0\,t_{\rm sp}=0.6\,\rm Myr$, $\gtrsim 80$ per cent of the cloud mass is still in the dense-gas phases, while only a very small percentage remains in non-radiative models. In both cases, the compact models end up with slightly higher percentages of dense gas, implying a mild correlation with the initial density distribution. Note also that the mass fractions in radiative models do not directly represent mass loss as in the non-radiative models, as most of the mass accounted for in our diagnostics comes from recondensation, following efficient mixing (as revealed by panel 7a in the same figure).

\subsubsection{Hot gas entrainment}
\label{Ambient_entrainment}
Panels 7b and 7c of Figure \ref{Figure7} also show that the mass content of dense gas in an outflow not only depends on how much cloud gas becomes mixed and precipitates back onto a colder phase due to cooling, but that it also depends on how much hot ambient gas becomes shock-compressed and entrained into the cold component via pressure-driven condensation. The thin lines in these panels reveal that $\gtrsim 20$ per cent of the dense gas in the outflow is entrained (initially hot) ambient gas rather than mixed cloud gas. While we were unable to follow the evolution of these models for longer than $0.6\,\rm Myr$, the trend of these curves also suggests that this percentage may even increase as time progresses and more hot gas becomes compressed and entrained. Studying the evolution of such gas with larger-domain simulations in the future is warranted. Entrained ambient gas may explain the prevalence of a dense-gas component in observed galactic outflows, as also pointed out by recent studies by \citealt*{2018MNRAS.480L.111G,2020MNRAS.499.4261S,2020MNRAS.498.2391N,2021MNRAS.501.1143K}.

\subsection{Cloud gas dynamics and global dense-gas entrainment}
\label{subsec:Entrainment}

\subsubsection{Acceleration}

\begin{figure*}
\begin{center}
  \begin{tabular}{c c c c c}
       \multicolumn{1}{l}{\hspace{-2mm}8a) sole-k8-M10 \hspace{+2.5mm}$t_0$} & \multicolumn{1}{c}{$0.5\,t_{\rm sp}=0.10\,\rm Myr$} & \multicolumn{1}{c}{$1.8\,t_{\rm sp}=0.36\,\rm Myr$} & \multicolumn{1}{c}{$3.0\,t_{\rm sp}=0.60\,\rm Myr$} & $\frac{M_{\rm mc}}{M_{\rm mc,0}}$\\      
       \hspace{-0.40cm}\resizebox{40mm}{!}{\includegraphics{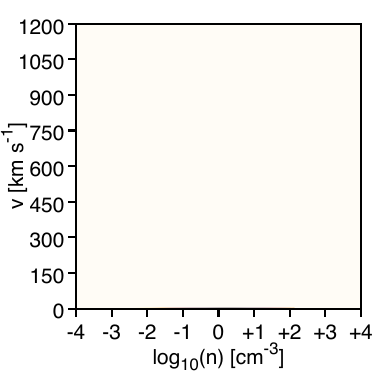}} & \hspace{-0.4cm}\resizebox{40mm}{!}{\includegraphics{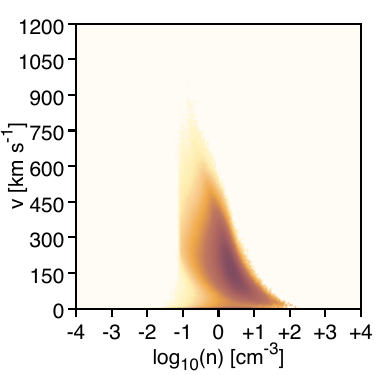}} & \hspace{-0.4cm}\resizebox{40mm}{!}{\includegraphics{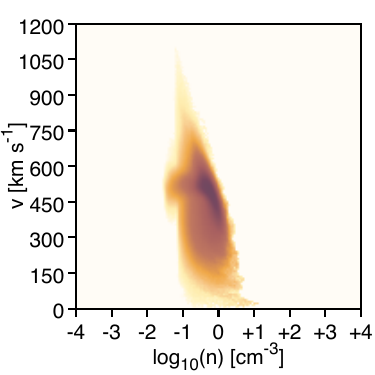}} & \hspace{-0.4cm}\resizebox{40mm}{!}{\includegraphics{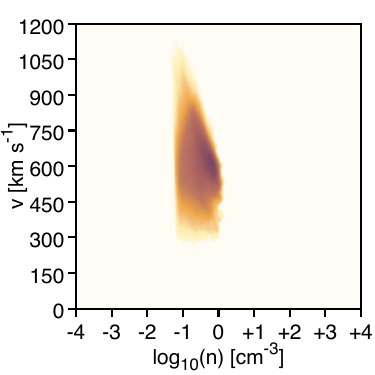}} &
\hspace{-0.2cm}\resizebox{9mm}{!}{\includegraphics{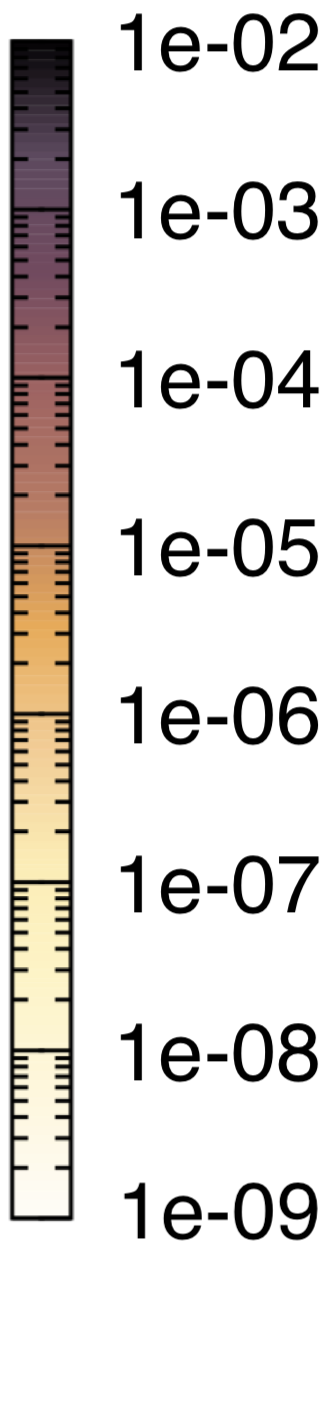}}\\
       \multicolumn{1}{l}{\hspace{-2mm}8b) sole-k8-M10-rad \hspace{+2.5mm}$t_0$} & \multicolumn{1}{c}{$0.5\,t_{\rm sp}=0.10\,\rm Myr$} & \multicolumn{1}{c}{$1.8\,t_{\rm sp}=0.36\,\rm Myr$} & \multicolumn{1}{c}{$3.0\,t_{\rm sp}=0.60\,\rm Myr$} & $\frac{M_{\rm mc}}{M_{\rm mc,0}}$\\    
       \hspace{-0.40cm}\resizebox{40mm}{!}{\includegraphics{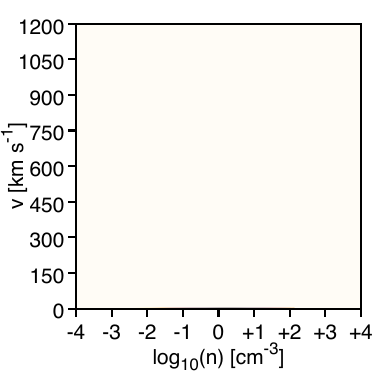}} & \hspace{-0.4cm}\resizebox{40mm}{!}{\includegraphics{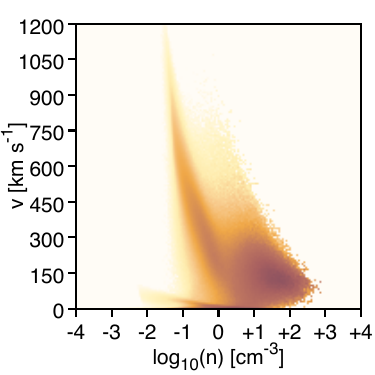}} & \hspace{-0.4cm}\resizebox{40mm}{!}{\includegraphics{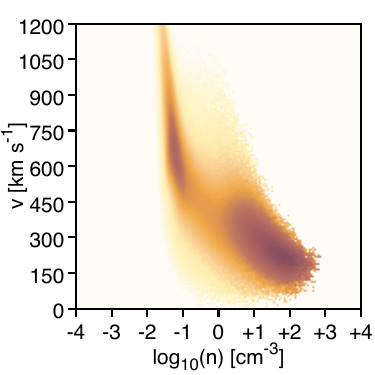}} & \hspace{-0.4cm}\resizebox{40mm}{!}{\includegraphics{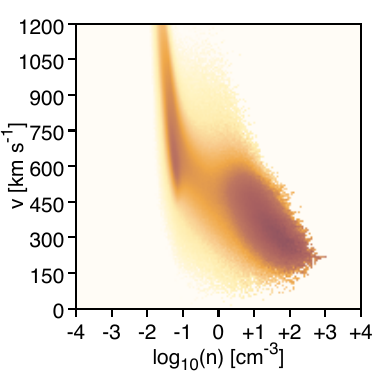}} &
\hspace{-0.2cm}\resizebox{9mm}{!}{\includegraphics{mass-bar-2.png}}\\
       \multicolumn{1}{l}{\hspace{-2mm}8c) comp-k8-M10 \hspace{+2.5mm}$t_0$} & \multicolumn{1}{c}{$0.5\,t_{\rm sp}=0.10\,\rm Myr$} & \multicolumn{1}{c}{$1.8\,t_{\rm sp}=0.36\,\rm Myr$} & \multicolumn{1}{c}{$3.0\,t_{\rm sp}=0.60\,\rm Myr$} & $\frac{M_{\rm mc}}{M_{\rm mc,0}}$\\       
       \hspace{-0.40cm}\resizebox{40mm}{!}{\includegraphics{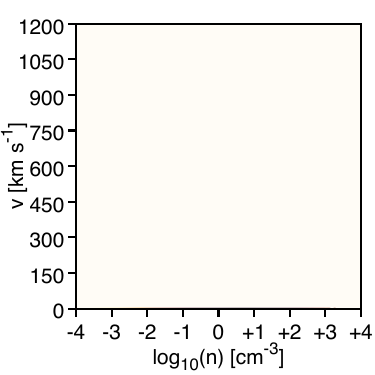}} & \hspace{-0.4cm}\resizebox{40mm}{!}{\includegraphics{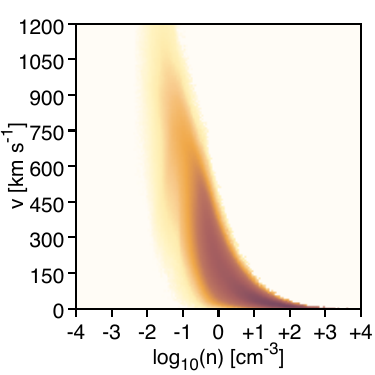}} & \hspace{-0.4cm}\resizebox{40mm}{!}{\includegraphics{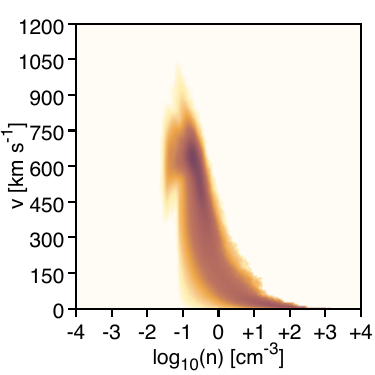}} & \hspace{-0.4cm}\resizebox{40mm}{!}{\includegraphics{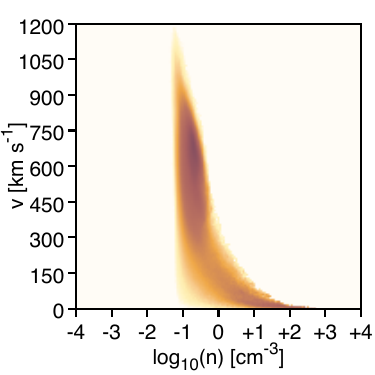}} &
\hspace{-0.2cm}\resizebox{9mm}{!}{\includegraphics{mass-bar-2.png}}\\
       \multicolumn{1}{l}{\hspace{-2mm}8d) comp-k8-M10-rad \hspace{+2.5mm}$t_0$} & \multicolumn{1}{c}{$0.5\,t_{\rm sp}=0.10\,\rm Myr$} & \multicolumn{1}{c}{$1.8\,t_{\rm sp}=0.36\,\rm Myr$} & \multicolumn{1}{c}{$3.0\,t_{\rm sp}=0.60\,\rm Myr$} & $\frac{M_{\rm mc}}{M_{\rm mc,0}}$\\  
       \hspace{-0.40cm}\resizebox{40mm}{!}{\includegraphics{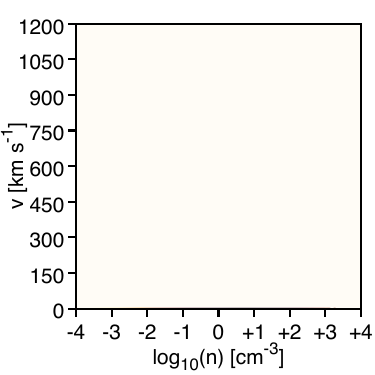}} & \hspace{-0.4cm}\resizebox{40mm}{!}{\includegraphics{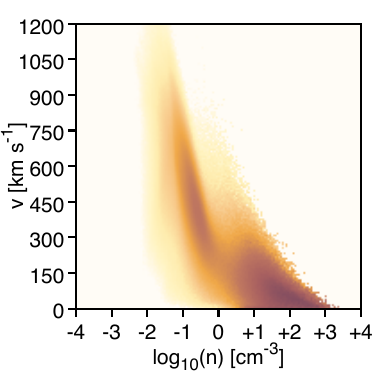}} & \hspace{-0.4cm}\resizebox{40mm}{!}{\includegraphics{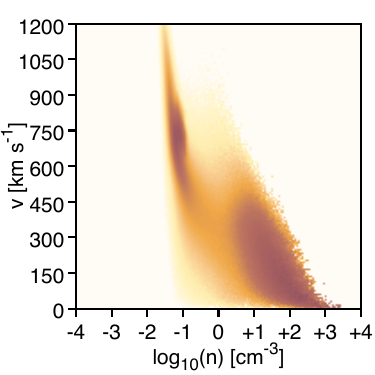}} & \hspace{-0.4cm}\resizebox{40mm}{!}{\includegraphics{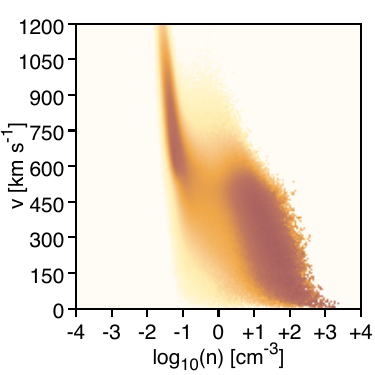}} &
\hspace{-0.2cm}\resizebox{9mm}{!}{\includegraphics{mass-bar-2.png}}\\
  \end{tabular}
  \caption{Mass-weighted phase diagrams showing the 2D velocity--number density distribution of cloud material in different models, sole-k8-M10 (panel 8a), sole-k8-M10-rad (panel 8b), comp-k8-M10 (panel 8c), and comp-k8-M10-rad (panel 8d), at four different times. In adiabatic models most of the gas at the end of the evolution is diffuse and acquires speeds $>400\,\rm km\,s^{-1}$ at $t=3.0\,t_{\rm sp}=0.60\,\rm Myr$. In radiative models there is fast-moving diffuse gas with speeds $>700\,\rm km\,s^{-1}$ at $t=3.0\,t_{\rm sp}=0.60\,\rm Myr$, and also slower dense gas with speeds $>100\,\rm km\,s^{-1}$ in solenoidal models and $>10\,\rm km\,s^{-1}$ in compressive models at $t=3.0\,t_{\rm sp}=0.60\,\rm Myr$. In general, cloud gas entrainment is more efficient in solenoidal models than in compressive models.} 
  \label{Figure8}
\end{center}
\end{figure*}

\begin{table*}\centering
\caption{Columns 1-3: Temperature bins used for studying different gas phases of cloud material in the outflow (see Figures \ref{Figure9}, \ref{Figure10}, \ref{Figure11}, and \ref{FigureB1}). Columns 4-8: Diagnostics measured at $t=3.0\,t_{\rm sp}=0.60\,\rm Myr$ for the compact model sole-k8-M10-rad. Columns (9-13): Same diagnostics, but measured for the porous model comp-k8-M10-rad. Columns 4-5 and 9-10 show the average $\pm$ 1$\sigma$ values of the distances and speeds (the average values can also be seen in panels 9a and 9b of Figure \ref{Figure9}). Columns 6 and 11 show the LOS velocity dispersions (also shown in panel 9c of Figure \ref{Figure9}). Columns 7-8 and 12-13 show the mass fractions, normalised to the total multicloud layer mass (also shown in panels 11a and 11b of Figure \ref{Figure11}).}
\hspace{-0.2cm}\begin{adjustbox}{max width=\textwidth}
\begin{tabular}{c c c c c c c c c c c c c}
\hline
&  &  & \multicolumn{5}{c}{\textbf{sole-k8-M10-rad} at $t=3.0\,t_{\rm sp}=0.60\,\rm Myr$} &  \multicolumn{5}{c}{\textbf{comp-k8-M10-rad} at $t=3.0\,t_{\rm sp}=0.60\,\rm Myr$}  \\ 
\textbf{(1)} & \textbf{(2)} & \textbf{(3)} & \textbf{(4)} & \textbf{(5)} & \textbf{(6)} & \textbf{(7)} & \textbf{(8)} & \textbf{(9)} & \textbf{(10)} & \textbf{(11)} & \textbf{(12)} & \textbf{(13)}\\ \vspace{+0.1cm}
\textbf{Medium} & \textbf{Identifier} & \textbf{Temperature} & ${\langle d_{\rm y}\rangle}_{\rm T}$ & ${\langle v_{\rm y}\rangle}_{\rm T}$ & $\delta_{{v_{\rm z}}_{\rm T}}$ & $\frac{V_{{\rm mc}_{\rm T}}}{V_{\rm mc}}$ & $\frac{M_{{\rm mc}_{\rm T}}}{M_{\rm mc}}$ & ${\langle d_{\rm y}\rangle}_{\rm T}$ & ${\langle v_{\rm y}\rangle}_{\rm T}$ & $\delta_{{v_{\rm z}}_{\rm T}}$ & $\frac{V_{{\rm mc}_{\rm T}}}{V_{\rm mc}}$ & $\frac{M_{{\rm mc}_{\rm T}}}{M_{\rm mc}}$ \\ 
&  &  & $[\rm pc]$ & $[\rm km\,s^{-1}]$  & $[\rm km\,s^{-1}]$ & & & $[\rm pc]$ & $[\rm km\,s^{-1}]$  & $[\rm km\,s^{-1}]$ & &  \\  \hline
All & mc & All & $156\pm32$ & $381\pm135$ & $24$ & $1$ & $1$ & $122\pm58$ & $333\pm195$ & $24$ & $1$ & $1$ \\
Molecular & MM-H$_2$ & $T\leq10^2\,\rm K$ & $147\pm26$ & $373\pm122$ & $21$ & $0.001$ & $0.017$ & $93\pm48$ & $302\pm190$ & $21$ & 0.001 & $0.012$ \\
Cold Neutral & CNM-H$_{\rm I}$ & $10^2\,\rm K<T\leq5\times10^2\,\rm K$ & $146\pm24$ & $303\pm67$ & $12$ & $0.004$ & $0.319$ & $90\pm50$ & $185\pm125$ & $11$ & $0.002$ & $0.247$ \\
Warm Neutral & WNM-H$_{\rm I}$ & $5\times10^2\,\rm K<T\leq5\times10^3\,\rm K$ & $152\pm29$ & $343\pm79$ & $17$ & $0.005$ & $0.157$ & $120\pm53$ & $279\pm125$ & $16$ & $0.003$ & $0.141$ \\
Warm Ionised & WIM-H$_{\alpha}$ & $5\times10^3\,\rm K<T\leq3\times10^4\,\rm K$ & $158\pm33$ & $390\pm92$ & $23$ & $0.032$ & $0.362$ & $129\pm55$ & $335\pm132$ & $21$ & $0.021$ & $0.406$\\
Hot Ionised & HIM & $3\times10^4\,\rm K<T\leq 10^6\,\rm K$ & $164\pm33$ & $455\pm105$ & $37$ & $0.039$ & $0.073$ & $137\pm54$ & $403\pm141$ & $33$ & $0.024$ & $0.080$\\
Hot & HM-$X_{\rm ray}$ & $T>10^6\,\rm K$ & $186\pm42$ & $689\pm176$ & $51$ & $0.920$ & $0.073$ & $157\pm57$ & $668\pm183$ & $45$ & $0.950$ & $0.114$\\
\hline
\end{tabular}
\end{adjustbox}
\label{Table3}
\end{table*}

Next, we study the ram pressure acceleration of diffuse and dense cloud gas in non-radiative and radiative multicloud systems. Figure \ref{Figure8} presents mass-weighted phase diagrams of speed versus number density of cloud gas at four different times in non-radiative and radiative (compact and porous) models. In general we confirm that cloud gas has different dynamical properties in non-radiative and radiative cloud models. In non-radiative models most of the momentum at late stages ($t=3.0\,t_{\rm sp}=0.60\,\rm Myr$) is concentrated in gas with number densities between $\sim 0.1$--$1\,\rm cm^{-3}$, while in radiative models we find bi-modality with the cloud mass distributed in a low-density phase with densities $\lesssim0.1\,\rm cm^{-3}$ and a high-density phase with densities $\sim 1$--$10^3\,\rm cm^{-3}$. Gas at intermediate densities is generally out of thermal equilibrium, and so it is short-lived as it can either be further heated up to the hot phase or cooled down to the more stable warm and cold, high-density phases.\par

Combined, the panels in Figures \ref{Figure5} and \ref{Figure8} reveal that the outflow in radiative models has a multi-phase structure (similar to that captured in larger-scale simulations of supernova-driven outflows, e.g., see \citealt{2008ApJ...674..157C,2015MNRAS.454..238W,2020ApJ...895...43S}), and that it is different than the relatively warm-to-hot outflow produced in non-radiative models (e.g., see \citealt{2002ApJ...576..832P,2012MNRAS.425.2212A}; \citetalias{2020MNRAS.499.2173B}). In general, cloud gas in non-radiative models is faster than in radiative models. Since cooling of warm, mixed gas is efficient in radiative models, there is more dense gas available in the cold phases. Gas with larger column densities is more difficult to accelerate via direct momentum transfer. However, since most of such dense gas in radiative models precipitates from the warm and hot phases, it retains some of its momentum. Therefore, the average dense-gas speed in radiative models can be in the range between $\sim 100$--$500\,\rm km\,s^{-1}$ at $t=3.0\,t_{\rm sp}=0.60\,\rm Myr$.\par

Figure \ref{Figure8} also shows that compact (solenoidal) multicloud systems in both non-radiative and radiative models are faster than porous (compressive) models, which have dense cores with more inertia. This result implies that entrainment of dense gas is a function of both the thermodynamical interplay between radiative processes and the initial density structure in these multicloud systems. Dense gas entrainment is more efficient in compact models than in porous models. For instance, in compact models all gas has speeds $\gtrsim 100\,\rm km\,s^{-1}$ at $t=3.0\,t_{\rm sp}=0.60\,\rm Myr$, while in porous models it spans speeds that start at much lower values, $\sim 10\,\rm km\,s^{-1}$. Thus, in order to better understand the dynamics of the different gas phases in our radiative outflow models, we separate the cloud gas using different temperature bins (see definitions in Table \ref{Table3}).\par

\subsubsection{Evolution of gas in different temperature bins}
\label{sec:tempbins}
\begin{figure}
\begin{center}
  \begin{tabular}{l}
    9a) Travelled distance vs. time\\
    \hspace{-0.40cm}\resizebox{80mm}{!}{\includegraphics{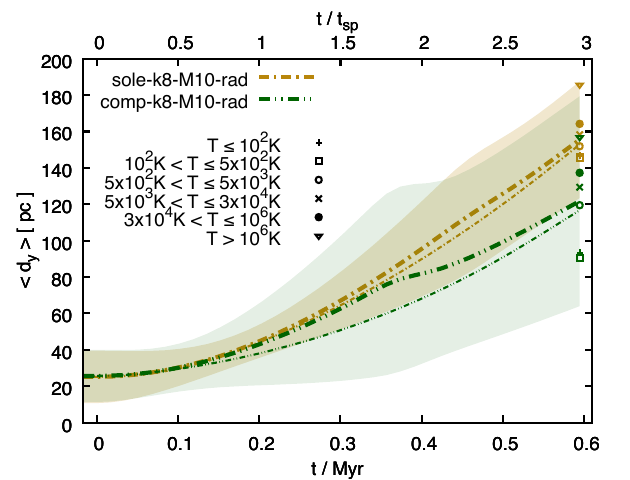}}\\
    9b) Bulk speed vs. time\\
    \hspace{-0.40cm}\resizebox{80mm}{!}{\includegraphics{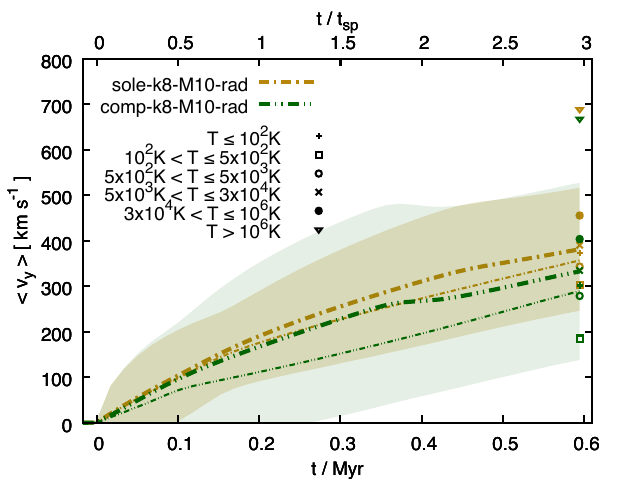}}\\
    9c) Velocity dispersion along the LOS vs. time\\
    \hspace{-0.40cm}\resizebox{80mm}{!}{\includegraphics{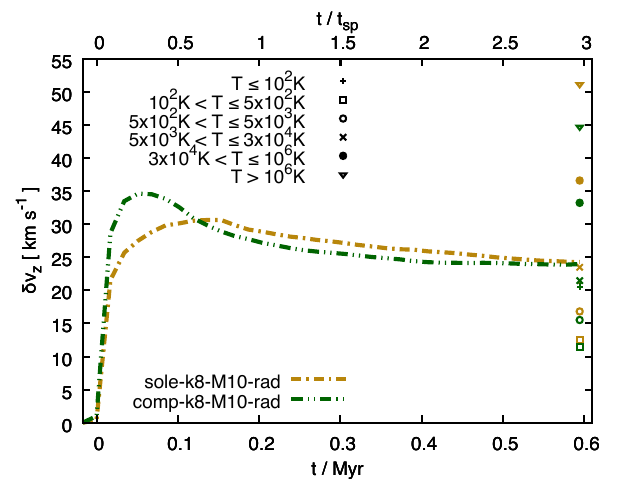}}\\
  \end{tabular}
  \caption{The distance travelled by the centre of mass (panel 9a), the mass-weighted bulk speed (panel 9b), and the line-of-sight velocity dispersion (LOS, i.e., along the $Z$ axis) of the multicloud layer as a function of time. These are all mass-weighted quantities. In general, the porous (compressive) model is slower than the compact (solenoidal) model, but it has a richer overall kinematics. In the two upper plots, the thick lines represent average values for the whole cloud, the shaded areas cover the $1\sigma$ limits, and the thin lines show gas denser than the conventional, $\rho_{\rm cloud,0}/3$. Also the points displayed at $t=0.6\,\rm Myr$ show where gas with different temperatures lies on these plots at the end of the simulations.} 
  \label{Figure9}
\end{center}
\end{figure}

Figure \ref{Figure9} reports the distance, mass-weighted speed, and velocity dispersion of cloud material in our radiative multicloud models (see Table \ref{Table2}). The thick lines and the shadowed areas around them show the average parameters and the $1\sigma$ standard deviations, respectively, for the whole cloud material. The thin lines show the same parameters for cloud gas only with densities above the conventional density threshold of $\rho_{\rm cloud,0}/3$, so that the values can be readily compared to previous studies (e.g., \citealt{2015ApJ...805..158S,2020ApJ...892...59C}). In addition, Figure \ref{Figure9} also displays with points the final distances, speeds, and velocity dispersions of cloud material in different temperature bins as data points at $t=0.6\,\rm Myr$. These temperature bins are chosen to reflect (to a first approximation) the typical temperatures of several ISM gas phases (see Table \ref{Table3}). In addition, we note that we did not follow the gas chemistry and did not carry out radiative transfer, which we leave for future work.\par


In panel 9a of Figure \ref{Figure9} we show the distance that the centre of mass of the multicloud layer has travelled as a function of time. It is clear that these multicloud systems can travel $\gtrsim 120\,\rm pc$ downstream from their starting positions within $0.6\,\rm Myr$. If we assume the systems were originally near the galactic mid-plane, this result implies that they can readily travel into the disc-halo transition region at $\sim 200\,\rm pc$, which is similar to the scale height of cold and warm neutral medium. The distance is approximately a parabolic function of time as expected for a constant acceleration. This is confirmed in panel 9b of Figure \ref{Figure9}, which shows the cloud bulk speed as function of time. The speed is approximately linearly increasing with time, although a slight flattening of the slope is visible. What is remarkable is that the speed of cloud material is already quite high, with values $\approx 100$--$500\,\rm km\,s^{-1}$ at the end of the simulations. Recall that the upper bound is already within a factor of two of e.g. the escape velocity of the Milky Way (of $\sim 900\,\rm km\,s^{-1}$; see \citealt{2004ApJ...613..326M}) even in the mid-plane close to the Galactic centre, and, given its trend, it is expected to continue increasing. The data points in panel 9b of Figure \ref{Figure9} show the mass-weighted vertical velocities of the various gas phases. The hot gas is, of course, the fastest component with a speed of $\approx$700~$\rm km\,s^{-1}$ at the end of the simulation ($t=0.6\,\rm Myr$). But even the warm and cold components with $T\approx 10^{4}\,\rm K$ and $T\approx 10^{2}~\rm K$ already have velocities of $\approx 300$--$400\,\rm km\,s^{-1}$ and $\approx 200$--$300\,\rm km\,s^{-1}$, respectively, by this time (see a summary in Table \ref{Table3}).\par

How are the warm and cold phases accelerated to such high speeds? Our radiative simulations indicate that mixing between ambient and cloud gas is very efficient ($f_{\rm mix}\gtrsim80$ per cent; see panel 7a of Figure \ref{Figure7}). Therefore, the warm ($T\approx 10^{4}\,\rm K$) and cold ($T\approx 10^{2}\,\rm K$) phases are not directly accelerated by ram pressure. Instead, Figures \ref{Figure7}, \ref{Figure8}, and \ref{Figure9} show that these gas phases precipitate and acquire momentum from both the mixed gas (consistent with the acceleration mechanism proposed by \citealt{2020ApJ...895...43S}, in which mixing redistributes momentum) and also the hot ambient gas (consistent with the mass-growth scenario proposed by \citealt{2018MNRAS.480L.111G}, in which ambient hot gas is entrained/accreted into the dense phase).

\subsection{Multi-phase wind properties}
\label{subsec:Multi-phase}

\begin{figure*}
\begin{center}
  \begin{tabular}{c c c c c c c l}
       \multicolumn{8}{l}{\hspace{+0.0cm}10a) sole-k8-M10-rad}\\   
       \multicolumn{1}{c}{All} & \multicolumn{1}{c}{MM - H$_2$} & \multicolumn{1}{c}{CNM - H$_{\rm I}$} & \multicolumn{1}{c}{WNM - H$_{\rm I}$} & \multicolumn{1}{c}{WIM - H$_{\alpha}$} & \multicolumn{1}{c}{HIM} & \multicolumn{1}{c}{HM - $X_{\rm ray}$} & $N_{\rm mc}$\\
       \hspace{-0.25cm}\resizebox{26mm}{!}{\includegraphics{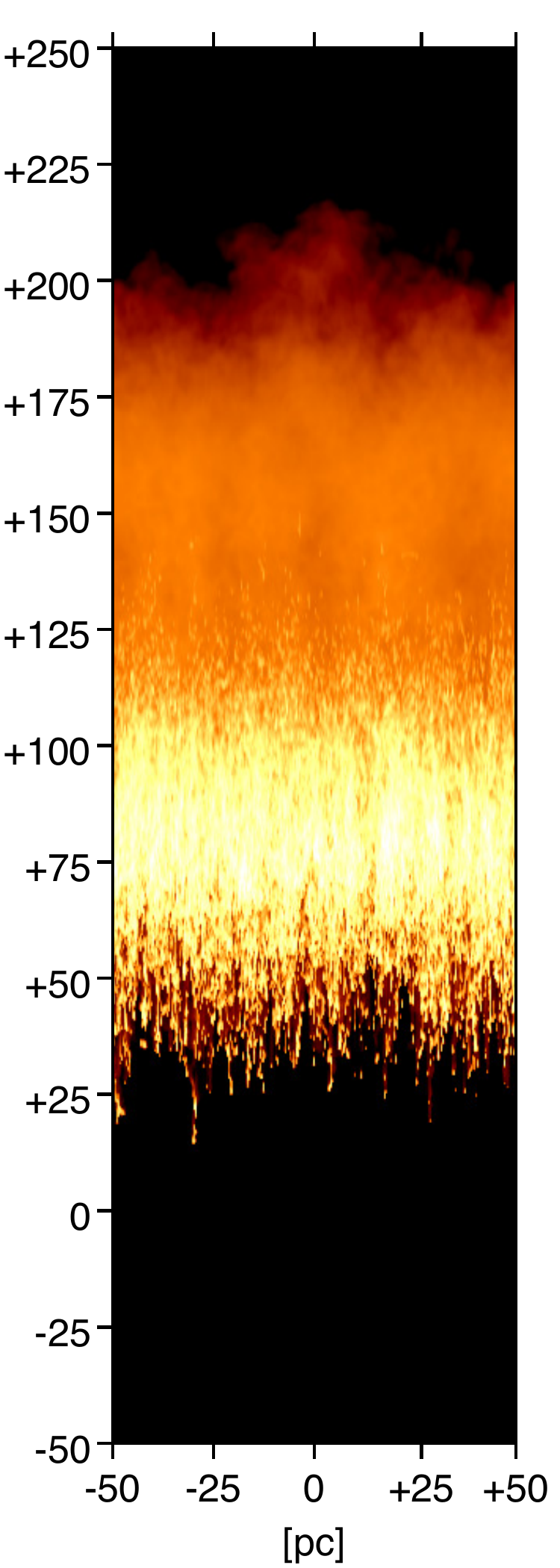}} & \hspace{-0.55cm}\resizebox{26mm}{!}{\includegraphics{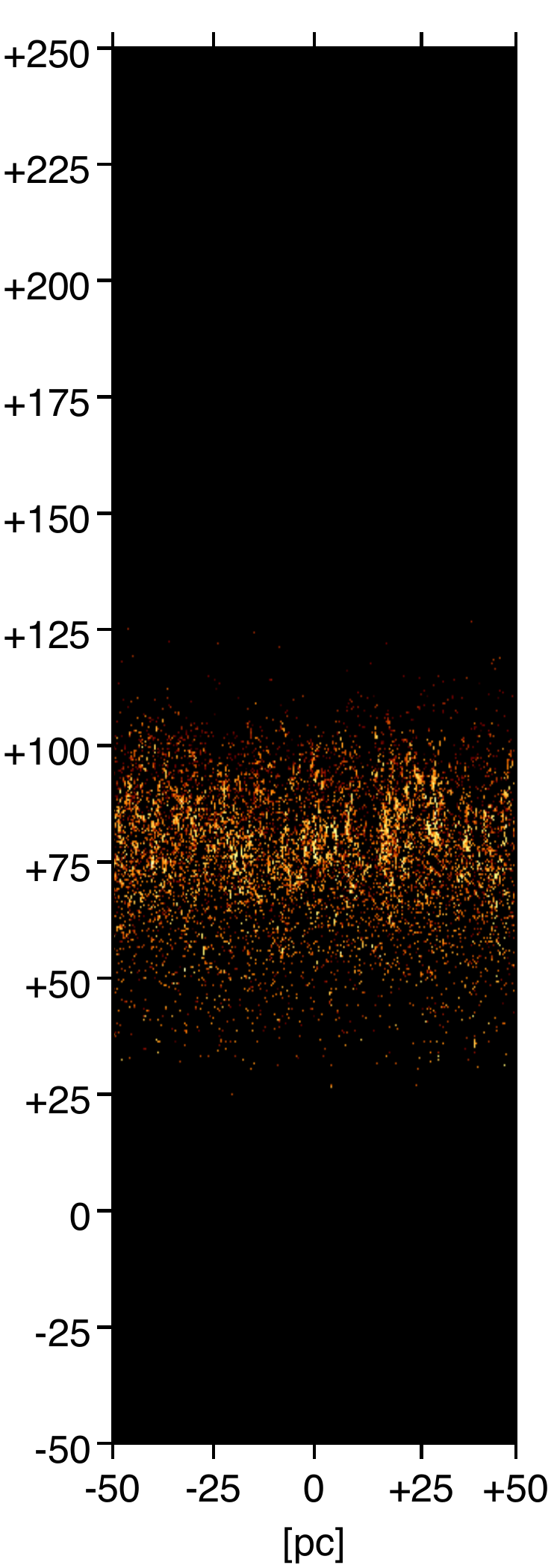}} & \hspace{-0.55cm}\resizebox{26mm}{!}{\includegraphics{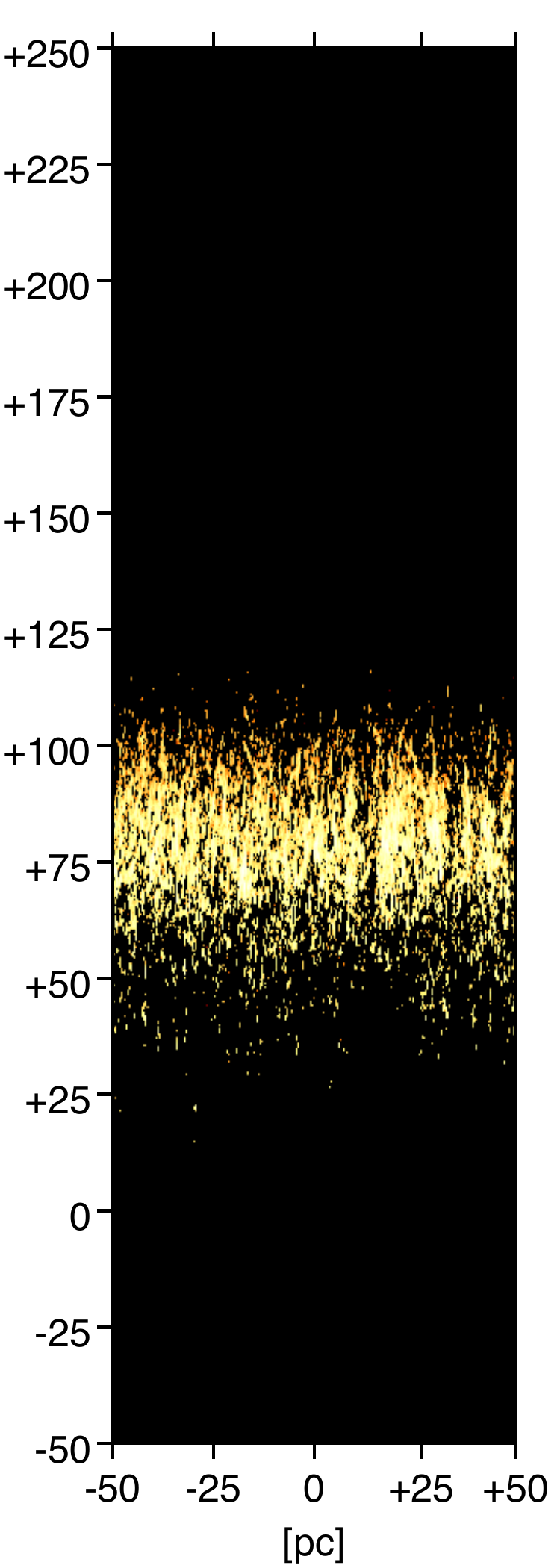}} & \hspace{-0.55cm}\resizebox{26mm}{!}{\includegraphics{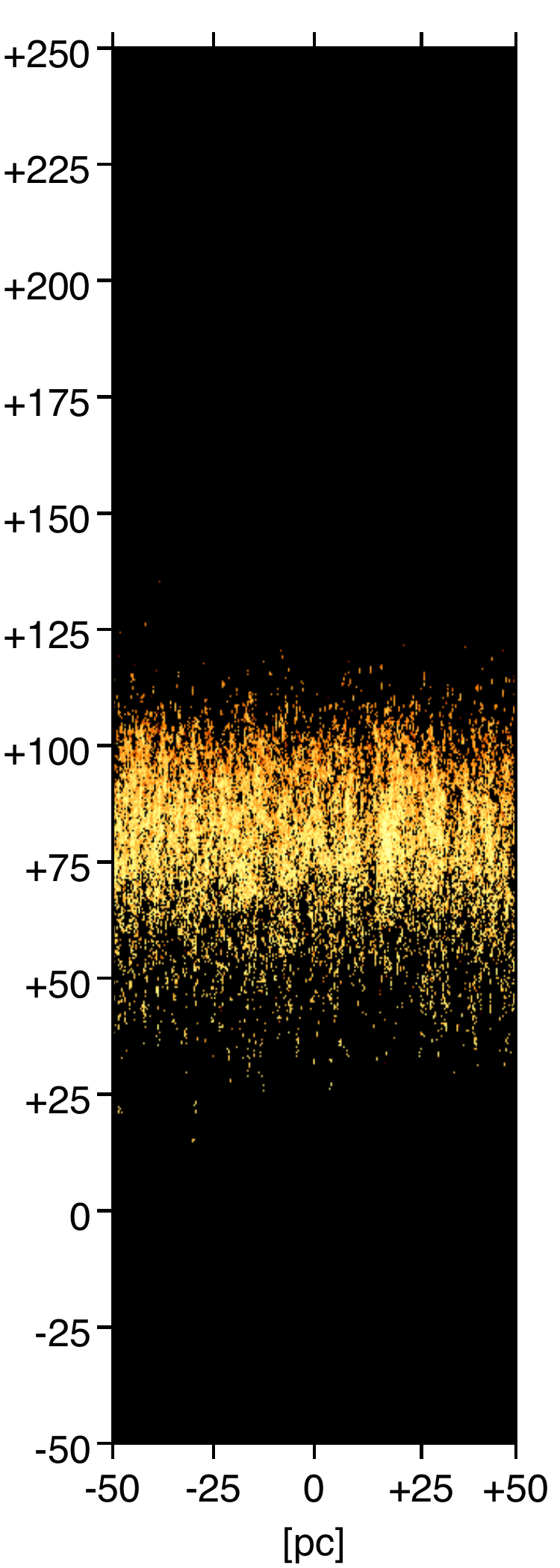}} & \hspace{-0.55cm}\resizebox{26mm}{!}{\includegraphics{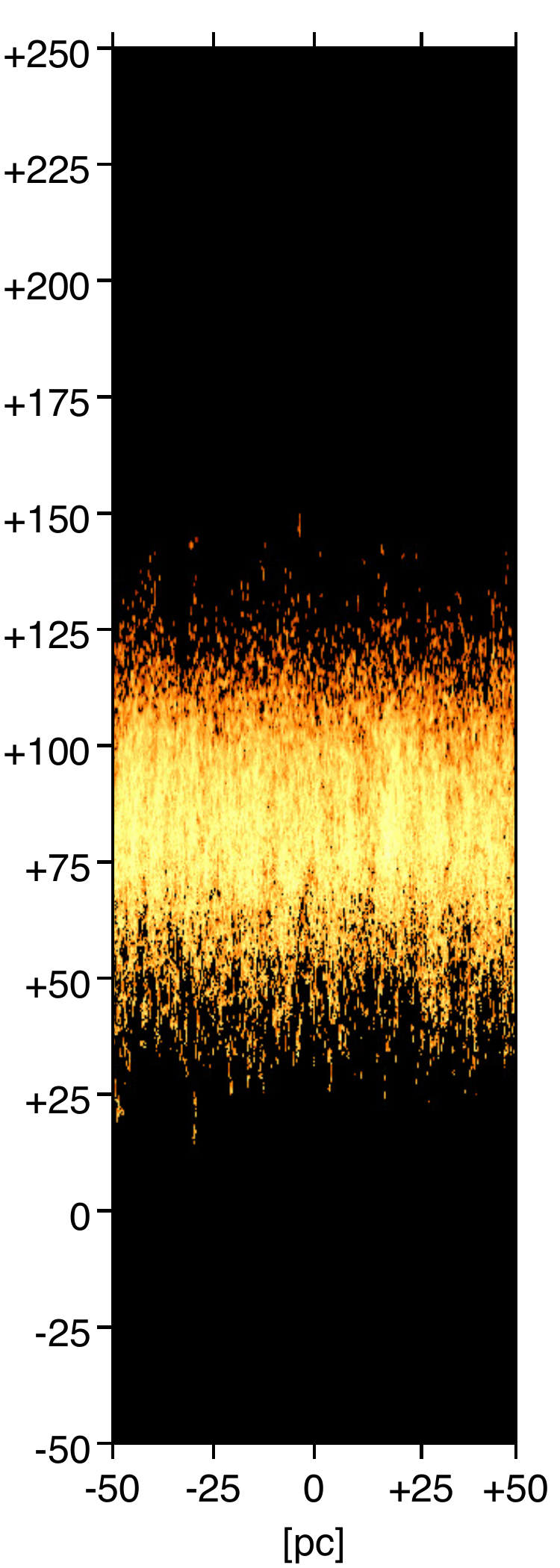}} & \hspace{-0.55cm}\resizebox{26mm}{!}{\includegraphics{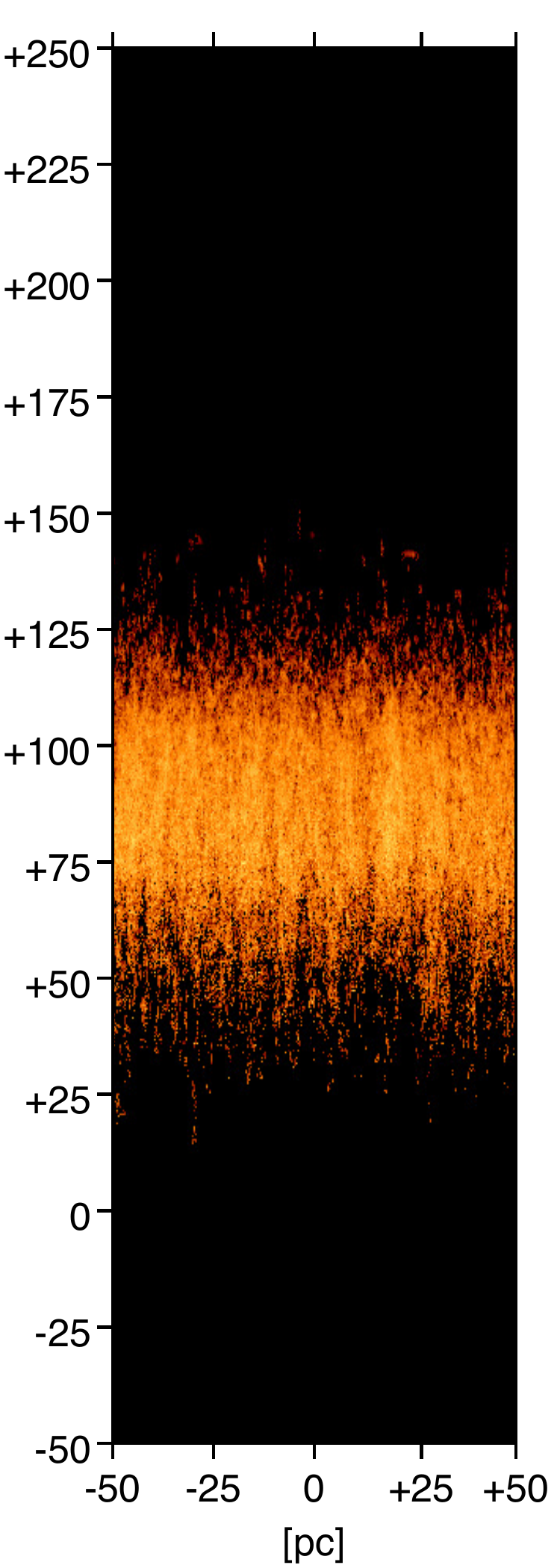}} &
\hspace{-0.55cm}\resizebox{26mm}{!}{\includegraphics{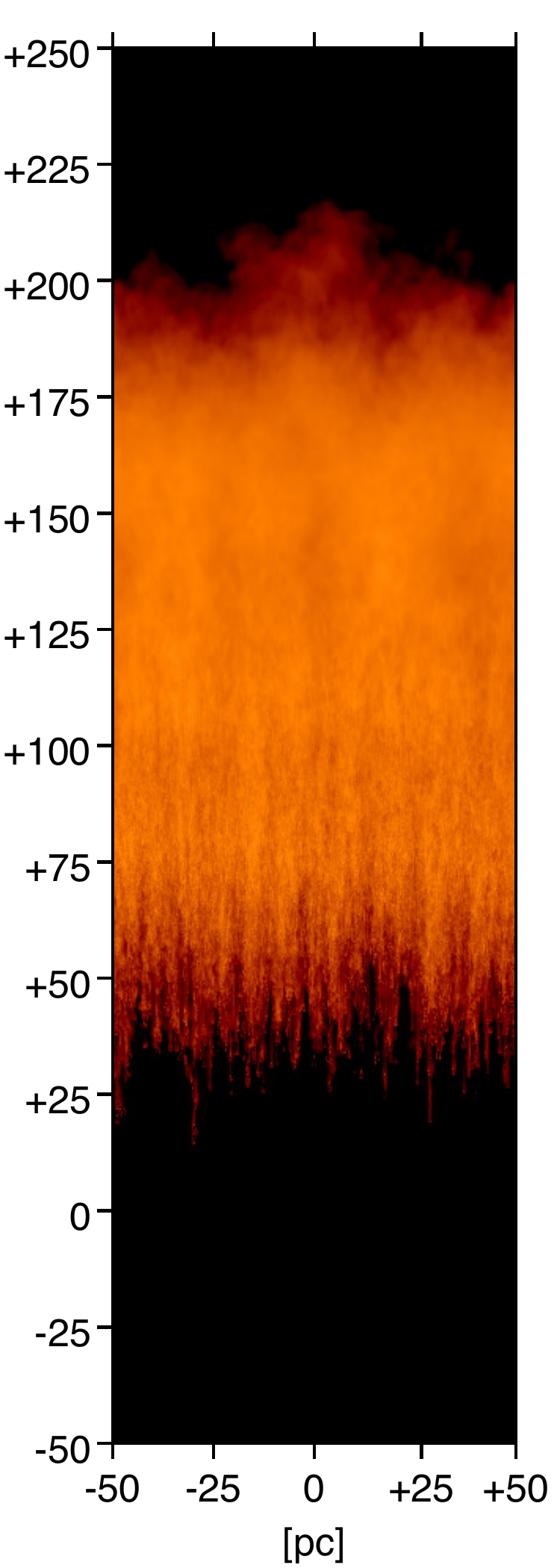}} &
\hspace{-0.40cm}\resizebox{11mm}{!}{\includegraphics{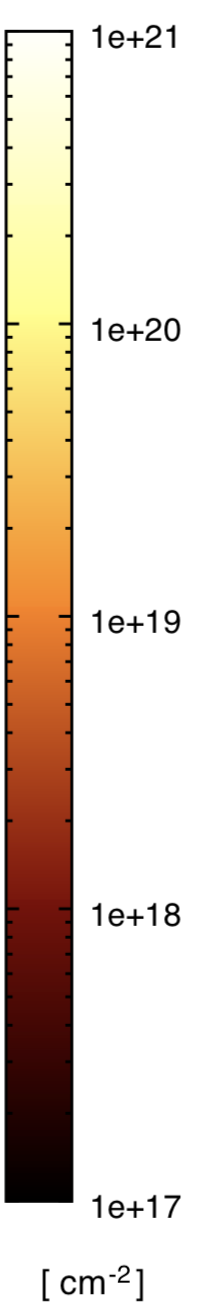}}\\
       \multicolumn{8}{l}{\hspace{+0.0cm}10b) comp-k8-M10-rad}\\    
       \multicolumn{1}{c}{All} & \multicolumn{1}{c}{MM - H$_2$} & \multicolumn{1}{c}{CNM - H$_{\rm I}$} & \multicolumn{1}{c}{WNM - H$_{\rm I}$} & \multicolumn{1}{c}{WIM - H$_{\alpha}$} & \multicolumn{1}{c}{HIM} & \multicolumn{1}{c}{HM - $X_{\rm ray}$} & $N_{\rm mc}$\\
       \hspace{-0.25cm}\resizebox{26mm}{!}{\includegraphics{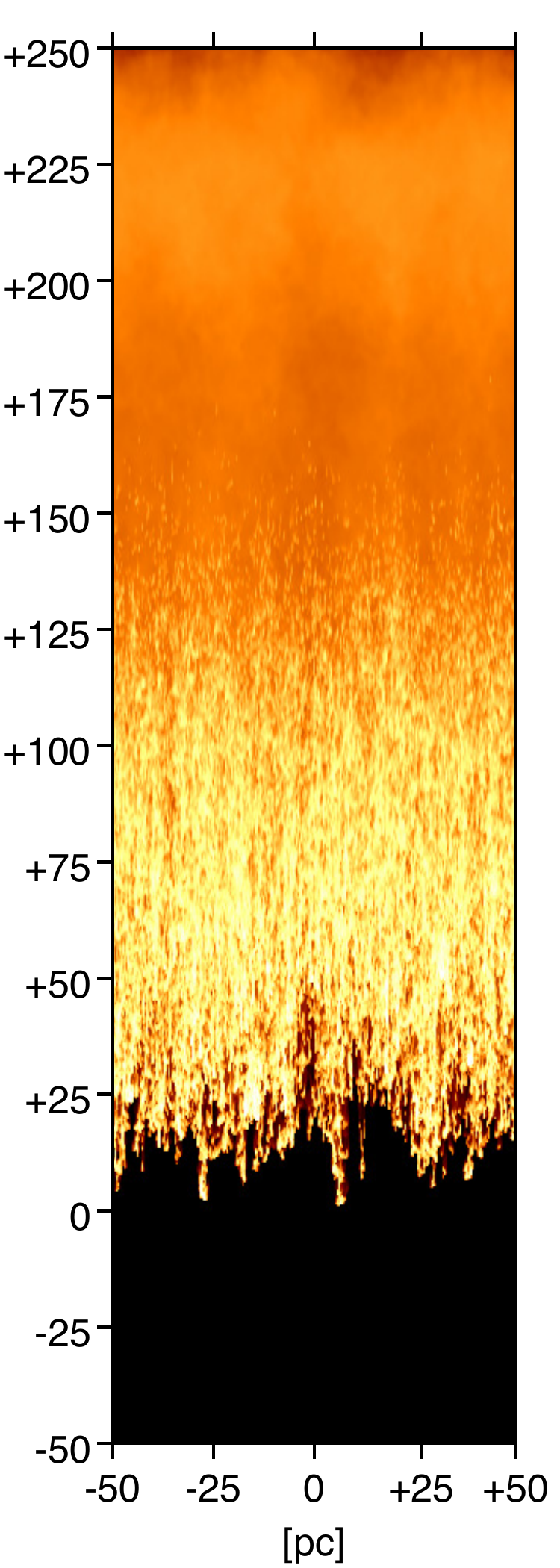}} & \hspace{-0.55cm}\resizebox{26mm}{!}{\includegraphics{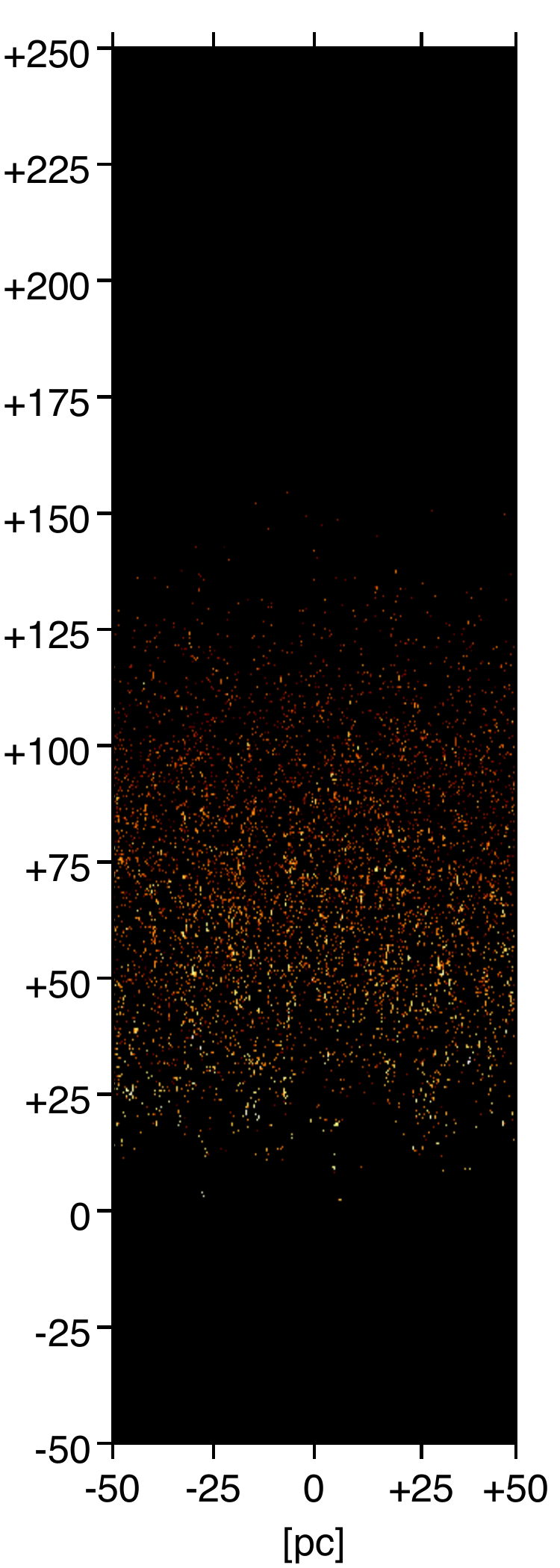}} & \hspace{-0.55cm}\resizebox{26mm}{!}{\includegraphics{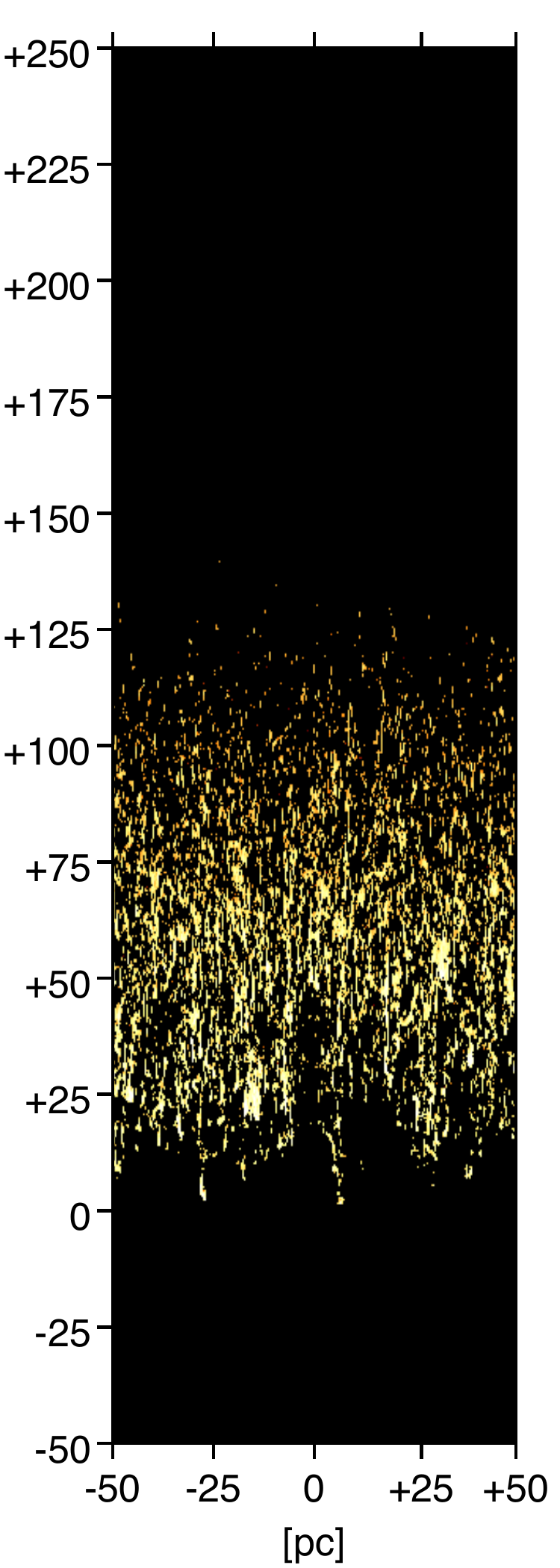}} & \hspace{-0.55cm}\resizebox{26mm}{!}{\includegraphics{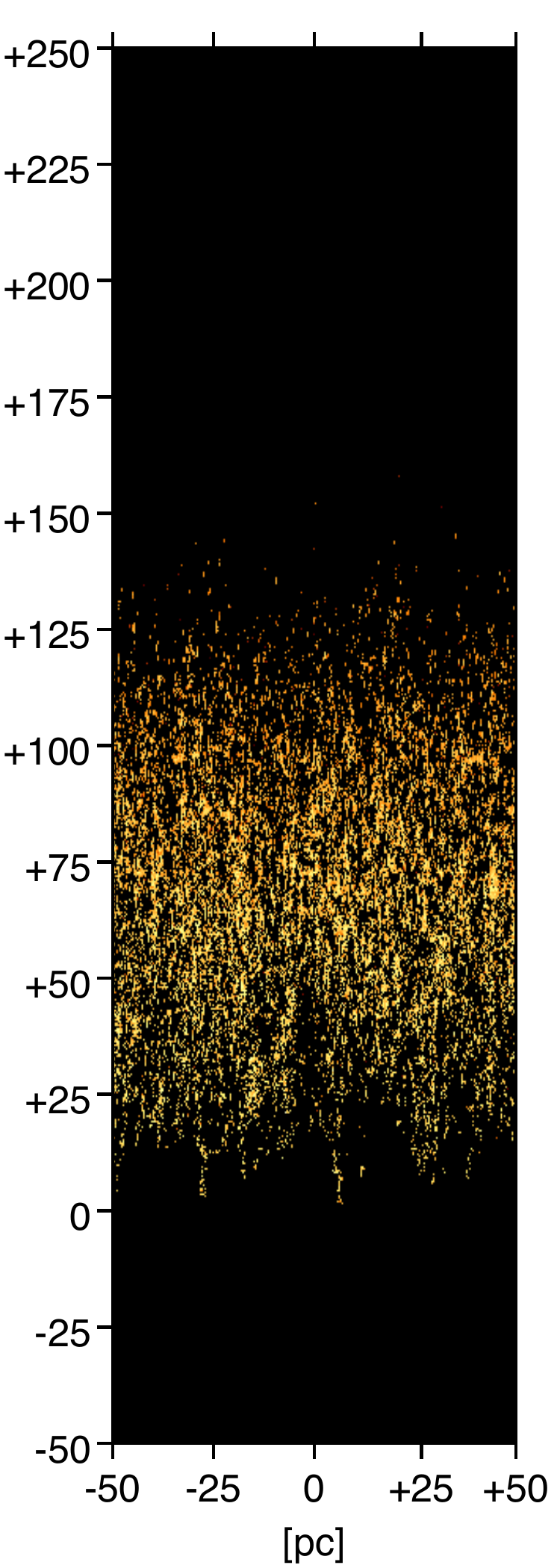}} & \hspace{-0.55cm}\resizebox{26mm}{!}{\includegraphics{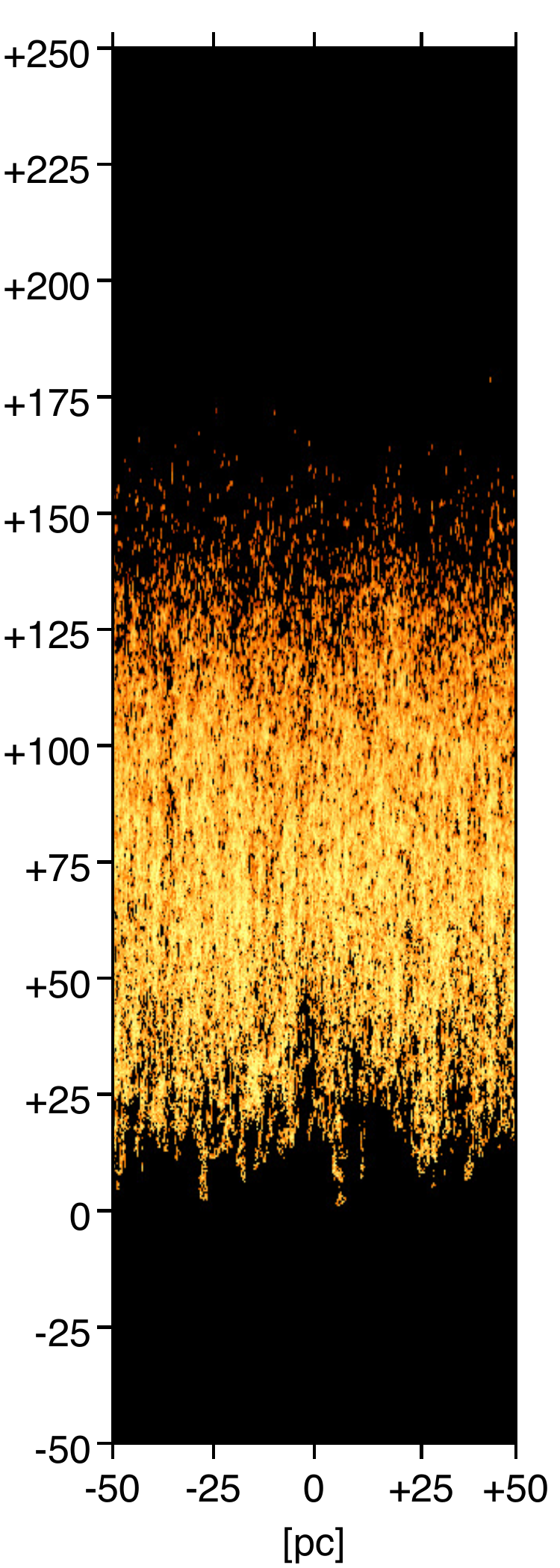}} & \hspace{-0.55cm}\resizebox{26mm}{!}{\includegraphics{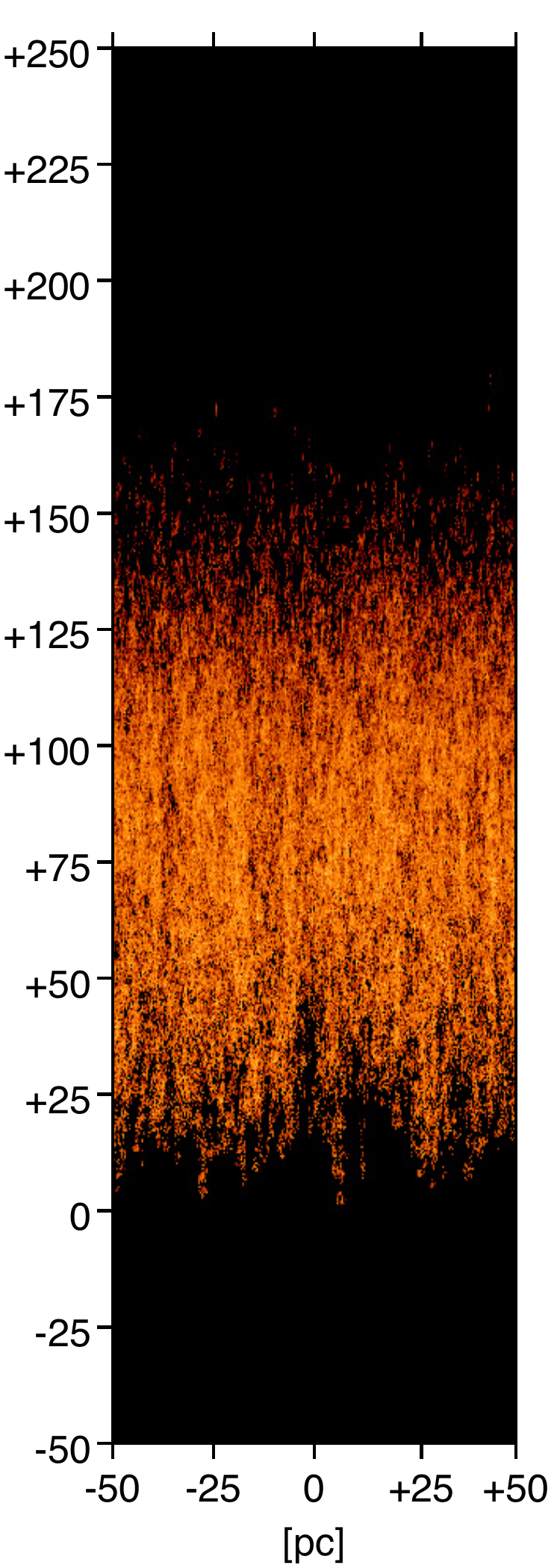}} &
\hspace{-0.55cm}\resizebox{26mm}{!}{\includegraphics{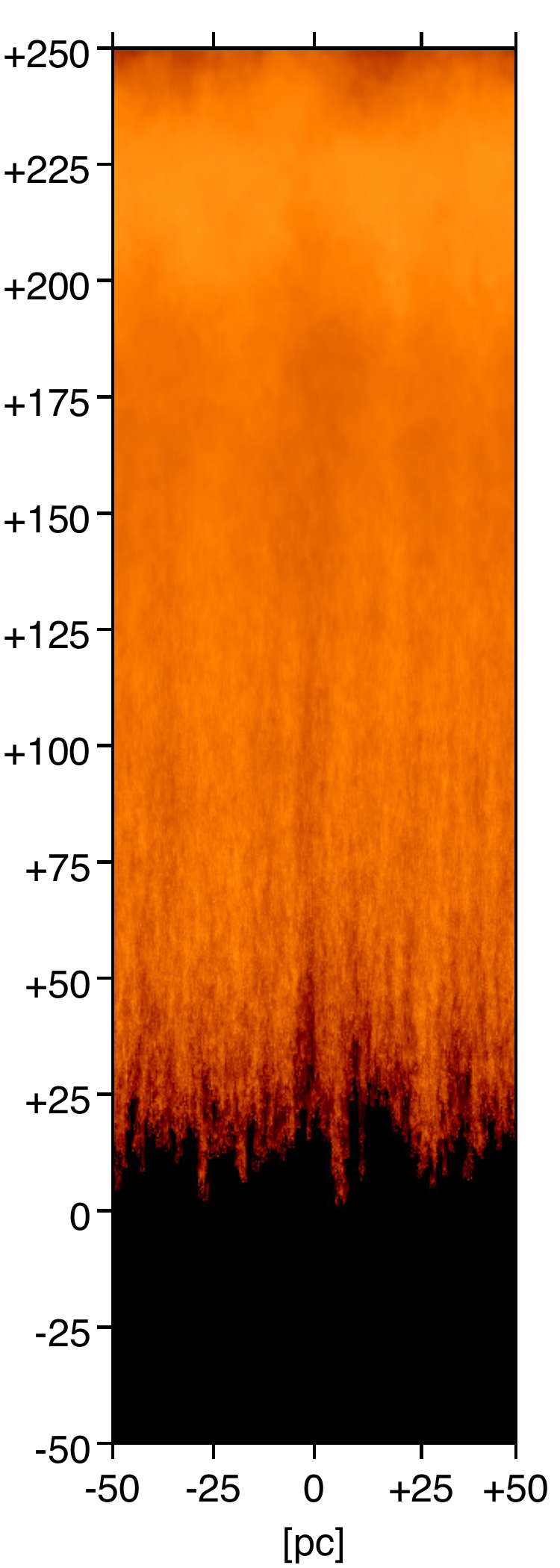}} & \hspace{-0.40cm}\resizebox{11mm}{!}{\includegraphics{barra.png}}\\
  \end{tabular}
  \caption{Column number density ($N_{\rm mc}$) maps of cloud gas in our radiative models, sole-k8-M10-rad (top panel) and comp-k8-M10-rad (bottom panel), for different temperature ranges (see Table \ref{Table3}) at $t=1.8\,t_{\rm sp}=0.36\,\rm Myr$ (a later snapshot is shown in Appendix \ref{AppendixB}). The very hot phase of the outflow ($T>10^6\,\rm K$) is volume filling and outruns all the other warm ($T\sim10^4\,\rm K$) and cold ($T\sim10^2\,\rm K$) phases. The warm and cold phases of the outflow have low volume filling factors and high 2D covering fractions, and they spatially coincide with each other, which indicates they have similar momenta. Note that for illustration purposes we only show cloud material in these panels. Therefore, the volume filling factor and 2D covering fraction of the hottest phase of the outflow (with $T>10^6\,\rm K$) would be even larger had we considered the contribution from the ambient gas, which is also hot.}
  \label{Figure10}
\end{center}
\end{figure*}

Figure \ref{Figure9} also reports the line-of-sight (LOS) velocity dispersion of cloud material as a whole and in different temperature bins at $t=0.60\,\rm Myr$ (see points and a summary in Table \ref{Table3}). Hot gas with temperatures $>10^6\,\rm K$ has higher velocity dispersions ($\delta_{v_z}>40\,\rm km\,s^{-1}$) than the warm and cold gas phases. The warm gas phase with temperatures between $5\times10^2$ and $3\times10^4\,\rm K$ has velocity dispersions $15$--$25\,\rm km\,s^{-1}$, while the cold gas phase with temperatures $<5\times10^2\,\rm K$ has velocity dispersions $<15\,\rm km\,s^{-1}$. These values can be compared to the full-width-at-half-maximum ($\rm \Delta v_{\rm FWHM}\approx2.355\delta_{v_z}$) velocity dispersions reported in observations of outflows for these different gas phases. To improve this comparison, we also compute the column number densities for gas in different temperature bins in our simulations.\par

Figure \ref{Figure10} reports maps of the column number density for the whole cloud material (see leftmost column), and for cloud gas in each of the above-mentioned temperature bins. These projections correspond to a time of $t=0.36\,\rm Myr$ in both compact (top panel) and porous (bottom panel) multicloud models. We show the same figure for $t=0.60\,\rm Myr$ in Appendix \ref{AppendixB}. The panels confirm our results above. The hottest $X_\mathrm{ray}$-emitting gas overshoots the warm and cold phases, moving faster and reaching larger distances. This gas is easily accelerated and does not cool as efficiently as gas with temperatures between $10^4\,\rm K$ and $10^6\,\rm K$ and densities between $0.01\,\rm cm^{-3}$ and $1\,\rm cm^{-3}$, which is subjected to strong cooling. In addition, most of the gas in other warm (ionised and atomic) and cold (atomic and molecular) phases is spatially coincident, confirming that such gas is not directly accelerated by the post-shock flow, but rather forms in situ from hot and warm gas undergoing fast cooling. If gas would be accelerated, we would expect to see the much colder gas phases lagging behind the warmer phases, but since they precipitate from the warm and hot phases respectively, they have similar momenta and form in-situ at overlapping locations.\par

\begin{figure}
\begin{center}
  \begin{tabular}{l}
    11a) Volume filling factors vs. time\\
    \hspace{-0.40cm}\resizebox{80mm}{!}{\includegraphics{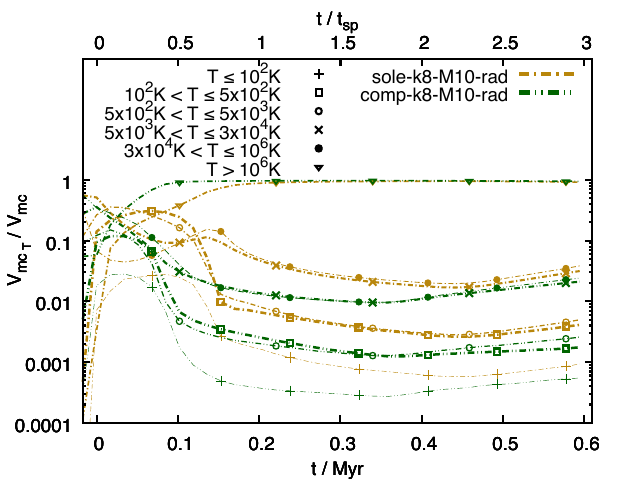}}\\
    11b) Mass fractions vs. time\\
    \hspace{-0.40cm}\resizebox{80mm}{!}{\includegraphics{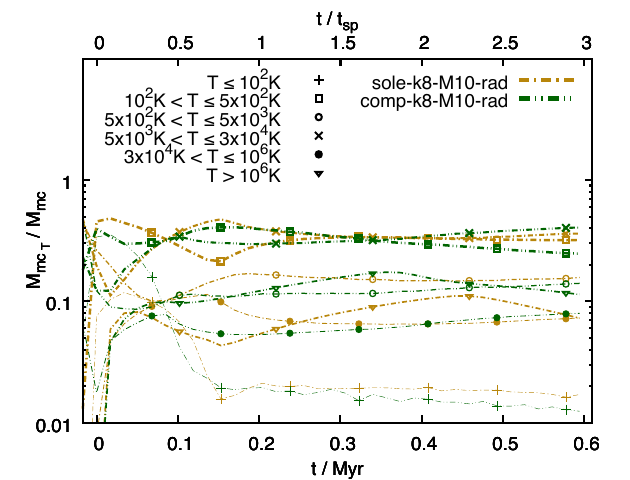}}\\
  \end{tabular}
  \caption{Evolution of the volume filling factors (panel 11a) and mass fractions (panel 11b) of cloud gas at different temperatures, normalised with respect to the total volume and the total mass of cloud material, respectively. While $>90$ per cent of the cloud gas volume is occupied by hot gas, $\sim 5$ per cent by ionised gas, and $<1$ per cent by cold gas, hot gas only contains $\sim 15$ per cent of the cloud gas mass with the warm and cold phases representing between $\sim 30$ per cent and $\sim 40$ per cent of the total cloud mass each.} 
  \label{Figure11}
\end{center}
\end{figure}

Figure \ref{Figure11} reports the volume filling factors and mass fractions computed for cloud gas in different temperature bins and normalised with respect to the total volume and the total mass of cloud material. At $t=0.60\,\rm Myr$, Figure \ref{Figure11} indicates that hot gas with $T>10^6\,\rm K$ occupies $>90$ per cent of the cloud gas volume, but only represents $<15$ per cent of the cloud gas mass in both porous and compact systems. This implies that hot ($X_\mathrm{ray}$-emitting) gas is volume filling in outflows, as has also been assumed by e.g. \citealt{2019Natur.567..347P}, who studied the $X_\mathrm{ray}$ chimneys associated with the Galactic centre wind. In such outflows, hot gas would be even more volume filling than $90$ per cent because the ambient gas (which is not included in Figure \ref{Figure11}) is also hot and would contribute to the $X_\mathrm{ray}$ emission too.\par

On the other hand, warm and cold gas with $T\leq3\times10^4\,\rm K$ occupy less than $5$ per cent of the outflow's volume, but they contain $\sim 80$ per cent of the outflow's mass. Indeed, in the warm phase H${\alpha}$-emitting gas with $5\times10^3\,\rm K<T\leq 3\times10^4\,\rm K$ occupies $5$ per cent of the outflow's volume and contains $\sim40$ per cent of its mass, while the warm neutral medium with $5\times10^2\,\rm K<T\leq 5\times10^3\,\rm K$ occupies $0.2$--$0.5$ per cent of the outflow's volume and contains $<20$ per cent of its mass. In the cold phase, the cold neutral medium with $10^2\,\rm K<T\leq 5\times10^2\,\rm K$ occupies only $0.2$--$0.4$ per cent of the outflow's volume and contains $30$ per cent of its mass, while molecular gas has volume filling factors $<0.1$ per cent and has $<2$ per cent of the outflow's mass. The latter, however, may be affected by the $10^2$~K floor that we imposed in these simulations for our cooling and heating function. It is possible that if this temperature threshold for cooling is removed from the models, there would be more gas in the molecular component than the value we report here. Therefore, this should be considered as a lower limit for the molecular content of an outflow. Despite this, the fact that we find gas with temperatures $\lesssim 10^2\,\rm K$ may explain the presence of molecular phases in observed galactic outflows (e.g., see \citealt{2019ApJ...881...43K,2019ApJ...885L..32D,2020Natur.584..364D,2020ApJ...905...85S}).\par

Note also that Figures \ref{Figure3}, \ref{Figure10}, and \ref{Figure11} show that while the warm and cold gas phases (which are composed of cloudlets and filaments) in the outflow have very low volume filling factors, their 2D covering fractions are quite high when projected along the $Z$ axis. This is in agreement with the results presented in \cite{2020MNRAS.491.5056L}. Investigating how optical depths can affect these results is beyond the scope of this paper, but would be an interesting topic to follow up in future work.

\section{An application to the atomic H\,{\sc i} outflow in the Galactic centre}
\label{sec:Application}

\subsection{H\,{\sc i} dynamics}
In order to show how our simulations can be used to study features observed in cold outflows in galaxies, we discuss some of their implications for the local Galactic Centre (GC) outflow. The closest evidence of the existence of multi-phase galactic winds comes from multi-wavelength observations of the GC (e.g., see \citealt{1998ApJS..118..455O,2000ApJ...540..224S,2003ApJ...582..246B,2010ApJ...708..474L,2010ApJ...724.1044S,2013Natur.493...66C,2021A&A...646A..66P}). In this paper, we will solely focus on the results from $\rm {H}\,${\scriptsize{I}} surveys of the area surrounding the GC (e.g., see \citealt{2012ApJS..199...12M,2018ApJ...855...33D}), but our results could also be applied to studies of Galactic $\rm {H}\,${\scriptsize{I}} in general, e.g. to study the vertical filaments in the Milky Way's disc reported by \citealt{2020A&A...642A.163S} or foreground emission for extragalactic sources (e.g., see \citealt{2019MNRAS.489.3778D}). The studies of $\rm {H}\,${\scriptsize{I}} emission in the GC reveal a population of $\sim 200$ high-velocity $\rm {H}\,${\scriptsize{I}} clouds at high latitudes ($\sim 0.3$--$1.5\,\rm kpc$) above and below the Galactic plane (see also \citealt{2013ApJ...770L...4M,2020ApJ...888...51L}). These warm clouds (with temperatures $\sim 4\times10^3\,\rm K$) have been interpreted as entrained material in a much hotter GC wind (with temperatures $\sim10^7\,\rm K$), as they do not exhibit the kinematic patterns associated with Galactic rotation.\par

The origin of the multi-phase GC wind itself is still debated between scenarios that favour either sustained star formation (\citealt{2000ApJ...540..224S,2003ApJ...582..246B,2012MNRAS.423.3512C}) or past AGN activity (\citealt{2012ApJ...756..181G,2020ApJ...894..117Z}) as causing the outflow. However, a challenge in either scenario is explaining the presence of high-latitude $\rm {H}\,${\scriptsize{I}} gas moving at estimated speeds of $\sim 200$--$300\,\rm km\, s^{-1}$ in hot gas (\citealt{2013ApJ...770L...4M}). As mentioned earlier, direct entrainment of dense gas is difficult as even radiative clouds cannot survive ablation and disruption by the wind ram pressure and dynamical instabilities for long enough to reach large distances. However, cloud disruption processes also generate warm mixed gas, which mass-loads the wind and can cool very efficiently. Our simulations show that in multicloud systems, shocked cloud material that becomes mixed can condense back onto the cold phase and even sustain a multi-phase outflow by accreting material from the hot wind. Thus, in our models the colder and denser phases of the GC outflow, such as the $\rm {H}\,${\scriptsize{I}} clouds, would be a natural result of re-condensation of warm mixed gas, originally stripped from the clouds by dynamical instabilities, combined with some gas that precipitates from the originally hot component.\par

\begin{figure*}
\begin{center}
  \begin{tabular}{c c c l c c c l}
        \multicolumn{4}{l}{12a) sole-k8-M10-rad}\vspace{+0.1cm} & \multicolumn{4}{l}{12b) comp-k8-M10-rad}\vspace{+0.1cm} \\  
        \multicolumn{1}{c}{GBT - H\,{\sc i}} & \multicolumn{1}{c}{ATCA - H\,{\sc i}} & \multicolumn{1}{c}{ASKAP - H\,{\sc i}} & $N_{\rm H\,{\scriptstyle I}}$ &  \multicolumn{1}{c}{GBT - H\,{\sc i}} & \multicolumn{1}{c}{ATCA - H\,{\sc i}} & \multicolumn{1}{c}{ASKAP - H\,{\sc i}} & $N_{\rm H\,{\scriptstyle I}}$ \\    
\resizebox{26mm}{!}{\includegraphics{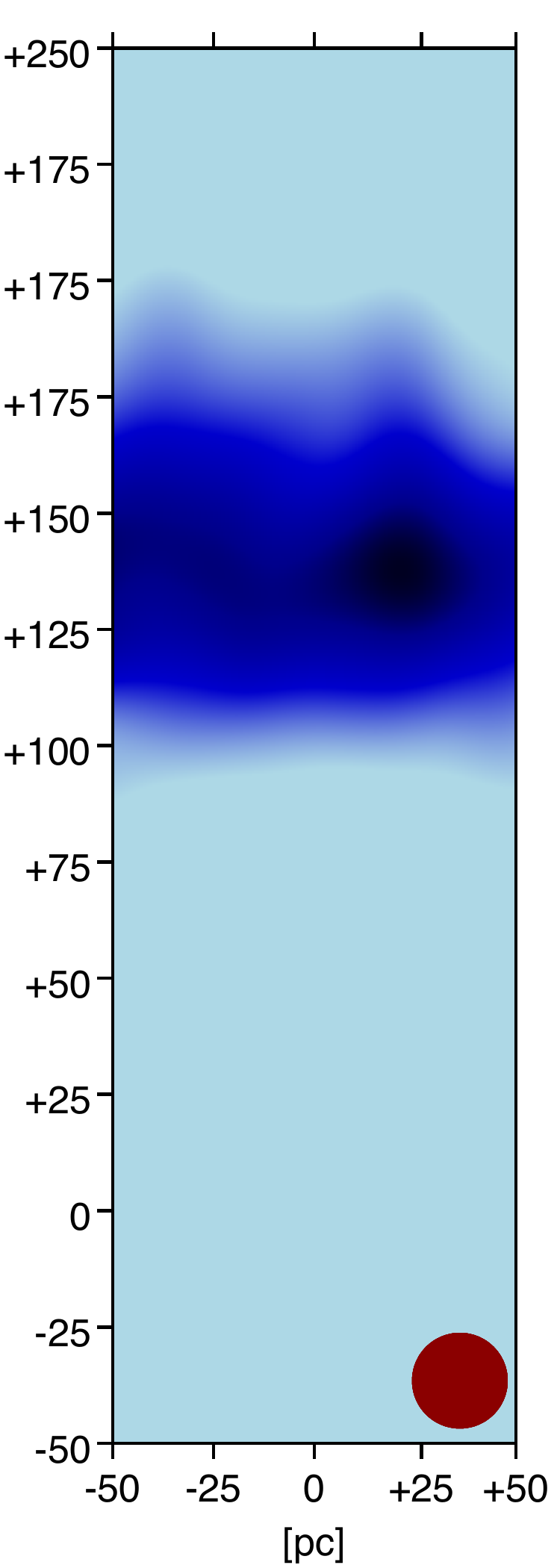}} & \hspace{-0.55cm}\resizebox{26mm}{!}{\includegraphics{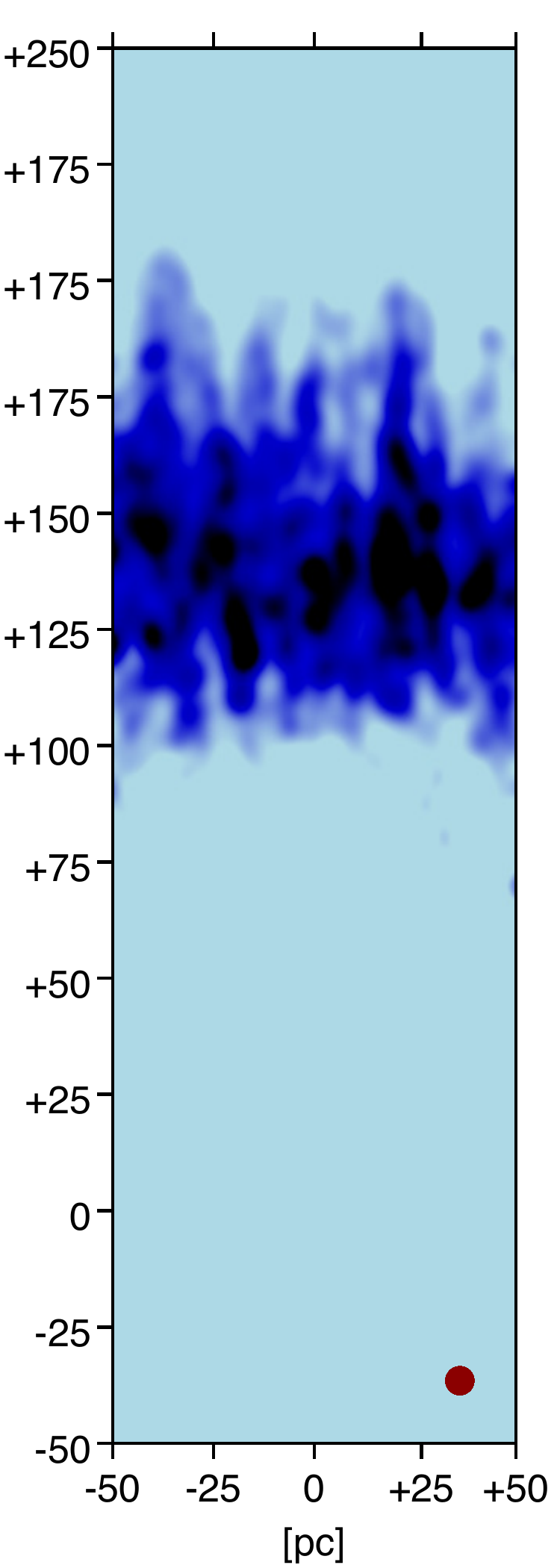}} & \hspace{-0.55cm}\resizebox{26mm}{!}{\includegraphics{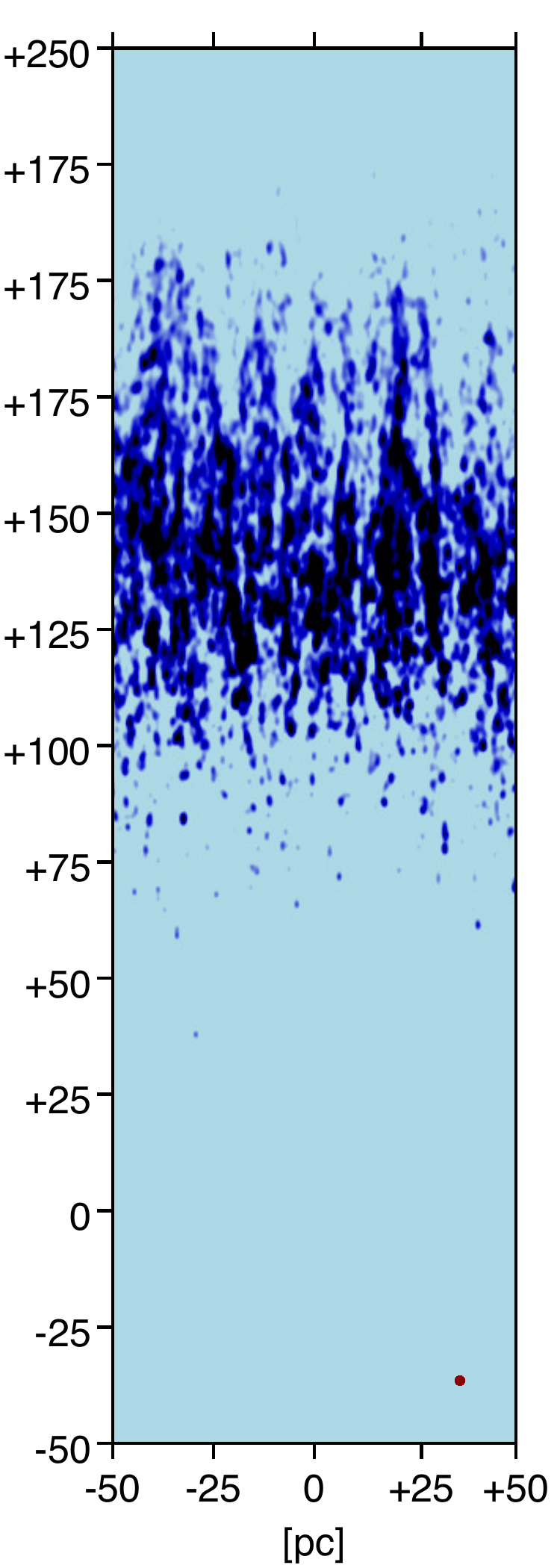}} &
\hspace{-0.40cm}\resizebox{11mm}{!}{\includegraphics{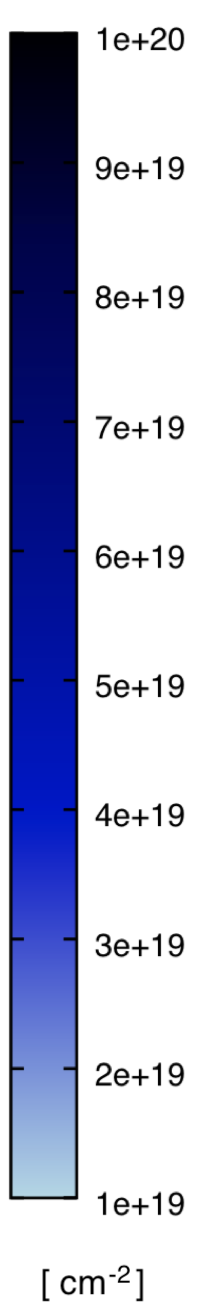}} & 
\resizebox{26mm}{!}{\includegraphics{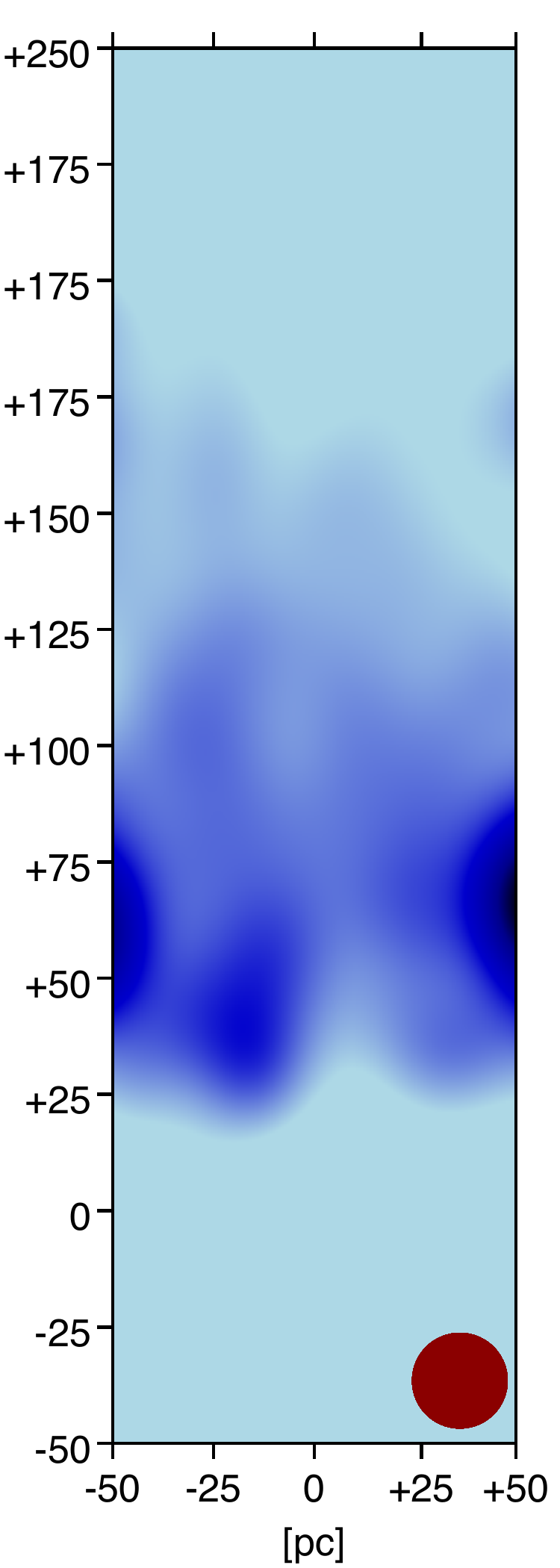}} & \hspace{-0.55cm}\resizebox{26mm}{!}{\includegraphics{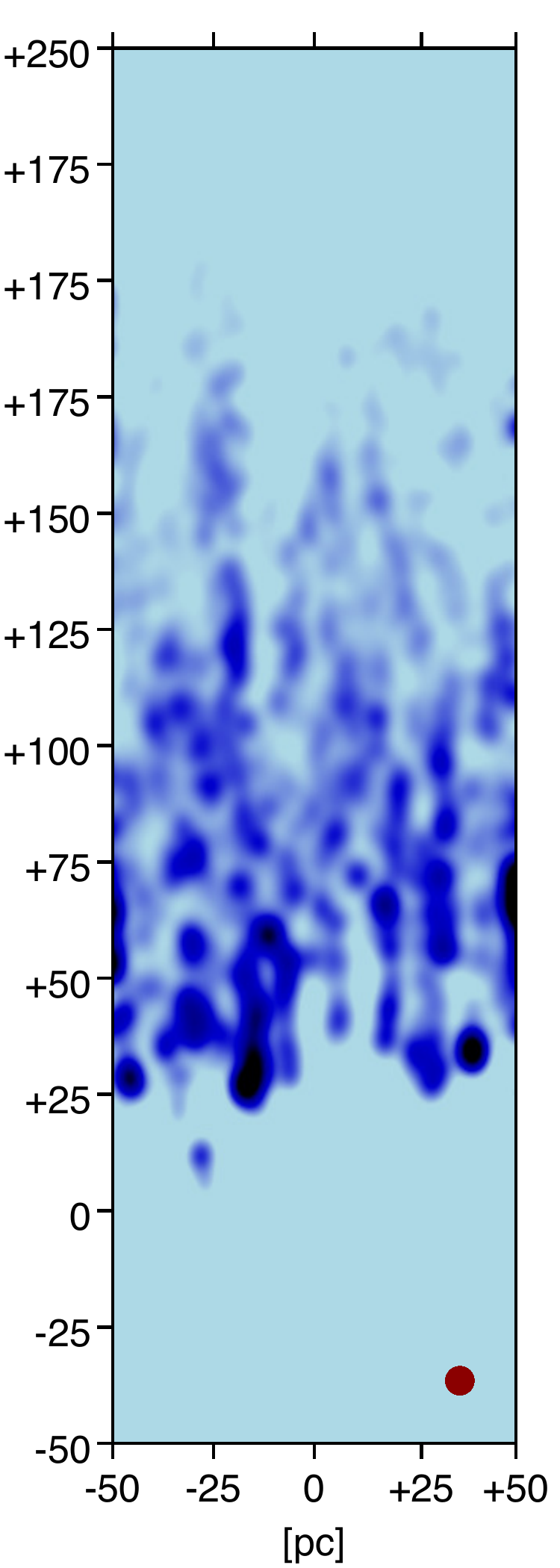}} & \hspace{-0.55cm}\resizebox{26mm}{!}{\includegraphics{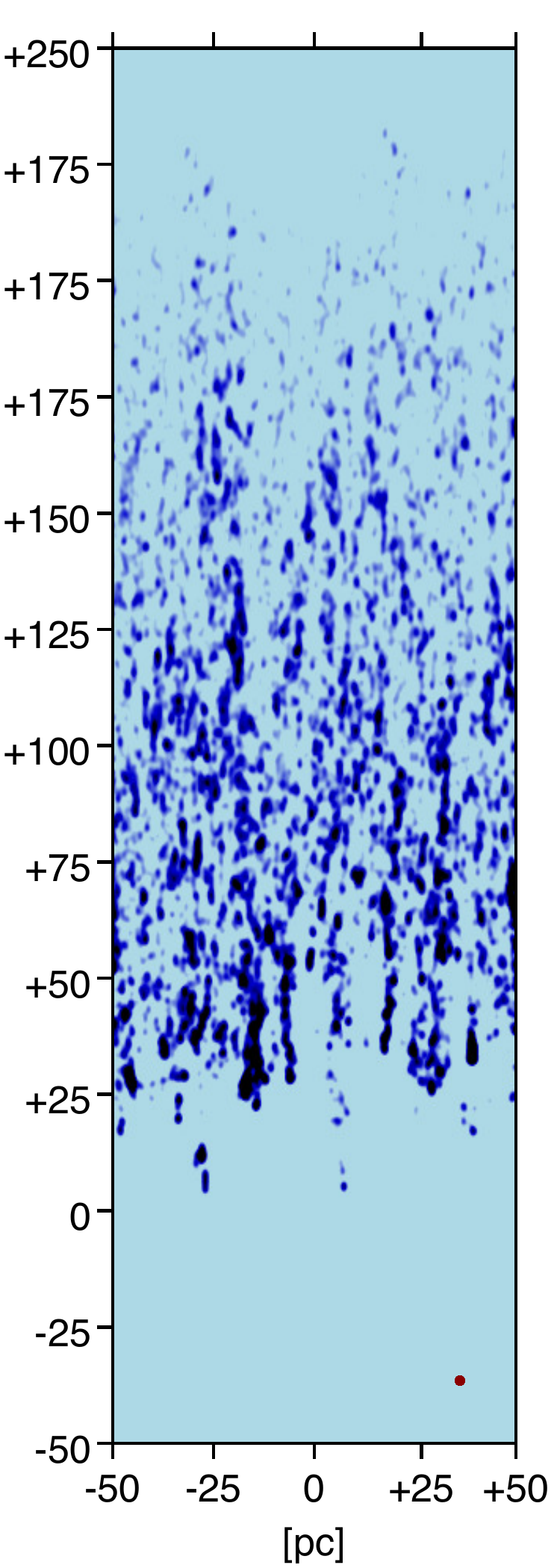}} &
\hspace{-0.40cm}\resizebox{11mm}{!}{\includegraphics{barra2.png}}\\
\multicolumn{4}{l}{\resizebox{80mm}{!}{\includegraphics{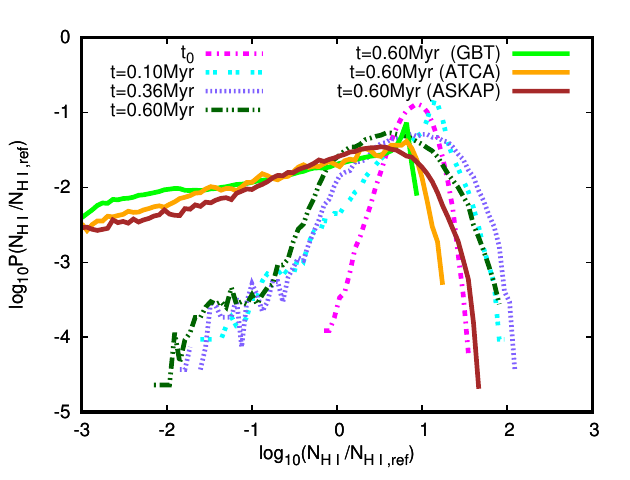}}} & \multicolumn{4}{r}{\resizebox{80mm}{!}{\includegraphics{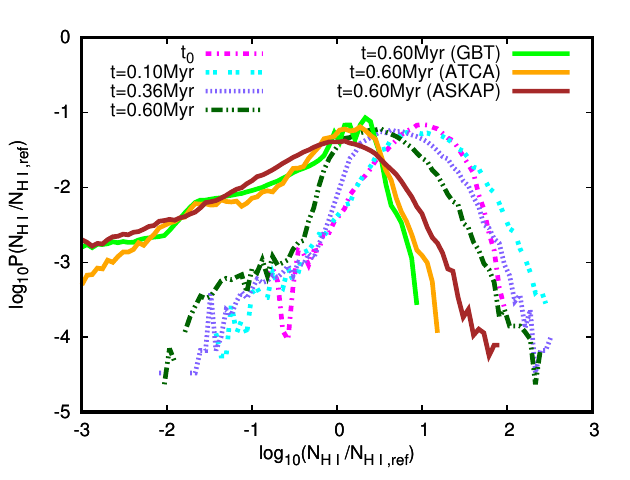}}} \\
  \end{tabular}
  \caption{Synthetic column number density maps of cold and warm H\,{\sc i} gas, $N_{\rm H\,{\scriptstyle I}}$, at $t=0.3\,t_{\rm sp}=0.60\,\rm Myr$ (top panels), and column number density histograms, normalised with respect to $N_{\rm H\,I, ref}=10^{19}\,\rm cm^{-2}$, as a function of time and surveying telescope (bottom panels), for both compact solenoidal (left column) and porous compressive (right column) multicloud systems. We assume a sensitivity threshold of $N_{\rm H\,I}=10^{18}\,\rm cm^{-2}$ for the top panels, and consider all the H\,{\sc i} gas for the bottom panels. The beam size of each telescope is represented by a red circle in the bottom right corner of the projections. Overall, H\,{\sc i} column number densities between $10^{19-21}\,\rm cm^{-2}$ should be expected in observed systems where warm and cold gas precipitates from warmer mixed phases via fast cooling.} 
  \label{Figure12}
\end{center}
\end{figure*}

In this context, we now compare some of the results in our models with the observed properties of the $\rm {H}\,${\scriptsize{I}} clouds in the GC. For this analysis we select our simulation outputs at the latest time, $t=0.3\,t_{\rm sp}=0.60\,\rm Myr$. We consider cloud gas with temperatures $10^2\,\rm K<T\leq 5\times10^3\,\rm K$ as $\rm {H}\,${\scriptsize{I}} gas, and for simplicity we assume that the optical depth of the gas is 0, i.e., that we can see all the $\rm {H}\,${\scriptsize{I}} emitting gas along the line of sight. Panels 9a and 9b in Figure \ref{Figure9} indicate that within that time such gas reaches average distances $140\,\rm pc<\langle d\rangle<150\,\rm pc$ and acquires average bulk speeds $300\,\rm km\,s^{-1}<\langle v_y\rangle<340\,\rm km\,s^{-1}$ in compact solenoidal models, and $90\,\rm pc<\langle d_y\rangle<120\,\rm pc$ and $180\,\rm km\,s^{-1}<\langle v_y\rangle <260\,\rm km\,s^{-1}$ in porous compressive models, respectively.\par

While we do not know at what stage of their evolution we observe the atomic clouds in the GC outflow or their initial positions, these values imply that $\rm {H}\,${\scriptsize{I}} gas can definitely be found at least $100\,\rm pc$ beyond the initial site of shock-multicloud interactions. Similarly, our results indicate that such gas can reach average speeds $\sim 200-300\,\rm km\,s^{-1}$. To know whether or not this gas continues accelerating and how far it can travel, we would need to extend the vertical extent of our simulation domains. However, the simulations presented here show that re-condensation can explain the presence of high-velocity dense gas (at least) near the base of the GC outflow.\par

\subsection{H\,{\sc i} properties}
Our simulations show that $\rm {H}\,${\scriptsize{I}}-emitting gas has line-of-sight velocity dispersions $\lesssim 16\,\rm km\,s^{-1}$, i.e., $\Delta v_{\rm FWHM}\lesssim 37\,\rm km\,s^{-1}$ (see panel 9c in Figure \ref{Figure9}) and average column number densities of $\bar{N_{\rm H_{\rm I}}}\approx 5\times10^{19}\,\rm cm^{-2}$ (see Figures \ref{Figure10} and \ref{Figure12}). While our models represent only small 3D sections of an outflow and we assume that the optical depth is negligible, these values are in good agreement with those found in the GC atomic gas, i.e., $3\lesssim\Delta v_{\rm FWHM}\lesssim 31\,\rm km\,s^{-1}$ and $N_{\rm H\,{\scriptstyle I}}=10^{18}$--$10^{21}\,\rm cm^{-2}$ (see \citealt{2013ApJ...770L...4M,2018ApJ...855...33D}).\par

In Figure \ref{Figure12} we show maps of the hydrogen column number densities, $N_{\rm H\,{\scriptstyle I}}$, projected along the $Z$ direction in the compact solenoidal (panel 12a) and porous compressive (panel 12b) multicloud models. The maps correspond to the latest time in our simulation, i.e., $t=0.3\,t_{\rm sp}=0.60\,\rm Myr$, and they have been smoothed so that the resolution in the synthetic maps reflects the beam size of different radio telescopes (see Table 1 in \citealt{2020ApJ...888...51L}, and \citealt{2013PASA...30....3D}). The dark red circle represents the beam size of the Green Bank telescope (GBT, with an angular resolution of $9\farcm 1$), the Australia Telescope Compact Array (ATCA, with an angular resolution of $2\farcm 4$), and the Australian Square Kilometre Array Pathfinder (ASKAP, with an angular resolution of $0\farcm 5$), which have been and are being used to survey H\,{\sc i} in the GC region. Additionally, our maps are accompanied by their respective histograms below. The histograms show the shapes of the column number density PDFs expected for different telescopes and also at distinct times in the evolution of the systems.\par

The volume-weighted $N_{\rm H\,{\scriptstyle I}}$ PDFs show that as time progresses the distributions become broader with the peaks moving rather slowly towards lower $N_{\rm H\,{\scriptstyle I}}$ values. In general the average $\bar{N}_{\rm H\,{\scriptstyle I}}$ remains within the $10^{19}-10^{20}\,\rm cm^{-2}$ range. After smoothing the projections to account for the different beam sizes of the radio telescopes, the PDFs in all cases become even broader and biased towards lower $N_{\rm H\,{\scriptstyle I}}$ values. The peak values also move towards lower values, thus indicating that the real $N_{\rm H\,{\scriptstyle I}}$ column densities are potentially higher than what can be calculated from current observations.\par

Similarly, our synthetic $N_{\rm H\,{\scriptstyle I}}$ maps show that the seemingly-diffuse $\rm {H}\,${\scriptsize{I}} gas observed with, e.g., GBT has an intricate substructure, which is only captured with interferometry-based instruments. ATCA can resolve that substructure better, but underestimates the maximum $N_{\rm H\,{\scriptstyle I}}$, while ASKAP would perform better and would also reduce the bias towards low $N_{\rm H\,{\scriptstyle I}}$ values. Interestingly, the density structure of the $\rm {H}\,${\scriptsize{I}} phase in the outflow also retains some information of the initial density structure in the multicloud system. Figure \ref{Figure12} reveals that compact (solenoidal) systems are more spatially cohesive than their more vertically extended porous (compressive) counterparts.

\section{Resolution effects and limitations}
\label{sec:Resolution}
In this section we discuss how the numerical resolution we employ for these simulations influences the results. For the non-radiative models presented in \citetalias{2020MNRAS.499.2173B}, we found that the global properties of the shock-swept flow is well captured at relatively low resolution. For example, 8~cells per cloudlet radius was sufficient to capture the dynamical evolution, mass losses, and some of the trends of variables related to the generation of vorticity and turbulence, which were also the most affected by resolution. Therefore, we concluded that the standard resolution of 16 cells per cloudlet radius we used for those models was sufficient to ensure convergence. Such resolutions are at least a factor of 4 lower than the resolution we need to capture the evolution of individual, isolated shock-swept cloudlets, which is of at least 64 cells per cloud radius (e.g, see \citealt{2015ApJ...805..158S,2016MNRAS.455.1309B,2016MNRAS.457.4470P}). The reason for this is that in a multicloud medium, the intra-layer dynamical interactions (e.g. cloud coalescence, see \citealt{2019ApJ...876L...3W}; and cloud-cloud collisions, see \citealt{2020MNRAS.499.4918A}) are dominant and they are well captured by global averages (although in our current set-ups we are unable to disentangle the contribution of coalescence from collisions when cloudlets merge), while in the case of single clouds the external interactions are dominant and therefore the diagnostics are more affected by the resolution we choose for the simulations.\par

\begin{figure*}
\begin{center}
  \begin{tabular}{c c c}
     \multicolumn{1}{l}{13a) sole-k8-M10-rad-lr} & \multicolumn{1}{l}{13b) sole-k8-M10-rad} & \multicolumn{1}{l}{13c) sole-k8-M10-rad-hr} \\
    \hspace{-0.4cm}\resizebox{52mm}{!}{\includegraphics{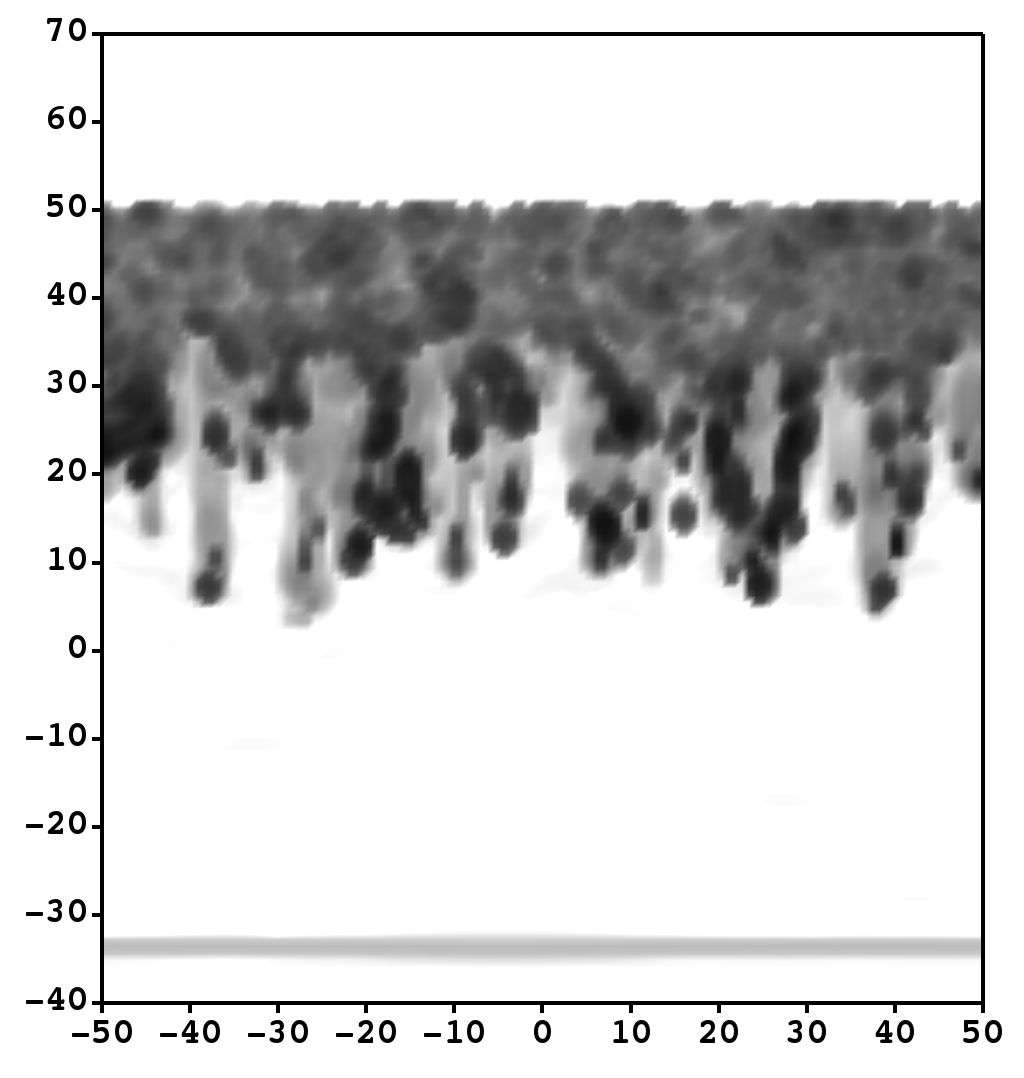}} & \hspace{-0.9cm}\resizebox{52mm}{!}{\includegraphics{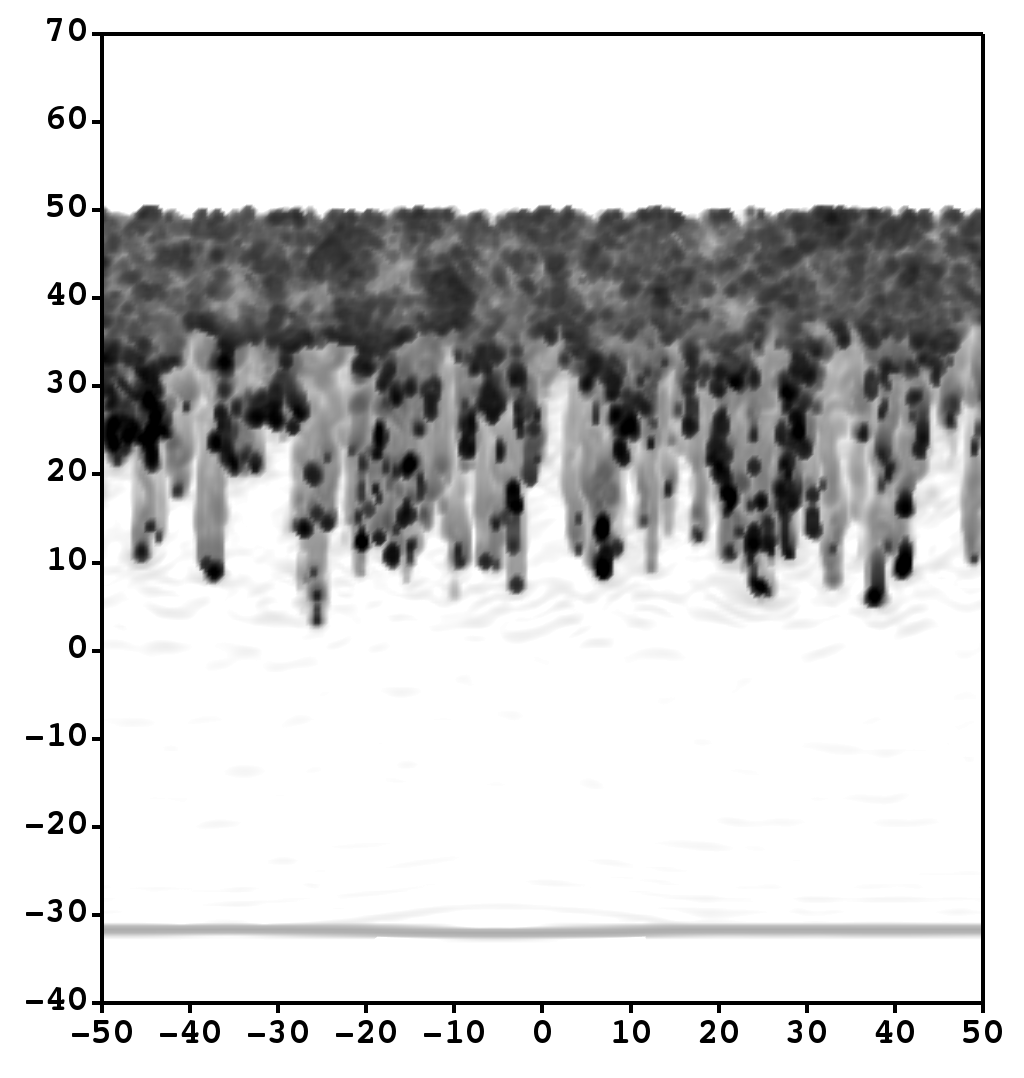}} & \hspace{-0.7cm}\resizebox{52mm}{!}{\includegraphics{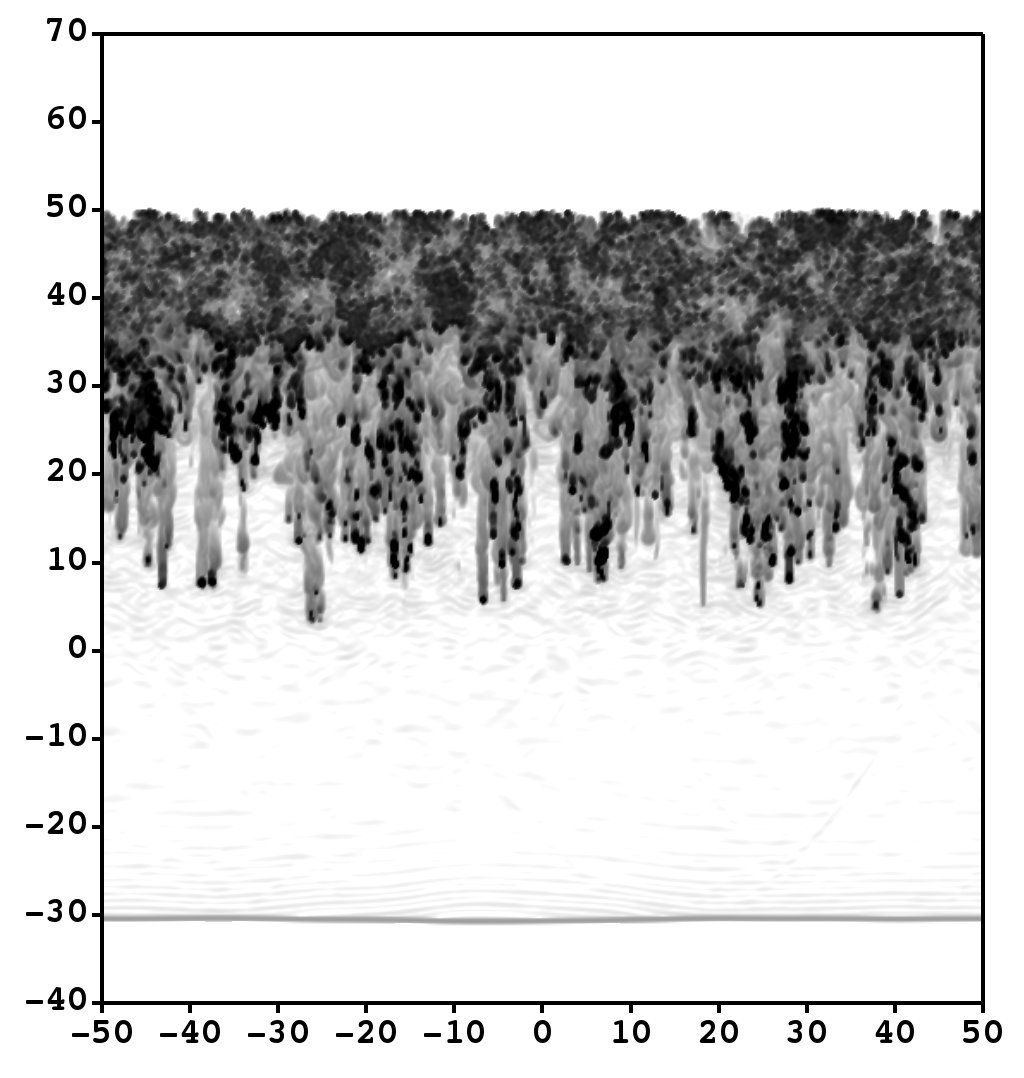}} \hspace{-0cm}\resizebox{12.5mm}{!}{\includegraphics{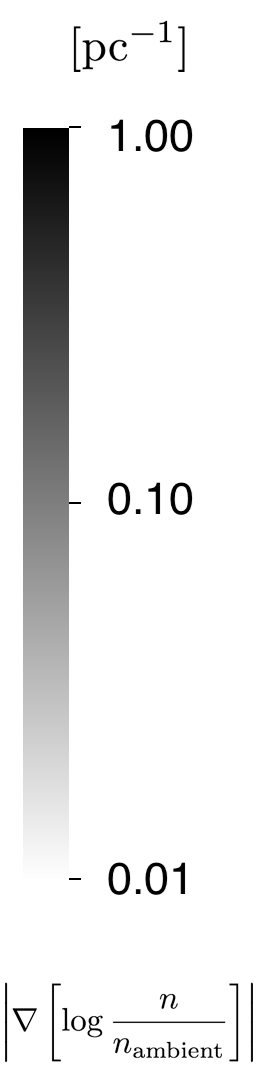}}\\
     \multicolumn{1}{l}{13d) comp-k8-M10-rad-lr} & \multicolumn{1}{l}{13e) comp-k8-M10-rad} & \multicolumn{1}{l}{13f) comp-k8-M10-rad-hr} \\
    \hspace{-0.4cm}\resizebox{52mm}{!}{\includegraphics{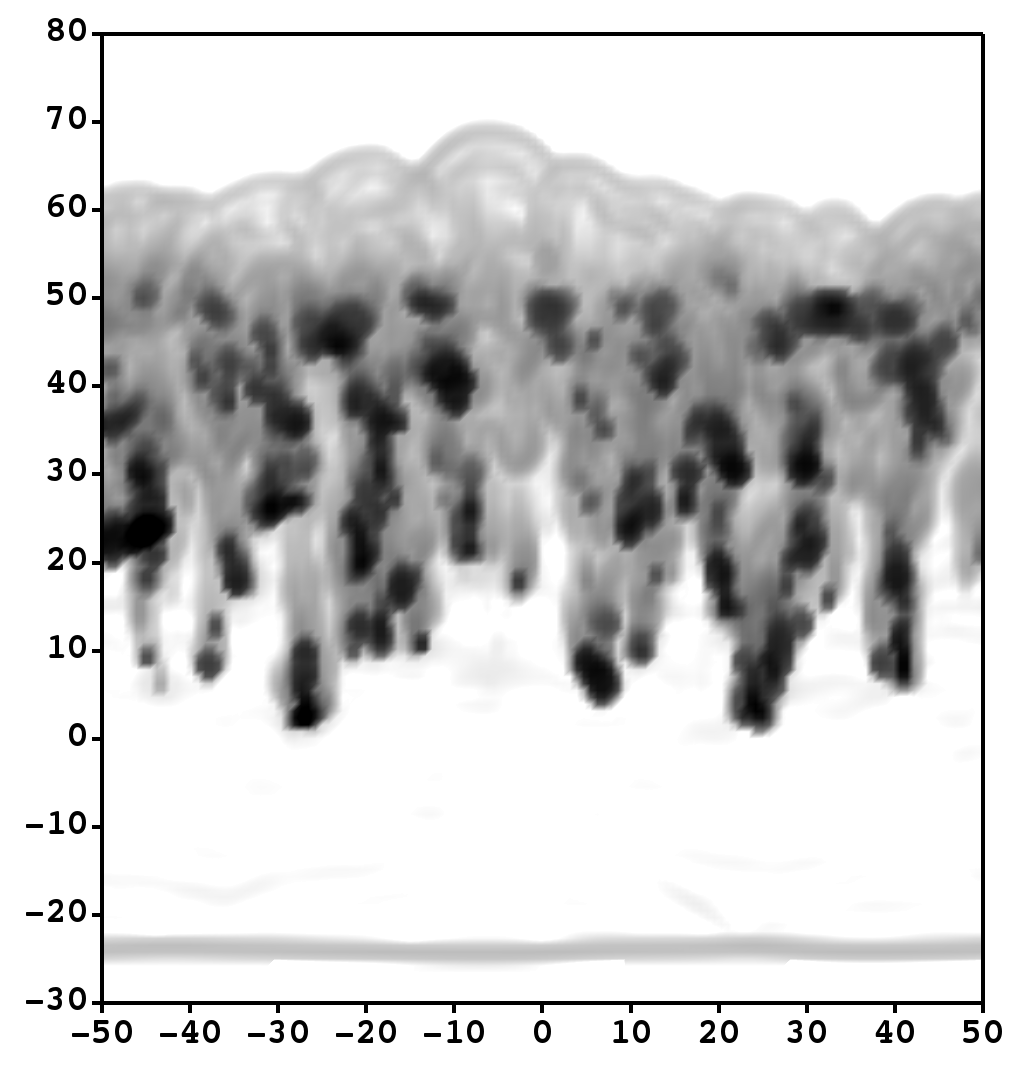}} & \hspace{-0.9cm}\resizebox{52mm}{!}{\includegraphics{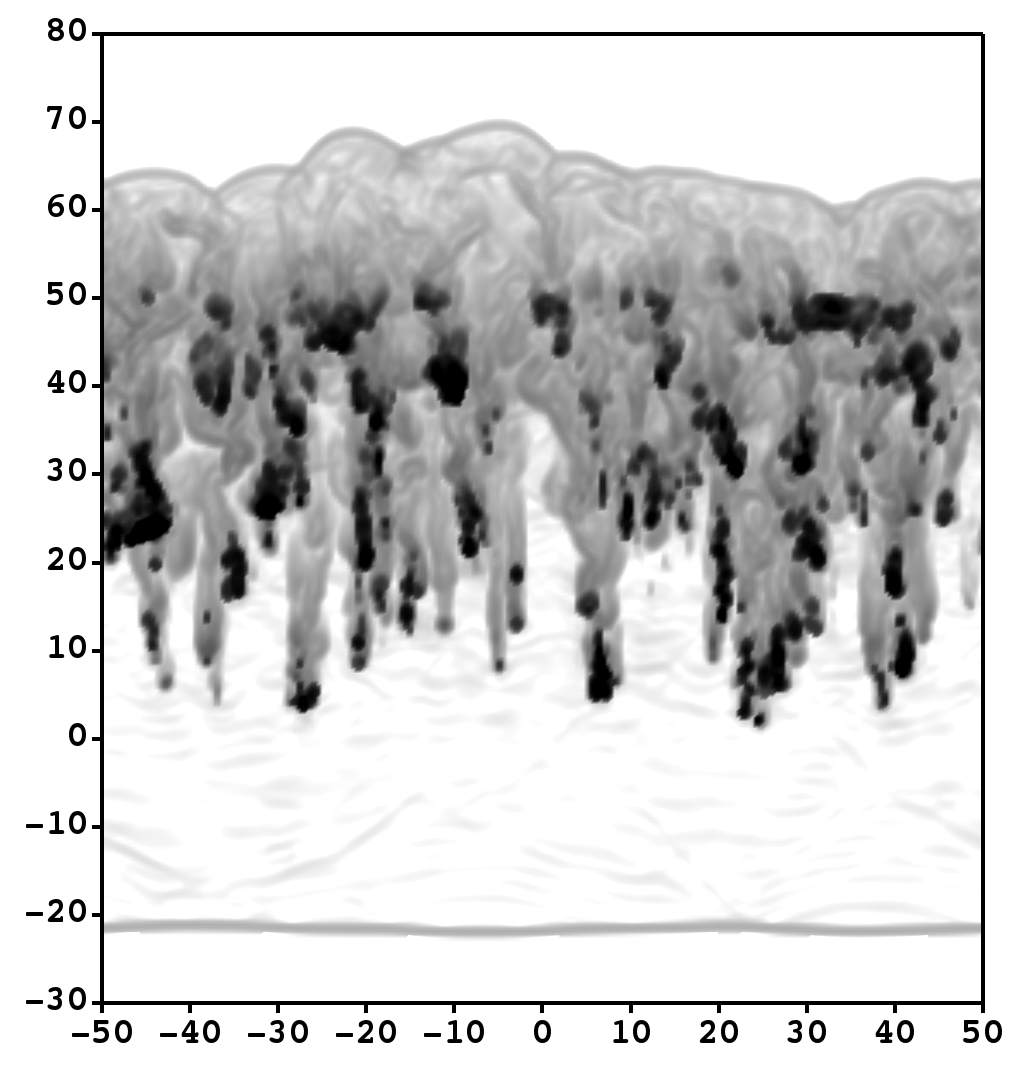}} & \hspace{-0.7cm}\resizebox{52mm}{!}{\includegraphics{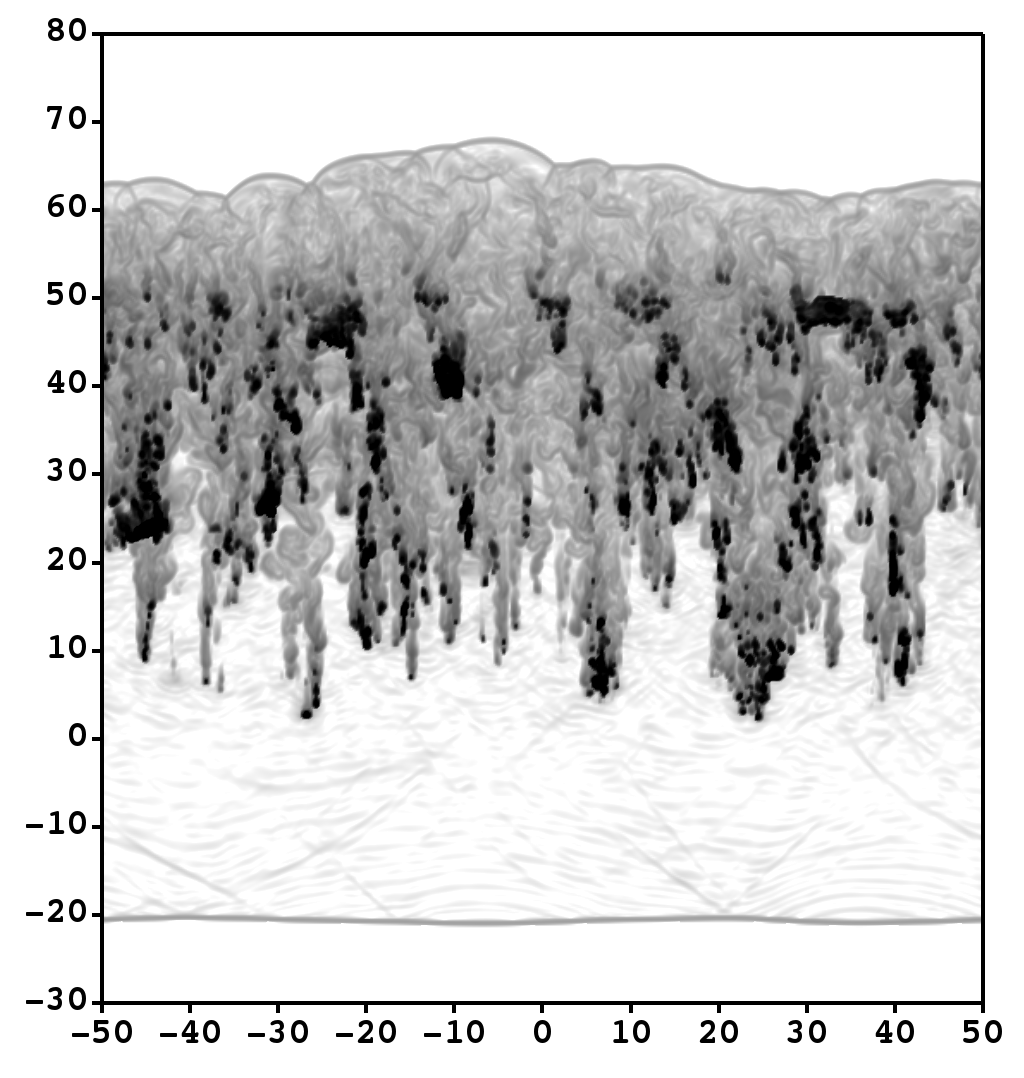}} \hspace{-0cm}\resizebox{12.5mm}{!}{\includegraphics{bar_grad.png}}\\
    \multicolumn{1}{l}{13g) Thermal pressure} & \multicolumn{1}{l}{13h) Volume filling factor} & \multicolumn{1}{l}{13i) Mixing fraction}\\
    \hspace{-0.4cm}\resizebox{60mm}{!}{\includegraphics{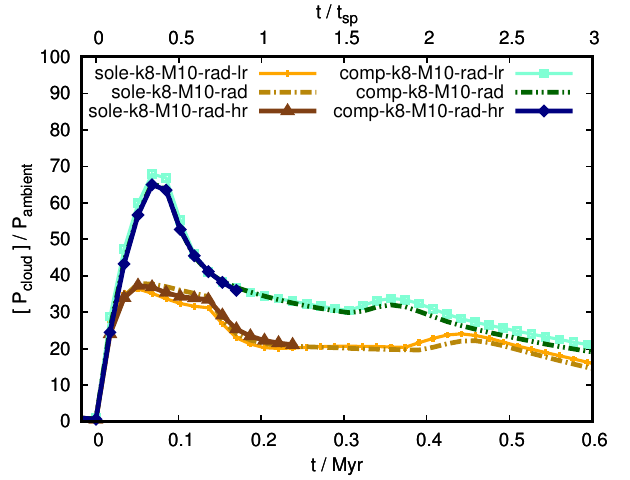}} & \hspace{-0.4cm}\resizebox{60mm}{!}{\includegraphics{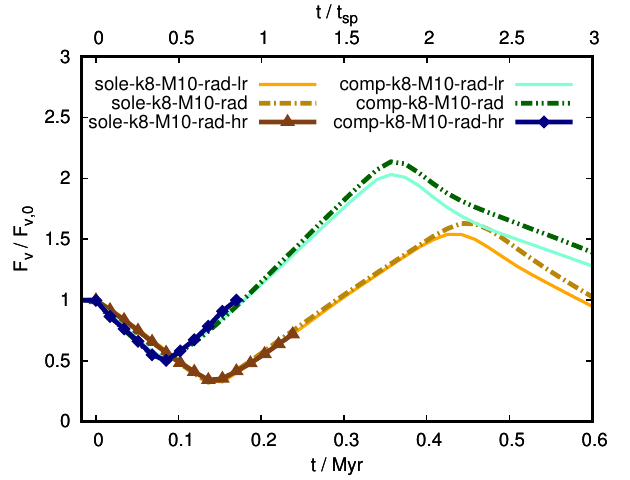}} & \hspace{-0.4cm}\resizebox{60mm}{!}{\includegraphics{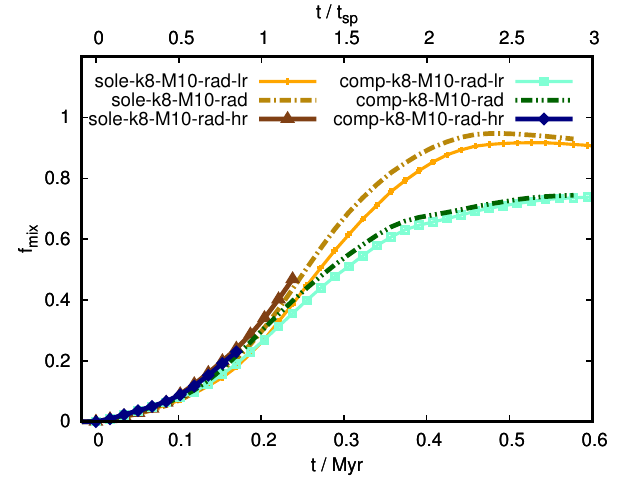}} \\
    \multicolumn{1}{l}{13j) Velocity dispersion} & \multicolumn{1}{l}{13k) Bulk speed} & \multicolumn{1}{l}{13l) Dense gas mass fraction}\\
    \hspace{-0.4cm}\resizebox{60mm}{!}{\includegraphics{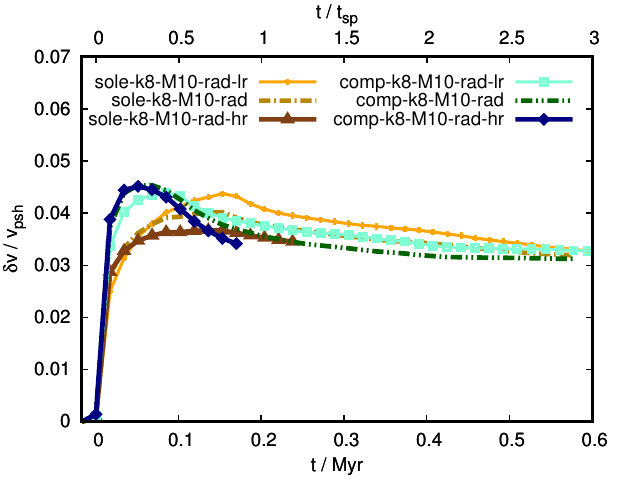}} & \hspace{-0.4cm}\resizebox{60mm}{!}{\includegraphics{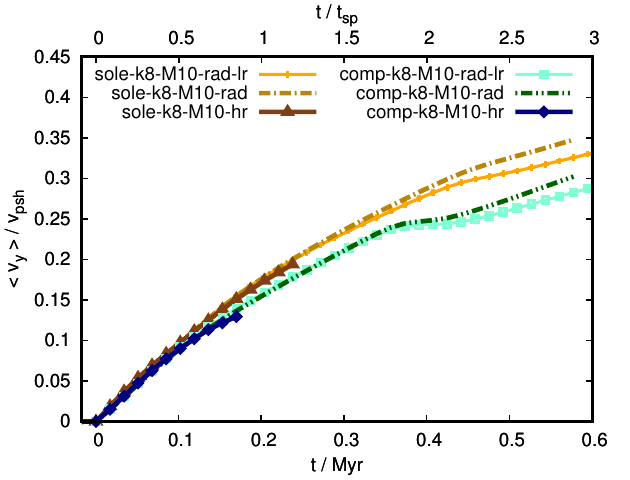}} & \hspace{-0.4cm}\resizebox{60mm}{!}{\includegraphics{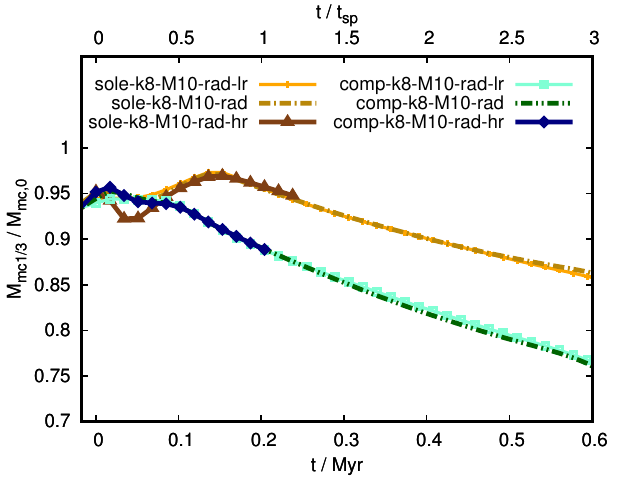}} \\
  \end{tabular}
  \caption{Numerical resolution study of our models. The top two rows show the Schlieren images of compact solenoidal and porous compressive models at increasing resolutions (8, 16, and 32 cells per cloudlet radius, see Table \ref{Table1}), and the two bottom rows show the evolution of six diagnostics: thermal pressure, volume filling factor, mixing fraction, velocity dispersion, bulk speed, and mass fraction of dense gas, respectively. Overall the evolution of the shocked multicloud layer is well captured at all resolutions, but small-scale instabilities and the cooling-induced fragmentation of dense gas are increasingly captured as we increase the resolution.}
  \label{Figure13}
\end{center}
\end{figure*}

To study if the above result holds for radiative multicloud models, we present a set of compact and porous shock-multicloud models at three different resolutions (8, 16, and 32 cells per cloudlet radius; see Table \ref{Table1}). In radiative models, cooling introduces additional length scales in the problem, e.g., the cooling and shattering lengths (see Section \ref{subsec:Diagnostics}). While the global cooling length is resolved in all our models, the shattering length is unresolved even by our highest-resolution models. Such models require time-consuming computations, and thus we were unable to run them beyond $\sim 0.2\,\rm Myr$. However, in both cases this time-scale is sufficient to carry out a meaningful comparison, which we expect to hold for the rest of the evolution of these systems. Panels 13a and 13b of Figure \ref{Figure13} show the Schlieren images of compact solenoidal (top row) and porous compressive (second row) models. In both cases increasing the resolution allows the gas to further fragment, thus showing a more intricate substructure at high resolutions. Similarly, the reflected and internal shock discontinuities become thinner with increasing resolution. In terms of turbulence generation, higher-resolution models capture intra-layer dynamical instabilities in greater detail and smaller-scale eddies are resolved.\par

Despite these expected differences, these panels also show that the overall shape and structure of the shocked multicloud layer is very similar at all resolutions. Thus, while individual fragmented cloudlets are not fully resolved at low resolution, the structure of the multicloud layer as a whole is well captured, even at the lowest resolution of 8 cells per cloudlet radius. The diagnostic quantities displayed in panels 13c and 13d show the same trend. Variables that depend on the generation of turbulence and small-scale eddies, such as the mixing fractions, mass fractions, and the velocity dispersions show slightly larger deviations when increasing the resolution, while variables that represent the dynamics, pressure, and volume of the layer are well captured at all resolutions. Given the evolution of these variables, we conclude that the dynamics and evolution of radiative multicloud layers can be well captured even at resolutions of 8 cells per cloudlet radius, while capturing the small-scale structure and shattering processes inside the multicloud layer does require higher resolutions, which should at least resolve the shattering length.\par

Regarding limitations, the simulations presented in this paper do not include magnetic fields, which are known to affect the dynamics and survival of dense gas (e.g., see \citealt{2018MNRAS.473.3454B,2020ApJ...892...59C}). We will study magnetohydrodynamical (MHD) scenarios in paper III (the reader is referred to \citealt{2014MNRAS.444..971A} for an earlier study on adiabatic shock-multicloud systems with magnetic fields). We have also neglected thermal conduction and self-gravity. Thermal conduction has been shown to delay the destruction of clouds (e.g., see \citealt{2017MNRAS.470..114A}) at the expense of acceleration (e.g., see \citealt{2016ApJ...822...31B}), while self-gravity would become important in regions of strong compression (e.g., see \citealt*{2014MNRAS.444.2884L}). We leave the analysis of these ingredients for future work.

\section{Conclusions}
\label{sec:Conclusions}
We have reported a new set of 3D shock-multicloud models that account for radiative cooling and heating at temperatures between $10^2\,\rm K$ and $10^7\,\rm K$. Our models represent 3D subsections of large-scale galactic outflows where a shock with ${\cal M}_{\rm shock}=10$ propagates across multicloud layers with compact and porous log-normal density fields, characteristic of solenoidal and compressive supersonic (${\cal M}_{\rm turb}\sim 5.5$) turbulence, respectively. We find that:

\begin{itemize}
    \item The ability of clouds to radiate energy leads to the compression of dense gas originally in the cloud layer as well as to the fast precipitation of warm, mixed gas, which would otherwise form the shells seen in non-radiative models (e.g., see \citealt{2012MNRAS.425.2212A}). Cloud gas is heated and eroded by the passage of the shock, but then rapidly cools down downstream.
    
    \item The interplay between heating and cooling creates a complex and long-lasting, multi-phase flow with a rain-like filamentary morphology, which is akin to the outflow structure produced by larger-scale simulations of starburst discs (e.g., see \citealt{2008ApJ...674..157C,2020ApJ...895...43S}). The multi-phase flow is supported by the lifecycle of dense-gas cloudlets, which break up, mix, condense, fragment again, and merge into larger fragments, which then repeat the cycle again. Such outflow structure is absent in non-radiative models, which are characterised by fast-moving, highly-turbulent shells of warm and hot gas containing only a few dispersed, low-momentum, dense cloudlets (see \citetalias{2020MNRAS.499.2173B} for further details).
    
    \item In radiative models, hot gas with temperatures $\gtrsim 10^6\,\rm K$ outruns the warm and cold phases, which reach thermal equilibrium near two temperatures, $\approx 10^4\,\rm K$ and $\approx 10^2\,\rm K$, respectively. For a post-shock flow moving at $1080\,\rm km\,s^{-1}$, we find that the hot, warm, and cold gas phases reach speeds of $\sim 700\,\rm km\,s^{-1}$, $\sim 300$--$400\,\rm km\,s^{-1}$, and $\sim 200$--$300\,\rm km\,s^{-1}$, respectively, at $t=3.0\,t_{\rm sp}=0.60\,\rm Myr$ (see Section~\ref{subsec:Entrainment}).
    
    \item Rather than being ram-pressure accelerated, the warm and cold phases in the outflow continuously precipitate and acquire momentum either from shocked, mixed cloud gas that resides out of thermal equilibrium or from the hot wind itself (see Section~\ref{subsec:Multi-phase}). We find that $\approx 20$ per cent of the total amount of dense gas at $t=3.0\,t_{\rm sp}=0.60\,\rm Myr$ comes from the hot ambient gas, while the rest is recycled mixed gas (originally from the multicloud layer, see Section~\ref{subsection:mixing}). Thus, radiative cooling leads to the replenishment of dense gas in the cold outflow, while heating prevents runaway cooling and allows different phases to coexist. The entrainment/accretion of hot ambient gas into the cold flow leads to a $\sim 20$-per-cent mass growth of dense gas, in agreement with recent studies of single-cloud systems, e.g. see \citealt{2018MNRAS.480L.111G,2020MNRAS.492.1970G}. Both cooling of mixed gas and accretion of hot ambient gas occur and may explain the pervasiveness of dense gas in observed galactic outflows.
    
    \item The volume filling factor of the hot gas phase in the outflow is higher than that of the warm and cold phases, but most of the mass is concentrated in dense gas clouds and filaments with warm ($\sim 10^4\,\rm K$) and cold ($\sim 10^2\,\rm K$) temperatures. Hot gas occupies $>90$ per cent of the volume of outflowing gas, but it only contains $\sim 15$ per cent of its mass. On the other hand, the warm and cold gas phases only occupy $\sim 5$ per cent of the volume, but they have $\sim 80$ per cent of its mass and a high 2D covering fraction when projected along a line-of-sight (in agreement with \citealt{2020MNRAS.491.5056L}).
    
    \item Similarly to \citetalias{2020MNRAS.499.2173B} we find that radiative outflows also contain some imprints of the initial density structure of multicloud systems. While the density PDFs of radiative models evolve into similar bi-modal distributions, the vertical extent, travelled distance, and dense-gas entrainment properties differ in compact solenoidal and porous compressive models. Porous multicloud layers result in more extended vertical outflows than compact multicloud layers, but dense gas is more efficiently produced in the latter. Thus, compact multicloud layers facilitate the entrainment of dense gas as their outflows reach larger distances and velocities than their porous counterparts.
    
    \item Our simulations have important implications for the H\,{\sc i} outflow observed in the GC (\citealt{2013ApJ...770L...4M,2020ApJ...888...51L}). We show that H\,{\sc i} gas, with column number densities $N_{\rm H\,{\scriptstyle I}}=10^{19}$--$10^{-21}\,\rm cm^{-2}$, can travel at least $\sim 100$--$200\,\rm pc$, reach average speeds $\sim 200$--$300\,\rm km\,s^{-1}$, and acquire FWHM velocity dispersions $\sim 20$--$37\,\rm km\,s^{-1}$ within $t=3.0\,t_{\rm sp}=0.60\,\rm Myr$. A molecular gas counterpart is also present, and the trends seen towards the end of our simulations also suggest that H\,{\sc i} gas could move even farther away. Investigating the terminal velocities of these components would require studying these systems in larger computational domains.
    
    \item While the development of dynamical instabilities and the process of cooling-induced fragmentation of cloudlets inside the multicloud layers occurs at increasingly smaller length scales as we double the numerical resolution, we find that the global dynamical evolution of shocked multicloud layers is well captured even at resolutions of $8$ cells per cloud radius. Thus, we confirm that our standard resolution of $16$ cells per cloud radius is adequate for studying shock-multicloud systems, and also that our diagnostics show better convergence properties in multicloud systems than in single-cloud systems.
    
\end{itemize}

Overall, our simulations show that cold gas with number densities between $\sim 1$--$10^3\,\rm cm^{-3}$ and temperatures of $\sim 10^2$--$10^4\,\rm K$ can coexist with hot gas with number densities $\lesssim0.1\,\rm cm^{-3}$ and temperatures of $\gtrsim10^6\,\rm K$ in an evolving three-phase outflow. Studying the role of magnetic fields in non-radiative and radiative shock-multicloud models will be topic of the next paper in this series.





\section*{Acknowledgements}
We thank the anonymous referee for providing insightful comments on this manuscript. WBB is supported by the Deutsche Forschungsgemeinschaft (DFG) via grant BR2026/25. WBB thanks H. D\'enes at ASTRON for reading this manuscript, and also thanks for support from the National Secretariat of Higher Education, Science, Technology, and Innovation of Ecuador, SENESCYT. ES was supported by NSF grant AST-1715876. CF acknowledges funding provided by the Australian Research Council (Discovery Project DP170100603 and Future Fellowship FT180100495), and the Australia-Germany Joint Research Cooperation Scheme (UA-DAAD). AYW is partially supported by the Japan Society for the Promotion of Science (JSPS) KAKENHI grant 19K03862. The authors gratefully acknowledge the Gauss Centre for Supercomputing e.V. (\url{www.gauss-centre.eu}) for funding this project by providing computing time (via grant pn34qu) on the GCS Supercomputer SuperMUC-NG at the Leibniz Supercomputing Centre (\url{www.lrz.de}) and on the GCS Supercomputer JUWELS at the J\"ulich Supercomputing Centre (JSC) under projects 16072 and 19590. We further acknowledge the Australian National Computational Infrastructure for grant~ek9 in the framework of the National Computational Merit Allocation Scheme and the ANU Merit Allocation Scheme. This work has made use of the VisIt visualisation software (\citealt{HPV:VisIt}), and the gnuplot program (\url{http://www.gnuplot.info}). We also thank the developers of the {\sevensize PLUTO} code for making it available to the community. The Starlink software (\citealt{2014ASPC..485..391C}) is currently supported by the East Asian Observatory.

\section*{Data availability}
The data underlying this article will be shared on reasonable request to the corresponding author.




\bibliographystyle{mnras}
\bibliography{RAD_paper} 




\appendix
\section{Clump analysis}
\label{AppendixA}
In this Appendix we discuss the evolution of the general number of cloudlets in the radiative multicloud density distributions analysed in this paper. This analysis complements the description on the number and size of cloudlets provided in Sections \ref{subsec:InitialConditions} and \ref{subsec:CoolingHeating}. In Section \ref{subsec:InitialConditions} we defined $N_{\rm cloudlet,k_{\rm min}}\approx 256$ and $r_{\rm cloudlet,k_{\rm min}}\approx 6.3\,\rm pc$, where $k_{\rm min}=8$, for both compact solenoidal and porous compressive models. These numbers represent the largest number of perturbations or `cloudlets' in the multicloud layers and their respective sizes, but the density fields employed in our simulations are fractal, and therefore they contain substructure at various length scales (or wavenumbers). To study how the general number of cloudlets, $N_{\rm cloudlet,k}$, and their respective sizes, $r_{\rm cloudlet,k}$, for other wavenumbers evolve, we search for clumps in the computational domain with sizes that are larger than those defined by two wavenumbers, $k\sim10$ and $k\sim 20$. For this, we used a customised version of the PyCupid project\footnote{See: \url{https://pycupid.readthedocs.io/en/latest/index.html}}, which is part of the Starlink project (\citealt{2014ASPC..485..391C}) and contains wrappers to the `clumpfind' algorithm developed by \citealt{1994ApJ...428..693W}.\par

\begin{figure}
\begin{center}
  \begin{tabular}{l}
    \hspace{+0.5cm}$N_{\rm cloud, k}$ vs. time\\
    \hspace{-0.40cm}\resizebox{80mm}{!}{\includegraphics{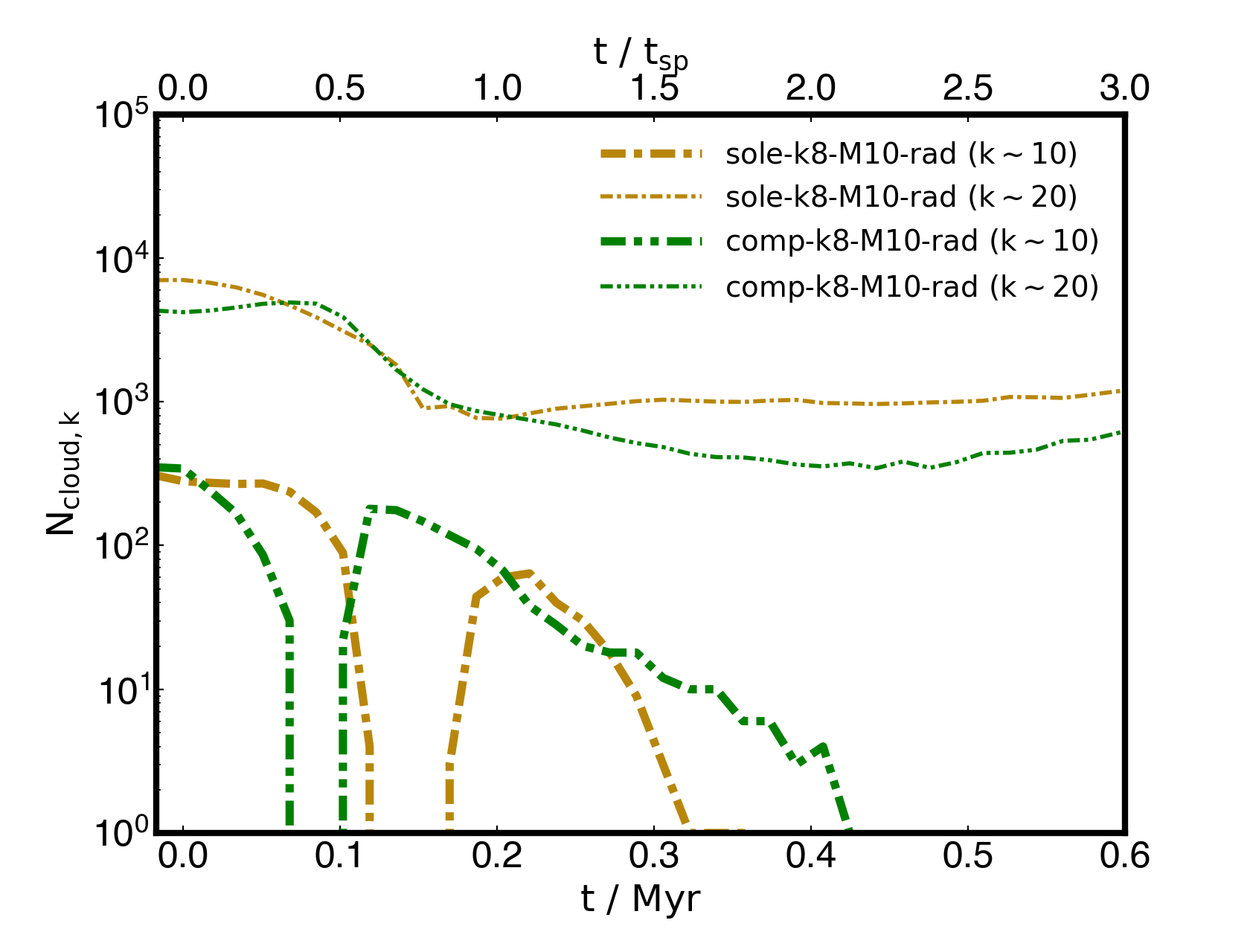}}\\
  \end{tabular}
  \caption{Time evolution of the number of cloudlets, $N_{\rm cloudlet,k}$, with sizes larger than those expected for $k\sim10$ (thick lines) and $k\sim 20$ (thin lines). For comparison we show both radiative multicloud models, sole-k8-M10-rad (compact solenoidal) and comp-k8-M10-rad (porous compressive).} 
  \label{FigureA1}
\end{center}
\end{figure}

To separate the main cloudlets from the density fields we calculated and used the rms value of the initial density fields as background `noise' and defined different contouring levels starting from $1/10$th of the initial mean cloud density and going up to the maximum density in the fields. Using these predefined levels, the algorithm searches for topologically connected structures by defining contours from the highest level to the lowest level and tagging connected regions with different ID numbers. At the end, we exclude tagged substructures that are smaller than a certain predefined volume, which for this analysis corresponds to $k\sim10$ and $k\sim 20$. Using the former value implies we exclude small-scale structure and only detect large-scale structure, while using the latter implies that we also detect smaller `cloudlets' in the distribution. By analysing the evolution of both $N_{\rm cloudlet,k}$, we can study the cloudlet population and their sizes in more detail. In Figure \ref{FigureA1} we show the evolution of $N_{\rm cloudlet,k}$ as a function of time, for compact and porous models. This figure shows that large-scale cloudlets ($k\sim10$) are fragmented by cooling and eroded by the forward shock, so they disappear by the end of the shock crossing phase, then re-emerge during the cloudlet expansion and shock re-acceleration phase, and disappear again during the turbulence generation phase.\par

\begin{figure}
\begin{center}
  \begin{tabular}{c c c}
 \multicolumn{1}{c}{$t=0$} & \multicolumn{1}{c}{$1.8\,t_{\rm sp}=0.36\,\rm Myr$} & \multicolumn{1}{c}{$3.0\,t_{\rm sp}=0.60\,\rm Myr$}\\
       \multicolumn{3}{l}{\hspace{-2mm}a) sole-k8-M10-rad ($k\sim 10$)}\\
       \hspace{-0.3cm}\resizebox{27mm}{!}{\includegraphics{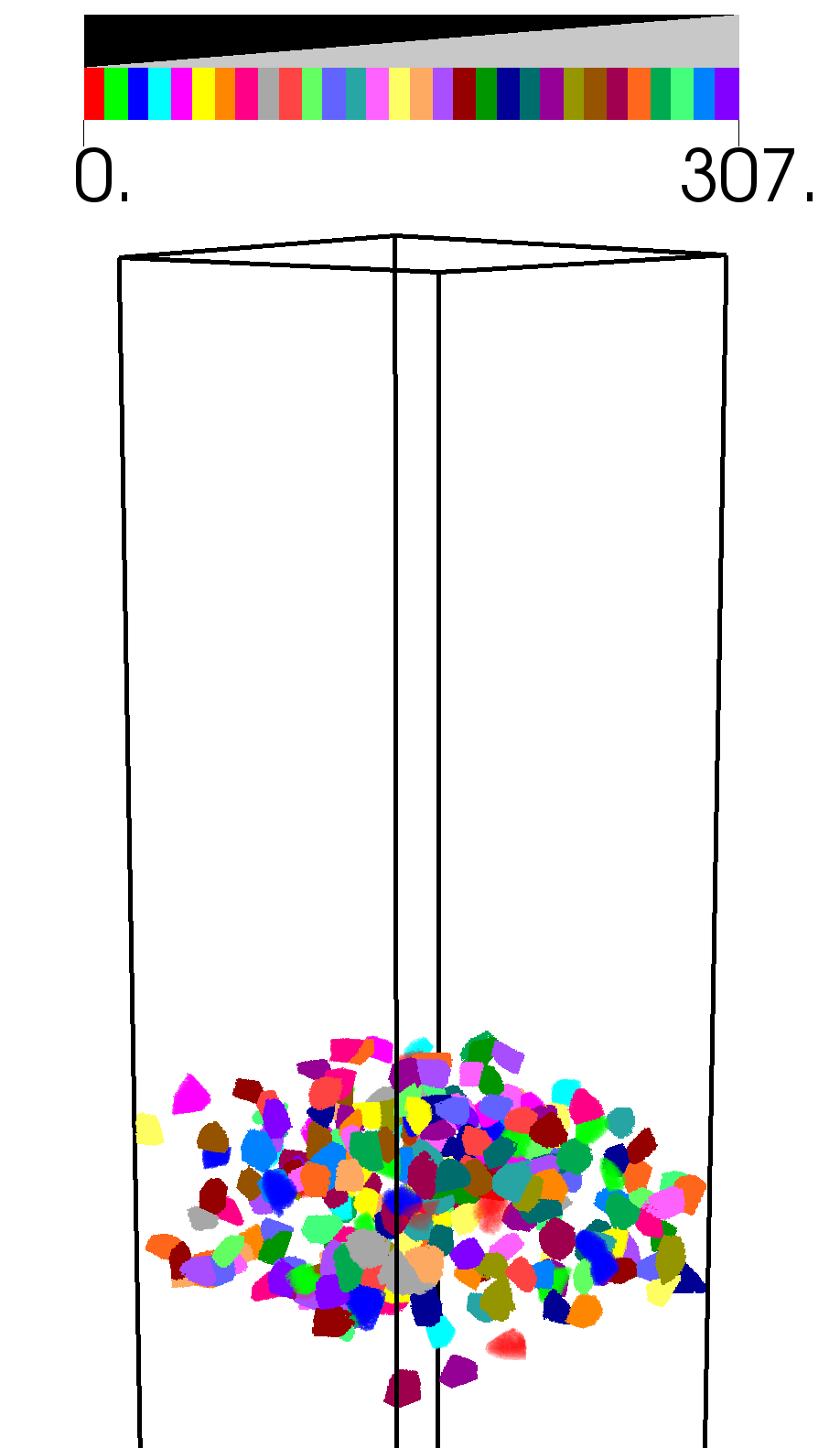}} & \hspace{-0.3cm}\resizebox{27mm}{!}{\includegraphics{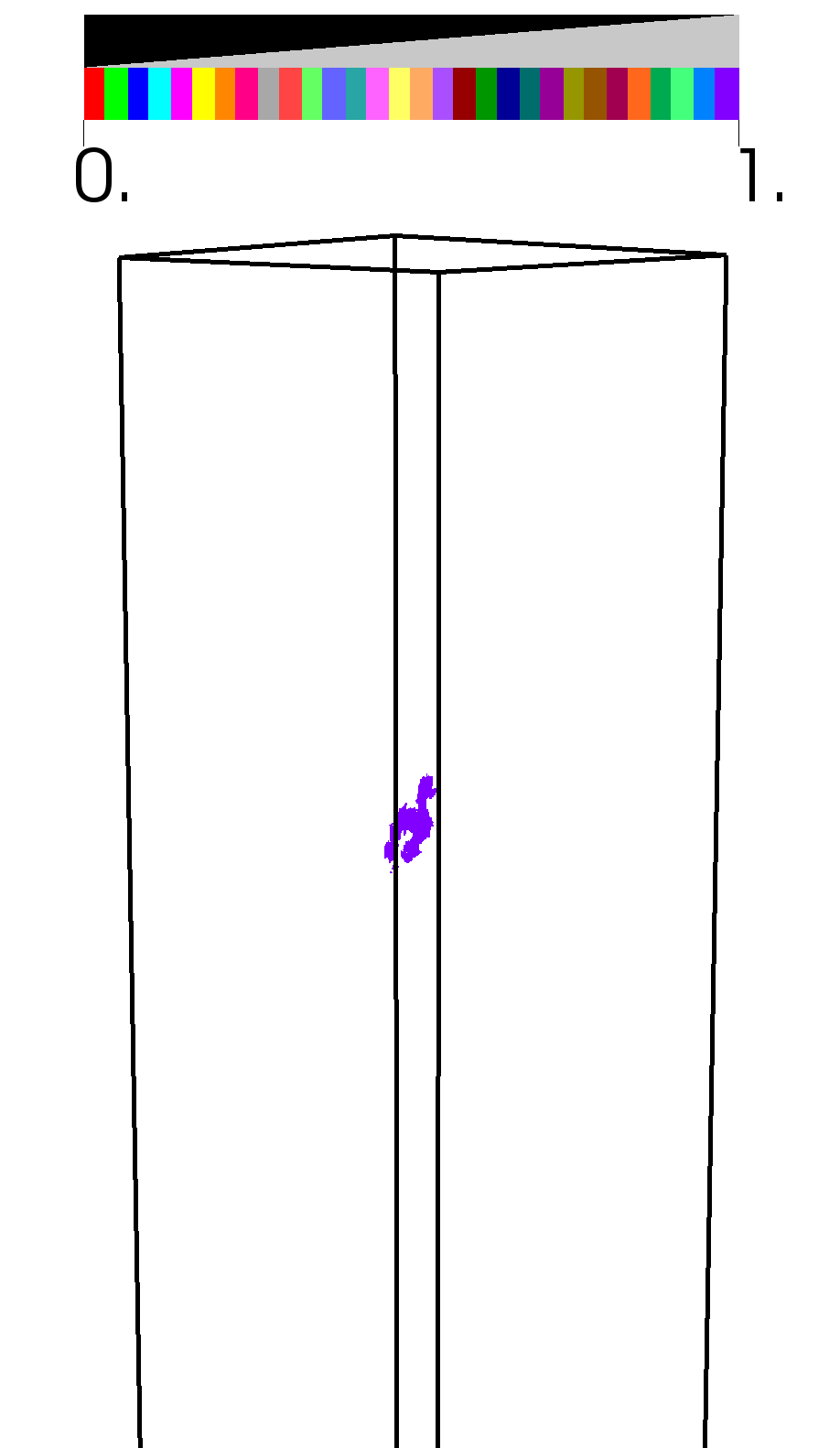}} & \hspace{-0.3cm}\resizebox{27mm}{!}{\includegraphics{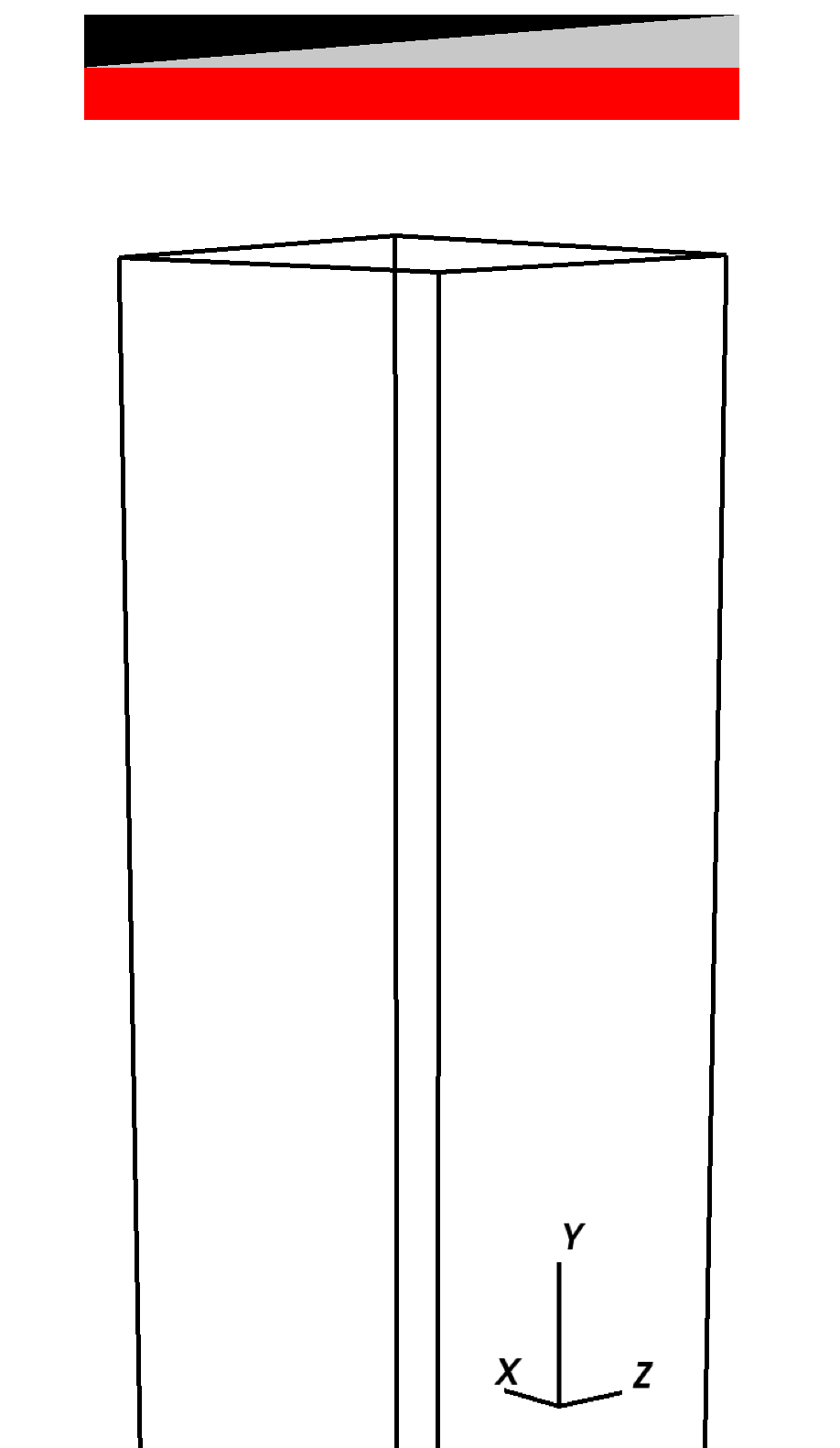}}\\
       \multicolumn{3}{l}{\hspace{-2mm}b) sole-k8-M10-rad ($k\sim 20$)}\\        
       \hspace{-0.3cm}\resizebox{27mm}{!}{\includegraphics{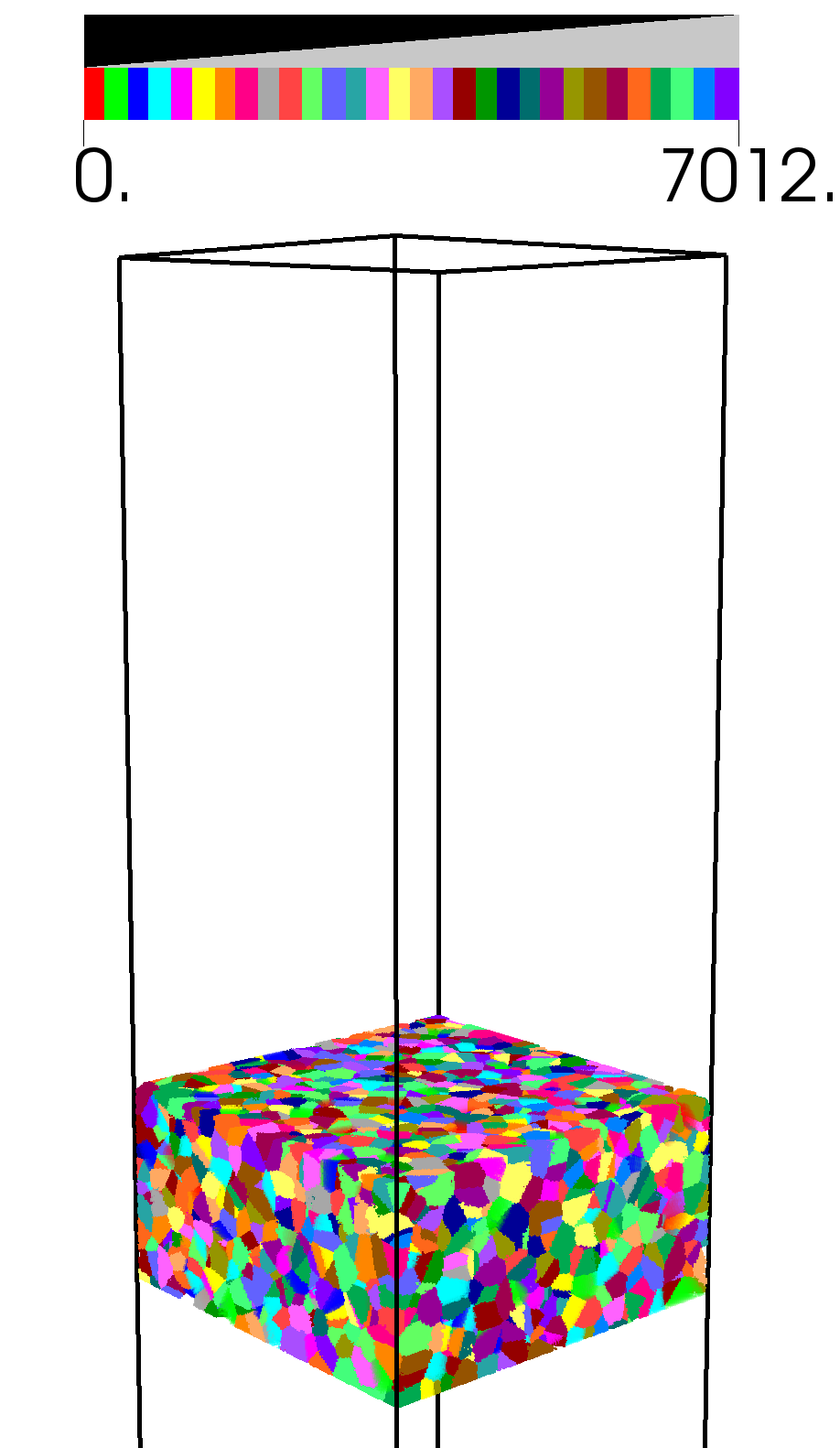}} & \hspace{-0.3cm}\resizebox{27mm}{!}{\includegraphics{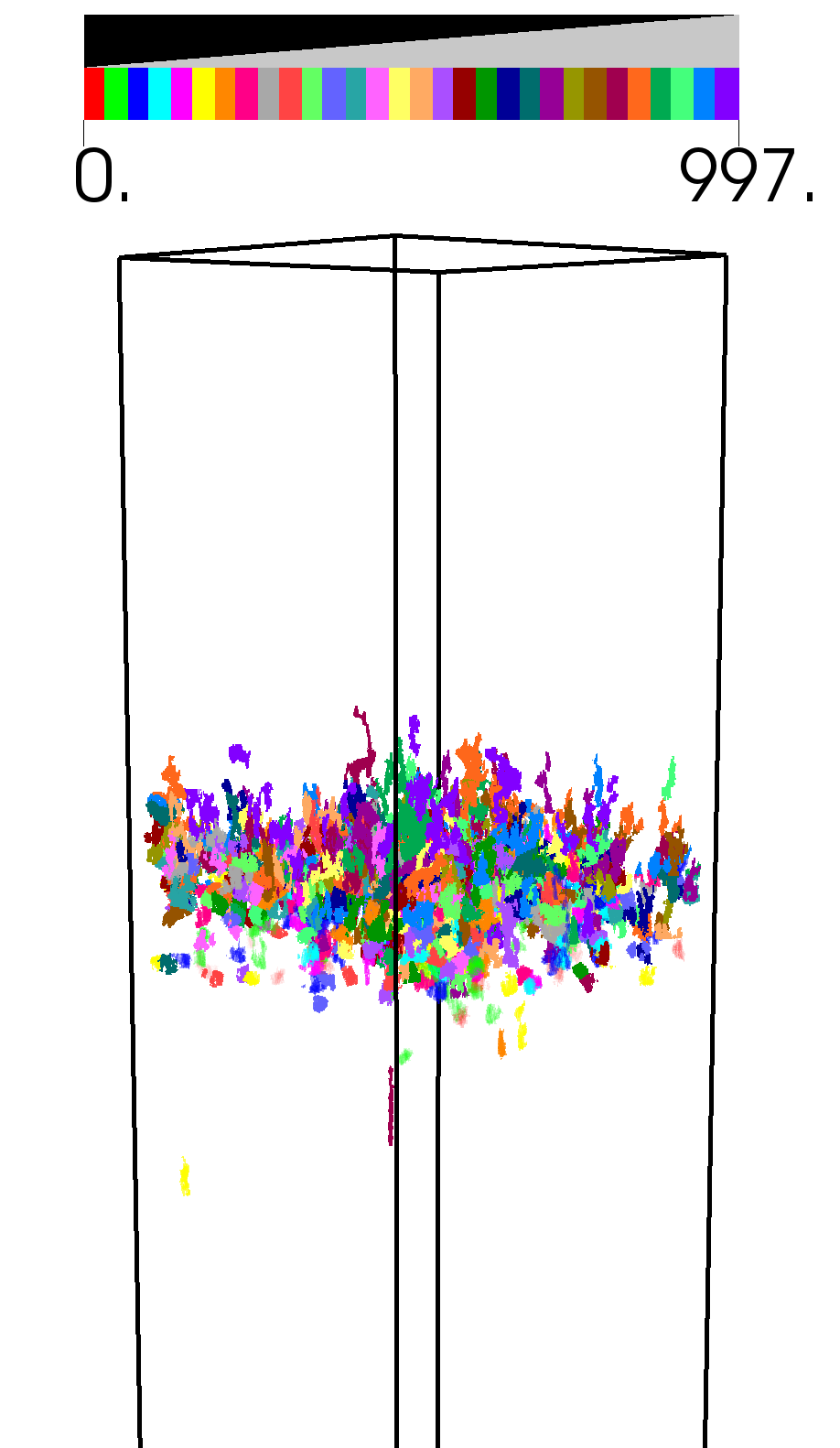}} & \hspace{-0.3cm}\resizebox{27mm}{!}{\includegraphics{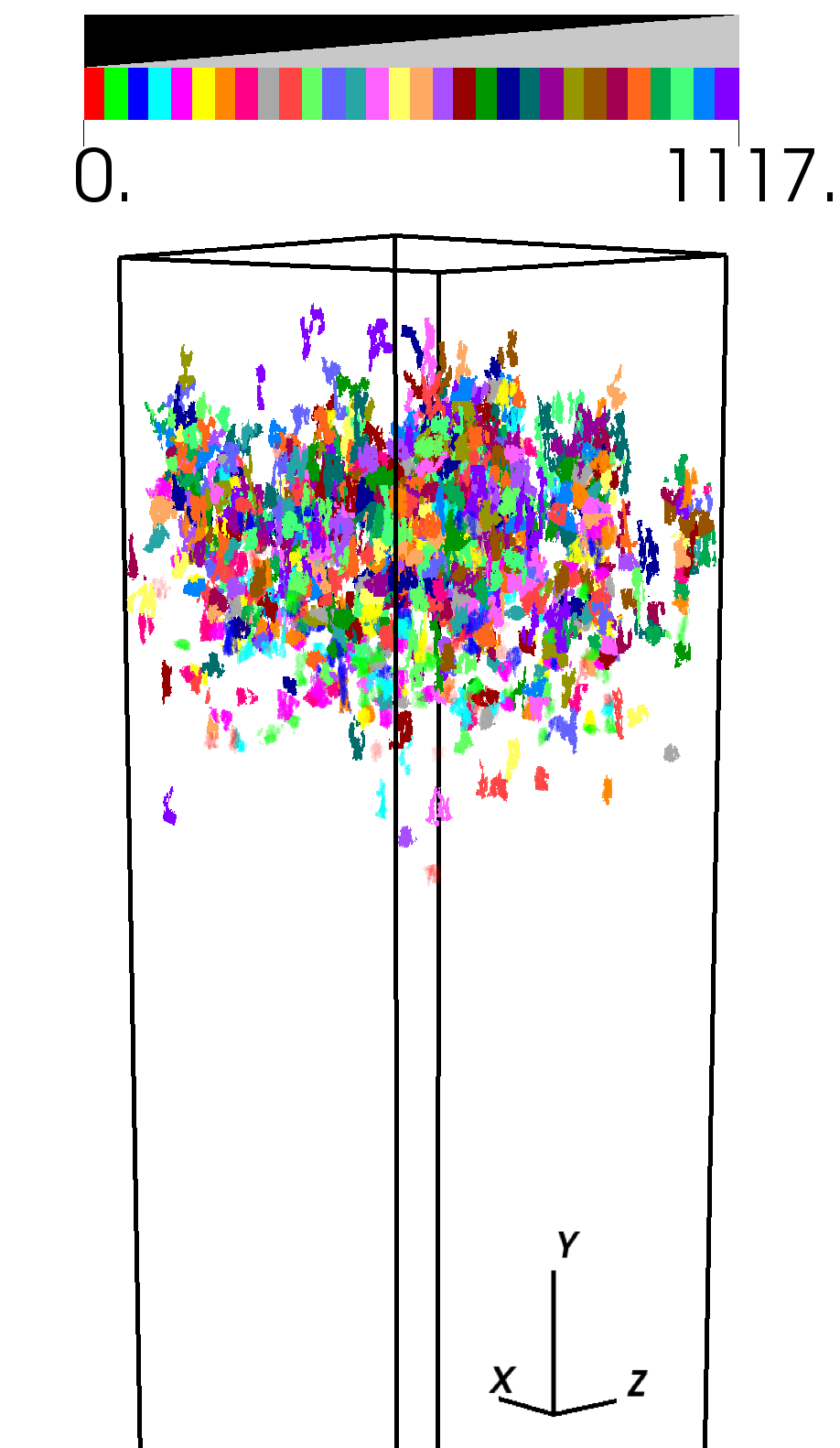}}\\
       \multicolumn{3}{l}{\hspace{-2mm}c) comp-k8-M10-rad ($k\sim 10$)}\\   
       \hspace{-0.3cm}\resizebox{27mm}{!}{\includegraphics{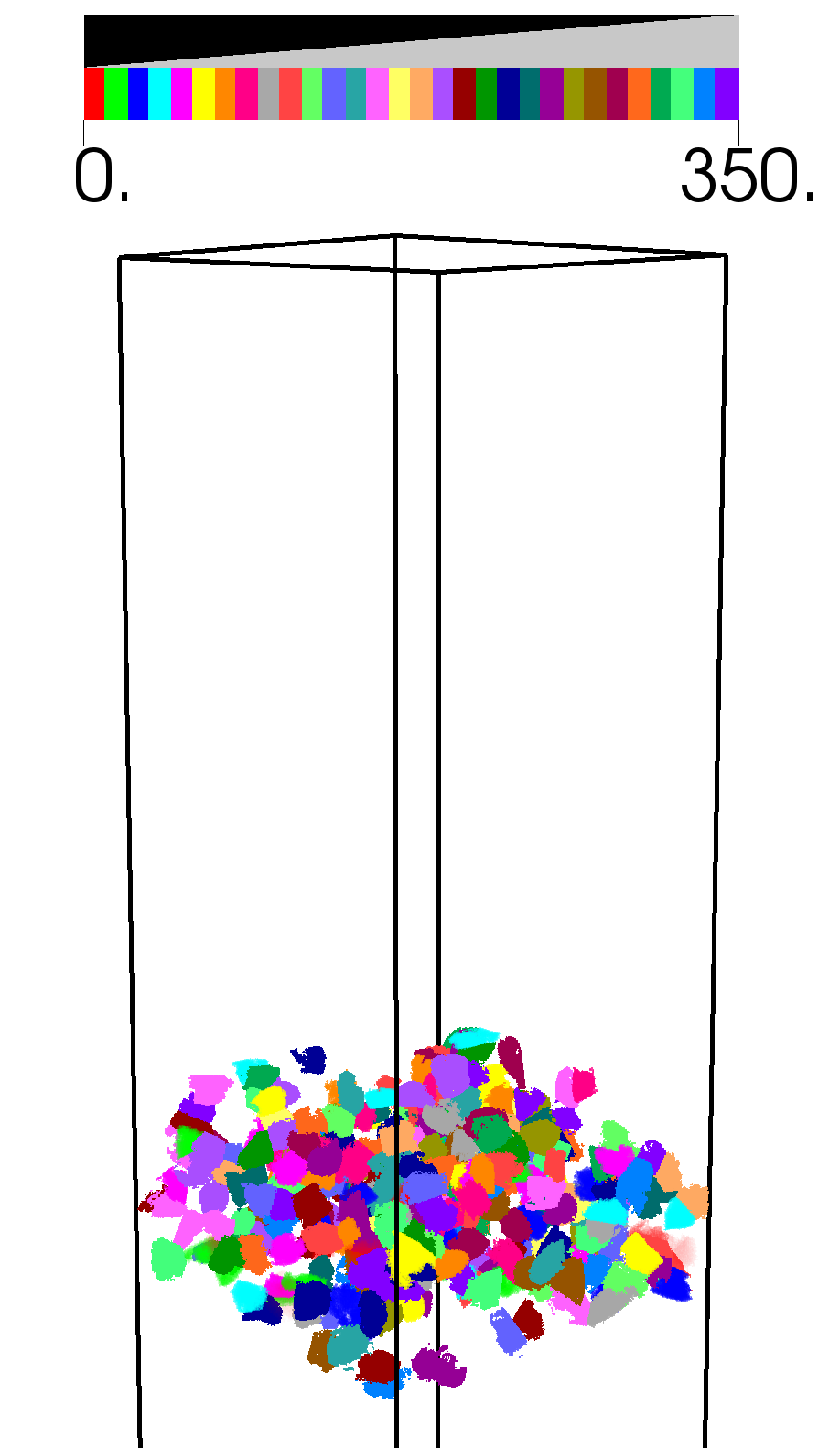}} & \hspace{-0.3cm}\resizebox{27mm}{!}{\includegraphics{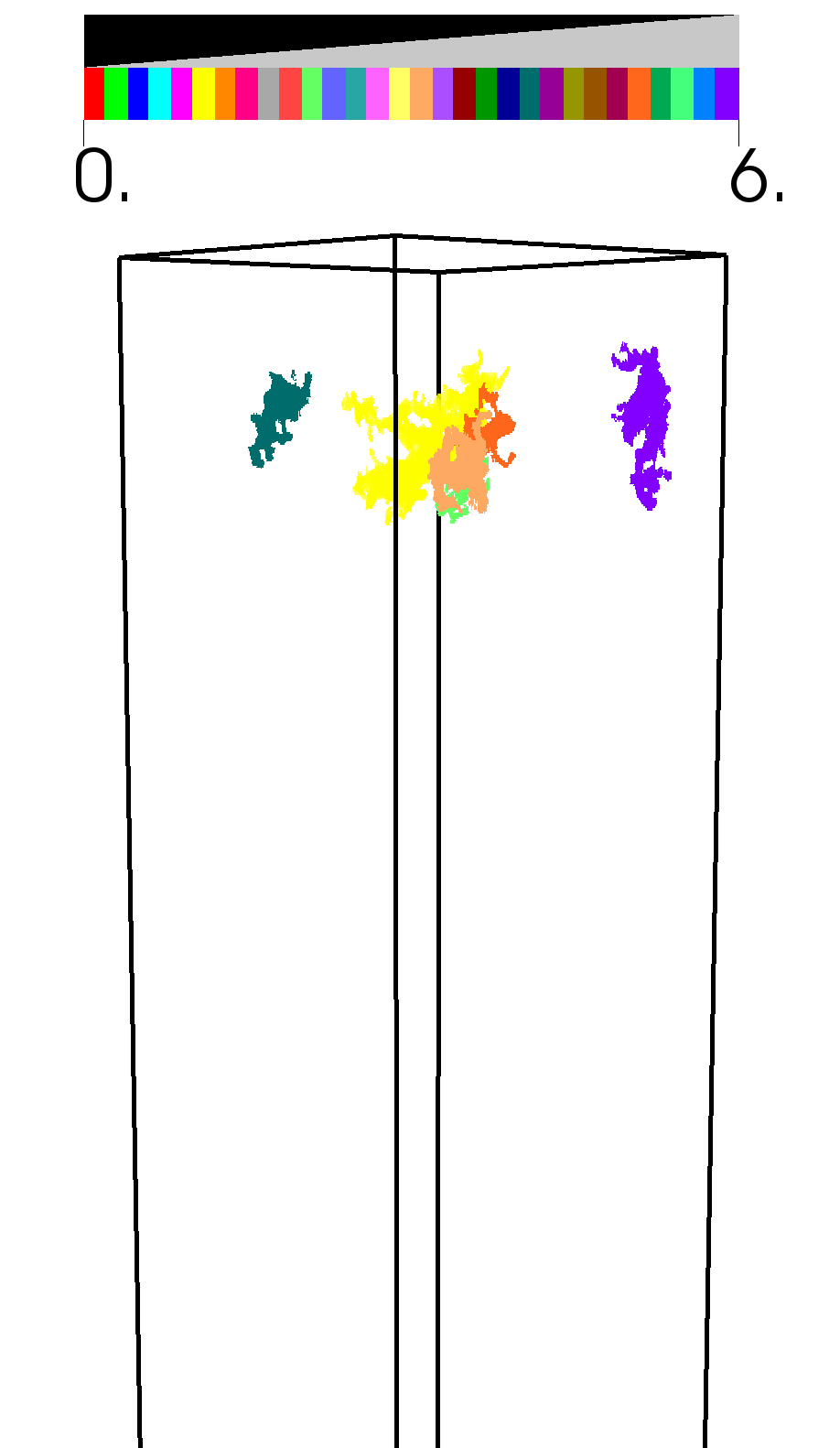}} & \hspace{-0.3cm}\resizebox{27mm}{!}{\includegraphics{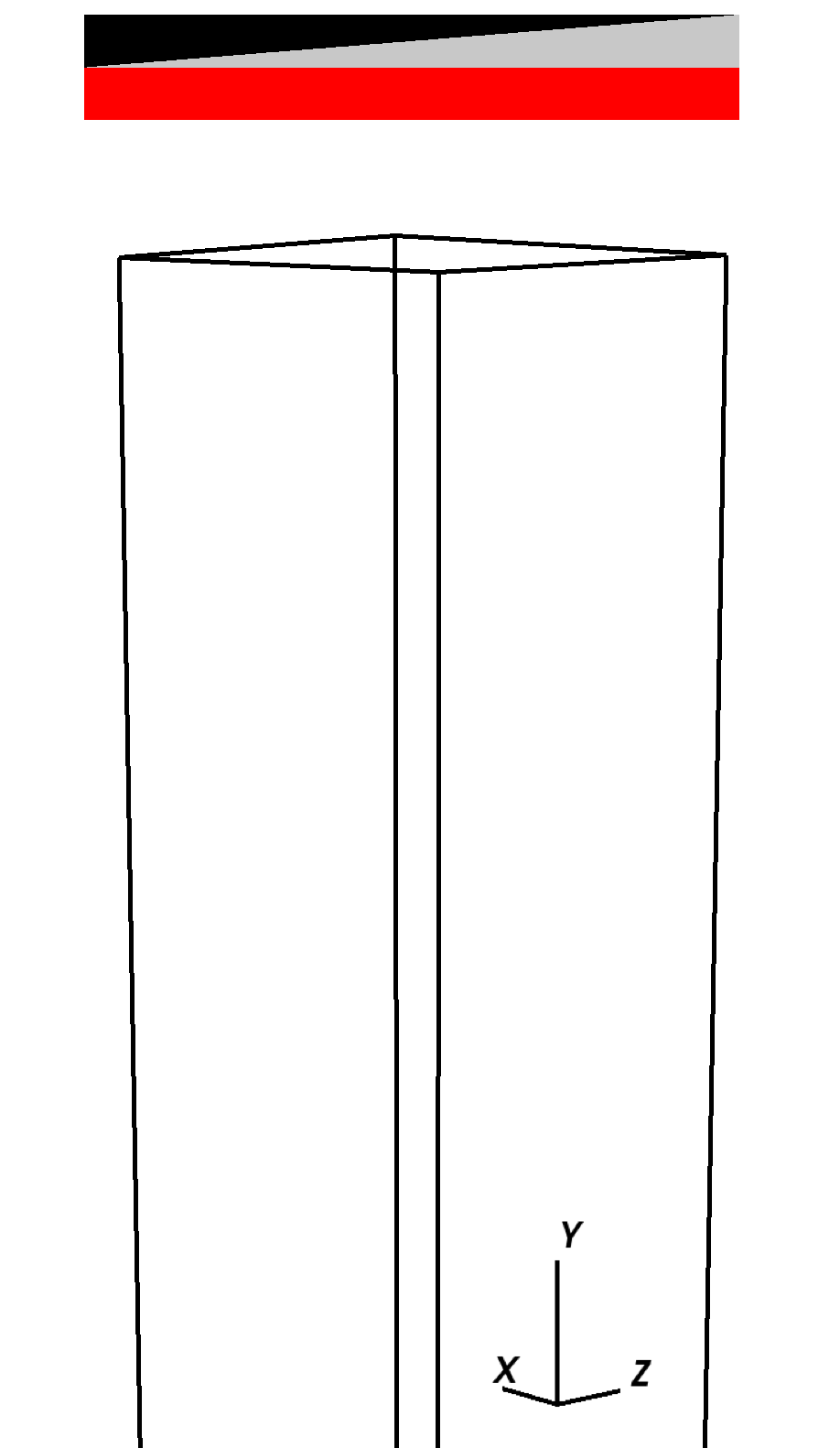}}\\  
       \multicolumn{3}{l}{\hspace{-2mm}d) comp-k8-M10-rad ($k\sim 20$)}\\   
       \hspace{-0.3cm}\resizebox{27mm}{!}{\includegraphics{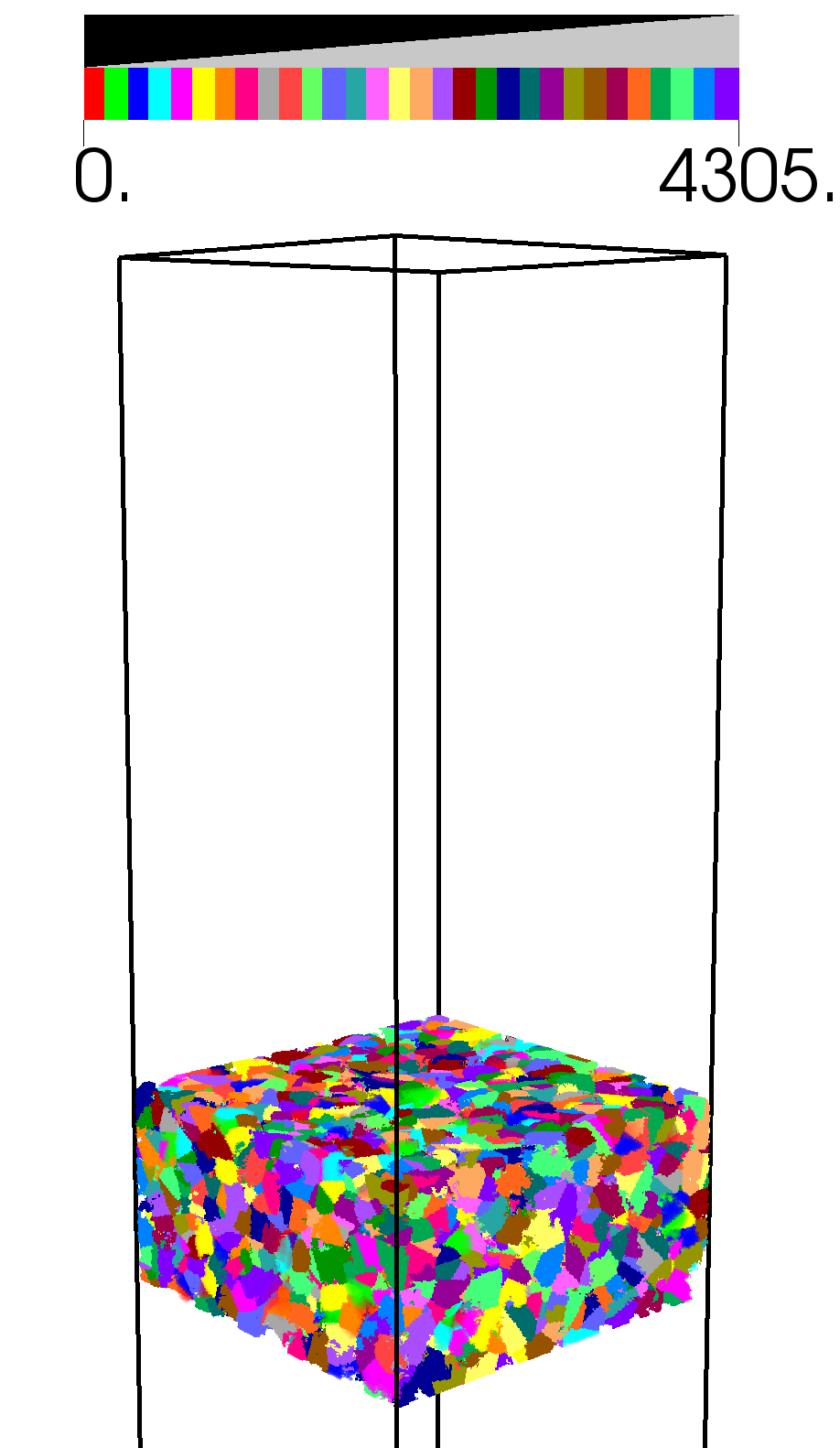}} & \hspace{-0.3cm}\resizebox{27mm}{!}{\includegraphics{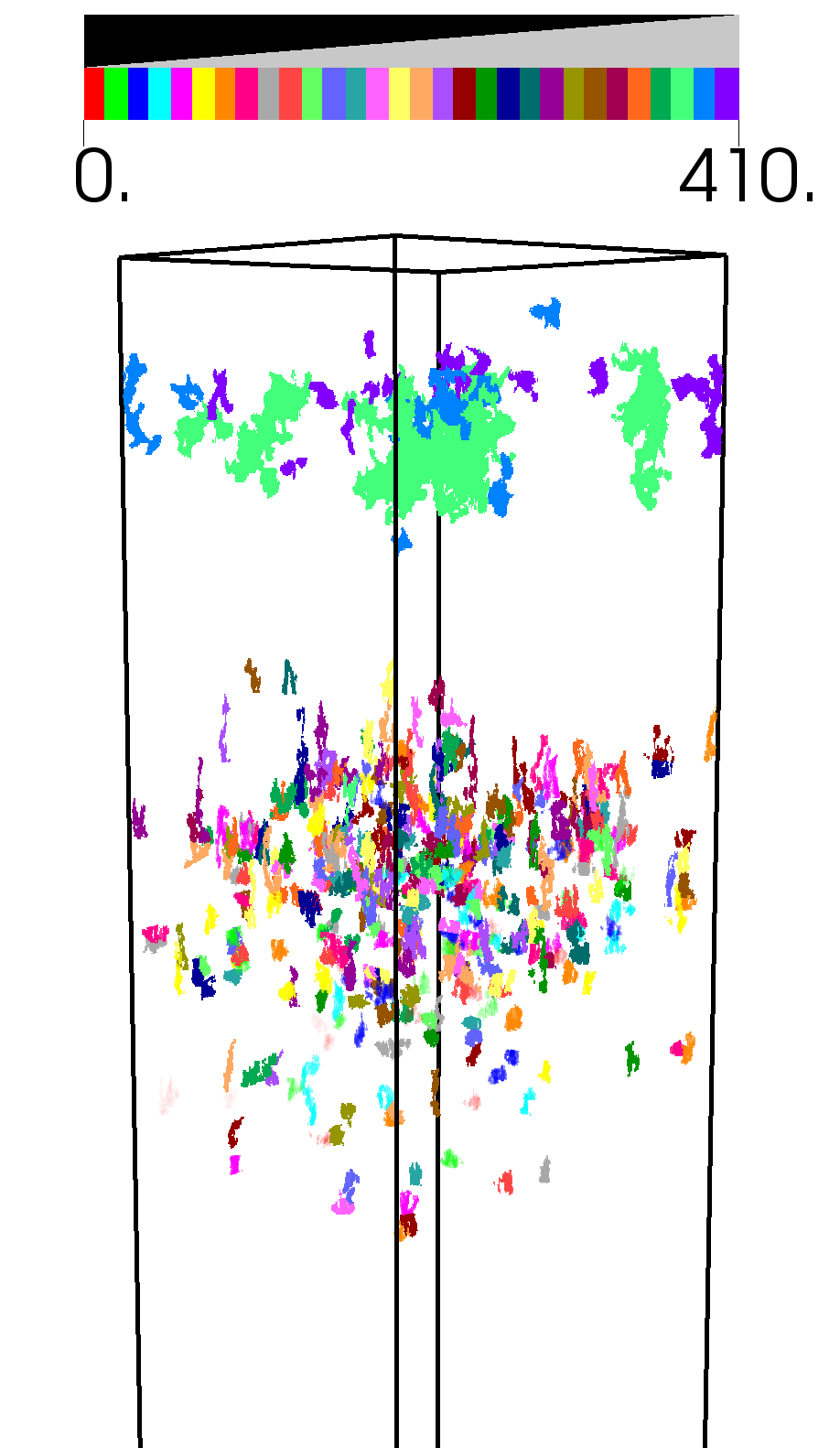}} & \hspace{-0.3cm}\resizebox{27mm}{!}{\includegraphics{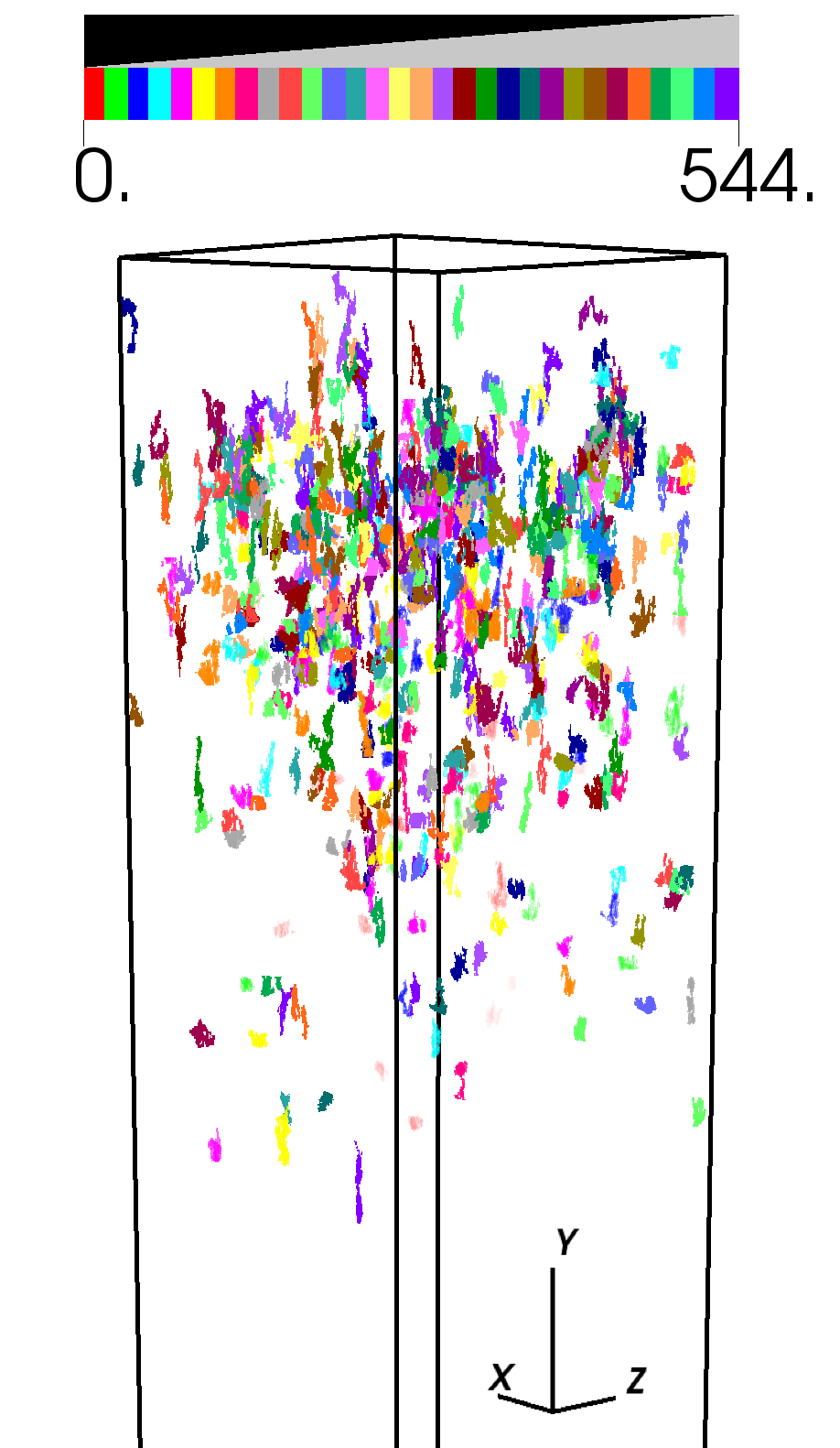}} \vspace{-0.1cm}\\
  \end{tabular}
  \caption{3D renderings showing the evolution for $t/t_{\rm sp}\leq3.0$ of the evolution of the number and 3D distribution of cloudlets, $N_{\rm cloudlet,k}$, for different radiative models, compact solenoidal (panels a and b) and porous compressive (panels c and d), and different $k$. The colour bar shows the clump id number in a linear scale, from 0 to $N_{\rm cloudlet,k}$.} 
  \label{FigureA2}
\end{center}
\end{figure}

The number and 3D distribution of cloudlets for these two cases ($k\sim10$ and $k\sim 20$) and for both radiative multicloud models are displayed in Figure \ref{FigureA2}. These panels show the cloudlet distribution at three different times, $t_0$, $t=1.8\,t_{\rm sp}=0.36\,\rm Myr$, and $t=3.0\,t_{\rm sp}=0.60\,\rm Myr$. The population of small-scale cloudlets ($k\sim20$) also changes with time, decreasing during the shock crossing phase, but remaining constant during the shock re-acceleration and turbulence generation phases. The passage of the shock and post-shock flow rapidly erodes small-scale gas, but cooling continuously fragments the gas, thus maintaining a population of small cloudlets and filaments with typical sizes between $\sim 2$--$8\,\rm pc$ in the directions transverse to the shock normal and lengths between $\sim 4$--$10\,\rm pc$ in the direction parallel to the shock normal (see Section \ref{subsec:CoolingHeating}). Figure \ref{FigureA2} also shows that porous compressive layers result in more vertically extended outflows. The properties of these cloudlet populations will be discussed in more detail in a future paper.

\section{Cold dense gas at late times}
\label{AppendixB}

\begin{figure*}
\begin{center}
  \begin{tabular}{c c c c c c c l}
       \multicolumn{8}{l}{\hspace{+0.0cm}10a) sole-k8-M10-rad}\\   
       \multicolumn{1}{c}{All} & \multicolumn{1}{c}{MM - H$_2$} & \multicolumn{1}{c}{CNM - H$_{\rm I}$} & \multicolumn{1}{c}{WNM - H$_{\rm I}$} & \multicolumn{1}{c}{WIM - H$_{\alpha}$} & \multicolumn{1}{c}{HIM} & \multicolumn{1}{c}{HM - $X_{\rm ray}$} & $N_{\rm mc}$\\
       \hspace{-0.25cm}\resizebox{26mm}{!}{\includegraphics{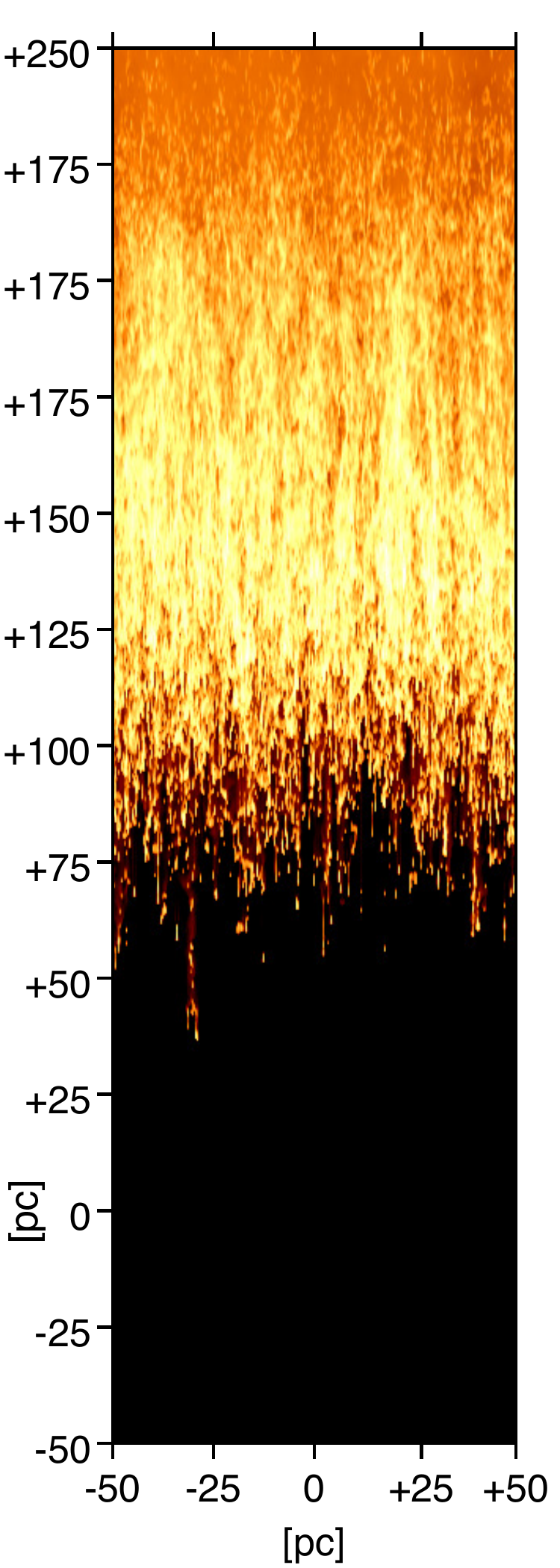}} & \hspace{-0.55cm}\resizebox{26mm}{!}{\includegraphics{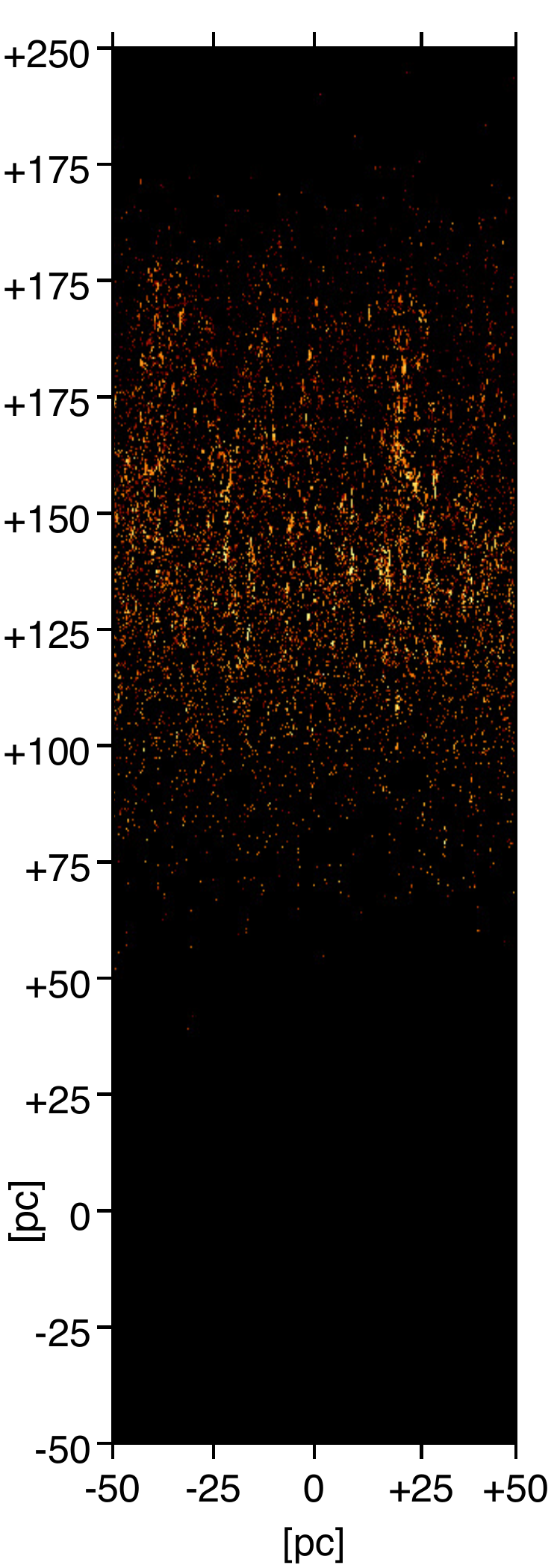}} & \hspace{-0.55cm}\resizebox{26mm}{!}{\includegraphics{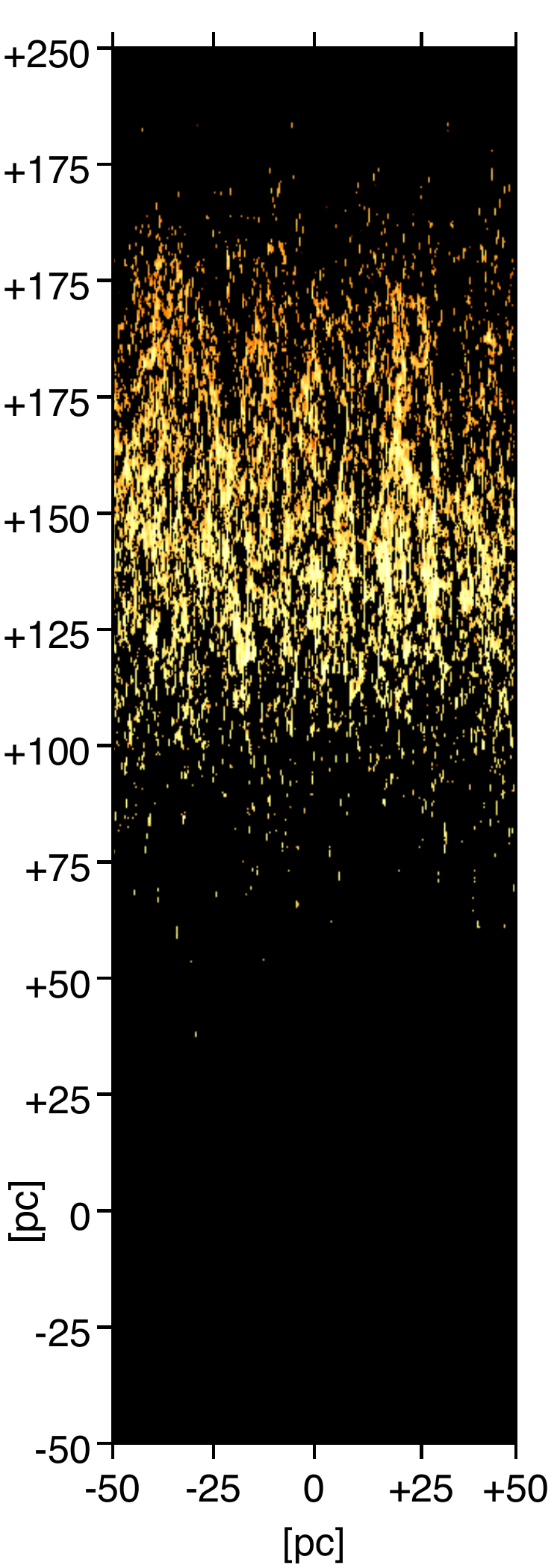}} & \hspace{-0.55cm}\resizebox{26mm}{!}{\includegraphics{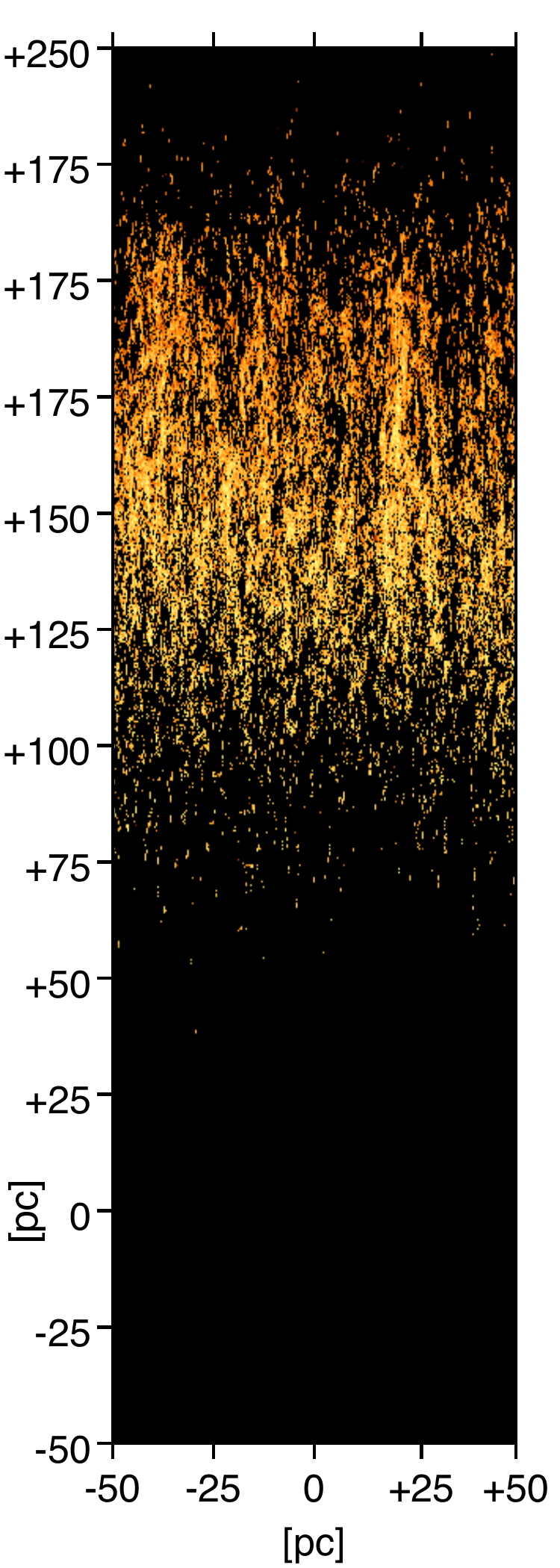}} & \hspace{-0.55cm}\resizebox{26mm}{!}{\includegraphics{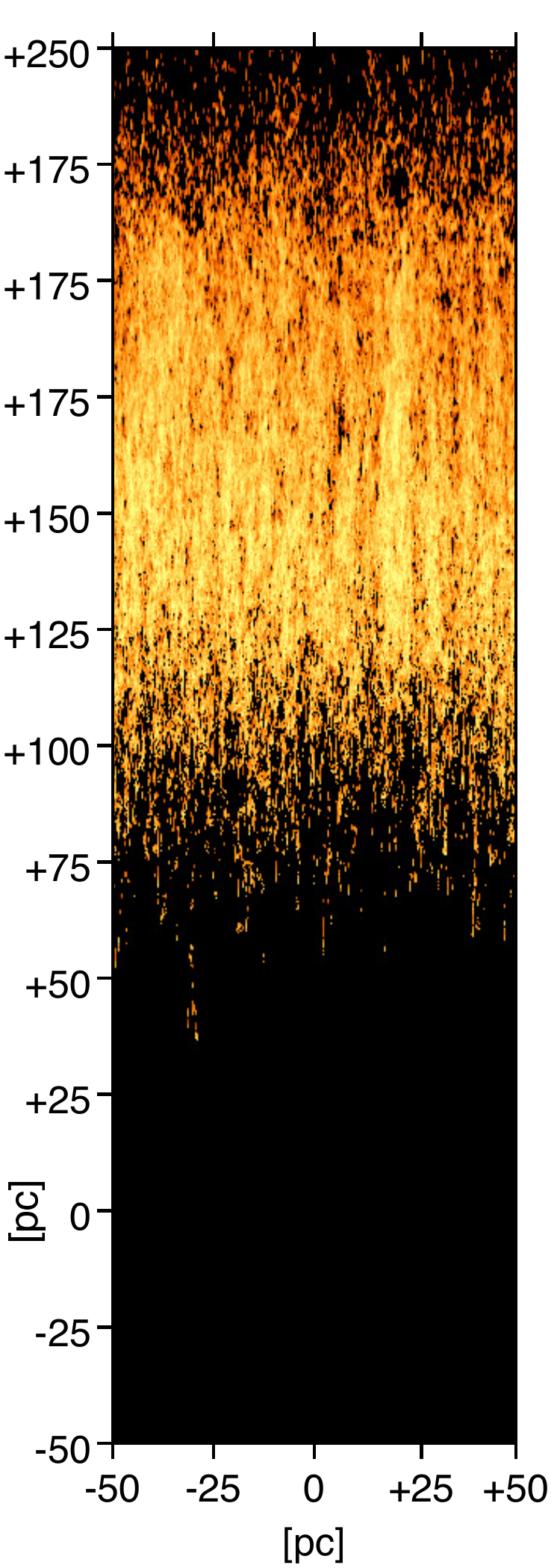}} & \hspace{-0.55cm}\resizebox{26mm}{!}{\includegraphics{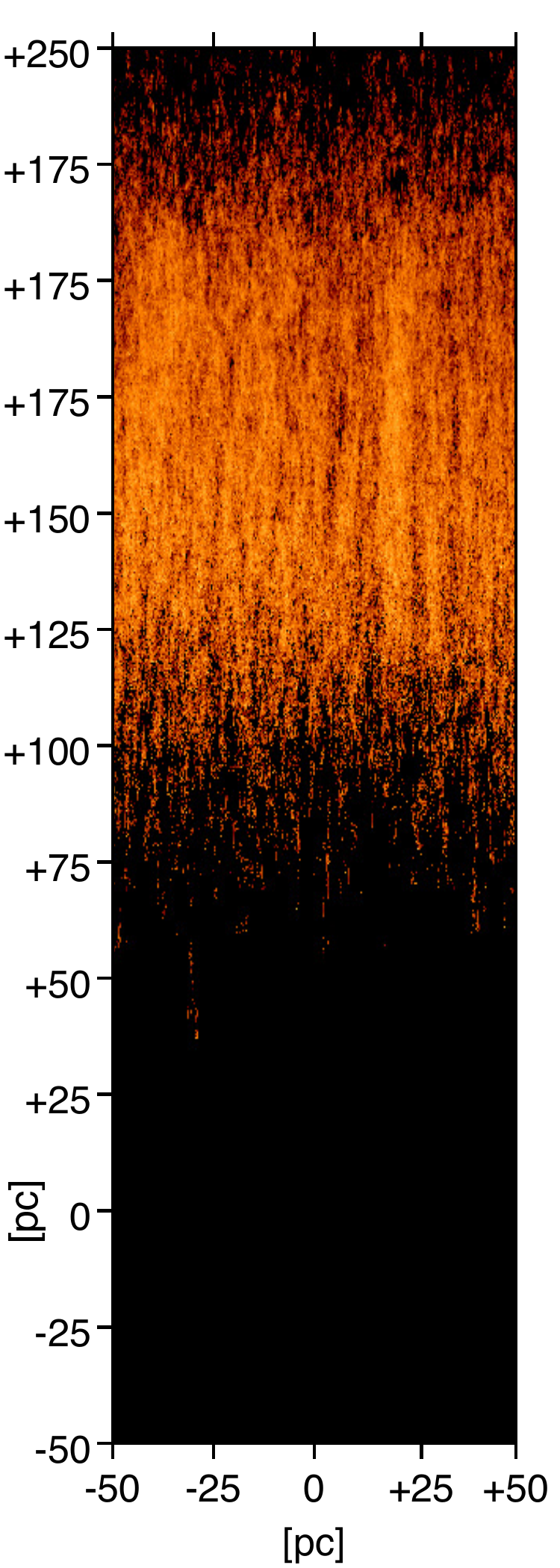}} &
\hspace{-0.55cm}\resizebox{26mm}{!}{\includegraphics{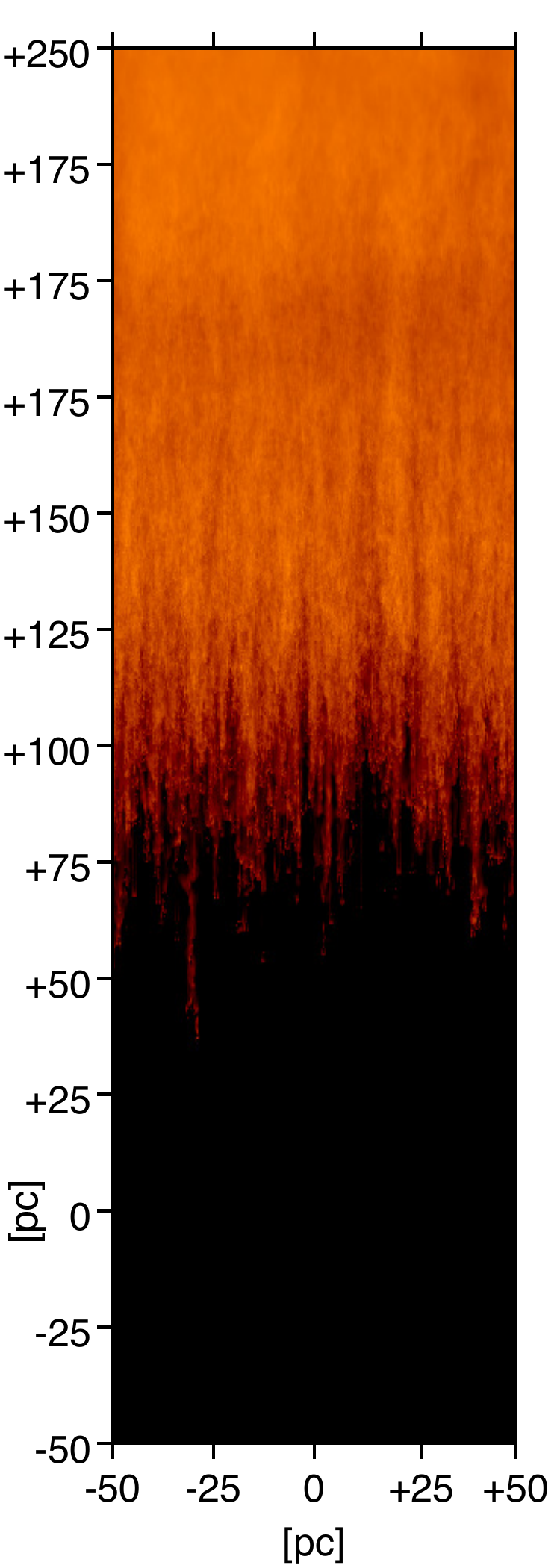}} &
\hspace{-0.40cm}\resizebox{11mm}{!}{\includegraphics{barra.png}}\\
       \multicolumn{8}{l}{\hspace{+0.0cm}10b) comp-k8-M10-rad}\\    
       \multicolumn{1}{c}{All} & \multicolumn{1}{c}{MM - H$_2$} & \multicolumn{1}{c}{CNM - H$_{\rm I}$} & \multicolumn{1}{c}{WNM - H$_{\rm I}$} & \multicolumn{1}{c}{WIM - H$_{\alpha}$} & \multicolumn{1}{c}{HIM} & \multicolumn{1}{c}{HM - $X_{\rm ray}$} & $N_{\rm mc}$\\
       \hspace{-0.25cm}\resizebox{26mm}{!}{\includegraphics{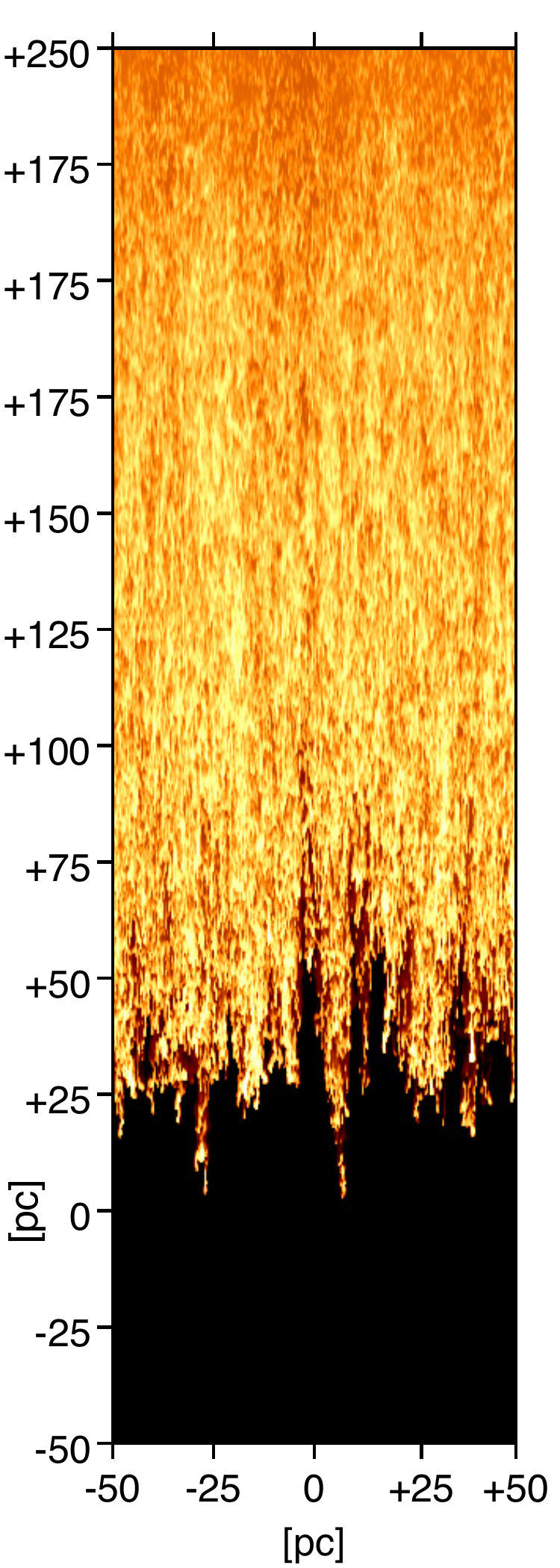}} & \hspace{-0.55cm}\resizebox{26mm}{!}{\includegraphics{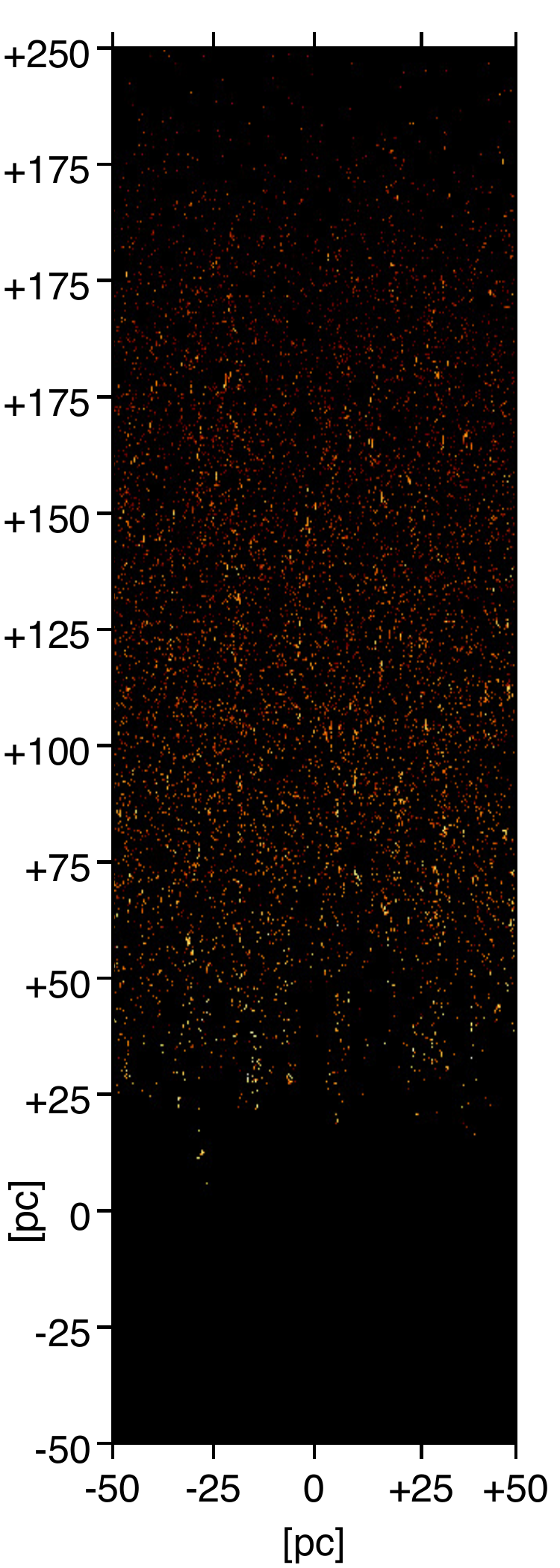}} & \hspace{-0.55cm}\resizebox{26mm}{!}{\includegraphics{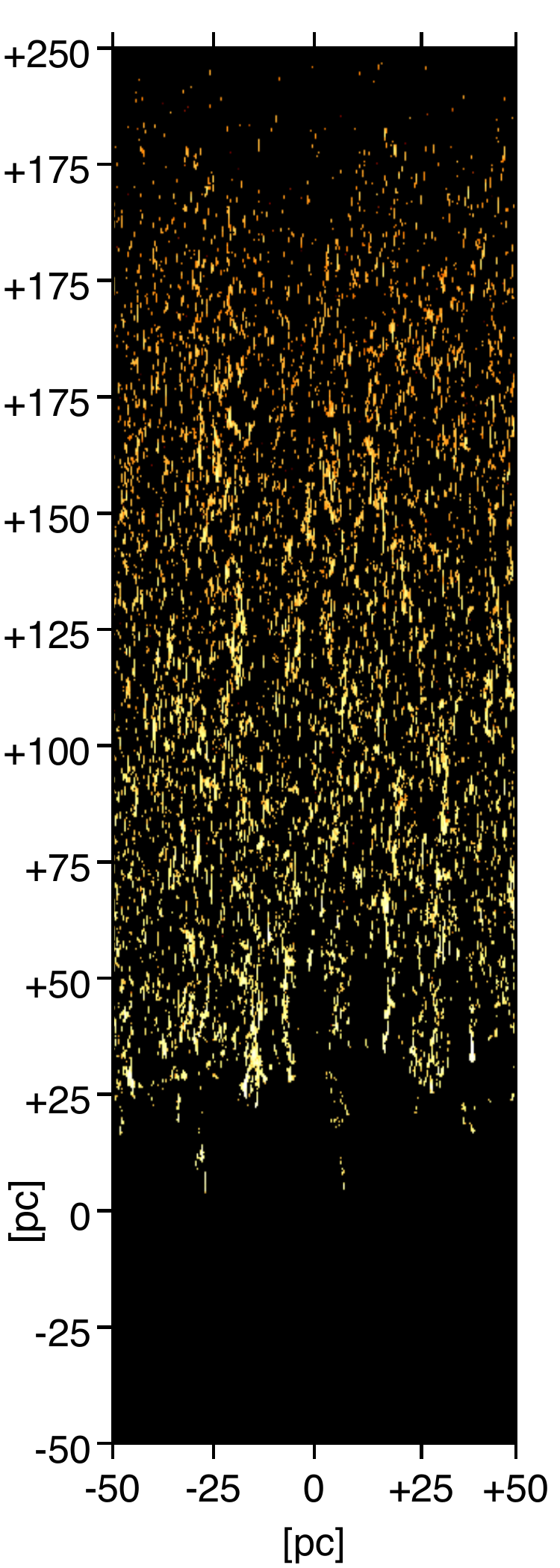}} & \hspace{-0.55cm}\resizebox{26mm}{!}{\includegraphics{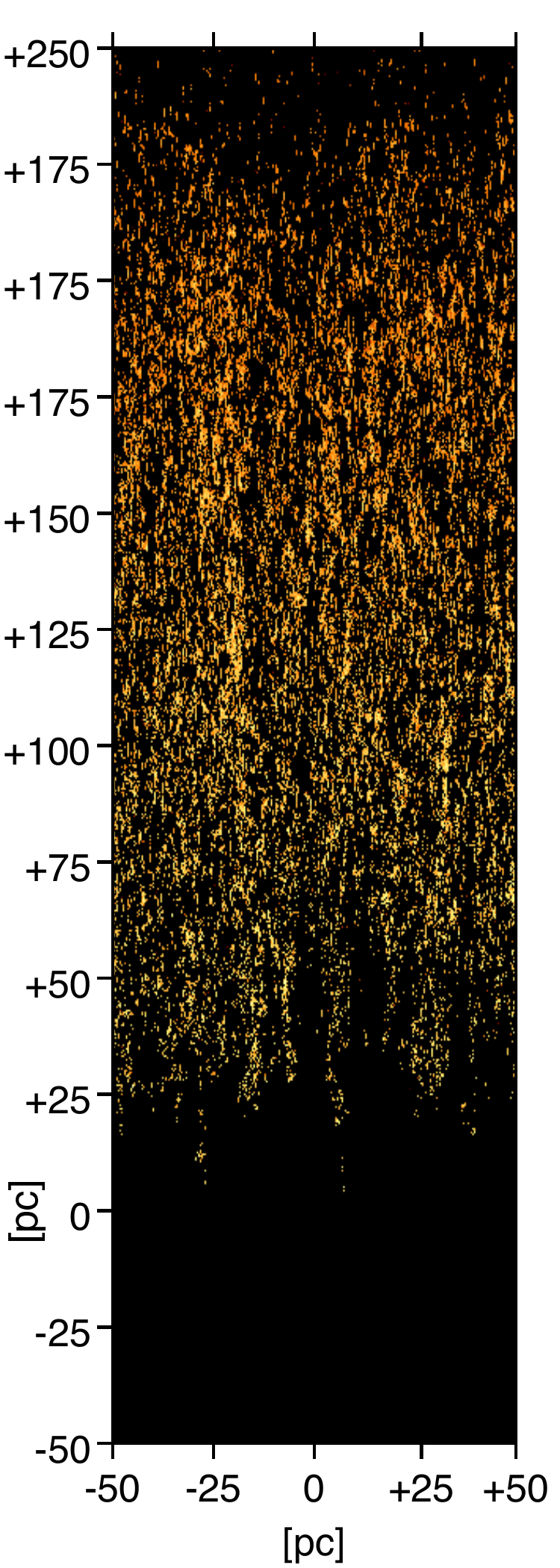}} & \hspace{-0.55cm}\resizebox{26mm}{!}{\includegraphics{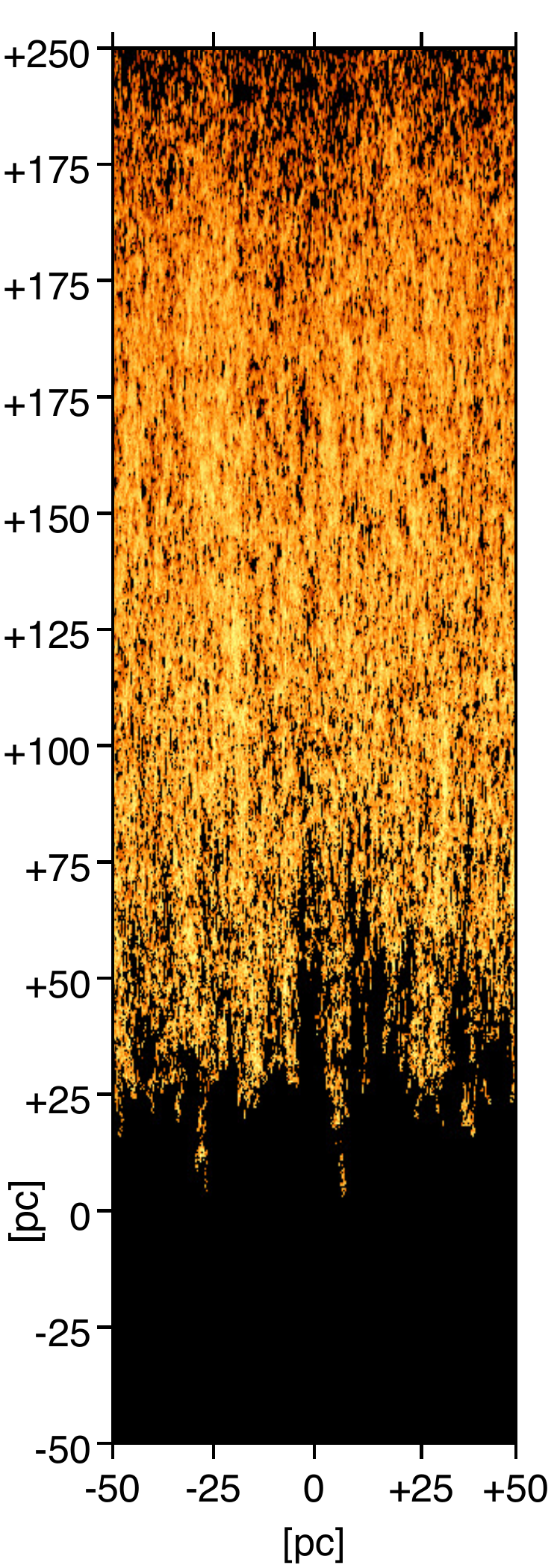}} & \hspace{-0.55cm}\resizebox{26mm}{!}{\includegraphics{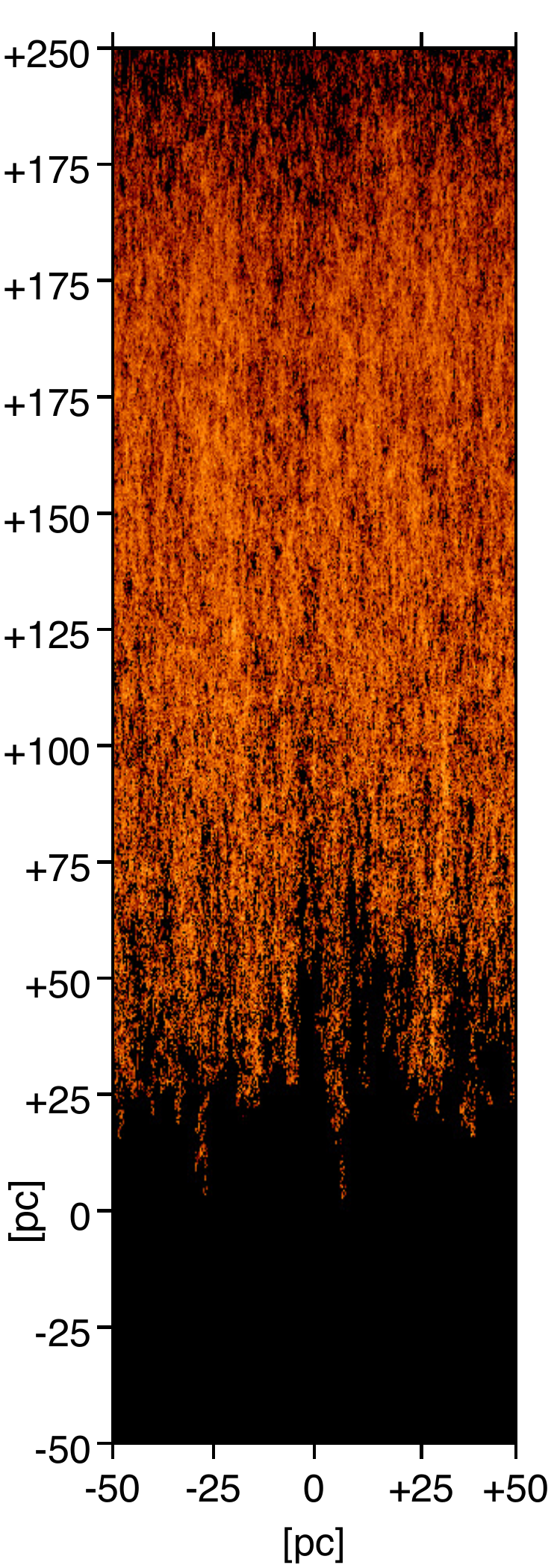}} &
\hspace{-0.55cm}\resizebox{26mm}{!}{\includegraphics{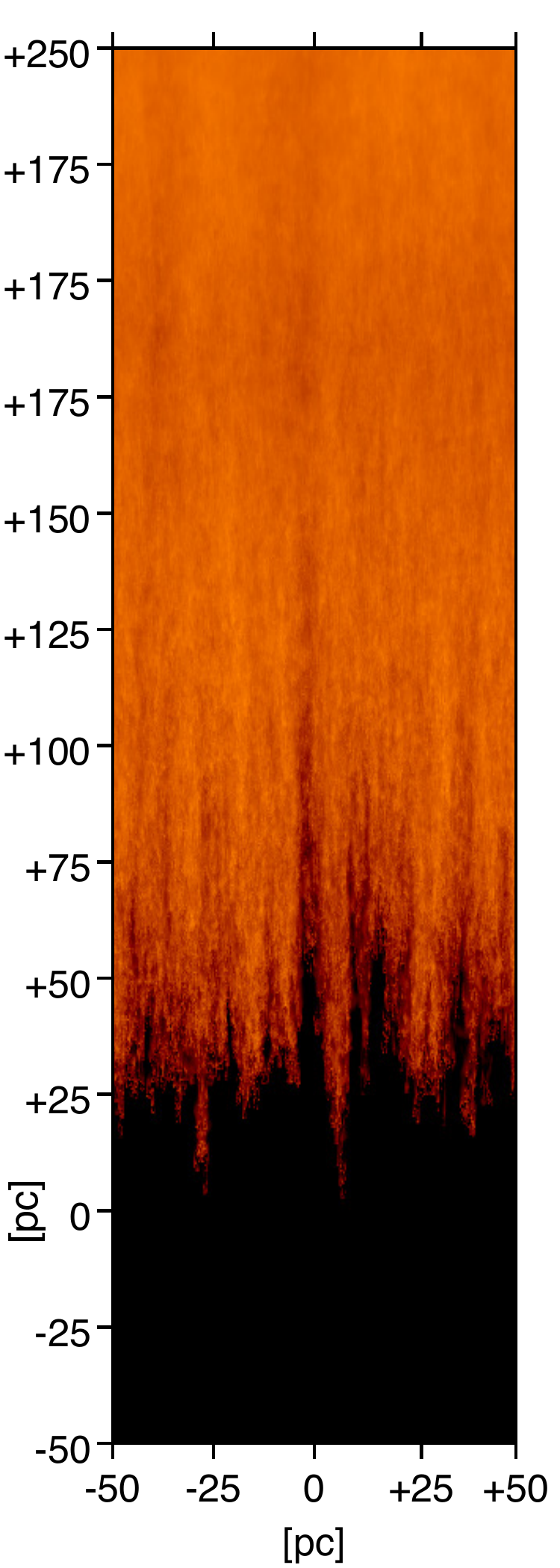}} & \hspace{-0.40cm}\resizebox{11mm}{!}{\includegraphics{barra.png}}\\
  \end{tabular}
  \caption{Same as Figure \ref{Figure10}, but for the latest time in our simulations. The column density maps of cloud material correspond to $t=1.8\,t_{\rm sp}=0.60\,\rm Myr$.}
  \label{FigureB1}
\end{center}
\end{figure*}

The panels in Figure \ref{FigureB1} show the column number density ($N_{\rm mc}$) maps of cloud gas in our radiative multicloud models, sole-k8-M10-rad (top panel) and comp-k8-M10-rad (bottom panel), for different temperature ranges (see Table \ref{Table3}) at $t=1.8\,t_{\rm sp}=0.60\,\rm Myr$. These projections show that, in both models, most dense cold gas is beyond the location of the initial multicloud layer at late times, and also that porous compressive cloud layers result in more vertically extended outflows than their compact solenoidal counterparts, as explained in Section \ref{subsec:Evolution}.

\bsp	
\label{lastpage}
\end{document}